%% file: PhysRep_All.tex
\documentclass[3p,sort&compress]{elsarticle}
\usepackage{amsfonts,amssymb,amsmath}
\usepackage{graphicx}
\usepackage{bm}
\usepackage{enumerate}
\usepackage{color}
\usepackage{todonotes,hyperref,url}
\usepackage{gensymb}

\journal{Physics Reports}

\begin{document}

\begin{frontmatter}

\title{\small{published as Zou et al, Physics Reports, 787, 1–97 (2019). DOI:10.1016/j.physrep.2018.10.005}\\ \vspace{12pt}
\Large{Complex network approaches to nonlinear time series analysis}}

\author[ECNU]{Yong Zou\corref{corrauthor}}
\ead{yzou@phy.ecnu.edu.cn}
\address[ECNU]{Department of Physics, East China Normal University, Shanghai 200062, China}

\author[MD,PIK]{Reik V. Donner\corref{corrauthor}}
\cortext[corrauthor]{Corresponding author}
\ead{reik.donner@pik-potsdam.de}
\address[MD]{Department of Water, Environment, Construction and Safety, Magdeburg--Stendal University of Applied Sciences, Breitscheidstra{\ss}e 2, 39114 Magdeburg, Germany}
\address[PIK]{Potsdam Institute for Climate Impact Research (PIK) -- Member of the Leibniz Association, Telegrafenberg A31, 14473 Potsdam, Germany}

\author[PIK]{Norbert Marwan}
\ead{marwan@pik-potsdam.de}

\author[PIK,SW]{Jonathan F. Donges}
\ead{donges@pik-potsdam.de}
	\address[SW]{Stockholm Resilience Centre, Stockholm University, Kr\"aftriket 2B, 114 19 Stockholm, Sweden}

\author[PIK,UK,HU,RU1,RU2]{J\"urgen Kurths}
\ead{kurths@pik-potsdam.de}
 \address[UK]{Institute for Complex Systems and Mathematical Biology, University of Aberdeen, Aberdeen AB243UE, United Kingdom}
 \address[HU]{Department of Physics, Humboldt University Berlin, Newtonstra{\ss}e 15, 12489 Berlin, Germany}
 \address[RU1]{Department of Control Theory, Nizhny Novgorod State University, Gagarin Avenue 23, 606950 Nizhny Novgorod, Russia}
 \address[RU2]{Institute of Applied Physics of the Russian Academy of Sciences, 603950 Nizhny Novgorod, Russia}

\begin{abstract}

In the last decade, there has been a growing body of literature addressing the utilization of complex network methods for the characterization of dynamical systems based on time series. While both nonlinear time series analysis and complex network theory are widely considered to be established fields of complex systems sciences with strong links to nonlinear dynamics and statistical physics, the thorough combination of both approaches has become an active field of nonlinear time series analysis, which has allowed addressing fundamental questions regarding the structural organization of nonlinear dynamics as well as the successful treatment of a variety of applications from a broad range of disciplines. In this report, we provide an in-depth review of existing approaches of time series networks, covering their methodological foundations, interpretation and practical considerations with an emphasis on recent developments. After a brief outline of the state-of-the-art of nonlinear time series analysis and the theory of complex networks, we focus on three main network approaches, namely, phase space based recurrence networks, visibility graphs and Markov chain based transition networks, all of which have made their way from abstract concepts to widely used methodologies. These three concepts, as well as several variants thereof will be discussed in great detail regarding their specific properties, potentials and limitations. More importantly, we emphasize which fundamental new insights complex network approaches bring into the field of nonlinear time series analysis. In addition, we summarize examples from the wide range of recent applications of these methods, covering rather diverse fields like climatology, fluid dynamics, neurophysiology, engineering and economics, and demonstrating the great potentials of time series networks for tackling real-world contemporary scientific problems. The overall aim of this report is to provide the readers with the knowledge how the complex network approaches can be applied to their own field of real-world time series analysis.
\end{abstract}


\end{frontmatter}
\clearpage
\tableofcontents
\clearpage

\input{Chapter01_Introduction/Chapter01_Introduction.tex}

\input{Chapter02_CompNetworkT/Chapter02_CompNetworkT.tex}

\input{Chapter03_RecurrenceNt/Chapter03_RecurrenceNt.tex}

\input{Chapter04_VisibilityGt/Chapter04_VisibilityGt.tex}

\input{Chapter05_TransitionNt/Chapter05_TransitionNt.tex}

\input{Chapter06_Applications/Chapter06_Applications.tex}

\input{Chapter07_Software/Chapter07_Software.tex}

\input{Chapter08_Discussion/Chapter08_Discussion.tex}

\section*{Acknowledgements}
The reported development of complex network frameworks for nonlinear time series analysis has been a community effort, to which various colleagues have been actively contributing. We particularly acknowledge the contributions and discussions by Jan H. Feldhoff, Zhongke Gao, Shuguang Guan, Jobst Heitzig, Zonghua Liu, Michael Small, Marco Thiel, and Marc Wiedermann.

Financial support of this work has been granted by the National Natural Science Foundation of China (grant nos. 11872182, 11835003), the Natural Science Foundation of Shanghai (Grant No. 17ZR1444800), the German Research Association (DFG) via the IRTG 1740 ``Dynamical phenomena in complex networks'' and DFG projects no.~MA4759/8 and MA4759/9, the European Union's Horizon 2020 Research and Innovation Programme under the Marie Sk{\l}odowska-Curie grant agreement no.~691037 (project QUEST), the German Federal Ministry for Education and Research (BMBF) via the Young Investigators Group CoSy-CC$^2$ (grant no.~01LN1306A), the Stordalen Foundation (via the Planetary Boundaries Research Network PB.net), the Earth League's EarthDoc network and the Leibniz Association (project DOMINOES). Numerical computations have been performed using the software packages \texttt{pyunicorn} \cite{Donges2015} and the CRP Toolbox for MATLAB, which are available at \url{https://tocsy.pik-potsdam.de}.

Parts of this report have been based on text from an earlier book chapter \cite{Donner2015RPBook}. We acknowledge the corresponding permission of the publisher of that chapter (Springer Nature) to reuse the corresponding material in this work. In a similar spirit, we are grateful to the publishers of several of our previous works for their permission to reuse several key figures in this review.
\clearpage

\input{Chapter09_Appendix/Chapter09_Appendix.tex}

\bibliography{PhysRep2018N,refBooks,rpN,physrep_reik.bib}

\end{document}

%% file: Chapter01_Introduction/Chapter01_Introduction.tex
\begin{table}
\caption{Nomenclature and abbreviations used in the manuscript \label{tab:symabbr}}
\centering
\begin{tabular}{|p{1.3cm}@{\extracolsep{\fill}}p{7.8cm} @{\extracolsep{\fill}}p{1.3cm}@{\extracolsep{\fill}}p{5.6cm}|}\hline
\multicolumn{2}{|l}{ \textbf {Nomenclature}}  & \multicolumn{2}{l|}{\textbf {Abbreviations}} \\[5pt]
$\mathbf{A}$     & Adjacency matrix                                                               &   ACF     & auto-correlation function \\ 
$b_p$                & betweenness centrality of vertex $p$                                &  AR	&	auto regressive 	 \\
$c_{p}$              & closeness centrality of vertex $p$                                     &  B. P.  &	 before present \\
$\mathcal{C}$    & global clustering coefficient 						  &   COPTN &	 cross ordinal pattern transition network \\
$\mathcal{C}_{p}$    & local clustering coefficient of vertex $p$ 			 &   CRP	&	cross recurrence plots	\\
$\mathcal{D}$    	& network diameter							& 	DHVG	& dfference horizontal visibility graph     \\
$\hat{D}_{\mathcal{C}} $              &      clustering dimension			&	DVG	& directed visibility graph\\
$\hat{D}_{\mathcal{T}} $              &      transitivity dimension			&	ER        & Erd\"os R\'enyi random network   \\
$\Delta k_p$    & excess degree of vertex $p$						&	fBm       & fractional Brownian motion  \\
$\Delta_{rel} k_p$    & relative excess degree of vertex $p$			&	fGn        & fractional Gaussian noise\\
$\Delta t $         & sampling time  								 &   FNN      & false nearest neighbors\\
$\delta(\cdot)$   & delta function ($\delta(x) = \{ 1 | x = 0; 0 | x \neq = 0 \}$)  &  GS        & generalized synchronization \\
$\varepsilon$    & radius of neighborhood							  &   HVG      & horizontal visibility graph\\
$e_p$       & local efficiency of vertex $p$ 							  &   IRN	&	inter-system recurrence network \\
$E$   	 & edge set										&	ISN   &	international sunspot number\\
$\mathcal{E}$    & global efficiency 								 &   JOPTN  &	 joint ordinal pattern transition network  \\
$F(k)$   &  cumulative degree distribution function 		 			&   JRN	&	joint recurrence network	\\
$\gamma$   	 & power law exponent							&	KLD	  & Kullback-Leibler divergence \\
$H$   	 & Hurst exponent									&	KS  		 & Kolmogorov-Smirnov test	\\
$k_p$       & degree of vertex $p$  								  &  Ma    &	  million years \\
$l_{pq}$              & shortest path between vertices $p$ and $q$                    &  ODP  &	 ocean drilling program \\
 $\mathcal{L}$      & average path length 							&     OP        & ordinal pattern\\
$\lambda$     & Lyapunov exponent; exponential scaling factor  		&   OPTN	&   ordinal pattern transition network \\
$m$        & embedding dimension 								  &  PDF      & probability density function \\
$\mu$    & coupling strength									&	PS        & phase synchronization  \\
$N$      & length of time series 									&     RGG &	random geometric graph 	 \\
$\pi$               & ordinal pattern 			 						  &  RN       & recurrence network  \\
$p(x) $              & probability density function of $x$ 					  &  RP        & recurrence plot \\
$RR$    & recurrence rate 										& 	RQA      & recurrence quantification analysis  \\
$\mathcal{R}$      & assortativity coefficient 						&     SF         & scale free \\
$r$			&	cross correlation coefficient 					&	SSA	  &  sunspot area \\
$\rho $              &      edge (link) density of a network 				&	SSN      & sunspot numbers \\
$\sigma_{pq}$     & multiple shortest paths between vertices $p$ and $q$        &  SW        & small world\\
$\tau$      & embedding delay 									  &   UPO      & unstable periodic orbits \\
$\mathcal{T}$    & transitivity									&	VG        & visibility graph\\
$S$    	& Shannon entropy									&	&	  \\
$\Theta(\cdot)$   	 & Heaviside function 						&	&	 \\
$\mathbf{W}$    	& weighted adjacency matrix 					&	 &	 \\
$\Omega$    & average frequency			 					&	&	\\
$w_{pq}$   	      &       transition frequency from vertex $p$ to $q$		&	&	 \\
$\hat{x}$   	 &       estimator of $x$							&	&	 \\
$\left< x \right> $   	 &       average of $x$						&	&	 \\
$V$    	& vertex set										&	 &	\\
\hline
\end{tabular}
\end{table}

\clearpage

\section{Introduction}
Artificial Intelligence is generating data in new forms of complexity, leading to the new era of big data \cite{schonberger2013}. This brings big challenges for researchers from various fields working together to extract patterns or new structures from data of very high volume, high velocity, or high variety. Advanced interdisciplinary data analytics techniques help to capture the hidden structures amidst otherwise chaotic data points, including approaches from machine learning, data mining, statistics, natural language and text processing \cite{hurwitz2018}. In consequence, we transform the messy datasets into something that we can learn fast, which allows us making better and faster decisions. Among these processes, there is ample scope for developing new tools for data analysis. In the context of dynamical systems and statistical physics, such methods are often associated with concepts like complex networks \cite{Albert2002,Newman2003,Newmanbook2010} and complexity theory. 

In this report, we focus on some particular subfield that has attracted great interest in the last years -- the application of various approaches from complex network theory in the context of nonlinear time series analysis \cite{kantz1997,abarbanel1993,Sprott2003}. A time series is a sequence of data points indexed by the time of observations, which are made at successive, in many cases equally spaced points in time. Hence, time series data have a natural discrete temporal ordering. Examples of such time series cover a great variety of variables potentially relevant for everyday life, including (but not being limited to) the following areas: (i) weather conditions, like surface-air temperatures, sea level pressure, and wind speeds that are collected from meteorological stations or satellites; (ii) finance, e.g., the daily closing prices of stock market indices like
the Dow Jone Industrial Average, individual assets, or exchange rates; (iii) bio-medical conditions of humans, for instance, physiological and clinical data that are collected by electroencephalogram (EEG) monitoring or high resolution brain imaging techniques like magnetic resonance imaging (MRI) and computed tomography (CT). Time series analysis considers the study of the entire collection of observations as a whole instead of individual numerical values at several temporal instances. 

The natural temporal ordering makes time series analysis distinct from data analysis in the case of no natural ordering of the observations (for instance, cross-sectional studies of explaining people's wages by reference to their respective education levels, or spatial data analysis accounting for house prices by the location as well as the intrinsic characteristics of the houses). In order to discover hidden patterns of such more general large data sets from different sources, data mining tools have been proposed in the research field of computer science, which have also found many applications in the context of time series mining \cite{Keogh2003,FU2011,Aghabozorgi2015}. Here, time series mining focuses more on indexing, clustering, classification, segmentation, motif discovery, and forecasting \cite{Keogh2003,Aghabozorgi2015}. In the recent decade, there has been a considerable amount of rapid developments of data mining tools initiated by the advent of big data and cloud computing reflecting the increasing size and complexity of available datasets. One particular example of such developments is to design algorithms of high efficiency that can learn from and make predictions on the large data sets in terms of supervised or unsupervised learning methods \cite{hurwitz2018}. Furthermore, there is an emerging trend to combine complex network approaches with data mining tools, which provide many novel analysis concepts for discovering hidden pattern in large data sets \cite{Zanin2016}. The classification task of data mining allows for a rich representation of some complex systems, for instance, a meaningful reconstruction of functional networks from rather large data sets by choosing feature vectors of lower dimension \cite{Zanin2014a}. Hence, the application of feature selection algorithms provides a complementary understanding of the characteristics of network structures. There have been a few successful applications of mining tools for complex network analysis from synthetic and experimental data, in particular, related with disease classification \cite{Zanin2014a,Karsakov2017,Whitwell2018}. 

Despite the considerable practical relevance of time series mining algorithms, there has been practically no overlap with the subject of nonlinear time series analysis by means of complex network methods, which is the focus of this report. Unlike most established data mining techniques, the time series network approaches reviewed here are based on the dynamical systems theory \cite{Ott1993,kantz1997} and present themselves as state-of-the-art contributions to nonlinear time series analysis \cite{Bradley2015c}. From the viewpoint of complex network research, the topics reviewed in this paper can be regarded as successful applications of network theory to tackling synthetic as well as experimental series from a diversity of fields of applications. We will discuss potential generalizations of the different time series network approaches in the context of data mining tools wherever appropriate.
 
	\subsection{Nonlinear time series analysis}
	Time series analysis is essentially data compression \cite{Bradley2015c}. Given a time series, we interpret the underlying dynamical system by a few characteristic numbers that are computed from a large sample of measurements. Therefore, the reduced information as represented by these characteristic numbers must highlight some specific features of the system. Early approaches of time series analysis heavily relied on the linearity assumption on the underlying processes, for instance, autoregressive (AR) and moving average (MA) models, both of which result in almost exponentially decaying auto-correlation functions. However, it is by now well accepted that the dynamical laws governing nature or human activities are seldomly linear. Nonlinearity is everywhere, for example: (a) phase transitions (e.g., the melting of the ice of a glacier) are an important signature of nonlinearity in physical systems; (b) animals behave differently (e.g. hunting effort) during times of short food supply versus times of abundant food supply; (c) for many electronic devices (e.g., transistors) saturation velocity and current are well-known nonlinear phenomena; (d) in many engineering problems, controlling the system to operate at desired states introduces various forms of feedback mechanisms. Accordingly, the development of nonlinear time series analysis has been primarily driven by the needs to overcome the corresponding limitations of linear models and methods. 

	Nonlinear time series analysis is not as well established and far less well understood than its linear counterpart \cite{kantz1997}. The collection of ideas and techniques of nonlinear time series analysis originates from the fast development of dynamical systems theory or so-called ``chaos theory'', which explores system dynamics by a set of nonlinear difference equations or nonlinear ordinary differential equations. Techniques from chaos theory allow to characterize dynamical systems in which nonlinearities give rise to a complex temporal evolution, for instance, a sensitive dependence on initial conditions and strongly limited predictability. Importantly, this concept allows extracting information that cannot be resolved using classical linear techniques such as the power spectrum or spectral coherence.

	Since its early stages in the 1980s \cite{Packard1980}, numerous conceptual approaches have been introduced for studying the characteristic features of nonlinear dynamical systems based on observational time series \cite{abarbanel1993,kantz1997,Sprott2003}. The mathematical beauty of this analysis framework is that we characterize the invariant measure in phase space in a number of different ways. Generally speaking, we quantify the system from either geometric or dynamic perspectives. Important examples include, but are not limited to the correlation dimension (or, more generally, the spectrum of generalized fractal dimensions $D_q$ \cite{Grassberger1983PRL}) that has been suggested to characterize the geometry of chaotic attractors in phase space; the Lyapunov exponent as a measure for stability of dynamics with respect to infinitesimal perturbations; and the Kolmogorov-Sinai entropy (or other information theory measures) to quantify uncertainty about the future states of a chaotic trajectory. All these techniques have in common that they quantify certain dynamically invariant phase space properties of the considered system based on temporally discretized realizations of individual trajectories. 

	One typical task is to perform a precise system characterization from a single time series, which is, however, not the final goal of most time series analyses \cite{Bradley2015c}. Here, we give just a few examples that nonlinear time series analysis can contribute to: (1) system characterization from a single time series; (2) discrimination between a signal and some other signals; (3) quantification of various bifurcation transition scenarios to complex dynamics, including period doubling, band merging, more general examples of subtle changes like intermittency or other phenomena associated with chaos-to-chaos transitions, detection of general regime shifts or tipping points in essential dynamical properties; (4) testing for time series reversibility; (5) noise reduction and filtering; and (6) prediction of future time series values.

	The aforementioned nonlinear time series characteristics are based on univariate series, i.e., they can be applied to single signals measured upon individual dynamical systems. In contrast, bivariate measures are used to analyze pairs of signals measured simultaneously from two dynamical systems. There has been considerable interest in the study of the synchronization behavior of coupled chaotic systems, which have been observed in many physical and biological systems \cite{Boccaletti2002,Pikovsky_Kurths_synchr}. Thus, such bivariate time series analysis measures aim to detect and to distinguish transition forms from non-synchronized states to synchronization (for instance, paths to phase synchronization, lag synchronization, complete synchronization and generalized synchronization). In different synchronization scenarios, it is important to extract not only the coupling strength but also the direction of these couplings, i.e., identifying causal relationships between the studied sub-systems. Unraveling the governing functional interactions between sub-systems contained in a large network of complex connectivity topology remains a big challenge in modern nonlinear sciences \cite{Boccaletti2006,Stankovski2017}. Various methods have been proposed to extract the statistical associations from data, for instance, Pearson correlation, mutual information (including its time delayed version) \cite{Frenzel_prl_2007,vejmelka_pre_2008}, Granger causality \cite{Granger_1969,Dhamala_prl2008}, transfer entropy \cite{schreiber_prl2000,runge2012}, or methods for detecting coupling directions from time series data \cite{Quiroga_pre2000,Rosenblum2001,rosenblum_pre2002,smirnov_pre2005,Palus_pre_2007,Romano_pre_2007,bahraminasab_prl2008,Nawrath_prl_2010}. More generally, coupling functions can have various forms. We do not expand the corresponding discussion here, but refer the interested readers to several review papers on this topic \cite{Ding_Book_2007,Hlavackovaschindler2007,Stankovski2017}.

	Nonlinear time series analysis provides a powerful toolbox of methods that are useful for many applications, but also have some practical limitations. Some common problems originating from experimental measurements challenge the computations of nonlinear measures. For instance, most of the existing nonlinear methods are in practice only applicable to low-dimensional dynamical systems. In reality, very few real-world data sets are measured by perfect sensors operating on low-dimensional dynamics. One has to be aware of non-stationarity, proper choice of embedding parameters, dependence on finite data length (with possibly rather short time series in many real-world situations), effects of noise, or irregular sampling \cite{Bradley2015c}. Statistical concerns come also from algorithmic aspects requiring proper choice of parameters. For instance, scaling regimes should be pronounced for implementing linear line fitting to estimate the numerical values of dynamically invariant measures like fractal dimensions and Lyapunov exponents, the proper selection of which often influences the results significantly. In addition, computational complexity has to be well evaluated since it varies significantly among these measures. Currently, the choice of algorithmic parameters largely depends on the researchers' experience. 

	In this review, we demonstrate that complex network approaches can contribute many aspects that we have discussed above for nonlinear time series analysis. More importantly, complex network approaches can solve partially some fundamental and long standing problems not successfully addressed by other existing methods so far, yielding a more robust estimation of dynamical invariants, for instance, using the transitivity dimension and local clustering dimension of recurrence networks to approximate the fractal dimension of the system \cite{Donner2011b}; or by computing the mean out-degree of ordinal pattern transition networks or the associated network diameter performing similarly well as the Lyapunov exponent \cite{McCullough2015}. 
	
	\subsection{Complex network approaches}
	With the recent increase in available computational capacities and rising data volumes in various fields of science, complex networks have become an interesting and versatile tool for describing structural interdependencies between mutually interacting units \cite{Albert2002,Boccaletti2006,Costa2007,Newman2003}. Besides ``classical'' areas of research (such as sociology, transportation systems, computer sciences, or ecology), where these units are clearly (physically) identifiable, the success story of complex network theory has recently lead to a variety of ``non-conventional'' applications. 

	One important class of such non-traditional applications of complex network theory are \emph{functional networks}, where the considered connectivity does not necessarily refer to ``physical'' vertices and edges, but reflects statistical interrelationships between the dynamics exhibited by different parts of the system under study. The term ``functional'' was originally coined in neuroscientific applications, where contemporaneous neuronal activity in different brain areas is often recorded using a set of standardized EEG channels. These data can be used for studying statistical interrelationships between different brain regions when performing certain tasks, having the idea in mind that the functional connectivity reflected by the strongest statistical dependencies can be taken as a proxy for the large-scale anatomic connectivity of different brain regions \cite{Bullmore2009,Zhou2006,Zhou2007}. Similar approaches have been utilized for identifying dominant interaction patterns in other multivariate data sets, such as climate data \cite{Tsonis2004,Donges2009b,Donges2009a}.

\begin{figure}[htbp]
	\centering
	\includegraphics[width=\textwidth]{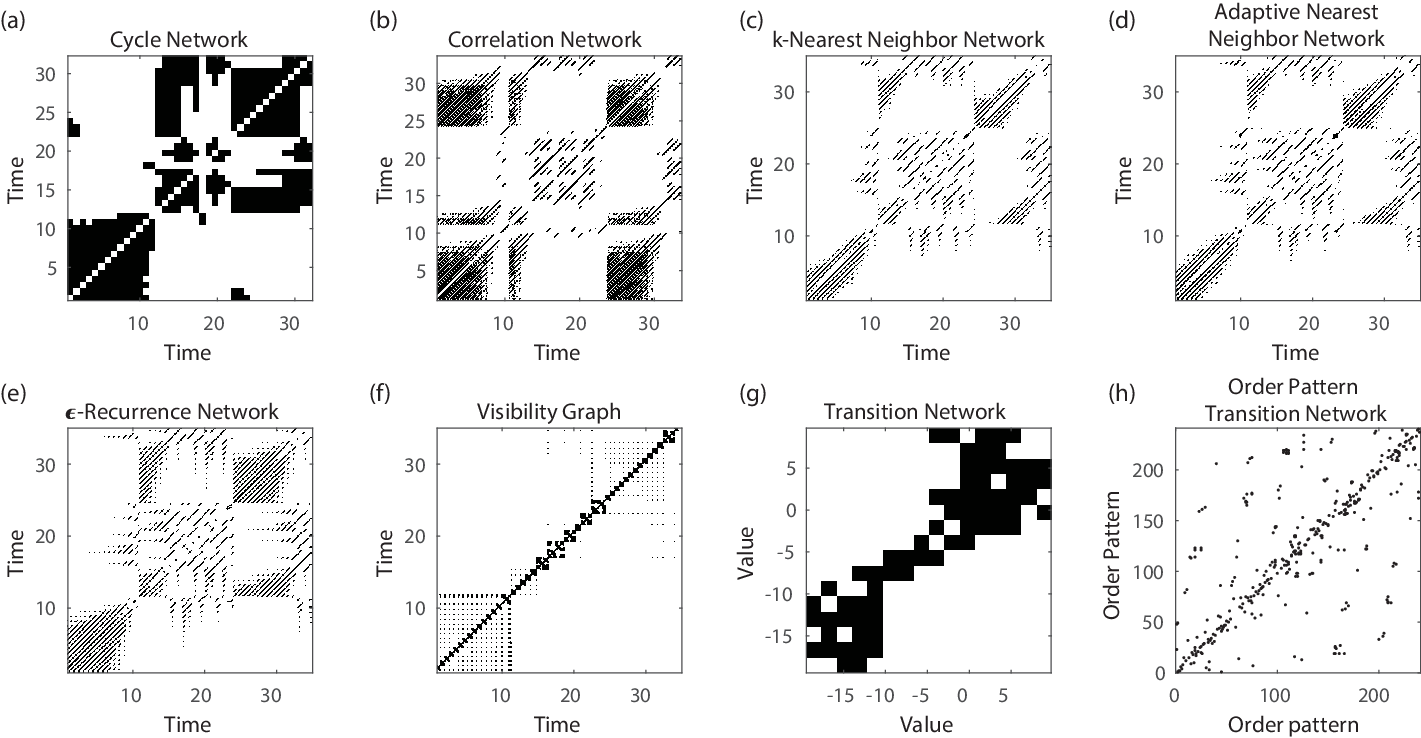} 
\caption{Adjacency matrices corresponding to different types of time series networks constructed from the $x$-coordinate of the Lorenz system: (a) cycle network ($N=40$, critical cycle distance in phase space $D_{max}=5$), (b) correlation network ($N=654$, embedding dimension $m=10$ with delay $\tau=3$), (c) $k$-nearest neighbor network {(asymmetric version)}, $N=675$, $m=3$, $\tau=3$, $k=10$, corresponding to a recurrence rate of $RR\approx 0.015$ using Euclidean norm, (d) adaptive nearest neighbor network $N=675$, $m=3$, $\tau=3$, (e) $\varepsilon$-recurrence network ($N=675$, $m=3$, $\tau=3$, $\varepsilon=2$, maximum norm), (f) visibility graph ($N=681$), and (g) coarse-graining based transition network (based on an equipartition of the range of observed values into $N=20$ classes of size $\Delta x=3.0$, minimum transition probability $p=0.2$ during 3 time steps), (h) ordinal pattern transition network ($N=240$ neglecting disconnected patterns, $m=6$, $\tau=3$). Modified from \cite{Donner2011}. } \label{fig:lorenz_adj-matrices}
\end{figure}

\begin{figure}[htbp]
	\centering
	\includegraphics[width=\textwidth]{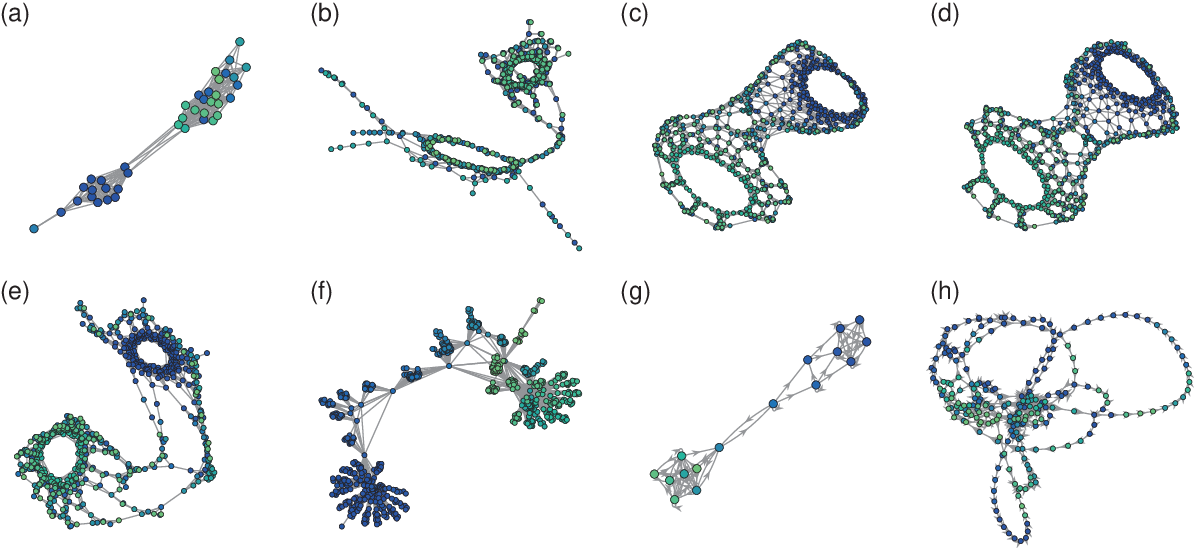} 
\caption{Graphical representation of the different complex networks based on the adjacency matrices shown in Fig.~\ref{fig:lorenz_adj-matrices}. The graphs have been embedded into an abstract two-dimensional space using a force directed placement algorithm~\cite{Battista1994}, which has been integrated into the {\tt {graph}} toolbox of Matlab. For panels (a)-(f), the vertex color indicates the temporal order of observations (from blue to green), for transition networks (panels (g, h)), colors correspond to the different partitions. Note that in panels (b), (g) and (h) some individual disconnected vertices have been removed from the corresponding network representations. Modified from \cite{Donner2011}. } \label{fig:lorenz_network}
\end{figure}

	Besides functional networks derived from multivariate time series, there have been numerous efforts for utilizing complex network approaches for quantifying structural properties of individual time series. By means of complex network analysis, the first step is to find a proper network representation for time series, i.e., an algorithm defining what network vertices and network edges are. To this end, several approaches have been proposed \cite{Zhang2006,Xu2008,Lacasa2008,Marwan2009,Donner2010a,Donner2011,McCullough2015}. Based on these network representations, the rich toolbox of complex network measures \cite{Boccaletti2006,Costa2007,Newman2003} provides various quantities that can be used for characterizing the system's dynamical complexity from a complex network viewpoint and allow discriminating between different types of dynamics~\cite{Donner2011}. More importantly, complementary features of dynamical systems (i.e., properties that are not captured by existing methods of time series analysis) can be resolved. In this report, we give an exhaustive review on complex network approaches for nonlinear time series analysis. To this end, we first provide an overall impression of various complex network representations for time series, which are illustrated in Figs.~\ref{fig:lorenz_adj-matrices},\ref{fig:lorenz_network} for the $x$-coordinate of one realization of the Lorenz system (Eq.~\eqref{eqlorenz}) with the parameters $r=28$, $\sigma=10$ and $\beta=8/3$ (sampling time $\Delta t=0.02$). In the following sections, we will expand the discussions on the reconstruction of these networks from given time series data and their resulting characteristics. Specifically, we will focus on some important transformation methods that have been widely applied to various artificial as well as real-world observational or experimental data, in particular, recurrence networks, visibility graphs, and transition networks. In addition to these main approaches, we will also discuss some algorithmic variants of these concepts and corresponding relevant network measures wherever appropriate. 

    In this review we provide in-depth discussions on complex network representations of individual or potentially interrelated time series, which distinctively differ from existing dara minimg tools from computer sciences \cite{Zanin2016} as discussed above in terms of the underlying motivation and methodology. Specifically, all network approaches discussed in this report provide different applications of complex network theory to nonlinear time series analysis. We do not further expand the discussion on the differences between time series networks and data mining tools to keep this report topically focused.

	\subsection{Transformations of time series into the complex network domain}
	In order to make time series accessible to complex network analysis techniques, we have to transform them into a proper network representation. At a first place, this requires an algorithm defining network vertices and edges. Depending on these definitions, there are at least three main classes of complex network approaches to the analysis of individual time series that will be put in the focus of this review (see Tab.~\ref{tab:methods}). These three types of methods are based on different rationales, i.e.,
\begin{enumerate}[(i)]
\item mutual statistical similarity or metric proximity between different segments of a time series (proximity networks),
\item convexity of successive observations (visibility graphs), and
\item transition probabilities between discrete states (transition networks).
\end{enumerate}

	The first important class of time series networks make use of similarities or proximity relationships between different parts of a dynamical system's trajectory \cite{Marwan2009,Donner2010a,Donner2011}, including such diverse approaches as cycle networks \cite{Zhang2006,Zhang2008e}, correlation networks \cite{Yang2008}, and phase space networks based on a certain definition of nearest neighbors \cite{Xu2008}. One especially important example of proximity networks are recurrence networks (RNs) \cite{Marwan2009,Donner2010a}, which provide a reinterpretation of recurrence plots \cite{marwan2007} in network-theoretic terms and are meanwhile widely applied in a variety of fields. 

	The second class are visibility graphs and related concepts, which characterize some local convexity or record-breaking property within univariate time series data \cite{Lacasa2008,Luque2009,Donner2012}. The standard visibility graph and its various variants have important applications, such as providing new estimates of the Hurst exponent of fractal and multi-fractal stochastic processes \cite{Lacasa2009,Ni2009} or statistical tests for time series irreversibility \cite{Donges2013,Lacasa2012}. 

	The third important class of network approaches are transition networks, which make use of ideas from symbolic dynamics and stochastic processes. Transforming a given time series into a transition network is a process of mapping the temporal information into a Markov chain to obtain a compressed or simplified representation of the original dynamics. More specifically, we first discretize the dynamics and then study the transition probabilities between the obtained groups in some Markov chain-like ways \cite{Nicolis2005}. Depending on the particular choice of partitions, we obtain different versions of transition networks. For instance, we can construct transition networks by threshold-based coarse graining of the underlying system's phase space \cite{Donner2011} or based on ordinal patterns \cite{McCullough2015,Kulp2016b}. 

	It may be interesting to note that both, proximity networks and transition networks, are somewhat related with the concept of recurrence in one way or the other \cite{Donner2011}. This is particularly evident for proximity networks, where connectivity is defined in a data-adaptive local way, i.e., by considering distinct regions with a varying center at a given vertex in either the phase space itself or an abstract metric space where (pseudo) distances measure similarities between states or sequences thereof. In contrast, for transition networks, the corresponding classes are rigid, i.e., determined by a fixed coarse-graining of the phase space, ordinal patterns, or other related symbolic approaches. In this regard, the distinction between both classes of time series network approaches closely resembles the duality between phase space based approaches of nonlinear time series analysis on the one hand, and symbolic time series analysis and related information-theoretic approaches, which may both be used for estimating similar dynamical invariants such as entropies and mutual information \cite{Balasis2013}.
\begin{table}[t]
\caption{Summary of the definitions of vertices and the criteria for the existence of edges in existing complex network approaches.}{
\resizebox{\columnwidth}{!}{%
\begin{tabular}{llll}
\hline
Method & Vertex & Edge & Directedness \\
\hline
Proximity networks & & \\
\textit{Cycle networks} & Cycle & Correlation or phase space distance between cycles & undirected \\
\textit{Correlation networks} & State vector & Correlation coefficient between state vectors & undirected \\
\textit{Recurrence networks} & & & \\
\quad \textit{$k$-nearest neighbor networks}& State (vector) & Recurrence of states (fixed neighborhood mass) & directed \\
\quad \textit{adaptive nearest neighbor networks}& State (vector)  & Recurrence of states (fixed number of edges) & undirected \\
\quad \textit{$\varepsilon$-recurrence networks} & State (vector) & Recurrence of states (fixed neighborhood volume) & undirected \\
\hline
Visibility graphs &  &  &  \\
\quad \textit{natural visibility graphs}& Scalar state & Mutual visibility of states & undirected \\
\quad \textit{horizontal visibility graphs}& Scalar state  & Horizontal mutual visibility of states & undirected \\
\hline
Transition networks &  &  &  \\
\quad \textit{threshold based networks}& Phase space partition & Temporal succession & directed \\
\quad \textit{ordinal pattern networks}& Ordinal patterns  & Temporal succession & directed \\
\hline
\end{tabular}
}
\normalsize
\label{tab:methods}}
\end{table}

	Among the three classes of methods listed above, the largest group of concepts is given by proximity networks, where the mutual closeness or similarity of different segments of a trajectory can be characterized in different ways. Consequently, there are various types of such proximity networks (see Tab.~\ref{tab:methods}). However, all these methods are characterized by two common general properties: Firstly, the resulting networks are invariant under relabeling of their vertices in the adjacency matrix. Hence, the topological characteristics of proximity networks yield nonlinear measures that are invariant under a permutation of their vertices. In this respect, these network-theoretic approaches are distinctively different from traditional methods of time series analysis where the temporal order of observations does explicitly matter. Secondly, we point out that especially proximity networks are spatial networks \cite{Barthelemy2011,Wiedermann2016}. In particular, recurrence networks are embedded in the phase space of the considered system, with distances being defined by one of the standard metrics (e.g., Euclidean, Manhattan, etc.), making them a specific type of random geometric graphs \cite{Donner2011b}. Similar considerations apply to other types of proximity networks as well. Moreover, also visibility graphs and related concepts can be viewed as spatially embedded networks, for which the one-dimensional time axis takes the role of a metric space in which the resulting network's vertices and edges are embedded.

\subsection{Outline of the report}
The remainder of this review is organized as follows: 

In Section \ref{sec:CompNetworkT}, we start with a brief introduction on complex network theory, mainly focusing on the characterization of the structural properties of networks based on the adjacency matrix. All relevant terminologies of network measures will be introduced in this section. We also discuss some concepts that are particularly important for transforming time series into network representations, particularly, the definitions of network vertices and edges. 

In Section \ref{sec:RecurrenceNt}, we focus on recurrence network approaches (RN). We will cover the theoretical background of Poincar\'e recurrences in dynamical systems and the popular visualization technique of recurrence plots \cite{marwan2007}. Furthermore, we summarize the current state of knowledge on the theoretical foundations and potential applications of RN approaches to nonlinear time series. We demonstrate that this type of time series networks naturally arise as random geometric graphs in the phase space of dynamical systems, which determines their structural characteristics and gives rise to a dimensionality interpretation of clustering coefficients and related concepts. Beyond the single-system case, we also provide a corresponding in-depth discussion of cross- and joint recurrence plots from the complex network viewpoint. Moreover, we discuss some recent ideas related to the utilization of multiplex and multilayer multivariate recurrence network-based approaches for studying geometric signatures of coupling and synchronization processes. 

In Section \ref{sec:VisibilityGt}, both the standard visibility graphs (VG) and horizontal visibility graphs (HVG) will be reviewed. We start with discussing the main variants of visibility algorithms applied in the context of time series analysis. Specifically, we summarize some conjectures of theoretical predictions of (H)VG properties in stochastic and deterministic processes. Some practical considerations when applying (H)VG analysis to experimental time series will be thoroughly discussed. In addition, we will discuss the generalization of (H)VG analysis from univariate to bi- and multivariate time series, for instance, multiplex (H)VGs. We further show that a decomposition of (H)VGs into time forward (outgoing) and backward (incoming) directions helps to test irreversibility of the underlying time series. 

In Section \ref{sec:TransitionNt}, we introduce the construction of transition networks by proper coarse graining of phase space and ordinal patterns. Specifically, the concept of ordinal pattern transition networks can be traced back to identifying ordinal patterns of time series \cite{Bandt2002}. We particularly review the ordinal pattern transition networks of \cite{McCullough2015} and their generalizations to multivariate time series \cite{Zhang2017b}, highlighting their great potential for studies of experimental observation data from climate sciences \cite{Eroglu2016}. 

In Section \ref{sec:Applications}, we review several applications of network approaches to different real-world time series. The following Section \ref{sec:Software} briefly summarizes existing software implementations, with a particular focus on the Python package \texttt{pyunicorn} that includes several methods from both, complex network theory and nonlinear time series analysis, including several of the approaches discussed in this review, and unites them in a high-performance, modular and flexible way \cite{Donges2015}. Finally, Section \ref{sec:Discussion} summarizes the main topics addressed in this report and puts them into a broader perspective. Specifically, we will outline a few important general directions for future research. We emphasize that applying complex network methods for time series analysis is still an emerging field, and that there are numerous relevant topics from both the theoretical and applied perspectives that still deserve further exploration.

%% file: Chapter02_CompNetworkT/Chapter02_CompNetworkT.tex
\section{Complex network theory} \label{sec:CompNetworkT}
In this section, we provide the brief introduction of the characterization of structural properties of a network, focusing on definitions, notations, and basic quantities that are often used to describe the topologies of networks reconstructed from time series. More comprehensive descriptions of complex networks can be found in a number of review articles \cite{Albert2002,Newman2003,Boccaletti2006,Costa2007} and books \cite{Cohenbook2010,Newmanbook2010}, which the reader may find useful to consult.

	\subsection{Basic concepts} \label{sec:basicCompNets}
	A complex network is often represented as a graph $G = (V, E)$ which consists of two sets $V$ and $E$, where $V$ is the set of vertices (nodes or points) of $G$, and $E$ is the set of edges (links or lines) representing pairs of connected elements of $V$ \cite{Costa2007}. Each vertex is identified by an integer index $p=1,\dots,N$, and each edge is identified by a pair $(p,q)$ connecting two vertices $p$ and $q$. A graph $G$ is called undirected if an edge from vertex $p$ to $q$ as denoted by $(p,q)$ is equivalent to the edge of $(q,p)$ from vertex $q$ to $p$, i.e., $(p,q)\in E \Leftrightarrow (q,p)\in E$. On the other hand, in a directed graph, typically $(p,q)\in E \nLeftrightarrow (q,p)\in E$. A graph may contain loops, i.e., edges from a vertex to itself, or multiple edges, i.e., pairs of vertices connected by more than one edge. 
    
    More generally, edges $(p,q)$ may be attributed additional weights $W_{pq}$. For convenience, one commonly defines $W_{pq}=0$ of $(p,q)\notin E$, In this case, a weighted directed graph can be completely described by its weight matrix $\mathbf{W}$ so that each entry $W_{pq}$ expresses the weight of the connection from vertex $p$ to vertex $q$. In this section, we only consider undirected and unweighted graphs for the sake of simplicity, since there is no need for discussing $W_{pq}$ in the context of most of the complex network approaches for time series analysis discussed in this review. A notable exception from this are transition networks, for which Weighted graphs will receive a special attention in Section \ref{sec:TransitionNt}. 
	
	An unweighted graph can be naturally constructed by applying a proper threshold $T$ to the elements of the weight matrix $\mathbf{W}$ of its weighted counterpart \cite{Costa2007}, resulting in the binary matrix $\mathbf{A}$. Specifically, we have $A_{pq} = 1$ if $W_{pq} > T$, otherwise, $A_{pq} = 0$. The resulting matrix $\mathbf{A}$ is called the \emph{adjacency matrix} of the resulting unweighted graph, and each nonzero element $A_{pq}$ of $\mathbf{A}$ indicates the presence of $(p,q)$ as a member of its edge set $E$. Further introduction of symmetry to $\mathbf{A}$, i.e., identifying $A_{pq} = A_{qp}$, is characteristic of an undirected graph. Such an undirected, unweighted graph is also called a simple graph.  
	
	Depending on the particular mappings for transforming a given time series into a complex network, the resulting adjacency matrix $\mathbf{A}$ often depends on some algorithmic parameters, for instance, the threshold value $\varepsilon$ of the recurrence network approach (Section~\ref{sec:RecurrenceNt}). More importantly, we often have some particular interpretations for network measures, for instance, in terms of the geometry of a dynamical system. In the following, we first introduce some general measures for characterizing some important aspects of network structures based on $\mathbf{A}$. More specific discussions of network measures in terms of the particular network transforming methods will be presented in later sections. 

	In addition to the concepts of vertices and edges, another important concept in complex network theory is the notion of paths. A \emph{path} between two specified vertices $p$ and $q$ is an ordered sequence of edges starting at $p$ and ending at $q$, with its \emph{path length} $l_{pq}$ given by the number of edges in this sequence. There are also various measures characterizing the structural properties of networks based on paths, which will be briefly reviewed here as well.  

	\subsection{Network characteristics} \label{sec:basictheoryCN}
		\subsubsection{Vertex characteristics}
		There are various measures to characterize the structures of a complex network, quantifying the importance of either a vertex or an edge in terms of a particular network property. The conceptually simplest measure characterizing the connectivity properties of a single vertex in a complex network is the \textit{degree} (or \textit{degree centrality})
\begin{equation} 
k_p=\sum_{q=1}^N A_{pq} ,
\label{eq:degree}
\end{equation}
\noindent
which simply counts the number of edges associated with a given vertex $p$. It is also convenient to introduce a normalized degree 
\begin{equation} \label{eq:localrho}
\rho_p = \frac{1}{N-1} k_p
\end{equation}
as the local connectivity density of $p$. Furthermore, a topological characterization of the graph $G$ as a whole can be obtained in terms of the degree distribution $p(k)$, defined as the probability that a vertex chosen uniformly at random has degree $k$ or, equivalently, as the fraction of vertices in the graph having degree $k$. Note that the variable $k$ assumes non-negative integer values. The degree distribution $p(k)$ is often used to classify complex networks, for instance, a scale-free network is characterized by $p(k) \sim k^{-\gamma}$, which will further discussed in Sec. \ref{sec:styleFacts}. Furthermore, one simple definition of a network entropy  is based on the degree distribution as $S = - \sum_{k} p(k) \log p(k)$, which can be computed straightforwardly \cite{Rashevsky1955,MacArthur1955}. A more recent survey of information theoretic measures based on different network partitions of complex topology has been presented in \cite{Dehmer2011}. 

		In order to characterize the density of connections among the neighbors of a given vertex $p$, we can utilize the \textit{local clustering coefficient}
\begin{equation}
  {\mathcal{C}}_p =\frac{1}{  {k}_p (  {k}_p -1)} \sum_{q,r=1}^N A_{pq}  A_{qr}  A_{rp} ,
\label{eq:locclustering}
\end{equation}
\noindent
which measures the fraction of pairs of vertices in the neighborhood of $p$ that are mutually connected. 

		While degree and local clustering coefficient characterize network structures on the local and meso-scale, there are further vertex characteristics that make explicit use of the concept of shortest paths and, thus, provide measures relying on the connectivity of the entire network. Two specific properties of this kind are the \textit{closeness} or \textit{closeness centrality}
\begin{equation}
  {c}_p =\left(\frac{1}{N-1}\sum_{q=1}^N   {l}_{pq}  \right)^{-1},
\label{eq:closeness}
\end{equation}
\noindent
which gives the inverse arithmetic mean of the shortest path lengths $l_{pq}$ between vertex $p$ and all other vertices $q\in V$, and the \textit{local efficiency}
\begin{equation}
  {e}_p =\frac{1}{N-1}\sum_{q=1}^N   {l}_{pq} ^{-1},
\label{eq:locefficiency}
\end{equation}
\noindent
which gives the inverse harmonic mean of these shortest path lengths. Notably, the latter quantity has the advantage of being well-behaved in the case of disconnected network components, where there are no paths between certain pairs of vertices (i.e., $  {l}_{pq}=\infty$). In order to circumvent divergences of the closeness due to the existence of disconnected components, it is convenient to always set ${l}_{pq}$ to the highest possible value of $N-1$ for pairs of vertices that cannot be mutually reached. Both $  {c}_p $ and $  {e}_p $ characterize the geometric centrality of vertex $p$ in the network, i.e., closeness and local efficiency exhibit the highest values for such vertices which are situated in the center of the networks. 

		Another frequently studied path-based vertex characteristic is the \textit{betweenness} or \textit{betweenness centrality}, which measures the fraction of shortest paths in a network traversing a given vertex $p$. Let $  {\sigma}_{qr}$ denote the total number of shortest paths between two vertices $q$ and $r$ and $  {\sigma}_{qr}(p)$ the multiplicity of these paths that include a given vertex $p$, betweenness centrality is defined as
\begin{equation}
  {b}_p =\sum_{q,r=1; q,r\neq p}^N \frac{  {\sigma}_{qr}(p)}{  {\sigma}_{qr}}.
\label{eq:betweenness}
\end{equation}
\noindent
It is commonly used for characterizing the importance of vertices for information propagation in networks. 

		\subsubsection{Edge characteristics}
		In contrast to vertices, whose properties can be characterized by a multitude of graph characteristics, there are fewer measures that explicitly relate to the properties of edges or, more general, pairs of vertices. One such measure is the \textit{matching index}, which quantifies the overlap of the network neighborhoods of two vertices $p$ and $q$:
\begin{equation}
  {m}_{pq} =\frac{\sum_{r=1}^N A_{pr}  A_{qr} }{  {k}_p +  {k}_q -\sum_{r=1}^N A_{pr}  A_{qr} }.
\label{eq:matching}
\end{equation}
\noindent

		While the concept of matching index does not require the presence of an edge between two vertices $p$ and $q$, there are other characteristics that are explicitly edge-based. To this end, we only mention that the concept of betweenness centrality $b_p$ can also be transferred to edges, leading to the \textit{edge betweenness} measuring the fraction of shortest paths on the graph traversing through a specific edge $(p,q)$:
\begin{equation}
  {b}_{pq} =\sum_{r,s=1; r,s\neq p,q}^N \frac{  {\sigma}_{rs}(p,q)}{  {\sigma}_{rs}},
\label{eq:edgebetweenness}
\end{equation}
\noindent
where $  {\sigma}_{rs}(p,q)$ gives the total number of shortest paths between two vertices $r$ and $s$ that include the edge $(p,q)$. If there is no edge between two vertices $p$ and $q$, we set $  {b}_{pq}=0$ for convenience. 
        
        		Finally, we mention the concept of network motifs \cite{Milo2002} as another edge-based way to obtain proper information on the meso-scale connectivity properties of a graph, which generalize the idea of the local clustering coefficient and are particularly useful for the study of directed networks. In this context, motifs are small connected subgraphs consisting of a small fixed number of vertices (typically, 3 or 4 due to their fastly increasing combinatorial variety and computational demanding), which represent specific local connection patterns that allow classifying real-world networks into superfamilies according to the relative frequencies of different motifs \cite{Milo2002}. In terms of time series networks, we focus on the frequency distribution of motifs which may serve as a sensitive indicator of specific type of network structures that are reconstructed by different methods, for instance, in Section \ref{sec:adaptiveRN}. 
		        
        		\subsubsection{Global network characteristics}
		Some, but not all useful global network characteristics can be derived by averaging certain local-scale (vertex) properties. Prominently, the \textit{edge density}
\begin{equation}
  {\rho} =\frac{1}{N}\sum_{p=1}^N   {\rho}_p =\frac{1}{N(N-1)} \sum_{p,q=1}^N A_{pq} 
\label{eq:edgedensity}
\end{equation}
is defined as the arithmetic mean of the degree densities of all vertices $\rho_p$ and characterizes the fraction of possible edges that are present in the network. 

		In a similar way, we consider the arithmetic mean of the local clustering coefficients $  {\mathcal{C}}_p $ of all vertices, resulting in the (Watts-Strogatz) \textit{global clustering coefficient}~\cite{Watts1998}
\begin{equation}
  {\mathcal{C}} =\frac{1}{N}\sum_{p=1}^N   {\mathcal{C}}_p 
= \frac{1}{N}\sum_{p=1}^N \frac{\sum_{q,r=1}^N A_{pq}  A_{qr}  A_{rp} }{  {k}_p (  {k}_p -1)}, 
\label{eq:globclustering}
\end{equation}
which measures the mean fraction of triangles that include the different vertices of the network. 

		Notably, in the case of a very heterogeneous degree distribution, the global clustering coefficient will be dominated by contributions from the most abundant type of vertices, the hubs. For example, for a scale-free network with $p(k)\sim k^{-\gamma}$, vertices with small degree will contribute predominantly, which can lead to an underestimation of the actual fraction of triangles in the network, since $  {\mathcal{C}}_p =0$ if $  {k}_p <2$ by definition. In order to correct for such effects, Barrat and Weigt~\cite{Barrat2000} proposed an alternative definition of the clustering coefficient, which is nowadays frequently referred to as \textit{network transitivity}~\cite{Boccaletti2006} and is defined as
\begin{equation}
  {\mathcal{T}} = \frac{\sum_{p,q,r=1}^N A_{pq}  A_{qr}  A_{rp} }{\sum_{p,q,r=1}^N A_{pq}  A_{rp} }.
\label{eq:transitivity}
\end{equation}
\noindent

	Finally, turning to shortest path-based characteristics, we define the \textit{average path length}
\begin{equation}
  {\mathcal{L}} =\frac{1}{N(N-1)} \sum_{p,q=1}^N   {l}_{pq}  = \frac{1}{N} \sum_{p=1}^N   {c}_p ^{-1}
\label{eq:apl}
\end{equation}
\noindent
as the arithmetic mean of the shortest path lengths between all pairs of vertices, and the \textit{global efficiency}
\begin{equation} 
  {\mathcal{E}} =\left(\frac{1}{N(N-1)} \sum_{p,q=1}^N   {l}_{pq} ^{-1} \right)^{-1} = \left( \frac{1}{N} \sum_{p=1}^N   {e}_p  \right)^{-1}
\label{eq:globefficiency}
\end{equation}
\noindent
as the associated harmonic mean. Notably, the average path length can be rewritten as the arithmetic mean of the inverse closeness, and the global efficiency as the inverse arithmetic mean of the local efficiency. Furthermore, based on shortest path length, one often defines the diameter of a network as the longest (maximum) of all the calculated shortest paths in a network, $\mathcal{D} = \max_{p,q} l_{pq}$. In other words, once the shortest path length from every vertex to all other vertices is calculated, the diameter $\mathcal{D}$ is the maximum of all the calculated path lengths. Certainly, the diameter is representative of the size of a network.       

	\subsection{Stylized facts of complex networks} \label{sec:styleFacts}
	Erd\"os and R\'enyi \cite{Erdos1959} introduced a model to generate random graphs consisting of $N$ vertices and $M$ edges. Starting with $N$ disconnected vertices, the network is constructed by the addition of $L$ edges at random, avoiding multiple and self connections. Another similar model defines $N$ vertices and a probability $p$ of connecting each pair of vertices. The latter model is widely known as the Erd\"os-R\'enyi (ER) model. For the ER model, in the large network size limit $(N \to \infty)$, the average number of connections of each vertex $\left < k \right>$ is given by $\left< k \right > = c (N - 1)$, where $c$ is fixed and often chosen as a function of $N$ to keep $\left < k \right >$ fixed. For this model, the degree distribution $p(k)$ is a Poisson distribution. 
	
	In regular hypercubic lattices in $d$ dimensions, the mean number of vertices one has to pass in order to reach an arbitrarily chosen vertex, grows with the lattice size as $N^{1/d}$. Conversely, in most real-world networks, despite of their often large size $N$, there is a relatively short path between any two vertices. This property, which is shared by many real-world networks, is the so-called small-world (SW) effect, that has been first described as the outcome of studies on social interrelationships, predominantly Milgram's famous chain-letter experiment in the 1960s \cite{Milgram1967}. In the spirit of the latter studies, the term ``SW effect'' originally denoted the fact that average shortest path lengths $\mathcal{L}$ (Eq.~\ref{eq:apl}) in social networks, but also other real-world networks, are much shorter than we would expect from random connectivity configurations. Given the importance of redundancy in such networks, Watts and Strogatz \cite{Watts1998} suggested including the presence of a high clustering coefficient $\mathcal{C}$ (i.e., higher than in random graphs) as a second criterion for identifying the small-world effect in real-world networks. According to their definition, small-world networks are characterized by having both a small value of $\mathcal{L}$, like random graphs, and a high clustering coefficient $\mathcal{C}$, like regular lattices. The generative model introduced by them (WS model) is based on a probabilistic rewiring of edges in a regular ring lattice (i.e., each existing edge is rewired uniformly at random with the same probability $c$) and is thus situated between an ordered finite lattice ($c=0$) and a random graph ($c=1$), presenting the small world property with short path length and high clustering coefficient at intermediate values of $c$. 

	Barab\'asi and Albert \cite{Barabasi199} showed that the degree distribution $p(k)$ of many real-world systems is characterized by a heavy-tailed distribution. Instead of the vertices of these networks having a random pattern of connections with a characteristic degree, as in the ER and WS models, some vertices are highly connected while others have few connections, with the absence of a characteristic degree. More specifically, the degree distribution has been found to follow a power-law for large $k$, $p(k) \sim k^{-\gamma}$. These networks are called scale-free (SF) networks, which are captured by a pronounced linear regime in the double logarithmic plot of $p(k)$. In order to model the emergence of such network structures, the BA model has been proposed which contains the two important ingredients of network growth and preferential attachment. A proper statistical identification of SF properties in real-world networks is a non-trivial task because of effects originating from finite size, intrinsic noise and finite sample size \cite{Clauset2009}. 
	
	In addition, a large number of real-world networks are correlated in the sense that the probability that a node of degree $k$ is connected to another node of degree, say $k'$, depends on $k$. This problem can be quantified by the average nearest neighbor degree of a vertex $p$, or, alternatively, the assortativity coefficient $\mathcal{R}$, i.e., the correlation coefficient between the degrees of all pairs of mutually connected vertices \cite{Newman2002}. In assortative networks, the vertices tend to connect to their connectivity peers ($\mathcal{R} > 0$), while in disassortative networks vertices with low degrees are more likely connected with highly connected ones ($\mathcal{R} < 0$). 
	
	In the following sections, the presence or absence of these stylized facts of complex networks will be discussed in the respective frameworks when introducing different network construction algorithms based upon possibly nonlinear time series. 
	
		\subsection{Multiplex and multilayer networks} \label{sec:multiplex}
			Many complex systems include multiple subsystems and layers of connectivity and they evolve, adapt and transform through internal and external dynamical interactions affecting the subsystems and components at both local and global scale. For example, the problem of information or rumor spreading on top of a social network like Facebook must take into account the intricate interactions on different levels \cite{Boccaletti2014}. In general, many interactions in social networks can be understood as a combination of interactions at different, independent levels, each representing different social scenarios such as family, friends, coworkers, etc. The actual relationships amongst the members of a social network must consider interactions mostly inside the levels and their influences on the other layers. An individual's behavior can be different in each level but it is conditioned by other levels. Multiplex and multilayer networks explicitly incorporate multiple levels of social interactions and have been successfully applied to the study of disease spreading and diffusion dynamics \cite{Gomez2013,Granell2013} or the evolution of cooperation in the presence of social dilemmas  \cite{Matamalas2015}. 
			
			Understanding and possibly predicting multi-scale and multi-component dynamics is a difficult challenge to complex systems theory \cite{Boccaletti2014}. In this context, the issues posed by the multi-scale modeling of both natural and artificial complex systems call for a generalization of the ``traditional" network theory, foremost including the development of a solid theoretical foundation and associated tools for studying multilayer and multi-component systems in a comprehensive fashion. 
		
			Here, we follow the formal definition of a {\textit{multilayer network}} \cite{Boccaletti2014} as a pair $\mathcal{M} = (\mathcal{G}, \mathcal{C})$ where $\mathcal{G} = \{G_{\alpha}; \alpha=1, \dots, M\}$ is a family of graphs $G_{\alpha} = (V_{\alpha}, E_{\alpha})$ and 
		\begin{equation}
		\mathcal{C} = \{ E_{\alpha\beta} \subseteq V_{\alpha} \times V_{\beta}; \alpha, \beta \in [1, 2, \dots, M], \alpha \neq \beta \}
		\end{equation}
		is the set of interconnections between nodes of different layers $G_{\alpha}$ and $G_{\beta}$ with $\alpha \neq \beta$. The elements of each $E_{\alpha}$ are called intra-layer connections of $\mathcal{M}$ as opposed to those of each $E_{\alpha\beta}$ with $\alpha\neq\beta$ that are called inter-layer connections. By using the multilayer network representation, we simultaneously consider edges that are located inside different layers and such that connect different layers. A {\textit{multiplex network}} is a special type of multilayer network in which each layer shares the same set of vertices, i.e., $V_{1} = \dots = V_M$, and the only possible type of interlayer connections are those in which a given node is connected to its counterparts in the other layers.  In other words, a multiplex network consists of a fixed set of vertices connected by different types of edges \cite{Boccaletti2014}. 
		
			The readers are referred to \cite{Boccaletti2014,Buldyrev2010} for a more thorough review on multilayer networks. Furthermore, it is important to remark that the concept of multilayer networks has been extended to other relevant notations, for instance, network of networks, interacting or interconnected networks, multidimensional networks, interdependent networks, multilevel networks, hypernetworks, etc., some of which are used as synonyms of each other. 	
	
		\subsection{Coupled networks}\label{sec:irn_measures}
	    		
                \subsubsection{Preliminaries}
			When describing multilayer and multiplex networks in Section \ref{sec:multiplex}, we have introduced a corresponding rather general framework. In the particular case of networks constructed from two or more possibly interdependent time series, different perspectives can be taken with respect to the coupled system. Under some conditions, it can be preferable to utilize the recently introduced framework of coupled or interdependent network analysis for a corresponding topological characterization \cite{Donges2011b,Wiedermann2013}. The study of coupled networks focuses on interrelationships between the different subnetworks, i.e., on a dependency scenario in which vertices in one network require connections to vertices in another subnetwork (e.g., in case of telecommunication networks and power grids) \cite{Buldyrev2010}. 
			
			Let us again consider an arbitrary undirected and unweighted simple graph $G=(V,E)$ with the adjacency matrix $\textbf{A}=\{A_{pq}\}_{p,q=1}^N$. Furthermore, let us assume that there is a given partition of $G$ with the following properties:
\begin{enumerate}
\item The vertex set $V$ is decomposed into $M$ disjoint subsets $V_\alpha \subseteq V$ such that $\bigcup_{\alpha=1}^M V_\alpha = V$ and $V_\alpha \cap V_\beta = \emptyset$ for all $\alpha \neq \beta$. The cardinality of $V_\alpha$ will be denoted as $N_\alpha$. 

\item The edge set $E$ consists of mutually disjoint sets $E_{\alpha\beta} \subseteq E$ with $\bigcup_{\alpha,\beta=1}^K E_{\alpha\beta} = E$ and $E_{\alpha\beta} \cap E_{\gamma\delta}=\empty$ for all $(\alpha,\beta) \neq (\gamma,\delta)$.

\item Let $E_{\alpha\beta}\subseteq V_\alpha \times V_\beta$. Specifically, for all $\alpha=1,\dots,M$, $G_\alpha=(V_\alpha,E_{\alpha\alpha})$ is the induced subgraph of the vertex set $V_\alpha$ with respect to the full graph $G$.
\end{enumerate}
Under these conditions, $E_{\alpha\alpha}$ comprises the (internal) edges within $G_\alpha$, whereas $E_{\alpha\beta}$ contains all (cross-) edges connecting $G_\alpha$ and $G_\beta$. 

			We are now in a position to study the interconnectivity structure between two subnetworks $G_\alpha, G_\beta$ on several topological scales drawing on the lineup of local and global graph-theoretical measures generalizing those used for single network characterization (see Section~\ref{sec:IntSRN}). In this context, local measures ${f}_p^{\alpha\beta}$ characterize a property of vertex $p \in V_\alpha$ with respect to subnetwork $G_\beta$, while global measures ${F}^{\alpha\beta}$ assign a single real number to a pair of subnetworks $G_\alpha, G_\beta$ to quantify a certain aspect of their mutual interconnectivity structure. Most interconnectivity characteristics discussed below have been originally introduced in \cite{Donges2011b}.

			\subsubsection{Vertex characteristics}
\begin{figure}
	\centering
	\includegraphics[scale=0.8]{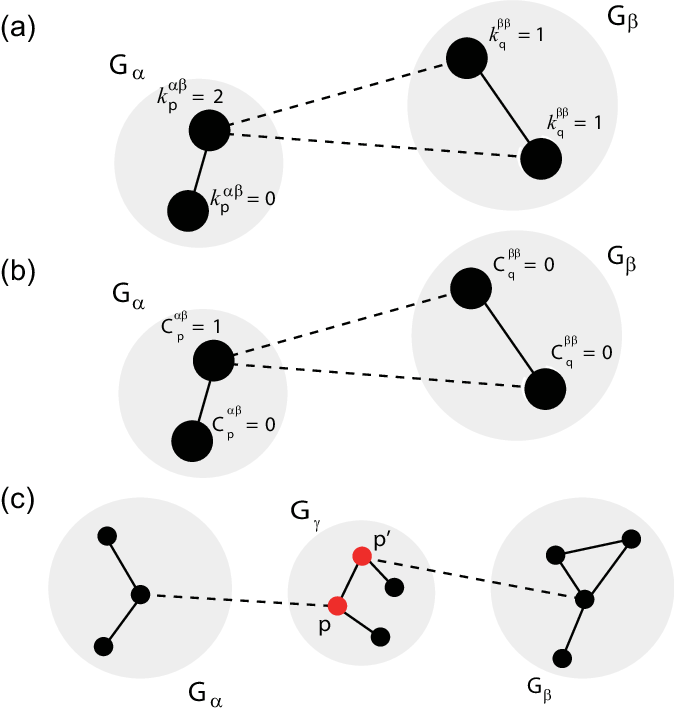} 
\caption{Schematic illustration of some characteristics of interdependent networks: (a) The cross-degree $k_p^{\alpha\beta}$ (Eq.~\eqref{eq:degree_cross}). In the example $\rho^{\alpha\beta}=0.5$. (b) The local cross-clustering coefficient $\mathcal{C}_p^{\alpha\beta}$ (Eq.~\eqref{eq:locclustering_cross}). In the example, the associated values are $\mathcal{C}^{\alpha\beta}=0.5$ and $\mathcal{C}^{\beta\alpha}=0$, whereas $\mathcal{T}^{\alpha\beta}=1$ and $\mathcal{T}^{\beta\alpha}=0$. (c) The cross-betweenness centrality $b_p^{\alpha\beta}$ (Eq. \eqref{eq:betweenness_cross}). In the example, $p,q\in V_\gamma$ (red) have a large cross-betweenness, whereas the remaining vertices $p\in V_\gamma\setminus\{p,q\}$ from subnetwork $G_\gamma$ do not participate in shortest paths between $G_\alpha$ and $G_\beta$ and therefore have vanishing values $b_r^{\alpha\beta}=0$. Modified from \cite{Donges2012PhD,Donges2011b}. } \label{fig:interacting_measures}
\end{figure}

			The \textit{cross-degree} (or \textit{cross-degree centrality})
\begin{equation}
{k}_p^{\alpha\beta} = \sum_{q \in V_\beta} A_{pq}
\label{eq:degree_cross}
\end{equation}
counts the number of neighbors of vertex $p \in V_\alpha$ within the subnetwork $G_\beta$, i.e., direct connections between $G_\alpha$ and $G_\beta$ (Fig.~\ref{fig:interacting_measures}A). Thus, this measure provides information on the relevance of $p$ for the network ``coupling'' between $G_\alpha$ and $G_\beta$. For the purpose of the present work, it is useful studying a normalized version of this measure, the \textit{cross-degree density}
\begin{equation}
{\rho}_p^{\alpha\beta} = \frac{1}{N_\beta} \sum_{q \in V_\beta} A_{pq} = \frac{1}{N_\beta} {k}_p^{\alpha\beta}.
\label{eq:locrho_cross}
\end{equation}

		As for the single network case, important information is governed by the presence of triangles in the network. Given two subnetworks, the \textit{local cross-clustering coefficient}
\begin{equation}
{\mathcal{C}}_p^{\alpha\beta} = \frac{1}{{k}_p^{\alpha\beta}({k}_p^{\alpha\beta} - 1)} \sum_{q,r \in V_\beta} A_{pq} A_{qr} A_{rp},
\label{eq:locclustering_cross}
\end{equation}
\noindent
measures the relative frequency of two randomly drawn neighbors $q,r\in V_\beta$ of $p\in V_\alpha$ are mutually connected (Fig.~\ref{fig:interacting_measures}B). For ${k}_p^{\alpha\beta}<2$, we define ${\mathcal{C}}_p^{\alpha\beta}=0$. In general, ${\mathcal{C}}_p^{\alpha\beta}$ characterizes the tendency of vertices in $G_\alpha$ to connect to clusters of vertices in $G_\beta$. 

		The \textit{cross-closeness centrality}
\begin{equation}
{c}_p^{\alpha\beta} = \left(\frac{\sum_{q \in V_\beta} l_{pq}}{N_\beta}\right)^{-1}
\label{eq:closeness_cross}
\end{equation}
\noindent
(where $l_{pq}$ is the shortest-path length between $p$ and $q$) characterizes the topological closeness of $p\in G_\alpha$ to $G_\beta$, i.e., the inverse arithmetic mean of the shortest path lengths between $p$ and all vertices $q\in V_\beta$. If there exist no such paths, $l_{pq}$ is commonly set to the maximum possible value $N-1$ given the size of $G$. As in the single network case, replacing the arithmetic by the harmonic mean yields the \textit{local cross-efficiency}
\begin{equation}
{e}_p^{\alpha\beta} = \frac{\sum_{q \in V_\beta} l_{pq}^{-1}}{N_\beta},
\label{eq:locefficiency_cross}
\end{equation}
\noindent
which can be interpreted in close analogy to ${c}_p^{\alpha\beta}$. 

		As a final vertex characteristic, we generalize the betweenness concept to the case of coupled subnetworks, which results in the \textit{cross-betweenness centrality}
\begin{equation}
{b}_p^{\alpha\beta} = \sum_{q\in V_\alpha,r\in V_\beta;q,r\neq p} \frac{{\sigma}_{qr}(p)}{{\sigma}_{qr}}.
\label{eq:betweenness_cross}
\end{equation}
\noindent
Note that ${\sigma}_{qr}(p)$ and ${\sigma}_{qr}$ are defined as in the case of a single network. The ${b}_p^{\alpha\beta}$ (Eq.~\eqref{eq:betweenness_cross}) measures the fraction of shortest paths between vertices from subnetworks $G_\alpha$ and $G_\beta$ that traverse the vertex $p\in V_\gamma$ (note that $G_\gamma$ can coincide here with $G_\alpha$ or $G_\beta$). Note that unlike the other vertex characteristics discussed above, in the case of ${b}_p^{\alpha\beta}$, we do not require $p$ belonging to $G_\alpha$ or $G_\beta$  (Fig.~\ref{fig:interacting_measures}C). The reason for this is that vertices belonging to any subnetwork may have a non-zero betweenness regarding two given subgraphs $G_\alpha$ and $G_\beta$, in the sense that shortest paths between $q\in V_\alpha$ and $r\in V_\beta$ can also include vertices in other subnetworks. 

		\subsubsection{Global characteristics}
		The density of connections between two subnetworks can be quantified by taking the arithmetic mean of the local cross-degree density (Eq.~\ref{eq:locrho_cross}), yielding the \textit{cross-edge density}
\begin{equation}
{\rho}^{\alpha\beta} = \frac{1}{N_\alpha N_\beta} \sum_{p \in V_\alpha, q \in V_\beta} A_{pq}.
\label{eq:globrho_cross}
\end{equation}
Since we consider here only undirected networks (i.e., bidirectional edges), $\rho^{\alpha\beta}$ is invariant under mutual exchange of the two considered subnetworks.

		The \textit{global cross-clustering coefficient}
\begin{equation}
{\mathcal{C}}^{\alpha\beta} = \left<{\mathcal{C}}_p^{\alpha\beta}\right>_{p \in V_\alpha} = \frac{1}{N_\alpha} \sum_{p \in V_\alpha, {k}_p^{\alpha\beta}>1} \frac{\sum_{q,r \in V_\beta} A_{pq} A_{qr} A_{rp}}{\sum_{q \neq r \in V_\beta} A_{pq} A_{rp}}
\label{eq:globclustering_cross}
\end{equation}
estimates the probability of vertices in $G_\alpha$ to have mutually connected neighbors in $G_\beta$. Unlike $\rho^{\alpha\beta}$ (Eq. \eqref{eq:globrho_cross}), the corresponding ``cross-transitivity'' structure is typically asymmetric, i.e., ${\mathcal{C}}^{\alpha\beta} \neq {\mathcal{C}}^{lk}$. As in the single network case, we need to distinguish ${\mathcal{C}}^{\alpha\beta}$ from the \textit{cross-transitivity}
\begin{equation}
{\mathcal{T}}^{\alpha\beta} = \frac{\sum_{p \in V_\alpha; q,r \in V_\beta} A_{pq} A_{qr} A_{rp}}{\sum_{p \in V_\alpha; q \neq r \in V_\beta} A_{pq}  A_{rp}},
\label{eq:transitivity_cross}
\end{equation}
for which we generally have ${\mathcal{T}}^{\alpha\beta}(\varepsilon) \neq {\mathcal{T}}^{lk}(\varepsilon)$ as well. Again, we have to underline that cross-transitivity and global cross-clustering coefficient are based on a similar concept, but capture distinctively different network properties as global versus mean local network features.

		Regarding the quantification of shortest path-based characteristics, we define the \textit{cross-average path length}
\begin{equation}
{\mathcal{L}}^{\alpha\beta} = \frac{1}{N_\alpha N_\beta} \sum_{p \in V_\alpha, q \in V_\beta} l_{pq} 
\label{eq:apl_cross}
\end{equation}
and the \textit{global cross-efficiency}
\begin{equation}
{\mathcal{E}}^{\alpha\beta} = \left( \frac{1}{N_\alpha N_\beta} \sum_{p \in V_\alpha, q \in V_\beta} l_{pq}^{-1} \right)^{-1} 
\label{eq:globefficiency_cross}
\end{equation}
\noindent
Unlike ${\mathcal{C}}^{\alpha\beta}$ and ${\mathcal{T}}^{\alpha\beta}$, ${\mathcal{L}}^{\alpha\beta}$ and ${\mathcal{E}}^{\alpha\beta}$ are (as shortest path-based measures) symmetric by definition, i.e., ${\mathcal{L}}^{\alpha\beta}={\mathcal{L}}^{\beta\alpha}$ and ${\mathcal{E}}^{\alpha\beta}={\mathcal{E}}^{\beta\alpha}$. In the case of disconnected network components, the shortest path length $d_{ij}$ is defined as discussed for the corresponding local measures.

		In the same spirit as mentioned above, other single network characteristics~\cite{Boccaletti2006,Costa2007} can be adopted as well for defining further coupled network measures. This includes measures characterizing edges or, more generally, pairs of vertices like edge betweenness or matching index, further global network characteristics (assortativity, network diameter and radius), mesoscopic structures (motifs), or even characteristics associated with diffusion processes on the network instead of shortest paths (e.g., eigenvector centrality or random walk betweenness). The selection of measures explicitly mentioned above reflects those characteristics which have already been utilized in studying the interdependence structure between complex networks in different contexts \cite{Donges2011b,Wiedermann2013}.

%% file: Chapter03_RecurrenceNt/Chapter03_RecurrenceNt.tex
\section{Recurrence networks in phase space} \label{sec:RecurrenceNt}
In this section, we introduce and discuss recurrence networks (RNs) together with similar types of phase space based complex network representations of time series as an alternative framework for studying recurrences in phase space from a geometric point of view. We start with introducing the notion of a recurrence plot (RP) \cite{Eckmann1987,marwan2007}, which provides the fundamental framework for the visual and dynamical analysis of individual dynamical systems. Subsequently, different types of related network representations are introduced together with a detailed discussion of the resulting network characteristics of recurrence networks and their interpretation. In this context, we highlight the capability of these networks to unveil geometric characteristics associated with the structural organization of the underlying system in its phase space, which distinctively differs from other recurrence based techniques like recurrence quantification analysis (RQA) \cite{zbilut92,trulla96}, recurrence time statistics, or estimation of dynamical invariants from RPs. Finally, we close this section by introducing and discussing different bi- and multivariate generalizations of the recurrence network concepts and their respective interpretations, focusing on recent extensions of cross-recurrence plots \cite{marwan2002,Zbilut1998}, joint recurrence plots \cite{romano2004} and multiplex recurrence plots \cite{Eroglu2018}.

	\subsection{Theoretical background}
		\subsubsection{Phase space and attractor reconstruction} \label{sec:attractorReconstruct}
		We start with a (possibly multivariate) time series $\{x_i\}_{i=1}^N$ with $x_i=x(t_i)$, which we interpret as a finite representation of the trajectory of some (deterministic or stochastic) dynamical system. For a discrete system (map), the sampling of the time series is directly given by the map, whereas for a continuous-time system, the time series values correspond to a temporally discretized sampling of a finite part of one trajectory of the system determined by some initial conditions together with the corresponding sampling rate.

        In the case of observation functions not representing the full variety of dynamically relevant components, we additionally assume that attractor reconstruction has been performed \cite{Fraser1986,kantz1997,Kennel1992,Takens1981}. More specifically, when given a scalar time series $\{x_i\}$ ($i=1,\dots,N$), we first convert the data into state vectors in some appropriately reconstructed phase space. A common method from dynamical systems theory to define such a phase space is time-delay embedding~\cite{Takens1981}. In fact, the concept of a phase space representation rather than a ``simple'' time or frequency domain approach is the hallmark of many methods of nonlinear time series analysis, requiring embedding as the first step. Here, we define $\vec{x}_i = (x_i, x_{i-\tau}, \cdots, x_{i-(m-1)\tau})$ to obtain an $m$-dimensional time-delay embedding of $x_i$ with embedding delay $\tau$ for obtaining state vectors in phase space~\cite{Takens1981}. It has been proven that for deterministic dynamical systems, the thus reconstructed phase space is topologically equivalent to the original space if $m > 2 D_F$, where $D_F$ is the fractal dimension of the support of the invariant measure generated by the dynamics in the true (but often at most partially observed) state space. Note that $D_F$ can be much smaller than the dimension of the underlying original (physical) phase space spanned by all relevant system variables.

		From a practical perspective, when analyzing a scalar time series of whatever origin, neither embedding dimension $m$ nor delay $\tau$ are known a priori. The false nearest-neighbors (FNN) method~\cite{Kennel1992} was introduced to derive a reasonable guess of how to choose $m$ based on studying whether or not proximity relations between state vectors are lost when the embedding dimension is successively increased. If a reasonable embedding dimension is found, all dynamically relevant coordinates of the system are appropriately represented, so that all proximity relationships are correct and not due to lower-dimensional projection effects.

		In a similar spirit, a delay $\tau$ may be appropriate when the statistical dependence between the components of the embedding vector $\vec{x}$ approaches zero. For example, this can be achieved by choosing $\tau$ corresponding to the first root of the auto-correlation function (ACF) of a time series. This minimizes the linear correlation between the components, but does not necessarily mean they are independent. However, the converse is true: if two variables are statistically independent they will be uncorrelated. Therefore, another well established strategy for determining $\tau$ is to use the first minimum of the time-delayed mutual information \cite{Fraser1986}.

		The aforementioned approaches to determining $m$ and $\tau$ commonly work well for data from deterministic dynamical systems. It is an important issue when dealing with experimental time series and we have to first check if appropriate embeddings are applicable. Practically, we need to show the dependence of any analysis on the embedding explicitly. Let us first assume in the following sections that a proper embedding has been obtained, before discussing the possible effects of embedding parameters on the reconstructed recurrence networks in Section \ref{subsubsec:embedding}.

		\subsubsection{Recurrences and recurrence plots}
		Recurrence of states, in the meaning that states become again arbitrary close to previous ones after some time, is a fundamental property of deterministic dynamical systems and is typical for nonlinear or chaotic systems \cite{Ott1993,poincare1890}. From the set of (original or reconstructed) state vectors representing a discrete sampling of the underlying system's trajectory (e.g., the chaotic attractor of a dissipative system), recurrences can be visualized by recurrence plots (RP) \cite{marwan2007}, originally introduced by Eckmann {\textit{et al.}} \cite{Eckmann1987}. The RP is a graphical representation of the corresponding recurrence matrix $\textbf{R}(\varepsilon)$ that is defined in the standard way as
\begin{equation} \label{eq:rpDefinition}
R_{ij}(\varepsilon)=\Theta(\varepsilon-\|\vec{x}_i - \vec{x}_j\|),
\end{equation}
\noindent
where $\|\cdot\|$ can be any norm in phase space (e.g., Manhattan, Euclidean, or maximum norm). For convenience, we will use the maximum norm in all following examples unless stated otherwise. A RP enables us to investigate the recurrences of the $m$-dimensional phase space trajectory through a two-dimensional graphical representation $R_{ij}$ in terms of black and white dots indicating recurrent and non-recurrent pairs of vectors, respectively. The algorithmic parameter $\varepsilon$ is a pre-defined threshold value which determines whether two state vectors are close or not. The effects of $\varepsilon$ on the resulting RP and their statistics will be further discussed in Section \ref{subsub:epsilon}.

\begin{figure}
	\centering
	\includegraphics[width=\textwidth]{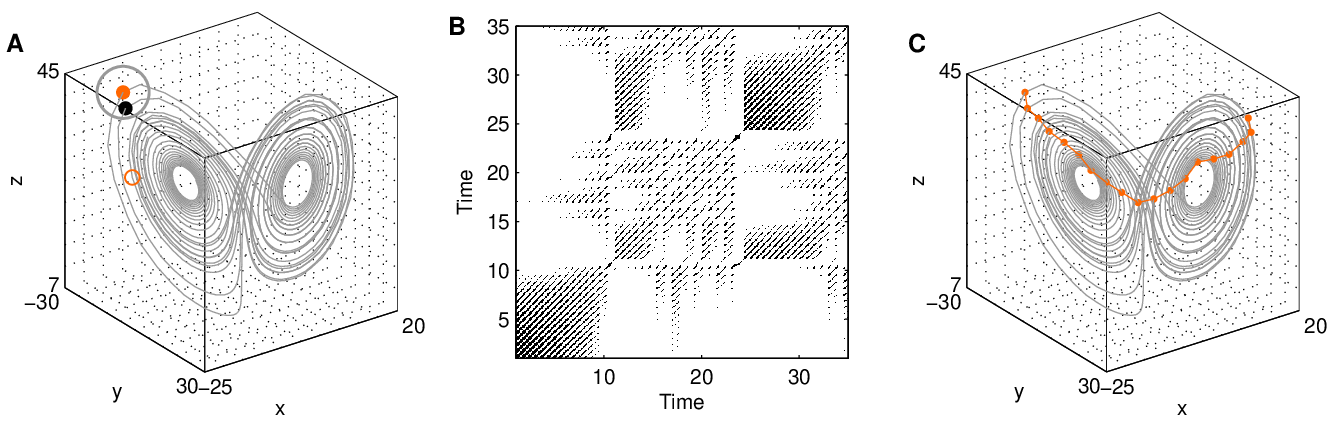}
\caption{Basic concepts beyond recurrence plots and the resulting recurrence networks (RNs), exemplified for one realization of the Lorenz system (Eq.~(\ref{eqlorenz})) using the same time series as in Fig. \ref{fig:lorenz_network}(e). (a) A state at time $i$ (red dot) is recurrent at another time $j$ (black dot) when the phase space trajectory visits its close neighborhood (gray circle). This is marked by value 1 in the recurrence matrix at $(i,j)$. States outside of this neighborhood (small open red circle) are marked with 0 in the recurrence matrix. (b) Graphical representation of the corresponding recurrence matrix (recurrence plot, Eq. \eqref{eq:rpDefinition}) and adjacency matrix of the RN (modulo main diagonal). (c) A particular path in the RN for the same system embedded in the corresponding phase space. Reproduced from \cite{Donner2011} with permission by World Scientific Publishing Co.. }
\label{lorenz_constr}
\end{figure}

The basic principle of a RP is illustrated in Fig.~\ref{lorenz_constr} for one realization of the Lorenz system (Eq. \eqref{eqlorenz}). Further more specific alternative definitions of recurrences add dynamical aspects, such as local rank orders or strictly parallel evolution of states (parallel segments of phase-space trajectory considered in iso-directional RPs~\cite{Horai2002}). For a more detailed overview, we refer to~\cite{marwan2007}.

RPs of dynamical systems with different types of dynamics exhibit distinct structural properties, which can be characterized in terms of their associated small-, medium- as well as large-scale features~\cite{marwan2007}. The study of recurrences by means of RPs has become popular with the introduction of recurrence quantification analysis (RQA)~\cite{zbilut92,marwan2002herz}. The initial purpose of this framework has been to introduce measures of complexity which distinguish between different appearances of RPs~\cite{marwan2008epjst}, since they are linked to certain dynamical properties of the studied system. RQA measures use the distribution of small-scale features in the RP, namely individual recurrence points as well as diagonal and vertical line structures. For instance, the recurrence rate (RR) simply counts the density of recurrence points of the matrix $\textbf{R}(\varepsilon)$ for a given threshold $\varepsilon$ (Eq.~\eqref{eq:rpDefinition}). RQA as a whole has been proven to constitute a very powerful technique for quantifying differences in the dynamics of complex systems and has meanwhile found numerous applications, e.g., in ecology \cite{facchini2007}, engineering \cite{litak2009c}, geo- and life sciences \cite{marwan2003climdyn,marwan2007pla}, or protein research \cite{Giuliani2002a,zbilut2004a}.  For a more comprehensive review on the potentials of this method, we refer to \cite{marwan2007,marwan2008epjst,webber2009}. In addition, we would like to remark that even dynamical invariants, like the $K_2$ entropy, mutual information, or fractal dimensions (i.e., the information and correlation dimensions $D_1$, $D_2$) can be efficiently estimated from RPs \cite{thiel2004a,marwan2007}. Moreover, RPs have also been successfully applied to study interrelations, couplings, and phase synchronization between dynamical systems~\cite{marwan2002,romano2004,romano2005,Romano2007,vanLeeuwen2009,Nawrath2010,marwan2013c}.

In order to highlight different domains of recurrences in the RPs, some algorithms have been proposed recently. For example, Pham {\textit{et al.}} introduced fuzzy recurrence plots, which determine an optimal relation between the observed states in phase space and a number of predefined clusters \cite{Pham2016}. This algorithm highlights the recurrence regions and, thus, provides possibly clearer visualization of the underlying recurrence structures. An alternative algorithm has been proposed in \cite{graben2013} to search for specific recurrence domains. Here, intersecting $\varepsilon$-balls around sampling points are merged into cells of a phase space partition, and a maximum entropy principle defines the optimal size of these intersecting balls. This data-adaptive algorithm for obtaining phase partitions has been found to perform better than techniques based on Markov chains, which require an ad hoc partition of the system's phase space. Along the same lines of research, another algorithm has been proposed recently in \cite{Costa2018} to capture the recurrence density structures in the RP.

	\subsection{Types of recurrence networks}
		\subsubsection{$\varepsilon$-recurrence networks}
\begin{figure}
	\centering
	\includegraphics[width=0.48\textwidth]{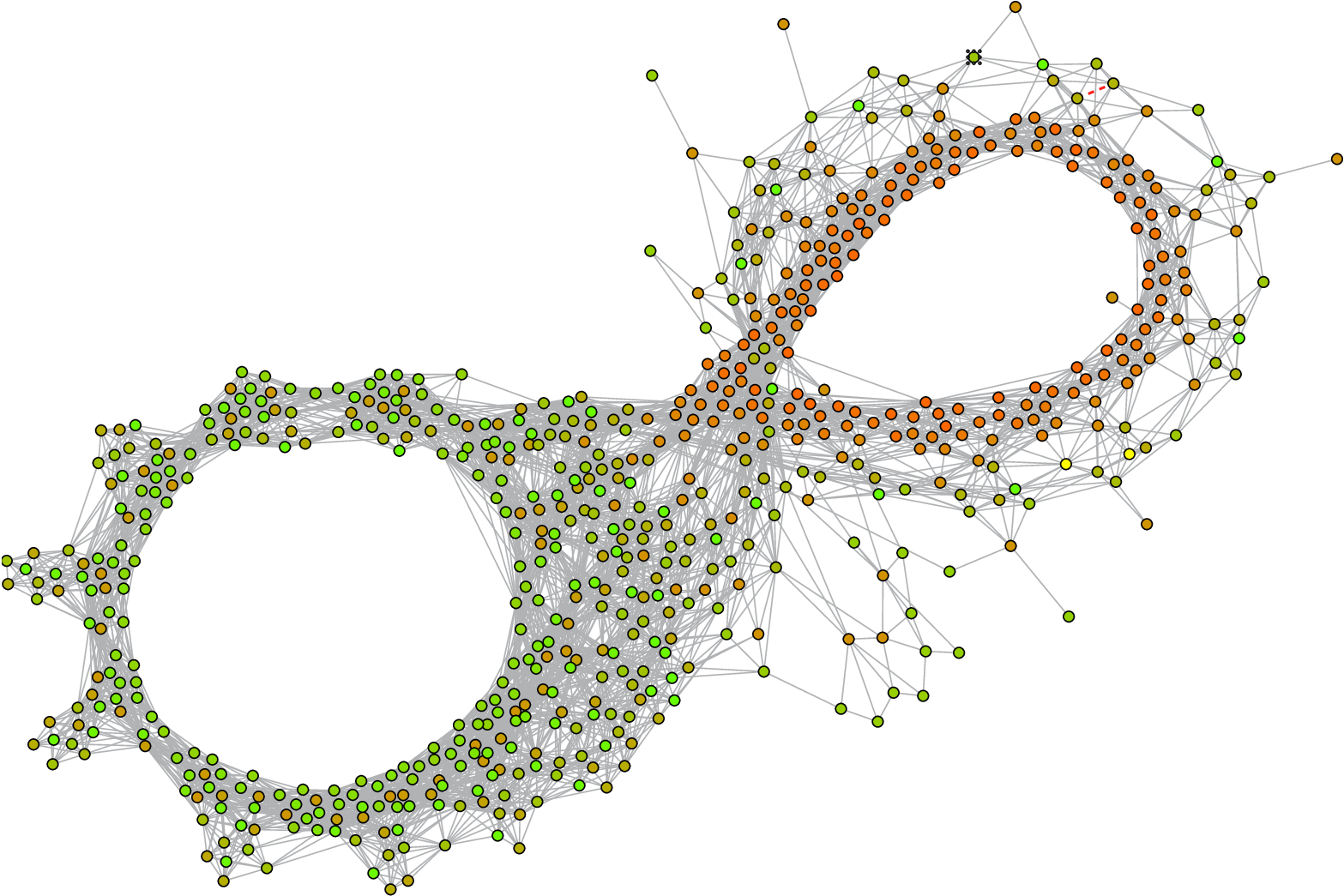}
\caption{A graphical representation of the Lorenz attractor based on the recurrence matrix represented in Fig.~\ref{lorenz_constr}. The color of the vertices corresponds to their temporal order (from orange to bright green). Reproduced from \cite{Donner2011} with permission by World Scientific Publishing Co.. }
\label{lorenz_net}
\end{figure}

		The crucial step of the recurrence network approach is to re-interpret the mathematical structure $\textbf{R}(\varepsilon)$ as the adjacency matrix $\mathbf{A}$ of some adjoint complex network embedded in phase space by setting
\begin{equation}
\mathbf{A}(\varepsilon)=\mathbf{R}(\varepsilon)-\mathbf{1}_N,
\label{eq:rn_definition}
\end{equation}
\noindent
where $\mathbf{1}_N$ is the $N$-dimensional identity matrix. The complex network defined this way is called \emph{$\varepsilon$-recurrence network (RN)}, as opposed to other types of proximity-based networks in phase space making use of different definitions of geometric closeness, e.g., considering $k$-nearest neighbors~\cite{Donner2011}, which will be further discussed below. Specifically, the sampled state vectors $\{x_i\}$ are interpreted as vertices of a complex network, which are connected by undirected edges if they are mutually close in phase space (i.e., describe recurrences). In the remainder of this review, we will adopt the time indices $i$, $j$, etc., of observations as vertex indices of the corresponding time series networks to highlight their equivalence whenever individual observations or state vectors coincide one-by-one with the vertices of the network representations (which is the case for recurrence networks, but also visibility graphs and related methods discussed later in this review). Notably, the binary matrix $\mathbf{A}(\varepsilon)$ retains the symmetry properties of $\mathbf{R}(\varepsilon)$, which implies that the RN is a \emph{simple graph}, i.e., a complex network without multiple edges or self-loops (note that $A_{ii}=0$ according to definition~(\ref{eq:rn_definition})). We show an example of such an unweighted  $\varepsilon$-RN network in Fig. \ref{lorenz_net}.

		One of the fundamental concepts of network theory is the notion of a {\textit{path}} (Section \ref{sec:basicCompNets}). In an $\varepsilon$-RN, a \emph{path} between two specified vertices $i$ and $j$ is an ordered sequence of edges starting at $i$ and ending at $j$, with its \emph{path length} $l_{ij}(\varepsilon)$ given by the number of edges in this sequence. An example of a path is shown in Fig. \ref{lorenz_constr}(c). In the context of RNs, we can thus understand a path as a sequence of mutually overlapping $\varepsilon$-balls $B_{\varepsilon}(x_i),B_{\varepsilon}(x_{k_1}),\dots,B_{\varepsilon}(x_{k_{l_{i,j}-1}}),B_{\varepsilon}(x_j)$, where $$B_{\varepsilon}(x)=\{y\,|\,\|x-y\|<\varepsilon\}$$ is an open set describing a volume with maximum distance $\varepsilon$ (measured in a given norm) from $x$, and $B_{\varepsilon}(x_i)\cap B_{\varepsilon}(x_{k_1})\neq\emptyset,\dots,B_{\varepsilon}(x_{k_{l_{i,j}-1}})\cap B_{\varepsilon}(x_j)\neq\emptyset$. Note that $\varepsilon$-balls refer to general (hyper-)volumes according to the specific norm chosen for measuring distances in phase space, e.g., hypercubes of edge length $2 \varepsilon$ in case of the maximum norm, or hyperballs of radius $\varepsilon$ for the Euclidean norm.

		Following these considerations, a \emph{shortest path} is a minimum sequence of edges (connecting vertices in mutually overlapping $\varepsilon$-balls) between two fixed vertices (state vectors) $i$ and $j$. Note that a shortest path does not need to be unique. In turn, due to the discrete character of a network, it is rather typical that there are multiple shortest paths between some specific pair of vertices. In what follows, the shortest path length will be denoted as ${l}_{ij}$, and the multiplicity of such shortest paths as ${\sigma}_{ij}$, following the corresponding general notation introduced in Section~\ref{sec:CompNetworkT}.

		We have to emphasize that the network-theoretic concept of a path on a given graph is distinctively different from the trajectory concept that records the dynamical evolution of a system \cite{Donner2010a}. Furthermore, RNs based on Eq.~(\ref{eq:rn_definition}) can be generalized by withdrawing the application of a specific threshold, which leads to weighted networks and unthresholded RPs (distance plots), respectively. For example, the unthresholded RP obtained from one trajectory of a given dynamical system may be re-interpreted as the connectivity matrix of a fully coupled, weighted network in phase space.

	 	Most of RN analysis has focused on the network representation using the adjacency matrix $\mathbf{A}(\varepsilon)$ (Eq.~\eqref{eq:rn_definition}) and the extraction of new network-theoretic measures, which will be reviewed below. We emphasize that $\mathbf{A}(\varepsilon)$ provides information on vertices and edges, but its graph structural layouts can take variable forms.  Network visualization is a non-trivial task and there are many tools for that purpose in computer science, e.g., spring-based layout systems, spectral layout method, tree layout algorithms, etc. These algorithms have been well integrated in many popular network visualization packages, i.e., \texttt{Mathematica}, \texttt{Gephi}, and \texttt{NetworkX}. The network visualization shown in Fig.~\ref{lorenz_net} has been created by the software package {\tt{GUESS}} using a force-directed placement algorithm. In \cite{Yang2013}, Yang {\textit{et al.}} proposed to use a spring-electrical model to explore the self-organized geometry of RNs. In this algorithm, they simulate the recurrence network as a physical system by treating the edges as springs and the nodes as electrically charged particles. Then, force-directed node placement algorithms are employed to automatically organize the network geometry by minimizing the system energy. It has been shown that this self-organized process recovers the attractor of a dynamical system, which provides important insights for attractor reconstruction from the adjacency matrix \cite{thiel2004b,hirata2008}.

		\subsubsection{$k$-nearest neighbor networks}
		Besides the recurrence definition based on a fixed distance threshold $\varepsilon$ in phase space (i.e., equal neighborhood volumes around all available state vectors), there are alternative ways for defining recurrences and, hence, RPs and RNs. For example, the original definition of a RP by Eckmann \textit{et~al.} \cite{Eckmann1987} makes use of $k$-nearest neighbors (i.e., a fixed probability mass of the considered neighborhoods). Re-interpreting the resulting recurrence matrix as the adjacency matrix of a complex network leads to a different type of RN \cite{Shimada2008}, typically referred to as \emph{$k$-nearest neighbor network}. Since in this definition, the neighborhood relation is not symmetric (i.e., $x_j$ being among the $k$ nearest neighbors of $x_i$ does not imply $x_i$ also being among the $k$ nearest neighbors of $x_j$), the resulting networks are in general directed graphs, and the local density of unidirectional edges (as opposed to bidirectional ones) is related to the gradient of the invariant density.

		\subsubsection{Adaptive neighbor networks} \label{sec:adaptiveRN}
		In order to circumvent the directedness of $k$-nearest neighbor networks, Xu \textit{et~al.} \cite{Small2009,Xu2008} proposed an algorithm for balancing the neighborhood relationships in such a way that they become symmetric again. The resulting networks embedded in phase space, sometimes also referred to as \emph{adaptive nearest neighbor networks} \cite{Donner2011}, are conceptually more similar to classical ($\varepsilon$-)RNs, but still exhibit somewhat different topological characteristics. In particular, this approach helps to understand the superfamily phenomena of time series, which concern the relative prevalence of motifs of the resulting networks. In particular, the motif distribution of adaptive nearest neighbor networks has been empirically shown to allow a discrimination between different types of dynamics in terms of a different motif ranking \cite{Xu2008}. Consequently, this approach has been mainly used for such discriminatory tasks, including applications to turbulence phenomena, instrumental music \cite{Donner2011}, fractional Brownian motions and multifractal random walks \cite{Liu2010a}.

		While these superfamily phenomena have been found in time series from various origins, no fundamental theories have been proposed so far in the literature to explain the corresponding empirical findings. Khol {\textit{et al.}} provide a heuristic explanation of superfamily phenomena by examining the dependence of the attractor dimension on motif prevalence \cite{Khor2016}. Since the constructed networks inherently capture the proximity of states, motifs represent specific arrangements of states in space, some of which are more or less likely to occur as dimension changes. Therefore, they found that the relative prevalence of motifs is strongly dependent on the local dimension of the space from which the state vectors are taken. Further evidence is given by identifying comparable superfamily phenomena in networks constructed from states randomly distributed in spaces of varying dimensions \cite{Khor2016}.

		\subsubsection{Algorithmic variants}
		A detailed discussion of the differences between $\varepsilon$-RNs, $k$-nearest neighbor and adaptive nearest neighbor networks introduced above has been recently provided in \cite{Donner2011}. While these three classes of time series networks exhibit very strong conceptual similarities (the same applies to the correlation networks \cite{Yang2008} discussed in Section~\ref{sec:correlationnetworks} if interpreting the correlation coefficient between two sufficiently high-dimensional state vectors as a generalized distance), the approach proposed by Li \textit{et~al.} \cite{Li2011a,Li2011b,Cao2014,Fan2015} can be understood as being derived from the RN idea. Here, for a set of $m$-dimensional embedding vectors, all mutual Euclidean distances are computed. Based on the maximum distance value $d_{max}(m)$, the threshold distance of a RN is taken as $\varepsilon(m)=d_{max}(m)/(N-1)$. This procedure is repeated for different $m$, and the critical value of the embedding dimension for which the resulting network gets completely disconnected is interpreted as a complexity index \cite{Cao2014}. However, it has not yet been demonstrated that this algorithmic approach has any conceptual benefits in comparison with the classical RN measures like transitivity (see Section~\ref{sec:transitivity}) obtained for a fixed embedding dimension. Moreover, we note that the maximum distance $d_{max}$ between embedding vectors may depend on the embedding dimension in some peculiar form, e.g., may be independent for the supremum norm while increasing monotonically for most other common norms \cite{Kraemer2018}. In this regard, it is commonly recommended to study fixed quantiles of the distance distribution function rather than fixed multiples of some location parameters (like mean, median or maximum) of the latter \cite{Kraemer2018}.

		Another conceptual approach loosely related to RNs provides the foundation of the frequency-degree mapping algorithm introduced by Li \textit{et~al.} \cite{Li2012}. Here, the resulting time series networks contain two types of edges: (i) temporal edges connecting subsequent points in time, and (ii) proximity edges containing observations of similar values, where similarity is defined by an initial grouping of the data into a discrete set of classes, and observed values being connected if and only if they belong to the same class. Note that this is conceptually related to the idea of coarse-graining based transition networks that will be discussed in Section~\ref{sec:cgtn}. Here the definition of a class is equivalent to a recurrence interval that is defined by amplitude quantization, for instance, the recurrence interval length $I = H/Q$ where $Q$ is the quantization level and $H$ is the amplitude range of the time series. Notably, the latter approach combines the classical recurrence idea and basic concepts of symbolic dynamics \cite{Daw2003} (see Section~\ref{sec:symbolic}). In this spirit, the resulting network's adjacency matrix is given as the recurrence matrix associated with a symbolic recurrence plot \cite{Donner2008,Faure2010,graben2013} plus a ``stripe'' around the matrix' main diagonal. The frequency-degree mapping algorithm has been successfully applied to characterizing signatures of various types of ventricular arrhytmias in human heart beat time series \cite{Li2012}, stock markets \cite{Cao2014}, and air quality indices \cite{Fan2015}.

		In order to highlight the recurrence domains in networks, the fuzzy recurrence network approach has been proposed in \cite{Pham2017}, which shares many similarities with fuzzy recurrence plots \cite{Pham2016}. Furthermore, a grammatical rewriting algorithm over the recurrence matrix has been proposed to search for recurrence domains in \cite{graben2013}, which presents a symbolic description of the recurrence properties of a time series. It is interesting to see that this algorithm yields an optimal symbolic recurrence representation revealing functional components of brain signals \cite{graben2013}. Note that the computation of the recurrence matrix is the first step of this grammatical algorithm.

		The computation time of a RN is proportional to $N^2$ where $N$ is the number of time points, which calls for more efficient algorithms for constructing RNs for long time series. In this case, on the other hand, we are more interested in the evolution of the RN over time. To this end, sliding window techniques are often suggested but require checking the dependence of the corresponding results on the window size \cite{Donges2011a,Donges2011}. Another idea is to perform coarse graining of the original RNs \cite{Costa2018}, which originates from the idea of meta-recurrence plots \cite{casdagli97}. In \cite{Iwayama2012}, the authors proposed to first divide the original long time series into short segments and RNs are then constructed for each piece. The next step is to build joint recurrence networks for each pair of windowed segments. Then, the long-term dynamics is characterized by the variations of the network properties computed for these meta-time series.

		\subsection{Complex network characteristics of RN} \label{sec:rn_measures}
		Based on the re-interpretation of the recurrence matrix $\mathbf{R}(\varepsilon)$ as the adjacency matrix of an adjoint RN, we can utilize the large toolbox of complex network measures (see Section~\ref{sec:basictheoryCN})~\cite{Albert2002,Boccaletti2006,Costa2007,Newman2003} for characterizing the structural organization of a dynamical system in its phase space. Notably, this viewpoint is complementary to other concepts of nonlinear time series analysis making use of RPs. For example, RQA characterizes the statistical properties of line structures in the RP, which implies an explicit consideration of dynamical aspects (i.e., sequences of state vectors) \cite{marwan2007}. In turn, RNs do not make use of time information, since network properties are generally invariant under vertex relabelling (i.e., permutations of the order of observations) \cite{Donner2010a}. In this spirit, RN analysis provides geometric instead of dynamical characteristics. This idea of a geometric analysis is similar to some classical concepts of fractal dimensions (e.g., box-counting or correlation dimensions), where certain scaling laws in dependence on the spatial scale of resolution (corresponding here to $\varepsilon$) are exploited. In turn, RN analysis can be performed (as RQA) using only a single fixed scale ($\varepsilon$) instead of explicitly studying scaling properties over a range of threshold values. We will further elaborate on this idea in Section~\ref{sec:transitivity}.

		The distinction between dynamical and geometric information implies that in case of RN analysis, the typical requirement of a reasonable (typically uniform) temporal sampling of the considered trajectory is replaced by the demand for a suitable spatial coverage of the system's attractor in phase space. Specifically, under certain conditions the latter could also be obtained by considering an ensemble of short trajectories instead of a single long one. If the trajectory under study is relatively densely sampled, trivial serial correlations can lead to a selection bias in the set of sampled state vectors; the latter could be avoided by reasonable downsampling. In the same context, the possibility of utilizing Theiler windows for removing edges representing short-term auto-correlations (e.g., recurrence points close to the main diagonal in the RP) should be mentioned as another strategy based on a somewhat different rationale~\cite{Donner2010a}. However, from a conceptual perspective, downsampling can provide an unbiased sampling of the attractor as long as the fixed sampling time does not correspond to any integer multiple of some of the system's natural frequencies. As an alternative, bootstrapping from the set of available state vectors provides another feasible option, which should be preferred if a sufficiently long time series is available. In general, numerical experiments and different applications suggest that stable estimates of RN characteristics can often already be obtained using a sample size of $N\sim\mathcal{O}(10^2\dots 10^3)$ data points~\cite{Donges2011,Donges2011a}.

		In Section~\ref{sec:basictheoryCN}, we have provided a general review on various network measures characterizing the structural properties of a complex network as denoted by the adjacency matrix $\mathbf{A}$. In this section, we further discuss the physical interpretations of these measures in terms of phase space properties as captured by RN representations. In what follows, we will denote some basic properties computed from a RN consisting of a finite number $N$ of state vectors as $\hat{f}$, pointing to the fact that they are estimated from a given sample of state vectors but shall characterize the entire trajectory of the system under study. In other words, we have specific finite sample estimates for Eqs.~\eqref{eq:degree}-\eqref{eq:apl}. Furthermore, we will discuss a corresponding continuous framework describing all network characteristics described below in terms of some fully analytical theory in Section \ref{sec:analyticRNtheory}. In order to focus the following discussion, we review only the possibly most relevant characteristics associated with RNs. More details including further measures can be found in \cite{Donges2012,Donner2010a}.

		When considering the quantitative characteristics of complex networks (see Section~\ref{sec:basictheoryCN}), different classifications of measures are possible \cite{Donner2010a,Donges2012}. First, we may distinguish measures based on the concept of graph neighborhoods from those making use of shortest path-based characteristics. This is not an exhaustive classification, since it potentially neglects other important network measures, e.g., such based on diffusion processes or random walks on the network. Second, network measures can be classified into such making use of local, meso-scale and global information. This scheme is widely equivalent to the first one in that local information refers to properties determined by the graph neighborhood of a given vertex, whereas global information takes contributions due to all vertices of the network into account, which is common for shortest path-based measures. Finally, we can differentiate between measures quantifying properties of single vertices, pairs of vertices, and the network as a whole. In the following, we will utilize the latter way of classification, since it appears most instructive from the applied point of view (i.e., we are commonly interested in either the local or the global geometry of an attractor or, more generally, some trajectory in phase space). Furthermore, we emphasize that in practice, the phase space properties captured by the estimates of RN characteristics are obtained for a particular value of $\varepsilon$. The effects of varying $\varepsilon$ on the resulting network statistics will be further discussed in Section \ref{subsub:epsilon}.

        In the following, we will adopt the notation of indexing each vertex of a RN with the time index $i$ (respectively, $j$, $k$, etc.) of the corresponding state vector $\vec{x}_i$ (etc.) instead of using the general vertex indices $p$, $q$, $r$, $s$, etc.\, used in Section~\ref{sec:basictheoryCN}. From here onward, we will adopt this notation whenever vertices can be uniquely identified with a point in time (respectively, the corresponding state or reconstructed state vector), which will also apply to the visibility graph based methods to be discussed in Section~\ref{sec:VisibilityGt}, but typically not to the transition networks reviewed in Section~\ref{sec:TransitionNt}. Moreover, some other types of proximity networks like cycle networks and correlation networks, which will be introduced in Section~\ref{sec:otherproxnets}, do not allow a unique identification of individual states (state vectors) with specific nodes of the networks, so that we will also adopt the more general index notation there.

			\subsubsection{Vertex characteristics}
\begin{figure}
	\centering
	\includegraphics[width=\columnwidth]{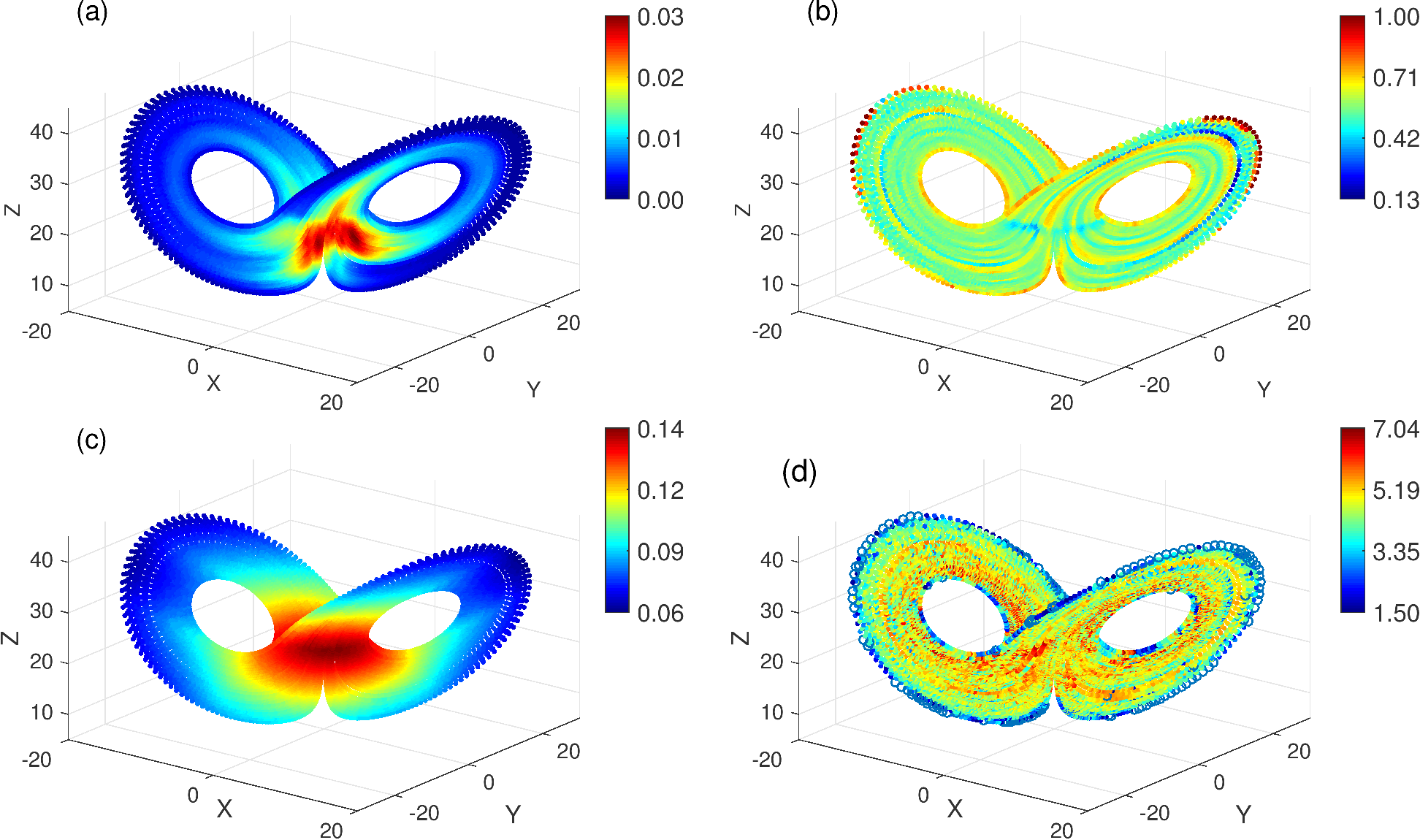}
\caption{Spatial distributions of vertex characteristics of the $\varepsilon$-RN obtained for the Lorenz system (Eq.~\ref{eqlorenz}) at the canonical parameters (using the maximum norm with $\hat{\rho}=0.01$, $N=20,000$ and sampling time $\Delta t=0.1$): (a) $\hat{k}_i(\varepsilon)$, (b) $\hat{\mathcal{C}}_i(\varepsilon)$, (c) $\hat{c}_i(\varepsilon)$, (d) $\hat{b}_i(\varepsilon)$. Modified from \cite{Donner2010a}. } \label{fig:local}
\end{figure}

        			From the perspective of recurrences, it is reasonable to replace the degree $\hat{k}_i(\varepsilon)$ (Eq.~\ref{eq:degree}) of a vertex by a normalized characteristic, the \textit{degree density} $\hat{\rho}_i(\varepsilon) =\frac{\hat{k}_i(\varepsilon)}{N-1}$, which corresponds to the definition of the local recurrence rate of the state $\vec{x}_i$. This means that $\hat{\rho}_i(\varepsilon)$ quantifies the density of states in the $\varepsilon$-ball around $\vec{x}_i$, i.e., the probability that a randomly chosen member of the available sample of state vectors is $\varepsilon$-close to $x_v$. An illustration of this fact for the Lorenz system is presented in Fig.~\ref{fig:local}(a); here, phase space regions with a high density of points (i.e., a high residence probability of the sampled trajectory) are characterized by a high degree density.

			The \textit{local clustering coefficient} (Eq.~\ref{eq:locclustering}) $\hat{\mathcal{C}}_i(\varepsilon)$ measures the fraction of pairs of vertices in the $\varepsilon$-ball around $\vec{x}_i$ that are mutually $\varepsilon$-close. For vertices with $\hat{k}_i(\varepsilon)<2$, we define $\hat{\mathcal{C}}_i(\varepsilon)=0$. It has been shown that the local clustering coefficient in a RN is associated with the geometric alignment of state vectors. Specifically, close to dynamically invariant objects such as unstable periodic orbits (UPOs) of low period, the dynamics of the system is effectively lower-dimensional, which results in a locally enhanced fraction of closed paths of length 3 (``triangles'') and, thus, a higher local clustering coefficient. The latter behavior is exemplified in Fig.~\ref{fig:local}(b) for the Lorenz system, where we recognize certain bands with higher values of $\hat{\mathcal{C}}_i$ corresponding to the positions of known low-periodic UPOs~\cite{Donner2010a}. The deeper reasons for this behavior will be further addressed in Section~\ref{sec:transitivity}.

			For estimating the \textit{closeness} (Eq.~\ref{eq:closeness}) $\hat{c}_i(\varepsilon)$ and \textit{local efficiency} (Eq.~\ref{eq:locefficiency}) $\hat{e}_i(\varepsilon)$, we again set $\hat{d}_{ij}$ to the highest possible value of $N-1$ for pairs of vertices that cannot be mutually reached. Both measures exhibit the highest values for vertices which are situated in the center of the RN (see Fig.~\ref{fig:local}(c) for an illustration for the Lorenz system).

			The \textit{betweenness} (Eq.~\ref{eq:betweenness}) $\hat{b}_i(\varepsilon)$ of nodes in a RN can be interpreted as indicating the local degree of fragmentation of the underlying attractor~\cite{Donner2010a}. To see this, consider two densely populated regions of phase space that are separated by a poorly populated one. Vertices in the latter will ``bundle'' the shortest paths between vertices in the former ones, thus forming geometric bottlenecks in the RN. In this spirit, we may understand the spatial distribution of betweenness centrality for the Lorenz system (Fig.~\ref{fig:local}(d)) which includes certain bands with higher and lower residence probability reflected in lower and higher betweenness values.

  		\subsubsection{Edge characteristics}
		The \textit{matching index} (Eq.~\ref{eq:matching}) $\hat{m}_{ij}(\varepsilon)$ quantifies the overlap of the network neighborhoods of two vertices $i$ and $j$. From the above definition, it follows that $\hat{m}_{ij}(\varepsilon)=0$ if $\|\vec{x}_i-\vec{x}_j\|\geq 2\varepsilon$. In turn, there can be mutually unconnected vertices $i$ and $j$ ($A_{ij}=0$) with $\varepsilon\leq \|\vec{x}_i-\vec{x}_j\|< 2\varepsilon$ that have some common neighbors and, thus, non-zero matching index. In the context of recurrences in phase space, $\hat{m}_{ij}(\varepsilon)=1$ implies that the states $\vec{x}_i$ and $\vec{x}_j$ are twins, i.e., share the same neighborhood in phase space \cite{thiel2006b}. In this spirit, we interpret $\hat{m}_{ij}(\varepsilon)$ measures the degree of twinness of two state vectors. Note that twins have important applications in creating surrogates for testing for the presence of complex synchronization~\cite{thiel2006b,Romano2009}.

		As the node betweenness of a RN, the associated \textit{edge betweenness} (Eq.~\ref{eq:edgebetweenness}) $\hat{b}_{ij}(\varepsilon)$ characterizes the local fragmentation of the studied dynamical system in its phase space.

		For the specific case of $\varepsilon$-RNs, we emphasize that there is no simple correspondence between matching index and edge betweenness, since both quantify distinctively different aspects of phase space geometry. Specifically, there are more pairs of vertices with non-zero matching index than edges, even though there are also pairs of vertices with $\hat{b}_{ij}(\varepsilon)>0$ but $\hat{m}_{ij}(\varepsilon)=0$ (i.e., there is an edge between $i$ and $j$, but both have no common neighbors). However, for those pairs of vertices for which both characteristics are non-zero, we find a clear anti-correlation \cite{Donner2010a}. One interpretation of this finding is that a large matching index typically corresponds to very close states in phase space; such pairs of RN vertices can in turn be easily exchanged as members of shortest paths on the network, which implies lower edge betweenness values. A similar argument may explain the coincidence between high edge betweenness and low non-zero matching index values.

		\subsubsection{Global network characteristics} \label{subsubsec:RNmeasureG}
		The \textit{edge density} (Eq.~\ref{eq:edgedensity}) $\hat{\rho}(\varepsilon)$. Notably, for a RN the edge density equals the recurrence rate $RR(\varepsilon)$ of the underlying RP. Strictly speaking, this is only true if the recurrence rate is defined such as the main diagonal in the RP is excluded in the same way as potential self-loops from the RN's adjacency matrix. It is trivial to show that $\hat{\rho}(\varepsilon)$ is a monotonically increasing function of the recurrence threshold $\varepsilon$: the larger the threshold, the more neighbors can be found, and the higher the edge density.

		The arithmetic mean of the local clustering coefficients $\hat{\mathcal{C}}_i(\varepsilon)$ of all vertices $i$ (Eq.~\ref{eq:globclustering}) defines the \emph{global clustering coefficient} $\hat{\mathcal{C}}(\varepsilon)$ in the usual way (see Section~\ref{sec:basictheoryCN}). Given our interpretation of the local clustering coefficient in a RN, $\hat{\mathcal{C}}(\varepsilon)$ can be interpreted as a proxy for the average local dimensionality of the dynamical system in phase space. Analogously, the \textit{network transitivity} (Eq.~\ref{eq:transitivity}) $\hat{\mathcal{T}}(\varepsilon)$ characterizes the effective global dimensionality of the system. In Section~\ref{sec:transitivity}, we will further expand this discussion by introducing the corresponding concepts of clustering and transitivity dimensions.

		The average path length (Eq.~\ref{eq:apl}) $\hat{\mathcal{L}}$ exhibits an inverse relationship with the recurrence threshold, since it approximates (constant) distances in phase space in units of $\varepsilon$~\cite{Donner2010a}. More specifically, the average phase space separation of states $\left<d_{ij}\right>$ serves as an $\varepsilon$-lower bound to $\hat{\mathcal{L}}$, namely,
\begin{equation}
\left<d_{ij}\right> \leq \varepsilon \hat{\mathcal{L}}. \label{avgL_eq}
\end{equation}
Interpreted geometrically, this inequality holds because $\hat{\mathcal{L}}$ approximates the average distance of states along geodesics on the RN (which can be considered as the geometric backbone of the attractor) in multiples of $\varepsilon$, while $\left<d_{ij}\right>$ gives the mean distance of states in $\mathbb{R}^m(\varepsilon)$ as measured by the norm $\|\cdot\|$.

	\subsection{Analytical theory of RN} \label{sec:analyticRNtheory}

	As we will demonstrate in the following, the properties of RNs can be also described analytically supporting their better understanding and, hence, applicability. For this purpose, we can exploit the formal equivalence of RNs and random geometric graphs (RGGs), a well-studied concept in graph theory and computational geometry. In this section, we motivate this equivalence and demonstrate how the variety of RN characteristics can be reformulated in the continuum limit $N\to\infty$ for any finite $\varepsilon$ \cite{Donges2012}. This framework allows gaining deep insights into the geometric organization of chaotic attractors by exploring the multitude of characteristics provided by complex network theory. Moreover, these analytical considerations will be extended to inter-system recurrence networks in Section \ref{sec:IntSRN}.

		\subsubsection{Preliminaries: random geometric graphs}
		Random geometric graphs \cite{Penrose2003} are based on a (finite) set of vertices randomly positioned in a $d$-dimensional ($d\in\mathbb{N}^+$) metric space according to some probability density function $p(\vec{x})$. In general, the connectivity among this set of vertices is taken to be distance-dependent, i.e., for two vertices $i$ and $j$, the probability of being connected in the RGG has the form $P(A_{ij}=1)=f(\|\vec{x}_i-\vec{x}_j\|)$ with some predefined function $f$, which is monotonically decreasing. As a consequence, spatially close vertices are more likely to be connected than distant ones. A particularly well studied special case is $f(\delta)=\Theta(\varepsilon-\delta)$ ($\delta$ denoting here the distance between any two points in the considered metric space as in Eq.~\eqref{eq:rpDefinition}), often referred to as RGG (in the strict sense). Notably, the latter definition has fundamental real-world importance (e.g., in terms of ad-hoc communication networks or, more general, contact networks) and matches that of the adjacency matrix of a RN (Eq.~\ref{eq:rn_definition}) if we identify $p(\vec{x})$ with the invariant density of the phase space object under study (e.g., some attractor in case of a dissipative system), and take the space in which the RGG is embedded as that spanned by the respective dynamical variables of the system. In this respect, for all following considerations, it is sufficient to restrict our attention to the support of $p(\vec{x})$ (respectively its closure), which is described by some manifold $S=\overline{\mbox{supp}(p)}$ embedded in the considered metric space (e.g., the attractor manifold).

		From a practical perspective, the spatial coverage of $p(\vec{x})$ by the RGG's vertices can be strongly affected by the sampling, leading to a spatial clustering of vertices if the sampling frequency is close to an integer multiple of the chaotic attractor's characteristic frequency. In such a situation, it is advantageous to follow alternative sampling strategies for $p(\vec{x})$. Note that for ergodic systems, sampling from one long trajectory, ensembles of short independent realizations of the same system, or directly from the invariant density should lead to networks with the same properties at sufficiently large $N$. In practice, generating the RGG/RN representation based on bootstrapping from the ensemble of available state vectors is often to be preferred over a regular sampling of a given trajectory, as discussed in Section~\ref{sec:rn_measures}.

		As outlined above, the importance of RGGs for the considerations on RNs is that some of their properties (like the degree distribution~\cite{Herrmann2003} or transitivity~\cite{Dall2002}) have been intensively studied for the generic case of a hard distance threshold in $f$ and arbitrary probability density functions $p(\vec{x})$ for metric spaces of various integer dimensions. For example, Hermann \textit{et~al.} \cite{Herrmann2003} give a closed-form expression of the degree distribution for arbitrary $p(\vec{x})$, whereas Dall and Christensen~\cite{Dall2002} provide a deep discussion of the transitivity properties of RGGs. Notably, the latter aspect has become particularly relevant in the interpretation of RN properties (see Section~\ref{sec:transitivity}) as well as those of some of their multivariate generalizations, as will be further discussed in Section~\ref{sec:IntSRN}.

		\subsubsection{Analytical description of $\varepsilon$-recurrence networks}
		By making use of the fact that RNs are a specific type of RGGs, all relevant graph-theoretical measures for RNs can be seen as discrete approximations of more general and continuous geometric properties of a dynamical system's underlying attractor characterized by a set $S$ together with an associated invariant density $p(\vec{x})$, $\vec{x}\in S$. This point of view allows obtaining deeper insights into the geometric meaning of the network quantifiers introduced in Section~\ref{sec:rn_measures} and enables us to establish surprising connections to other fields, e.g., the close relationship of transitivity measures like the local clustering coefficient and transitivity to the local and global fractal dimension of the dynamical system's attractor, respectively~\cite{Donner2011b} (see Section~\ref{sec:transitivity}). In the following, we review a corresponding analytical framework for general spatially embedded networks which is specifically tailored for defining continuous variants of the common discrete complex network characteristics \cite{Donges2012}.

		\paragraph{General setting}
		Let $S$ be a compact and smooth manifold with a non-vanishing continuous probability density function $p:S\to(0,\infty)$ with $\int_S d\vec{x}\ p(\vec{x}) = 1$. For the purpose of the present work, we identify $S$ with the set of points defining the attractor of a (dissipative) dynamical system. In case of chaotic attractors in time-continuous systems, we obtain a closure of the open attractive set by considering its union with the set of (infinitely many) unstable periodic orbits embedded in the attractor.

		Continuous analogs of the discrete complex network characteristics introduced in Section~\ref{sec:rn_measures} should be approximated by taking the limit $N\to\infty$ and $\varepsilon\to 0$ (note that the latter limit may not be assessible in the case of fractal sets $S$, which we will not further consider in the following). Here, ``continuous'' refers to a network with uncountably many vertices and edges, which is determined by the \emph{adjacency function}
\begin{equation}
A(\vec{x},\vec{y})=\Theta(\varepsilon-\|\vec{x}-\vec{y}\|)-\delta(\vec{x}-\vec{y})
\end{equation}
\noindent
for all $\vec{x},\vec{y}\in S$, which is a continuous analog of the adjacency matrix (Eq.~\ref{eq:rn_definition}). In the latter expression, $\delta(\vec{x}-\vec{y})=1$ if $\vec{x}=\vec{y}$, and $0$ otherwise.

\begin{figure}
\centering
\includegraphics[width=.50\columnwidth]{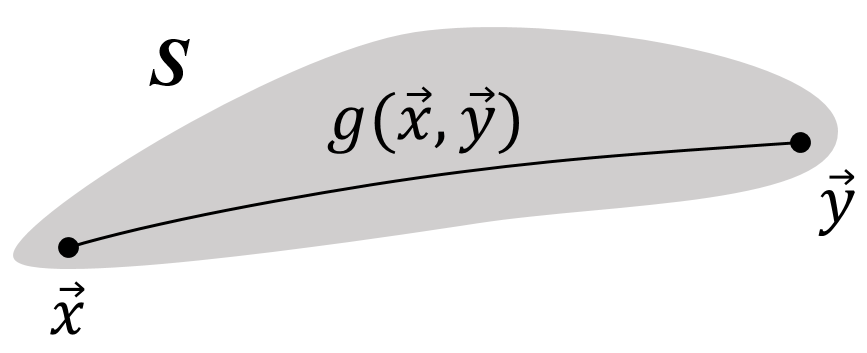}
\caption{Schematic illustration of a set $S$ (gray), where $g(\vec{x},\vec{y})$ denotes the geodesic distance between $\vec{x},\vec{y}\in S$ (after \cite{Donges2012}).}
\label{fig:manifold}
\end{figure}

		\paragraph{Shortest paths and geodesics}
		A large variety of complex network characteristics introduced in Section~\ref{sec:basictheoryCN} relies on the concept of shortest paths. Examples include closeness and betweenness centrality, local and global efficiency, and average path length. In the continuum limit, we consider a path in $S$ as a closed curve described by a properly parametrized function $\vec{f}:[0,1]\to S$, and define the associated path length
\begin{equation}
l(\vec{f}) = \sup_{n>0; \{t_i\}_{i=1}^n} \left.\left\{ \sum_{i=1}^n d(\vec{f}(t_{i-1}),\vec{f}(t_i)) \right| 0=t_0\leq t_1\leq\dots\leq t_n=1 \right\} \in[0,\infty]
\end{equation}
\noindent
where $d(\cdot)$ denotes some metric used for defining distances on $S$. The \textit{geodesic distance} between two points $\vec{x},\vec{y}\in S$, which serves as the analog of the shortest path length on a network, is then defined as (cf.~Fig.~\ref{fig:manifold}).
\begin{equation}
g(\vec{x},\vec{y}) = \inf_{\vec{f}} \left\{ l(\vec{f})\ |\ \vec{f}:[0,1]\to S,\ \vec{f}(0)=\vec{x},\ \vec{f}(1)=\vec{y} \right\}.
\end{equation}
\noindent
A path of length $g(\vec{x},\vec{y})$ is called a \emph{global geodesic} on $S$. Depending on the specific geometry of the considered set $S$, the multiplicity of global geodesics connecting $\vec{x}$ and $\vec{y}$ may differ, including no, one, or even infinitely many distinct global geodesics.

		Regarding a continuum limit for RNs, we note that shortest paths in such networks approximate global geodesics on the underlying invariant set $S$ in the limit of $\varepsilon\to 0$ and $N\to\infty$. Specifically, in the latter limit the shortest path length $l_{ij}(\varepsilon)$ between two points $\vec{x}_i, \vec{x}_j\in S$ behaves as
\begin{equation}
\varepsilon\ l_{ij}(\varepsilon) \to g(\vec{x}_i,\vec{x}_j)
\end{equation}
\noindent
independently of the chosen metric~\cite{Donges2012}.

		For defining betweenness centrality, we do not only require information on the lengths of global geodesics, but also their total multiplicity $\sigma(\vec{y},\vec{z};\varepsilon)$ as well as their multiplicity conditional on a third point $\vec{x}\in S$ being part of the curve, denoted as $\sigma(\vec{y},\vec{z}|\vec{x};\varepsilon)$ in the following. The definition of the latter quantity is, however, not unique for a given finite $\varepsilon$. Two possible, yet generally not equivalent expressions read~\cite{Donges2012}
\begin{eqnarray}
\sigma_1(\vec{y},\vec{z}|\vec{x};\varepsilon) &=& \sum_{k=1}^{\sigma(\vec{y},\vec{z};\varepsilon)} \int_0^1 dt\ \delta(\vec{f}_a(t)-\vec{x}) \label{eq:nrpaths_cont1} \\
\sigma_2(\vec{y},\vec{z}|\vec{x};\varepsilon) &=& \sum_{k=1}^{\sigma(\vec{y},\vec{z};\varepsilon)} \int_0^1 dt\ \Theta(\varepsilon-\|\vec{f}_a(t)-\vec{x}\|), \label{eq:nrpaths_cont2}
\end{eqnarray}
\noindent
where $\vec{f}_a(t)$ denotes the family of global geodesics between $\vec{y}$ and $\vec{z}$. Note that this family can have uncountably many members (to see this, consider, for example, the set of geodesics between the two poles on a sphere). In this case, the sum in Eqs.~(\ref{eq:nrpaths_cont1}) and (\ref{eq:nrpaths_cont2}) should be replaced by an integral. Furthermore, we emphasize that the $\varepsilon$-dependence in the multiplicities of shortest paths is implicit rather than explicit, since the chosen discretization level $\varepsilon$ can affect the effective ``shape'' of $S$ and, hence, the positions of possible edges in the considered space.

		\paragraph{Local (vertex-based) measures}
		The local vertex measures as introduced in Section \ref{sec:basictheoryCN} can be derived analytically \cite{Donges2012}. More specifically, the \emph{continuous $\varepsilon$-degree density}
\begin{equation} \label{eq:rhoAN}
\rho(\vec{x};\varepsilon)=\int_{B_{\varepsilon}(\vec{x})} d\mu(\vec{y})
\end{equation}
\noindent
gives the probability that a point $\vec{y}\in S$ randomly drawn according to $p$ falls into an $\varepsilon$-neighborhood $B_{\varepsilon}(\vec{x})=\{\vec{y}\in S\,|\, \|\vec{x}-\vec{y}\|<\varepsilon\}$ around $\vec{x}$. Its discrete estimator is given by the classical degree density $\hat{\rho}_i(\varepsilon)$. Here, we adopt the notion of an invariant measure, $d\mu(\vec{y})=p(\vec{y})d\vec{y}$ adjoint to the invariant density $p(\cdot)$ in order to shorten our notation.

		In order to quantify the density of closed paths of length 3 in the network, we consider the \emph{continuous local $\varepsilon$-clustering coefficient}
\begin{equation}
\mathcal{C}(\vec{x};\varepsilon)=\frac{\int\int_{B_{\varepsilon}(\vec{x})} d\mu(\vec{y})\ d\mu(\vec{z})\ \Theta(\varepsilon-\|\vec{y}-\vec{z}\|)}{\rho(\vec{x};\varepsilon)^2}.
\end{equation}
\noindent
This measure characterizes the probability that two points $\vec{y}$ and $\vec{z}$ randomly drawn according to $p$ from the $\varepsilon$-neighborhood of $\vec{x}\in S$ are mutually closer than $\varepsilon$. Its discrete approximation is provided by the classical local clustering coefficient $\hat{\mathcal{C}}_i(\varepsilon)$ (Eq.~\ref{eq:locclustering}).

		Let $\vec{y}\in S$ be drawn randomly according to $p$. For a fixed $\vec{x}\in S$, the \emph{continuous $\varepsilon$-closeness centrality}
\begin{equation}
c(\vec{x};\varepsilon)=\left(\int_S d\mu(\vec{y})\ \frac{g(\vec{x},\vec{y})}{\varepsilon}\right)^{-1}
\end{equation}
\noindent
and the \emph{continuous local $\varepsilon$-efficiency}
\begin{equation}
e(\vec{x};\varepsilon)=\int_S d\mu(\vec{y})\ \left(\frac{g(\vec{x},\vec{y})}{\varepsilon}\right)^{-1}
\end{equation}
\noindent
give the inverse expected geodesic distance and the expected inverse geodesic distance of $\vec{y}$ to some fixed $\vec{x}$, respectively. Hence, both measures quantify the geometric closeness of $\vec{x}$ to any other point in $S$ according to the probability density function $p$. By making use of RNs, they can be approximated by the classical closeness centrality $\hat{c}_i(\varepsilon)$ (Eq.~\ref{eq:closeness}) and local efficiency $\hat{e}_i(\varepsilon)$ (Eq.~\ref{eq:locefficiency}).

		Finally, the probability that a point $x$ lies on a randomly chosen global geodesic connecting two points $\vec{y},\vec{z}\in S$ according to $p$ is measured by the \emph{continuous $\varepsilon$-betweenness centrality}
\begin{equation}
b(\vec{x};\varepsilon)=\int\int_S d\mu(\vec{y})\ d\mu(\vec{z})\ \frac{\sigma(\vec{y},\vec{z}|\vec{x};\varepsilon)}{\sigma(\vec{y},\vec{z};\varepsilon)}.
\end{equation}
\noindent
Its discrete estimator is given by the standard RN betweenness centrality $\hat{b}_i(\varepsilon)$ (Eq.~\ref{eq:betweenness}) with the different possible expressions for $\sigma(\vec{y},\vec{z}|\vec{x};\varepsilon)$ (Eqs.~(\ref{eq:nrpaths_cont1},\ref{eq:nrpaths_cont2})) \cite{Donges2012}.

		\paragraph{Pairwise vertex and edge measures}
		The \emph{continuous $\varepsilon$-matching index}
\begin{equation}
m(\vec{x},\vec{y};\varepsilon)=\frac{\int_{B_{\varepsilon}(\vec{x})\cap B_{\varepsilon}(\vec{y})}d\mu(\vec{z})}{\int_{B_{\varepsilon}(\vec{x})\cup B_{\varepsilon}(\vec{y})}d\mu(\vec{z})}
\end{equation}
\noindent
quantifies the mutual overlap between the neighborhoods of two vertices $\vec{x},\vec{y}\in S$. In other words, $m(\vec{x},\vec{y};\varepsilon)$ is the probability that a point $\vec{z}\in S$ randomly chosen from $B_{\varepsilon}(\vec{x})$ according to $p$ is also contained in $B_{\varepsilon}(\vec{y})$ and vice versa. For $\vec{x}\to \vec{y}$, we have $B_{\varepsilon}(\vec{x})\to B_{\varepsilon}(\vec{y})$ and, consequently, $m(\vec{x},\vec{y};\varepsilon)\to 1$, whereas $m(\vec{x},\vec{y};\varepsilon)=0$ if $\|\vec{x}-\vec{y}\|>2\varepsilon$. As in the case of the other measures described above, $m(\vec{x},\vec{y};\varepsilon)$ can be approximated by the discrete RN matching index $\hat{m}_{ij}(\varepsilon)$ (Eq.~\ref{eq:matching}).

		Note again that $m(\vec{x},\vec{y};\varepsilon)$ does not require mutual $\varepsilon$-closeness between $\vec{x}$ and $\vec{y}$ (i.e., $\|\vec{x}-\vec{y}\|\in(\varepsilon,2\varepsilon)$ is possible). In contrast, the \emph{continuous $\varepsilon$-edge betweenness}
\begin{equation}
b(\vec{x},\vec{y};\varepsilon)=\int\int_S d\mu(\vec{z})\ d\mu(\vec{z}')\ \frac{\sigma(\vec{z},\vec{z}'|\vec{x},\vec{y};\varepsilon)}{\sigma(\vec{z},\vec{z}';\varepsilon)}
\end{equation}
\noindent
(with $\sigma(\vec{z},\vec{z}'|\vec{x},\vec{y};\varepsilon)$ denoting the number of global geodesics between $\vec{z}$ and $\vec{z}'$ containing both $\vec{x}$ and $\vec{y}$ under the condition $\|\vec{x}-\vec{y}\|\leq\varepsilon$, and $\sigma(\vec{z},\vec{z}';\varepsilon)$ the total number of global geodesics between $\vec{z}$ and $\vec{z}'$) is a measure whose discrete estimator $\hat{b}_{ij}(\varepsilon)$ (Eq.~\ref{eq:edgebetweenness}) is related to the presence of an edge between $\vec{x}_i$ and $\vec{x}_j$, i.e., $\|\vec{x}_i-\vec{x}_j\|< \varepsilon$. However, although this property has been originally introduced as an explicit edge property, it can be understood in a more general way as a two-vertex property such that $b(\vec{x},\vec{y};\varepsilon)$ measures the probability that two specific (not necessarily $\varepsilon$-close) points $\vec{x}$ and $\vec{y}$ both lie on a $p$-randomly drawn global geodesic connecting two points $\vec{z},\vec{z}'\in S$ \emph{and} are mutually closer than $\varepsilon$. Further generalizations towards $n$-point relationships are possible, but not instructive within the scope of this work.

		\paragraph{Global network measures}
		The \emph{continuous $\varepsilon$-edge density}
\begin{equation}
\rho(\varepsilon)=\int_S d\mu(\vec{x})\ \rho(\vec{x};\varepsilon)
\end{equation}
\noindent
is the $p$-expectation value of the continuous $\varepsilon$-degree density and approximated by the discrete edge density $\hat{\rho}(\varepsilon)$ of a RN (Eq.~\eqref{eq:edgedensity}).

		In the same spirit, the \emph{continuous global $\varepsilon$-clustering coefficient}
\begin{equation}
\mathcal{C}(\varepsilon)=\int_S d\mu(\vec{x})\ \mathcal{C}(\vec{x};\varepsilon)
\end{equation}
\noindent
is the $p$-expectation value of the continuous local $\varepsilon$-clustering coefficient. Its associated discrete estimator is the classical global (Watts-Strogatz) clustering coefficient $\hat{\mathcal{C}}(\varepsilon)$ (Eq.~\ref{eq:globclustering}). As an alternative measure characterizing geometric transitivity, we define the \emph{continuous $\varepsilon$-transitivity}
\begin{equation}
\mathcal{T}(\varepsilon)=\frac{\int\int\int_S d\mu(\vec{x})\, d\mu(\vec{y})\, d\mu(\vec{z})\, \Theta(\varepsilon-\|\vec{x}-\vec{y}\|)\, \Theta(\varepsilon-\|\vec{y}-\vec{z}\|)\, \Theta(\varepsilon-\|\vec{z}-\vec{x}\|)}{\int\int\int_S d\mu(\vec{x})\, d\mu(\vec{y})\, d\mu(\vec{z})\, \Theta(\varepsilon-\|\vec{x}-\vec{y}\|)\, \Theta(\varepsilon-\|\vec{z}-\vec{x}\|)},
\end{equation}
\noindent
which gives the probability that among three points $\vec{x},\vec{y},\vec{z}\in S$ randomly drawn according to $p$, $\vec{y}$ and $\vec{z}$ are mutually closer than $\varepsilon$ given they are both closer than $\varepsilon$ to $\vec{x}$. The corresponding discrete estimator is the RN transitivity $\hat{\mathcal{T}}(\varepsilon)$ (Eq.~\ref{eq:transitivity}).

		As examples of shortest path-based characteristics, we define the \emph{continuous $\varepsilon$-average path length}
\begin{equation}
\mathcal{L}(\varepsilon)=\int\int_S d\mu(\vec{x})\ d\mu(\vec{y})\ \frac{g(\vec{x},\vec{y})}{\varepsilon}
\end{equation}
\noindent
and the \emph{continous global $\varepsilon$-efficiency}
\begin{equation}
\mathcal{E}(\varepsilon)=\left(\int\int_S d\mu(\vec{x})\ d\mu(\vec{y})\ \left(\frac{g(\vec{x},\vec{y})}{\varepsilon}\right)^{-1}\right)^{-1},
\end{equation}
\noindent
which measure the expected geodesic distance and the inverse of the expected inverse geodesic distance, respectively, both measured in units of $\varepsilon$ between two points $x,y\in S$ drawn randomly according to $p$. Their discrete estimators are given by the classical RN average path length $\hat{\mathcal{L}}(\varepsilon)$ (Eq.~\ref{eq:apl}) and global efficiency $\hat{\mathcal{E}}(\varepsilon)$ (Eq.~\ref{eq:globefficiency}), respectively. Notably, we can reformulate $\mathcal{L}(\varepsilon)$ as the $p$-expectation value of the inverse continuous $\varepsilon$-closeness centrality,
\begin{equation}
\mathcal{L}(\varepsilon)=\int_S d\mu(\vec{x})\ c(\vec{x};\varepsilon)^{-1},
\end{equation}
\noindent
and $\mathcal{E}(\varepsilon)$ as the inverse $p$-expectation value of the continuous local $\varepsilon$-efficiency
\begin{equation} \label{eq:efficiencyAN}
\mathcal{E}(\varepsilon)=\left(\int_S d\mu(\vec{x})\ e(\vec{x};\varepsilon)\right)^{-1}.
\end{equation}

		\paragraph{Further characteristics}
		The selection of measures discussed above is far from being complete. Continuous versions of further complex network characteristics, such as assortativity, network diameter and radius, as well as network motifs are discussed in \cite{Donges2012}, where also some outlook on corresponding generalizations of other measures like eigenvector centrality or random walk betweenness have been given. Here, we restrict ourselves to the measures discussed above (Eqs.~(\ref{eq:rhoAN})-(\ref{eq:efficiencyAN})), since they have been most commonly used in recent applications of the RN framework.

	\subsection{General properties of recurrence networks}
		With the general RN framework (Section~\ref{sec:rn_measures}) and the associated analytical treatment of RNs (Section~\ref{sec:analyticRNtheory}) in mind, it is possible to study the properties of RNs as well as their multivariate generalizations from a solid theoretical basis. In the following, we will first discuss some general aspects of complex networks often found in real-world systems, such as small-world effects, the emergence of scale-free degree distributions, or assortative mixing (i.e., the tendency of vertices to connect with other vertices that exhibit a similar degree), regarding their presence or absence in RNs. Subsequently, we will turn to the transitivity characteristics of RNs, motivating their particular usefulness for detecting geometric signatures of qualitative changes in the dynamics of a single system.

       		\subsubsection{Degree distributions of RNs} \label{sec:scaling}
		A general analytical expression for the degree distribution $p(k)$ of a RGG has been given by Herrmann~\textit{et~al.}~\cite{Herrmann2003}. For this purpose, let us make the following assumptions: (i) The system under study is ergodic. (ii) The sampled trajectory is sufficiently close to its attractor, i.e., we exclude the presence of transient behavior. (iii) The sampling interval is co-prime to any possible periods of the system. If these three conditions are met, the vertices of the RN can be considered as being randomly sampled from the probability density function $p(\vec{x})$ associated with the invariant measure $\mu$ of the attractor~\cite{Eckmann1985}.

		For a RGG with an arbitrary $p(\vec{x})$, the degree distribution $p(k)$ can be derived from $p(\vec{x})$ in the limit of large sample size $N$ as
\begin{equation}\label{herrmann}
  p(k) = \int d\vec{x}\,p(\vec{x}) e^{-\alpha p(\vec{x})}(\alpha p(\vec{x}))^k/k!
\end{equation}
(representing an $n$-dimensional integral in case of an $n$-dimensional system) with $\alpha = \left<k\right> / \int d\vec{x}\,p(\vec{x})^2$~\cite{Herrmann2003}. In order to understand this relationship, note that for each $\vec{x}$, the probability that a sampled point falls into the $\varepsilon$-ball centered at $\vec{x}$ is approximately proportional to $p(\vec{x})$. Hence, the degree of a node at $\vec{x}$ has a binomial distribution. For sufficiently large $N$, the latter can be approximated by a Poissonian distribution with the parameter $\alpha p(\vec{x})$, leading to Eq.~(\ref{herrmann}).

\begin{figure}
	\centering
	\includegraphics[width=\columnwidth]{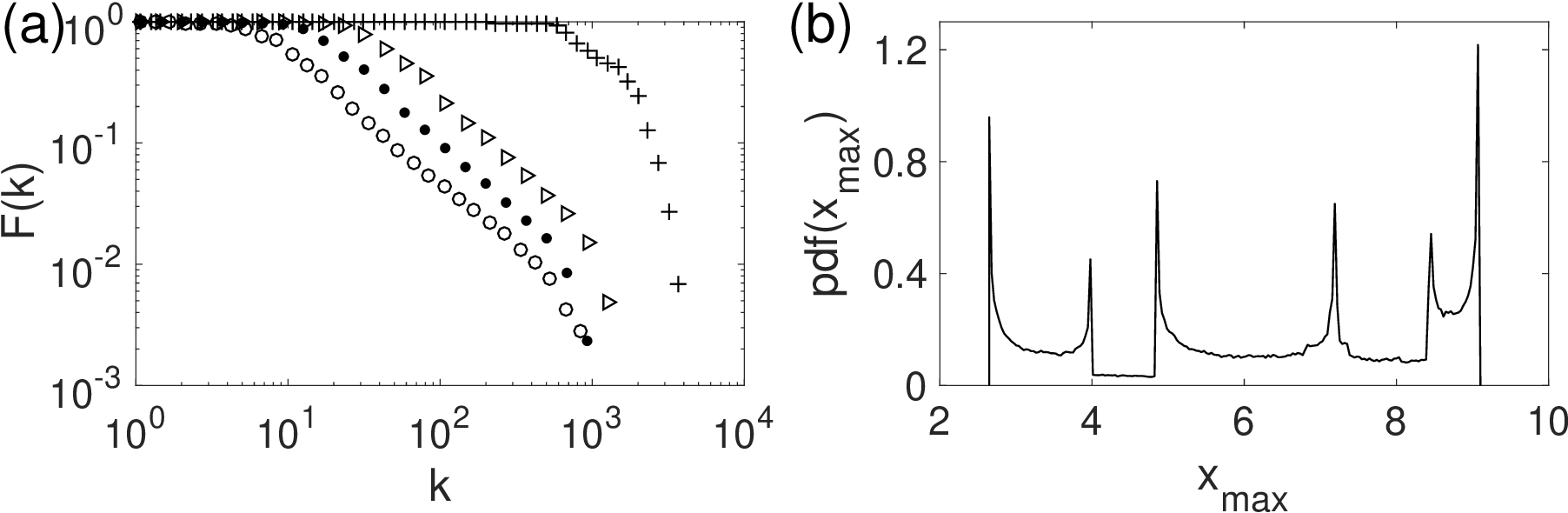}
	\caption{(a) Complementary cumulative distribution function $F(k)=\sum_{k'=k}^{\infty} p(k')$ for RNs obtained from the $x$-component of the first return map of the R\"ossler system (with $a=b=0.2$, $c=5.7$ in Eq.~\eqref{eq:roessler}) through the $y=0$ plane, using edge densities $\hat{\rho}_1 =0.02\%$ ($\circ$), $\hat{\rho}_2 =0.03\%$ ($\bullet$), $\hat{\rho}_3 =0.05\%$ ($\triangleright$), and $\hat{\rho}_4 =3\%$ (+). All curves have been obtained as mean values taken from $5$ independent realizations of the system with length $N=2\times 10^5$ and using the Euclidean norm. For $\hat{\rho}_1$ to $\hat{\rho}_3$, we find power-law behavior with a characteristic exponent of $\gamma=2.16\pm 0.03$, whereas no clear scaling region is found in the denser RN with edge density $\hat{\rho}_4$. (b) PDF of the $x$ values, where power-law shaped singularities are observed. Redrawn after \cite{Zou2012}.}
\label{fig:roessler_scaling}
\end{figure}

		The degree distribution $p(k)$ of RNs for a specific case of one-dimensional maps (i.e., the Logistic map), Eq.~(\ref{herrmann}) can be explicitly evaluated, leading to a general characterization of the conditions under which SF distributions can emerge in RNs \cite{Zou2010}, as shown in Fig.~\ref{fig:roessler_scaling}(a). When projecting higher-dimensional time-continuous systems to such one-dimensional maps by making use of appropriate (Poincar\'e) return maps, the corresponding considerations can be generalized to such systems, given the specific Poincar\'e surface is ``representative'' for the system's geometric structure (Fig.~\ref{fig:roessler_scaling}(b)). A detailed discussion has been presented in \cite{Zou2012}. To this end,  we only recall the main result that when the system's invariant density $p(\vec{x})$ exhibits a singularity with a power-law shape, Eq.~(\ref{herrmann}) implies that the resulting RN's degree distribution must also display a power-law in the limit $N\to\infty$ for sufficiently small $\varepsilon$. In turn, if $\varepsilon$ is chosen too large, the SF behavior cannot be detected anymore, since it is masked by too large neighborhoods of the points close to the singularity. Figure~\ref{fig:roessler_scaling} demonstrates the latter effect for the specific case of the R\"ossler system (Eq.~\ref{eq:roessler}).

		Notably, it is not trivial to provide an exhaustive characterization of the conditions under which SF distributions can emerge for higher-dimensional systems. As a consequence, generally applicable necessary and sufficient conditions for the presence of power-laws in the degree distributions of RNs have not been established so far. Based on the degree distribution $p(k)$, some higher-order statistics have been proposed in \cite{Jocob2016} quantifying heterogeneity properties of the connectivity.

		We note that in general complex systems, the emergence of power-laws is often associated with a hierarchical organization related to certain fractal properties. In contrast, for RNs it has been shown that the presence of power-laws is not directly related to some (global) fractal structure of the system, but rather the local shape of its invariant density. Consequently, although there are examples of dynamical systems where the scaling exponent of the degree distribution coincides well with the associated fractal dimension, there is no such relationship in general. It will be a subject of future studies under which conditions regarding the structural organization of the attractor, fractal structure and power-law singularities are sufficiently closely related so that the RN's degree distribution allows quantifying the system's fractal properties.

       		 \subsubsection{Small-world effect}
		Since small-world networks are characterized by both, a high clustering coefficient and short average path length, it is clear that RNs cannot obey small-world effects \cite{Donner2015RPBook,Jacob2017,Jacob2018}: although they may exhibit a high degree of transitivity (typically depending on the specific system under study), for any \emph{fixed} value of $\varepsilon$, the average path lengths can only take specific values, which become independent of the network size $N$ in case of sufficiently large samples. On the one hand, for any chosen pair of vertices $i$ and $j$ at positions $\vec{x}_i$ and $\vec{x}_j$, the shortest path length is bounded from below as $\hat{d}_{ij}\geq \lceil\|\vec{x}_i-\vec{x}_j\|/\varepsilon\rceil$ (respectively, the geodesic distance on the attractor $S$ divided by the recurrence threshold $\varepsilon$). Specifically, each shortest path length will converge to a finite value for $N\to\infty$. On the other hand, due to the finite diameter of chaotic attractors, the average path length $\hat{\mathcal{L}}(\varepsilon)$ cannot exceed a maximum value of $\lceil\max_{i,j}\{\|\vec{x}_i-\vec{x}_j\|\}/\varepsilon\rceil$ independent of $N$. Hence, the average path length is bounded from above by a value independent of $N$, which is distinct from the common behavior of SW networks ($\hat{\mathcal{L}}\sim \log N$) \cite{Watts1998}. Moreover, as another immediate consequence of the latter considerations, we observe that $\hat{\mathcal{L}}\sim\varepsilon^{-1}$~\cite{Donner2010a}. This implies that by tuning $\varepsilon$, it is possible to achieve any desired average shortest path length $\hat{\mathcal{L}}$; this fact notably reduces the explanatory power of this global network characteristic. Adding sufficient amount of noise or increasing the threshold $\varepsilon$ comparable to the attractor size, SW properties may be numerically observed for RNs \cite{Jacob2016a,Jacob2017}.

		\subsubsection{Assortative vs. disassortative mixing}
		Unlike SW effects and SF degree distributions, there are hardly any available results regarding the mixing properties of RNs. In general, RNs often obey a tendency towards showing assortative mixing (i.e., vertices tend to link to other vertices with similar degree), which is reasonable in situations where the invariant density $p(\vec{x})$ is continuous or even differentiable, which is supported by recent numerical results \cite{Donges2011,Donner2010a}.

		\subsubsection{Path-based characteristics}\label{sec:apl}
		One main field of application of RQA as well as other quantitative approaches to characterizing the distribution of recurrences in phase space (e.g., recurrence time statistics) is identifying and quantifying different degrees of dynamical complexity among realizations of the same system under different conditions (e.g., different values of the control parameter(s)), or even within a single time series given the system is non-stationary \cite{marwan2007}. While the line-based characteristics of RQA are founded on heuristic considerations (e.g., the higher the predictability of the observed dynamics, the longer the diagonal line structures off the main diagonal should be), we have argued in Section~\ref{sec:analyticRNtheory} that RNs have an analytical foundation in RGGs. Notably, the corresponding characteristics are based on the same binary structure (the recurrence matrix) as the RQA measures. Hence, both concepts allow deriving a similar kind of information, with the important difference being that RQA quantifies dynamical properties, whereas RNs encode topological/geometric characteristics. However, since both aspects are ultimately linked in the case of chaotic attractors, this general observation suggests that RN analysis is in principle suitable for characterizing dynamical complexity in the same way as other established concepts. Therefore, one natural question arises: How do RN measures perform in this task, and which of the multiple possible network measures are particularly suited for this purpose?

		The latter questions have been the main motivation behind much of the early work on RNs focussing on numerical studies of various paradigmatic model systems for low-dimensional chaos~\cite{Donner2010Nolta,Donner2011,Donner2010b,Donner2010a,Marwan2009,Zou2010,Zou2012c}. These studies suggest that for characterizing dynamical complexity, global network characteristics are conceptually easier to use and could provide potentially more stable and distinctive results than certain statistics over local network properties such as the distributions of vertex degrees~\cite{Zou2012b} or local clustering coefficients~\cite{Zou2012c}. Among the set of possible global RN measures, two properties have been found particularly useful: network transitivity $\hat{\mathcal{T}}$ and average path length $\hat{\mathcal{L}}$.

		Regarding $\hat{\mathcal{L}}$, the discriminatory skills concerning different degrees of dynamical complexity can be understood by the fact that for time-continuous systems, chaotic systems can display different degrees of spatial filling of the ``populated'' hyper-volume in phase space, i.e., a high (fractal) dimension of a chaotic attractor close to the (integer) dimension gives rise to a more homogeneous filling than lower ones, which has a natural geometric consequence for the possible path lengths between pairs of sampled state vectors on the attractor. However, it needs to be noted that quantifying dynamical complexity by means of $\hat{\mathcal{L}}$ suffers from two important drawbacks:

		On the one hand, the measure is not normalized and depends crucially on the choice of $\varepsilon$. Hence, working in different methodological settings (e.g., using fixed recurrence thresholds $\varepsilon$ versus fixed recurrence rates $RR=\hat{\rho}$) can provide potentially ambiguous results, since numerical values of $\hat{\mathcal{L}}$ cannot necessarily be directly compared with each other.

		On the other hand, the system's dynamical complexity influences the qualitative behavior of $\hat{\mathcal{L}}$ , which further depends on whether the system is a discrete map or time-continuous. In the latter case, a periodic orbit would result in a higher $\hat{\mathcal{L}}$ than a chaotic one, since a chaotic attractor is a ``spatially extended'' object in phase space on which there are ``shortcuts'' between any two state vectors connecting points corresponding to different parts of the trajectory~\cite{Donner2010a}. In turn, for discrete maps, a periodic orbit contains only a finite set of $p$ mutually different state vectors, so that for sufficiently low $\varepsilon$ and large $N$, the RN is decomposed into $p$ disjoint, fully connected components. In such a situation with not just single isolated vertices, but a completely decomposed network, a reasonable redefinition of $\hat{\mathcal{L}}$ would be summing up only over pairs of mutually reachable vertices in Eq.~(\ref{eq:apl}). Consequently, we approach the minimum possible value of $\hat{\mathcal{L}}=1$~\cite{Marwan2009}, whereas chaotic orbits typically lead to larger $\hat{\mathcal{L}}$.

		According to the above observations, there is no fully developed theoretical understanding and description of the influence of attractor dimensionality on the resulting $\hat{\mathcal{L}}$ beyond the general considerations presented in Section~\ref{sec:analyticRNtheory}. Corresponding further investigations might be an interesting subject for future studies.

       		 \subsubsection{Dimension characteristics by clustering and transitivity} \label{sec:transitivity}
		As mentioned in Section~\ref{sec:scaling}, the scaling exponent of a possible power-law degree distribution has no direct relationship to the fractal dimension of the system \cite{Zou2012}. In turn, such a relationship naturally exists when studying the corresponding integrated measure (i.e., the edge density $\hat{\rho}(\varepsilon)$) in terms of its scaling properties as the recurrence threshold is systematically varied. The latter approach has been extensively discussed in the literature in connection with the estimation of dynamical invariants from RPs~\cite{Faure1998,thiel2004a} and gives rise to estimates of the correlation dimension $D_2$. Notably, one of the classical approaches to estimating $D_2$ from time series data, the Grassberger-Procaccia algorithm~\cite{Grassberger1983PLA,Grassberger1983PRL}, makes use of the correlation sum, which can be easily formulated as a special case in terms of the recurrence rate or RN edge density \cite{Faure1998}.

		The relatively high computational complexity of the latter approaches to estimating the correlation dimension from a RP stems from the fact that a sequence of RPs for different values of $\varepsilon$ needs to be studied for obtaining a proper scaling relationship. In turn, as shown in~\cite{Donner2011b}, network transitivity $\hat{\mathcal{T}}$ provides an alternative approach to defining and estimating a different notion of fractal dimension. For this purpose, note that for a classical RGG embedded in some integer-dimensional metric space, the expected $\hat{\mathcal{T}}$ (which is numerically estimated as the ensemble mean over sufficiently many realizations of the stochastic generation of the RGG) is an analytical function of the dimension $m$, which decays (exactly when using the maximum norm, otherwise approximately) exponentially with $m$~\cite{Dall2002}. This analytical relationship can be generalized to attractor manifolds with non-integer fractal dimensions, which can in turn be estimated from the RN transitivity by inverting this function.

\begin{figure}
\centering
\includegraphics[width=\columnwidth]{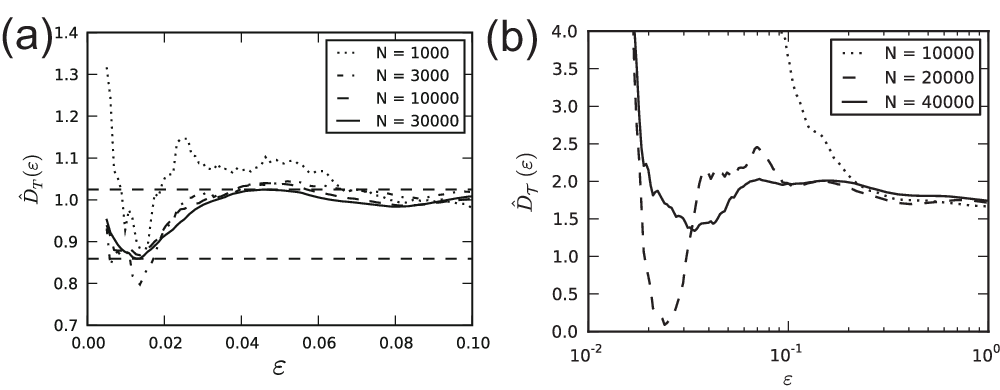}
\caption{(a) Transitivity dimensions $\hat{D}_{\mathcal{T}}$ of the H\'enon map (Eq.~\ref{eq:henon} for one realization with initial condition $(x, y) = (0, 0)$, the first 1000 iterations have been removed from the trajectory to avoid transient behavior) for different $N$. Dashed horizontal lines indicate numerical estimates $\hat{D}_{\mathcal{T}}^{u, l}$ (Eqs. (\ref{eq:dtu}, \ref{eq:dtl})). (b) Same for the R\"ossler system (Eq.~\ref{eq:roessler}) with different lengths $N$ (sampling time $\Delta t=0.05$, first part of the trajectory removed to avoid possible transient dynamics). Note the different scale on the $x$ axis. Modified from \cite{Donner2011b}. } \label{fig:roessler_transitivity}
\end{figure}

		\paragraph{Transitivity dimensions}
		For the general case, the idea formulated above leads to a pair of quantities referred to as upper and lower transitivity dimensions \cite{Donner2011b},
\begin{eqnarray}
D_{\mathcal{T}}^u &=& \limsup_{\varepsilon} \frac{\log(\mathcal{T}(\varepsilon))}{\log(3/4)}, \label{eq:dtu} \\
D_{\mathcal{T}}^l &=& \liminf_{\varepsilon} \frac{\log(\mathcal{T}(\varepsilon))}{\log(3/4)}, \label{eq:dtl}
\end{eqnarray}
\noindent
where the two definitions originate from the fact that certain systems (in particular, chaotic maps whose attractors form Cantor sets in at least one direction in phase space~\cite{Donner2011b}) can exhibit an oscillatory behavior between some upper and lower accumulation point of $\mathcal{T}(\varepsilon)$ as the recurrence threshold $\varepsilon$ is varied (Fig.~\ref{fig:roessler_transitivity}(a)). For systems without such fragmented structure, the upper and lower transitivity dimensions practically coincide, which allows estimating them from the sample RN transitivity with reasonable accuracy using only a single network instance with one suitably chosen value of $\varepsilon$. A detailed analytical investigation of the qualitatively different behavior of the RN transitivity for chaotic attractors with continuous and fragmented invariant densities in dependence on $\varepsilon$ will be subject of future work. Note that in the above definition, we do not explicitly consider a scaling behavior for $\varepsilon\to 0$, since the definition does not explicitly contain $\varepsilon$ (as it is the case for other classical notions of fractal dimensions), but makes use of normalized characteristics with a probabilistic interpretation (cf. Section~\ref{sec:analyticRNtheory}). In this spirit, the fraction on the right-hand side of the Eqs.~\eqref{eq:dtu} and \eqref{eq:dtl} is a well-defined object for each value of $\varepsilon$ (i.e., the specific scale under which the system is viewed) individually.
Figure~\ref{fig:roessler_transitivity} shows the behavior of the scale-dependent transitivity dimension estimate $\hat{D}_{\mathcal{T}}(\varepsilon)=\log(\hat{\mathcal{T}}(\varepsilon))/\log(3/4)$ for the H\'enon map (Eq.~\ref{eq:henon}) and the R\"ossler system (Eq.~\ref{eq:roessler}) for three different RN sizes. We note that very long realizations are typically required to numerically capture the local features of the chaotic attractor of the H\'enon map with reasonable confidence as shown in Fig.~\ref{fig:roessler_transitivity}(a). The larger $N$, the better the estimated values of this measure obtained for fixed $\varepsilon$ approach stationary values corresponding to the upper and lower transitivity dimensions. In contrast, for too small $N$, we observe significant deviations from the asymptotically estimated values, which becomes particularly important for $\varepsilon \to 0$.

		In the case of the R\"ossler system, we obtain similar results shown in Fig. \ref{fig:roessler_transitivity}(b).  We clearly recognize that $\hat{D}_{\mathcal{T}}(\varepsilon)$ assumes approximately stable (i.e., $N$- and $\varepsilon$-independent) values if the recurrence threshold is chosen sufficiently large. In general, there exist two limits that need to be taken into account: For too large recurrence rates, the RN characteristics lose their discriminatory skills, since too many edges are present masking subtle small-scale properties of the attractor \cite{Donner2011b,Donner2010b}. In turn, if $\varepsilon$ is too low (e.g., if $\hat{\rho}$ is below the RN's percolation threshold) \cite{Donges2012}, the network decomposes into mutually disjoint components, and the resulting network characteristics can become ambiguous. In the considered example of the R\"ossler system, this decomposition is mainly caused by the rare excursions of some cycles towards larger $z$ values, which give rise to a poorly populated region (low $p(\vec{x})$) of the attractor. In order to properly cover this part of the attractor for a given $\varepsilon$, many samples (i.e., a large network size $N$) are necessary. Otherwise, the edge density $\hat{\rho}$ starts saturating as $\varepsilon$ gets smaller (at least in the regime where most vertices close to the $z=0$ plane are still connected, cf.\, Fig.~\ref{fig:roessler_transitivity}(b)), and the transitivity dimension estimates strongly deviate from their expected values.

		Notably, the analytical relationship in Eqs.~(\ref{eq:dtu}), (\ref{eq:dtl}) between the effective (geometric) dimension of chaotic attractors and RN transitivity provides the theoretical justification and foundation for applying $\hat{\mathcal{T}}$ as a characteristic discriminating between high and low dynamical complexity of chaotic attractors. Unlike for $\hat{\mathcal{L}}$, the transitivity shows qualitatively the same behavior for discrete and time-continuous systems and is normalized, so that its values can be directly used as a quantitative measure of dynamical complexity associated with the effective geometric dimensionality and, hence, structural complexity of the attractor in phase space.

		\paragraph{Local clustering dimensions}\label{sec:transitivitylocal}
		With the same rationale as for the global network transitivity, we can make use of the local clustering properties of RNs for defining local measures of attractor dimensionality, referred to as upper and lower clustering dimensions~\cite{Donner2011b}:
\begin{eqnarray}
D_{\mathcal{C}}^u(\vec{x}) &=& \limsup_{\varepsilon} \frac{\log(\mathcal{C}(\vec{x};\varepsilon))}{\log(3/4)}, \\
D_{\mathcal{C}}^l(\vec{x}) &=& \liminf_{\varepsilon} \frac{\log(\mathcal{C}(\vec{x};\varepsilon))}{\log(3/4)}.
\end{eqnarray}
\noindent
Following the same argument as for the (global) transitivity dimensions, we do not need to consider the limit $\varepsilon\to 0$ here.

		With similar considerations regarding the possible existence of two distinct accumulation points of $\mathcal{C}(\vec{x})$ as $\varepsilon$ varies, we may utilize this framework for characterizing the point-wise dimension of chaotic attractors in a unique way without making explicit use of scaling characteristics as in the common point-wise dimensions~\cite{Donner2011b}. However, we need to keep in mind that the considered concept of (geometric) dimensionality is largely affected by the profile of the invariant density, e.g., the existence of sharp attractor boundaries or supertrack functions~\cite{Donner2010Nolta,Donner2011b,Donner2010a}. For example, if the attractor has distinct tips (e.g., in the case of the H\'enon system~\cite{Donner2011b,Donner2010a}), the geometric dimension at these points is effectively reduced to zero, which is reflected by $\hat{\mathcal{C}}_i=1$ for state vectors $\vec{x}_i$ sufficiently close to the tips. A similar behavior can be observed for the logistic map at the attractor boundaries and the supertrack functions~\cite{Donner2010Nolta,Donner2011b,Donner2010a}.

		The latter observations point to a prospective application of the local clustering properties of RNs. In case of chaotic attractors of time-continuous dynamical systems, it is known that an infinite number of unstable periodic orbits (UPOs) provide the skeleton of the chaotic dynamics and they are densely embedded in the attractor. The localization of such UPOs is, however, known to be a challenging task. Since UPOs are relatively weakly repulsive (from a practical perspective, those UPOs with low periods are typically least unstable), a trajectory getting close to the vicinity of an UPO will stay close to this orbit for some finite amount of time~\cite{Lathrop1989}. As a result, the dynamics close to UPOs is quasi one-dimensional, and state vectors sampled from the trajectories approximate some lower-dimensional (in the limiting case one-dimensional) subset of the attractor manifold. In such a case, the above theoretical considerations suggest that the local clustering coefficient $\hat{\mathcal{C}}_i$ of vertices $i$ close to low-periodic UPOs should be higher than the values typical for other parts of the chaotic attractor. This conceptual idea is supported by numerical results from~\cite{Donner2010b,Donner2010a} (cf.\, also the band structures with increased $\hat{\mathcal{C}}_i$ in Fig.~\ref{fig:local}(b)), but has not yet been systematically applied to the problem of UPO localization. Notably, the detection limit of UPOs should be ultimately determined by the recurrence threshold $\varepsilon$ in conjunction with the RN size $N$. Specifically, for every finite $\varepsilon>0$, there are infinitely many UPOs intersecting with the $\varepsilon$-neighborhood of some point $\vec{x}_i$ in phase space, whereas we will (for a finite sample of state vectors) only resolve the signatures of the least unstable orbits.

	\subsection{Practical considerations} \label{subsec:practicalRN}

The impact of several algorithmic parameters such as recurrence threshold $\varepsilon$, embedding parameters, sampling rate, or even the selection of variables in multi-dimensional systems has been extensively discussed in previous works~\cite{Donner2010b,Strozzi2009}, focusing mostly on deterministic systems, but also addressing stochastic ones recently. In the following, we will summarize the main findings that should guide corresponding methodological choices in practical applications of RN analysis.

		\subsubsection{Choice of recurrence rate or threshold} \label{subsub:epsilon}
		The most crucial algorithmic parameter of recurrence-based time series analysis is the recurrence threshold $\varepsilon$, which has been discussed extensively in the literature \cite{marwan2007,Donner2010b,Donges2012,Jacob2017}. The empirical choice of $\varepsilon$ often depends on time series embedded in phase space. Too small $\varepsilon$ causes very sparsely connected RNs with many isolated components; too large $\varepsilon$ results in an almost completely connected network. Several invariants of a dynamical system, e.g., the second-order R\'enyi entropy $K_2$  can be estimated by taking its recurrence properties for $\varepsilon \to 0$ \cite{Grassberger1983PLA,Grassberger1983PRL}, which suggests that for a feasible analysis of RNs, a low $\varepsilon$ is preferable as well. This is supported by the analogy to other types of complex networks based on spatially extended systems, where attention is usually restricted to the strongest links between individual vertices, i.e., observations from different spatial coordinates for retrieving meaningful information about relevant aspects of the systems' dynamics. In contrast, a high edge density, does not yield feasible information about the actually relevant structures, because these are hidden in a large set of mainly less important edges \cite{Donner2010b,Donges2012,Jacob2017}.

		As a consequence, only those states should be connected in a RN that are closely neighbored in phase space, leading to rather sparse networks. Following a corresponding rule of thumb confirmed for recurrence quantification analysis \cite{schinkel2008}, one common choice of $\varepsilon$ would be corresponding to an edge density $\rho \lesssim 0.05$ \cite{Marwan2009,Donner2010a}, which yields neighborhoods covering appropriately small regions of phase space. Note that since many topological features of recurrence networks are closely related to the local phase space properties of the underlying attractor~\cite{Donner2010a}, the corresponding information is best preserved for such low $\varepsilon$ unless the presence of noise requires higher $\varepsilon$ \cite{schinkel2008}.

		The heuristic criterion selecting $\varepsilon$ as the (supposedly unique) turning point of the plot of $\rho$ versus $\varepsilon$ \cite{Gao2009} is not generally applicable (as discussed in \cite{Donner2010b}). In particular, this heuristic criterion cannot attribute certain network features to specific \textit{small-scale} attractor properties in phase space \cite{Donner2010b}. Moreover, besides our general considerations supporting low $\varepsilon$, application of the turning point criterion can lead to serious pitfalls. We have to emphasize that various typical examples for both discrete and continuous dynamical systems are characterized by \textit{several} turning points. Depending on the particular types of signals from real measurements from civil engineering structures, a surrogate-assisted method for choosing an optimal threshold by searching for a turning point of a properly defined quality loss function might be a good solution \cite{Yang2015a}.

		For meaningfully estimating path-based and other higher-order structural properties of recurrence networks it is important that the recurrence network possesses a giant component and, hence, nearly all nodes are reachable from nearly all other nodes. At the same time, $\varepsilon$ should be as small as possible so that geometrical fine structure of the underlying attractor is still reflected in the recurrence network. Donges et al. propose to make use of this insight and suggest to set the recurrence threshold just above the percolation threshold $\varepsilon_c$ of the random geometric graph corresponding to the invariant density underlying the dynamical system under study \cite{Donges2012}. This approach allows to connect the problem of choice of recurrence threshold to insights from the theory of random geometric graphs \cite{Dall2002,penrose2003random,herrmann2003connectivity} and, more generally, spatial networks \cite{Gastner2006e,Barthelemy2011,Wiedermann2016} on the percolation threshold $\varepsilon_c$ or the more frequently critical mean degree $z_c(\varepsilon_c)$ in random geometric graphs. In this way, the method allows to make use of analytical results on $z_c$ that have been obtained for various geometries and invariant densities \cite{Dall2002}. These show that the percolation threshold $z_c=1$ of Erd\H"os-R\'enyi graphs is not a tight lower bound for random geometric graphs, i.e. also not in the case of RNs, but that the true critical mean degree tends to be larger due to spatial clustering effects. For example, \textit{Dall and Christensen}\cite{Dall2002} empirically find a scaling law
		\begin{equation}
		z_c(d) = z_c(\infty) + A d^{-\gamma}
		\end{equation}
		for the $d$-dimensional box $S=[0,1]^d$ with uniform probability density $p$, where $z_c(\infty) = 1$, $\gamma = 1.74$, and $A = 11.78$. Alternatively, $\varepsilon_c$ can be obtained from the available data point cloud by numerical methods, e.g., efficiently by $k$-d tree algorithms.

        		One of the problems preventing a uniform choice of $\varepsilon$ across different time series is that the size of the attractor after embedding is arbitrary. To overcome this, Jacob {\textit{et~al.}} \cite{Jacob2016b,Jacob2016a} proposed first to normalize the time series into a unit interval so that the size of the attractor gets rescaled into the unit cube $[0, 1]^m$ where $m$ is embedding dimension. Then, their choice of $\varepsilon$ has been based on empirical results from numerical computations such that the following two criteria are fulfilled \cite{Jacob2016b}: (a) the resulting RN has to remain mostly as ``one single cluster" (see also \cite{Donges2012}) and (b) the measures derived from the RN should uniquely represent the underlying attractor. The first condition fixes the lower bound for $\varepsilon$ which ensures that the network becomes fully connected. The second one fixes the upper bound. Furthermore, Jacob \emph{et~al.} showed that the above two conditions together provide an identical optimum $\varepsilon$ range for time series from all chaotic attractors. However, recent findings point to the fact that this range may still depend on the specific system under study and, more importantly, on the embedding dimension $m$ controlling the distribution of pairwise distances between state vectors in phase space, which should guide the corresponding threshold selection \cite{Kraemer2018}. Moreover, it should be noted that the above choice of the critical range of $\varepsilon$ is, in fact, conceptually related to the selection of a scaling region in the conventional nonlinear time-series analysis for computing dynamical invariants like the correlation dimension $D_2$ \cite{Grassberger1983PLA}, thereby relieving the potential advantage of RN properties providing scale-local characteristics related to such dynamical invariants \cite{Donner2011b}.

		The above strategies for choosing $\varepsilon$ based on normalizing the underlying time series or fixing the recurrence density help us to overcome the problem of sliding-window-based analyses of systems with varying amplitude fluctuations (as coming from different dynamical regimes or non-stationarities). However, in real-world applications, time series are usually not always smooth. When considering time series by studying their RN representations, extreme points (very high rises or falls in the fluctuations) in the time series could break the connected components in the network since the distance between an extreme point and other points would be larger than the threshold value \cite{Eroglu2014}. These unconnected components would cause problems for some complex network measures, since some of them need a connected network to be computed for the entire network. For example, even if we have just one node that is not connected to the network, the average path length will always be infinite for the entire network unless employing the artificial definition of $l_{ij}=N$ for vertices in disconnected components. In such a situation, the normalization method of Jacob \emph{et~al.} \cite{Jacob2016b,Jacob2016a} would result in non-optimal recurrence thresholds biasing the recurrence analysis. An even more important motivation for avoiding isolated components in a RN is that the RN provides a large amount of information about the dynamics of the underlying system, although it contains only binary information. To find a sufficiently small threshold $\varepsilon$ that fulfills the desired condition of connected neighborhoods, Eroglu {\textit{et al.}} proposed to use the connectivity properties of the network. In particular, here the value for $\varepsilon$ is chosen in such a way that it is the smallest one for the RN to be connected. The connectivity of a RN is measured by the second-smallest eigenvalue $\lambda_2$ of the Laplacian matrix associated to the RN's adjacency matrix \cite{Eroglu2014}. Note that this criterion for choosing $\varepsilon$ in an adaptive way shares the same idea as \cite{Jacob2016b,Lin2016}. In the special case where the phase space consists of several disjoint partitions, the method guaranteeing the connectivity might not be feasible, for instance, the support of the invariant density is not continuous when the control parameter is in the periodic or even chaotic regimes before the band merging crisis of the logistic map. Another example for such a behavior is the standard map, where there exist several spatially disjoint components of periodic dynamics in phase space \cite{Donner2010b,Zou2016d}).

		 Note that modular regions of RNs may correspond to a trajectory bundle, which can be associated with the existence of metastable states. Small variations of $\varepsilon$ may lead to very different modular structure in the associated RN, which poses a big challenge for identifying metastable states in real-world time series. Choosing an inadequate recurrence threshold can hide important geometric information related to the organization of a system in its phase space. In \cite{Vega2016}, it was suggested that an adequate recurrence threshold should lie in a range of values producing RNs with similar modular structures. This means that there is a range of recurrence thresholds for which the associated RNs describe reconstructed state space objects with equivalent topology. However, this region of values depends on the distribution of the particular time series data, which might not be uniform. Therefore, they define a filtration procedure to self-adaptively set an adequate $\varepsilon$ from RNs that are associated to a set of recurrence thresholds. The adequate $\varepsilon$ belongs to the subset of values in the filtration for which the modular structures of their associated RNs are the least dissimilar. Furthermore, in searching for metastable states \cite{Vega2016}, the authors suggested to compute modular similarity measures like the adjusted rank index, which may further help to identify an adequate recurrence threshold $\varepsilon$.
Finally, a more recent alternative approach to the threshold selection problem in RNs has been suggested by Wiedermann \emph{et~al.} \cite{Wiedermann2017} in terms of analyzing the statistical complexity of the resulting RNs based upon the Jensen-Shannon divergence between their mean random-walker entropy $\mathcal{S}^{RW}=\sum_{i} \log k_i /N\log(N-1)$ and that of ER random graphs. Specifically, they argued that the threshold could be chosen such that the resulting RN structure becomes most informative, which could be measured in terms of this specific complexity measure. However, it might be questioned if the most complex network structure also corresponds to the most informative one, e.g., when considering the selection of the corresponding edge density as a statistical model selection problem with the network's statistical complexity as the target function to be optimized and the number of links as an analog to the number of degrees of freedom that should be used for penalizing the target function. Additional theoretical work will be necessary to further address this problem.

		\subsubsection{Dependence on embedding parameters} \label{subsubsec:embedding}
Two other important algorithmic parameters of the RN approach are the parameters of time-delay embedding, i.e., embedding dimension $m$ and delay $\tau$. Note that alternative approaches of phase space reconstruction are possible, particularly so-called derivative embedding, which are however often harder to use and would just replace the embedding delay by other parameters possibly involved in more complex algorithms \cite{Lekscha2018}. For this reason, we will not further discuss these alternative approaches here.

		As mentioned in Section \ref{sec:attractorReconstruct}, our discussion so far has assumed that proper embeddings are available for the given time series. For instance, embedding dimension $m$ and delay $\tau$ could have been chosen properly by means of the FNN and ACF method, respectively, which commonly works well for data from deterministic dynamical systems. In turn, proper embeddings do not exist for non-stationary processes though they are more ubiquitous in real time series analysis, for instance, fractional Brownian motions (fBm) and related processes arising from an integration of stationary processes (e.g., fractional L\'evy motion, (F)ARIMA models, etc.). More specifically, we have to keep in mind that some severe conceptual problems may appear when applying them to non-stationary processes: First, finite estimates of $D_F$ are spurious due to the finite amount of data used. The latter result is reasonable since an infinite amount of data (i.e., the innovations at each time step) are necessary to fully describe the evolution of a stochastic process. Thus, from a conceptual perspective, the embedding dimension should be chosen infinitely large. In turn, finite $m$ will necessarily cause spurious results, since the full complexity of the system's (discrete) trajectory is not captured.

		On the other hand, the embedding delay $\tau$ is not considered in the mathematical embedding theorems for deterministic dynamical systems. Embeddings with the same embedding dimension $m$ but different $\tau$ are topologically equivalent in the mathematical sense~\cite{kantz1997}, but in reality a good choice of $\tau$ facilitates further analysis. If $\tau$ is small compared to the relevant internal time-scales of the system, successive elements of the delay vectors are strongly correlated. This leads to the practical requirement that the embedding delay should cover a much longer time interval than the largest characteristic time-scale that is relevant for the dynamics of the system. However, in fBm arbitrarily long time-scales are relevant due to the self-similar nature of the process \cite{Zou2015}. This makes finding a feasible value of $\tau$ a challenging (and, regarding formal optimality criteria, even theoretically impossible) task.

		We emphasize that in the case of non-stationary fBm, the fundamental concepts of phase space reconstruction and low-dimensional dynamics do not apply (not even approximately) anymore. Therefore, the corresponding RN results as have been presented in \cite{Liu2014} hold only for the particular choices of the algorithmic parameters (for instance, length of time series, embeddings etc), showing limited physical interpretations. In \cite{Zou2015}, it has been demonstrated that RN analysis can indeed provide meaningful results for stationary stochastic processes, given a proper selection of its intrinsic methodological parameters, whereas it is prone to fail to uniquely retrieve RN properties for intrinsically non-stationary stochastic processes like fBm. In cases of non-stationarity, a proper transformation is required to remove the particular type of non-stationarity from the data. This can be achieved by additive detrending, phase adjustment (de-seasonalization), difference filtering (incrementation) or other techniques, with the one mentioned last being the proper tool for the particular case of fBm transforming the original process into stationary fractional Gaussian noise (fGn). Further numerical results on the RN analysis for fGn will be presented in Section \ref{subsubsec:nonstation}.

		 \subsubsection{$\varepsilon$-dependence of RN properties}
\begin{figure}
  		\centering
  			\includegraphics[width=\columnwidth]{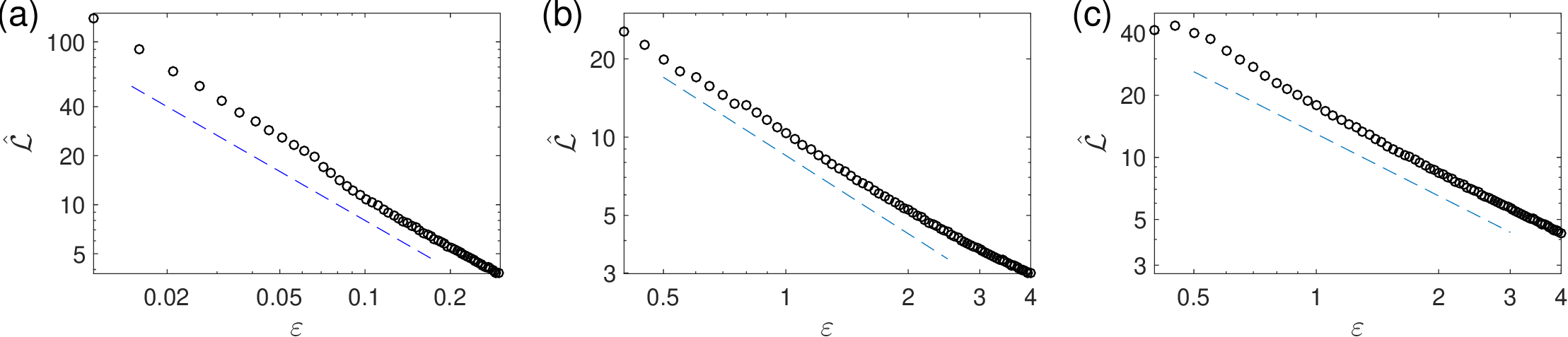} \\
  			\includegraphics[width=\columnwidth]{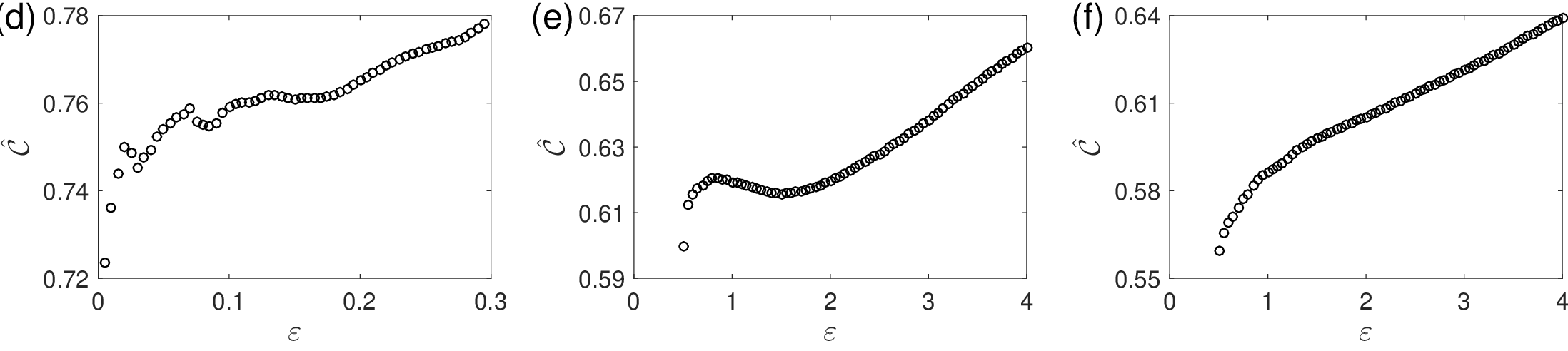}  \\
		\caption{\small Dependence of the average path length $\mathcal{L}$ (a-c) and the global clustering coefficient $\mathcal{C}$ (d-f) for the H\'enon map (Eqs.~\ref{eq:henon}), R\"ossler (Eqs.~\ref{eq:roessler}), and Lorenz system (Eqs.~\ref{eqlorenz}) (from left to right). Dashed lines in the plots on $\hat{\mathcal{L}}(\varepsilon)$ indicate the approximate presence of the theoretically expected $1/\varepsilon$ dependence of $\hat{\mathcal{L}}$. Reproduced from \cite{Donner2011}. }  \label{avg_path_all}
		\end{figure}

		 In order to evaluate the robustness of the topological properties of RNs, their dependence on the free parameter of the method, $\varepsilon$, has to be explicitly considered. In particular, we show in Fig.~\ref{avg_path_all} the effects of $\varepsilon$ on network measures $\hat{\mathcal{L}}$ and $\hat{\mathcal{C}}$. The scale dependence of $\hat{\mathcal{T}}$ has been briefly reviewed in Section \ref{sec:transitivity} and further results of $\hat{\mathcal{T}}$ and $\hat{\mathcal{R}}$ can be found in \cite{Donner2010a,Donner2011b}. In the following, we briefly summarize the main findings for the three model systems H\'enon map (Eqs.~\ref{eq:henon}), R\"ossler (Eqs.~\ref{eq:roessler}), and Lorenz system (Eqs.~\ref{eqlorenz}).

		 As we already discussed associated with Eq.~\eqref{avgL_eq} in Section~\ref{subsubsec:RNmeasureG}, there is an inverse relationship between $\hat{\mathcal{L}}$ and the threshold $\varepsilon$, which has been numerically confirmed in Fig.~\ref{avg_path_all}(a, b, c). However, for the global clustering coefficient $\hat{\mathcal{C}}$, the dependence on $\varepsilon$ is more complicated and depends on the specific properties of the considered system (Fig.~\ref{avg_path_all}(d, e, f)). In particular, while for too small $\varepsilon$, problems may occur, since the RNs will generally decompose into different disconnected clusters for a length $N$ of the considered time series, for intermediate threshold values, an approximately linear increase of $\hat{\mathcal{C}}$ with $\varepsilon$ seems to be a common feature of all three examples.

		\subsubsection{Stability and robustness against noise}
        		The results presented in Section~\ref{sec:transitivity} together with the numerical findings for model systems that will be presented in Section~\ref{sec:numex} show that RN approaches are able to clearly distinguish between periodic and chaotic dynamics under noise-free condition. However, in experimental time series, one is always confronted with measurement errors. Hence, it is necessary to analyze the influence of noise on the constructed RNs. In the framework of recurrence plots, choosing a larger $\varepsilon$ has been suggested to overcome the effect of additive noise \cite{thiel2002}, e.g., using a threshold $\epsilon$ that is at least 5 times larger than the standard deviation $\sigma$ of the observational Gaussian noise can yield reliable statistics. This criterion is based on an analytical computation of the probability of a recurrence point in the RP to be correctly recognized in the presence of observational noise. We suggest to use this criterion if weak observational noise is present as it has been found that the choice $\epsilon \sim 5\sigma$ is optimal for a wide class of processes \cite{thiel2002}.

		It is interesting to visualize noise effects on the reconstructed RNs, which have been presented in \cite{Jacob2016a}. In this work, the authors found numerically that RNs retain much of the information regarding the shapes of the attractor even with moderate addition of both, white and colored noise. Their numerical results suggest that the topology of RNs may completely change if the noise contamination level is above 50\% of signal-to-noise ratio. Furthermore, it has been shown in \cite{Subramaniyam2014} that the influence of noise on the clustering coefficient $\mathcal{C}$ can be minimized by an appropriate choice of $\rho(\varepsilon)$ (e.g., by setting $\rho(\varepsilon) > 0.02$), while the influence on the average path length $\mathcal{L}$ is independent of $\rho(\varepsilon)$. However, for noise levels greater than 40\% in case of $\mathcal{C}$ and 20\% in case of $\mathcal{L}$, the RN measures fail in distinguishing between noisy periodic dynamics and noisy chaotic dynamics. As the noise level reaches 50\% or more, the structural characteristics of RNs present a smooth transition to those of random geometric graphs \cite{Jacob2017}. In particular, Jacob {\textit {et~al.}} have numerically shown that these transitions hold for degree distributions, clustering coefficients and shortest path length \cite{Jacob2017}. The cross-over behavior of RNs towards random graphs has also been observed if $\varepsilon$ is increased towards the system size \cite{Jacob2017}. For large $\varepsilon$, the degree distributions tend to become Poissonian and both, clustering coefficient and shortest path length tend to be $1$, which are characteristics for random networks with very high edge density.

		By no means one can avoid noise effects when applying RN measures to discriminate different dynamical properties from experimental time series. In order to test the efficiency of RN measures as discriminating statistics, hypothesis testing using surrogate technique has been recently proposed \cite{Jacob2018}. For instance, a hypothesis that the data are derived from a linear stochastic process has been employed in \cite{Jacob2018}. The numerical results show that the clustering coefficient $\mathcal{C}$ is not a good discriminating measure if the data involves colored noise whereas the shortest path length $\mathcal{L}$ is effective in the presence of both white and colored noise. We have to emphasize that the specific choice of the hypothesis is fundamental to the interpretation of such results.

		More generally, the consideration of uncertainties plays a crucial role for experimental time series. There are various sources of uncertainties, including measurement errors, noise, irregularly distributed sampling times, etc. The importance of considering the type and magnitude of the uncertainties of an observable in time series analysis cannot be stressed enough. In \cite{Goswami2018}, Goswami {\textit{et~al.}} introduced a framework that merges the analysis of the measurements with that of their uncertainties, including uncertainties in the timing of observations, and shifts the focus from knowing the value of an observable at a given time to knowing how likely it is that the observable had a specific value at that time. In this case, since we consider time series with uncertainties, it is not possible to give a precise answer to the question whether time points $i$ and $j$ recurred, in the sense that we cannot answer this question with a $1$ or $0$ as in a traditional recurrence analysis. We estimate instead the probability that the observations at times $i$ and $j$ are contained in their respective $\varepsilon$-neighborhood. A further point of difference with traditional recurrence analysis is that, till date, there does not exist any meaningful way to embed a time series of probability densities, and one thus estimates the recurrence probabilities in the following without embedding. This novel framework of recurrence probabilities helps to track abrupt transitions in real time series with much improved statistical significance \cite{Goswami2018}. We emphasize that uncertainties bring a huge challenge to complex network approaches to nonlinear time series analysis in general. We will continue the corresponding discussions when reviewing the visibility graph methods in Section~\ref{secsec:VGpractical}.

	\subsection{Numerical examples}\label{sec:numex}
		RN approaches have been widely used for disentangling different dynamical regimes in different times of both, time-discrete and time-continuous dissipative systems \cite{Marwan2009,Donner2010a,Donner2010b,Donner2011,Donner2011b,Zou2010}. Specifically, as discussed above, discriminating qualitatively different types of dynamics can be achieved in terms of measures of complexity, dynamical invariants, or even structural characteristics of the underlying attractor's geometry in phase space. In the context of RN, local vertex-based network characteristics of time series can be visualized in the corresponding phase space as shown in Fig.~\ref{fig:local} \cite{Donner2010a}. For instance, phase spaces of discrete logistic and H\'enon maps, as well as the chaotic R\"ossler and Lorenz system have been color coded by vertex degrees $\hat{k}_i$, local clustering coefficient $\hat{\mathcal{C}}_i$ and betweenness $\hat{b}_i$, respectively \cite{Donner2010a,Donner2010Nolta,Donner2011,Donner2011b}. Furthermore, global network measures like transitivity $\hat{\mathcal{T}}$ or average path length $\hat{\mathcal{L}}$ have been applied to identify dynamical transitions in the logistic map when the control parameter is changed \cite{Marwan2009}. In addition to such stationary settings, the effect of drifting parameters on such characteristics has also been studied for classical model systems in terms of some sliding window analysis \cite{Donges2011}.

In the following, we will briefly review some ``non-classical'' numerical examples illustrating to which extent the aforementioned studies can be generalized to the contexts of higher-dimensional parameter spaces, Hamiltonian or even stochastic dynamics.

		\subsubsection{Parameter space in the R\"ossler system}
\begin{figure}
		\centering
			\includegraphics[width=\columnwidth]{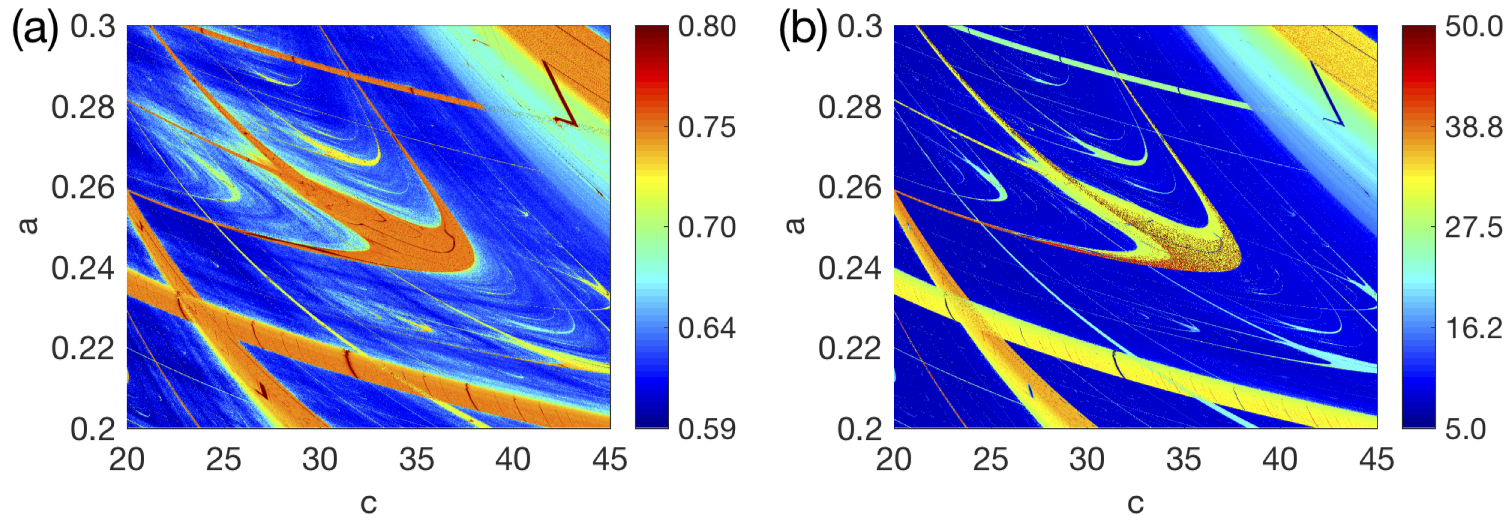}
		\caption{RN transitivity $\hat{\mathcal{T}}$ (A) and average path length $\hat{\mathcal{L}}$ (B) for a two-dimensional intersection ($a=b$) of the three-dimensional parameter space of the R\"ossler system (Eq.~\ref{eq:roessler}), displaying ``shrimp'' structures (i.e., self-similar periodic windows with complex shape). For details, see~\cite{Zou2010}.}
\label{fig:roessler_shrimp}
\end{figure}

		In order to further illustrate the performance of RN transitivity $\hat{\mathcal{T}}$ and average path length $\hat{\mathcal{L}}$ as tracers for qualitative changes in the dynamics of complex systems, we briefly recall results originally obtained by the authors of \cite{Zou2010}. In the latter work, the RN properties have been successfully used to discriminate between periodic and chaotic solutions in a two-dimensional subspace of the complete three-dimensional parameter space $(a,b,c)$ of the R\"ossler system. As Fig.~\ref{fig:roessler_shrimp} reveals, there are sequences of transitions between periodic and chaotic solutions. Specifically, we clearly see that the periodic windows are characterized by higher values of $\hat{\mathcal{T}}$ and $\hat{\mathcal{L}}$ than the chaotic solutions, which is in agreement with the general considerations in Sections~\ref{sec:apl} and \ref{sec:transitivity}. Specifically, for the periodic windows, we find $\hat{\mathcal{T}}$ close to $0.75$, the theoretical value for periodic dynamics (i.e., a system with an effective dimension of $1$).

		In a similar way, we may use the RN framework for capturing the signatures of qualitative changes in the attractor's shape and invariant density as a single control parameter is varied systematically. In a previous study using the R\"ossler system, the RN properties across the transition from the classical phase-coherent R\"ossler attractor to the non-coherent funnel regime have been investigated in \cite{Zou2012c}. Our results indicate that phase coherence -- in a similar spirit as fractal dimension -- can be characterized from a geometric rather than a dynamics viewpoint. However, as of today there is no single RN-based index for phase coherence that has been explicitly derived from theoretical considerations.

\subsubsection{Hamiltonian dynamics}

		In \cite{Zou2016d}, the validity of the RN approach for achieving the same goals in low-dimensional Hamiltonian systems has been demonstrated. Using the standard map as a paradigmatic example, RN analysis was applied to distinguish between regular and chaotic orbits co-existing in the same phase space. Specifically, it was shown (see below) that sticky orbits of the standard map (Eqs.~\ref{std_map_book}) can have a distinct geometric organization that can be detected reliably by RN analysis of relatively short time series (say, $N=1,000$ or $5,000$ points). Note that this model is probably the best-studied chaotic Hamiltonian map and can be interpreted as a Poincar\'e section of a periodically kicked rotor \cite{Lichtenberg_Lieberman_regular,Meiss_rmp_1992}.

		As in other Hamiltonian systems, the phase space of the standard map presents a complicated mixture of domains of chaotic trajectories coexisting with domains of regular ones. The regular component consists of both periodic and quasi-periodic trajectories, while the irregular one contains one or more chaotic orbits. A typical chaotic trajectory needs a long time to fill its corresponding domain in phase space. Due to the existence of periodic islands, once a chaotic orbit gets close to such an island, it can stay close to it and be almost regular in its motion for a rather long time. After this transient period it escapes again to the large chaotic sea. Such a long-term confinement of the trajectory close to the regular domain is commonly referred to as stickiness \cite{Karney_physicaD_1983,Meiss_rmp_1992}, which has been accepted as a fundamental property of many Hamiltonian systems.

\begin{figure}
	\centering
	\includegraphics[width=\columnwidth]{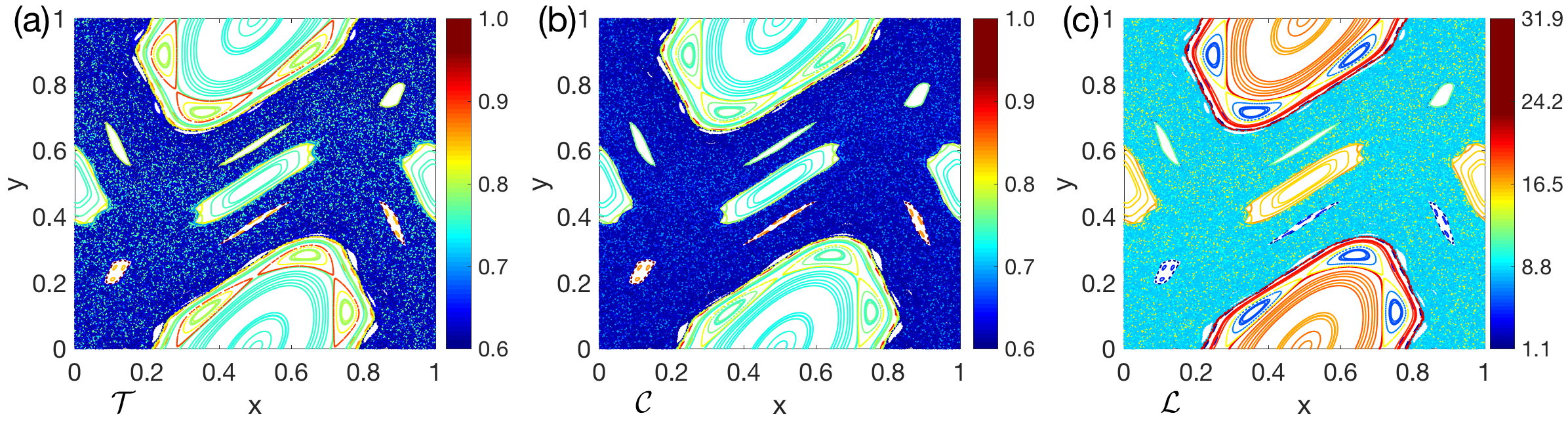}
\caption{\small {Phase space of the standard map (Eq.~\ref{std_map_book}) characterized by three RN measures for the standard map using fixed $RR=0.02$. (a) $\hat{\mathcal{T}}$, (b) $\hat{\mathcal{C}}$, (c) $\hat{\mathcal{L}}$. Reproduced from \cite{Zou2016d}. } \label{fig:sm_rec_rr}}
\end{figure}

The corresponding results for the three global RN measures $\hat{\mathcal{T}}, \hat{\mathcal{C}}$ and $\hat{\mathcal{L}}$ are shown in Figs.~\ref{fig:sm_rec_rr}. Here, we choose 200 initial conditions distributed randomly within the domain of definition of the standard map, $(x,y) \in [0, 1] \times [0, 1]$, and use the canonical parameter value of $\kappa = 1.4$ in Eqs.~\eqref{std_map_book}. All trajectories are computed for $5000$ time steps. Since we were aiming for a quantitative comparability of RN characteristics (which can depend on $RR$), we adaptively choose $\varepsilon$ such that the $RR$ has the same fixed value \cite{Zou2016d}. We observe that the overall structure of the phase space with its intermingled regular and irregular components is captured well by all three measures. Further dynamical and geometric measures have been discussed in \cite{Zou2016d}. Specifically, quasi-periodic trajectories are characterized by larger values of $\hat{\mathcal{T}}$ and $\hat{\mathcal{C}}$, while filling chaotic ones exhibit smaller $\hat{\mathcal{T}}$ and $\hat{\mathcal{C}}$ values (Fig.~\ref{fig:sm_rec_rr}(a, b)).

Unlike periodic or quasi-periodic orbits, chaotic trajectories can fill the complete domain of chaos (as $t\to\infty$). In turn, regular ones are distinct and mutually nested. Since the RN measure $\hat{\mathcal{L}}$ depends clearly on the size of the orbit, the corresponding pattern in Fig.~\ref{fig:sm_rec_rr}(c) is strongly influenced by the selection of a unique threshold $\varepsilon$ for all studied trajectories. In turn, when fixing $RR$ (as done here) the effect of different spatial distances on the estimated RN average path length $\hat{\mathcal{L}}$ is essentially removed (Fig.~\ref{fig:sm_rec_rr}(c)).

		\subsubsection{Non-stationary deterministic systems} \label{subsubsec:nonstation}
		While the aforementioned results have been obtained for stationary deterministic systems, i.e., independent realizations of the system at fixed parameter values, tracing temporal changes in the dynamical complexity of non-stationary systems is another interesting field of application with numerous examples in the real-world. Using model systems with drifting parameters such as the Lorenz \cite{Donges2011} or R\"ossler systems, it is possible to systematically evaluate the performance of RN characteristics in a sliding windows framework, underlining their capabilities for discriminating between qualitatively different types of dynamics and different degrees of complexity in non-stationary (transient) runs as well. For the example of a linearly drifting control parameters of the logistic map and the R\"ossler system (Eqs.~\eqref{eq:roessler}), Donges~{\textit{et al}} found that the values at which bifurcations between periodic and chaotic behavior occur in the non-stationary system do well coincide with the numerically estimated bifurcation points of the autonomous system, indicating that in the considered example, transient dynamics close to the bifurcation points does not play a major role as long as the considered RNs are still sufficiently large to obtain a reliable statistics.

        \subsubsection{Long-range correlated stochastic systems}

		Another category of non-stationary processes comprises long-term correlated stochastic dynamical systems, for instance, fractional Brownian motion (fBm) as already discussed in Section \ref{subsubsec:embedding}, which needs special care when applying recurrence based network analysis. In the case of non-stationary fBm, the fundamental concepts of phase space reconstruction and low-dimensional dynamics do not apply anymore \cite{Zou2015}. One solution to the problem could be transforming the process in a way so that it becomes stationary \cite{Zou2015}. In recent applications to non-stationary real-world time series~\cite{Donges2011,Donges2011a}, the authors have removed non-stationarities in the mean by removing averages taken within sliding windows from the data. In the particular case of fBm, the underlying stochastic process can be transformed into a stationary one by a first-order difference filter, i.e., by considering its increments $x_{i+1}- x_i$. The transformed series is commonly referred to as fractional Gaussian noise (fGn) in analogy with the classical Brownian motion arising from an aggregation of Gaussian innovations. Notably, fGn retains the long-range correlations and Gaussian probability density function (PDF) from the underlying fBm process.

		Because of its stationarity, for fGn the embedding parameters can be chosen more properly than for fBm. Following the discussion in Section~\ref{subsubsec:embedding}, we choose the embedding delay $\tau$ according to the decay of the ACF. For $H < 0.5$ (where $H$ is the Hurst exponent of the process), the estimated ACF drops to negative value at lag one resulting from subsequent values being negatively correlated for the anti-persistent process. Therefore, we choose $\tau = 1$ for $H < 0.5$. In contrast, for $H > 0.5$ we use an estimator of the de-correlation time (specifically, the delay $\tau_{0.1}$ at which the ACF drops below $0.1$) for selecting the embedding delay $\tau$, which increases with rising $H$ as one would expect since larger $H$ indicates a longer temporal range of correlations. The embedding dimension $m$ is chosen via the FNN method. Unlike for fBm, our results suggest that the optimal value $m$ rises with an increasing length of the time series. Hence, it is dominated by the effect of a finite sample size, since the proper theoretical embedding dimension for a stochastic process would in fact be infinite. Specifically, due to the finite sample size, we still find a vanishing FNN rate at a finite embedding dimension, which is probably related to a lack of proper neighbors when high dimensions are considered.

It has been numerically found for various deterministic chaotic systems that the RN characteristics transitivity $\hat{\mathcal{T}}$ and global clustering coefficient $\hat{\mathcal{C}}$ provide relevant information for characterizing the geometry of the resulting RNs. Here, we further demonstrate the application of RNs to fGn to unveil how the transitivity properties of RNs arising from stationary long-range correlated stochastic processes depend on the characteristic Hurst exponent. From the numerical perspective, we show the dependence of the results on the embedding dimension $m$ explicitly.
		\begin{figure}
			\centering
			\includegraphics[width=\columnwidth]{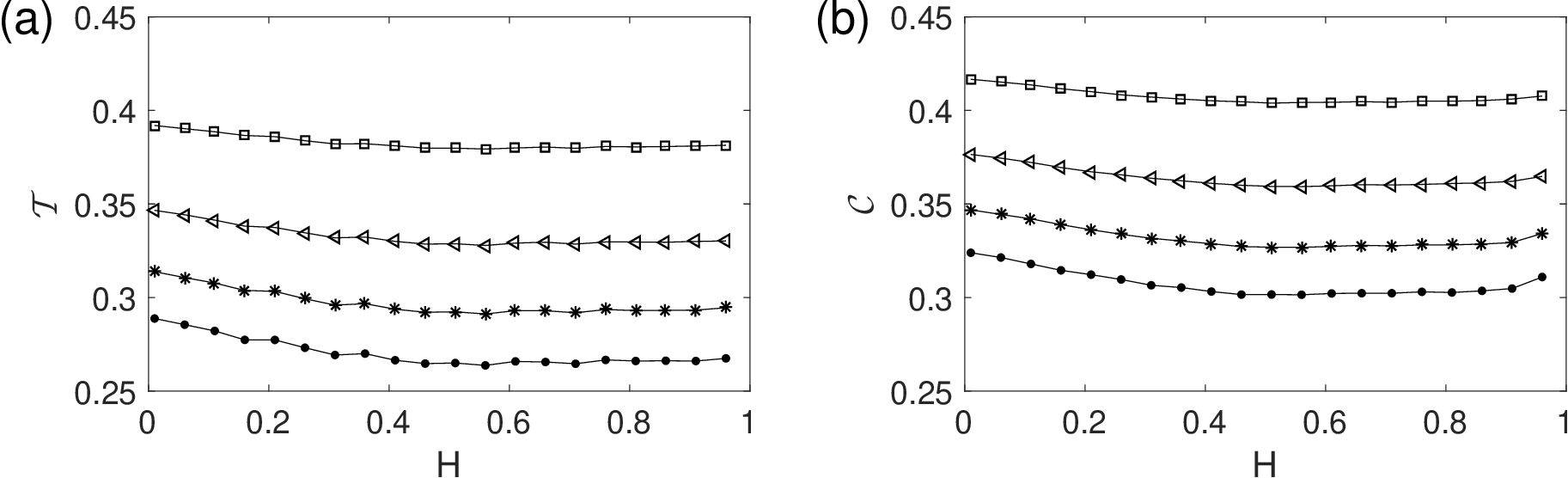}
		\caption{Dependence of (a) RN transitivity $\hat{\mathcal{T}}$ and (b) global clustering coefficient $\hat{\mathcal{C}}$ for fGn on the Hurst exponent $H$ for different embedding dimensions ($m=3$: $\square$, $m=4$: $\triangleleft$, $m=5$: $\ast$, $m=6$: $\bullet$), taken over 200 independent realizations and using a RN edge density of $\rho=0.03$. The embedding delay has been kept at the same value for all realizations with the same $H$ according to the de-correlation time $\tau_{0.1}$. In all cases, $N=2^{12}$. Reproduced from \cite{Zou2015}.  \label{fig:fgn_trans_H}}
\end{figure}

		For $H>0.5$, Fig.~\ref{fig:fgn_trans_H} shows that for a given embedding dimension $m$, both $\hat{\mathcal{T}}$ and $\hat{\mathcal{C}}$ do not depend much on $H$, which is expected since the $m$-dimensional Gaussian PDF of the process does not depend on $H$ \cite{Donges2012,Zou2015}. Some minor deviation from the constant values can be observed at $H$ close to 1, i.e., close to the non-stationary limit case represented by $1/f$-noise, which might be due to numerical effects \cite{Zou2015}.

		For $H<0.5$, both $\mathcal{T}$ and $\mathcal{C}$ rise with decreasing $H$. The reason for this behavior is that $\tau=1$ is recommended, but still not ``optimal'' embedding delay for anti-persistent processes. Specifically, the closer $H$ approaches 0, the stronger is the anti-correlation at lag one. This means that with the same embedding delay $\tau=1$, the smaller $H$ the stronger are the mutual negative correlations between the different components of the embedding vector. As a consequence, the state vectors do not form a homogeneous $m$-dimensional Gaussian PDF with independent components in the reconstructed phase space, but are stretched and squeezed along certain directions, so that the resulting geometric structure appears significantly lower-dimensional than $m$. More numerical considerations have been discussed in \cite{Zou2015}, for instance, systematical biases when $H$ is close to $0$ and due to a finite sample size $N$.

	\subsection{Multiplex recurrence networks}\label{sec:mrn}
	So far, RN approaches have been discussed in the framework of a single system. In the next three subsections, we focus on several different generalizations to multivariate analysis (cf.\, Sections~\ref{sec:multiplex} and \ref{sec:irn_measures}): multiplex recurrence networks, inter-system recurrence networks (ISRN) and joint recurrence networks (JRN), the last two which are based on cross-recurrence plots and joint recurrence plots, respectively.

\begin{figure}
			\centering
			\includegraphics[width=\columnwidth]{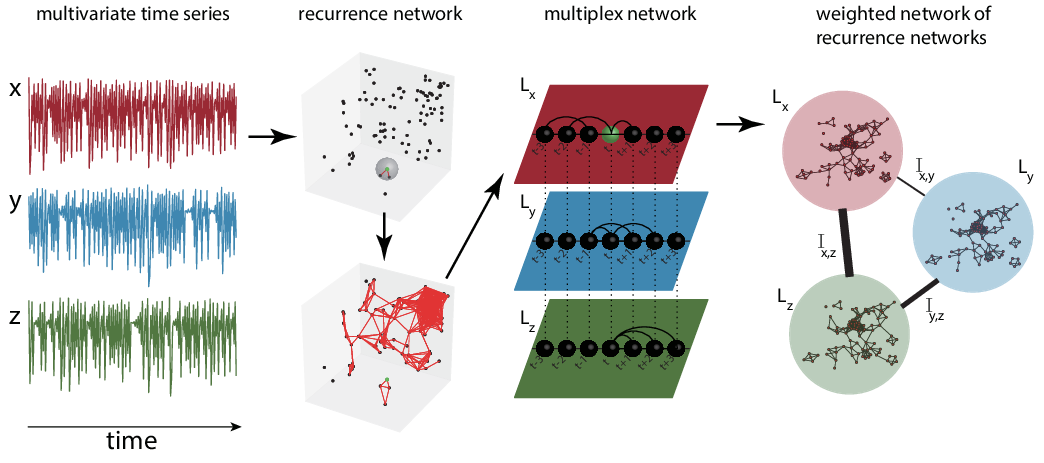}
		\caption{\small{Construction of multiplex recurrence networks for multivariate time series. Reproduced from \cite{Eroglu2018} with permission. } \label{fig:multiRN}}
		\end{figure}

		Let us start with the construction of multiplex recurrence networks, as schematically illustrated in Fig. \ref{fig:multiRN}. If in a multilayer network of $M$ layers, each layer has the same set of vertices and the connections between layers are only between a node and its counterpart in the other layers, we call such a network a ``multiplex". In \cite{Eroglu2018}, Eroglu {\textit{et al.}} proposed to construct multiplex recurrence networks from multivariate time series. In the framework of visibility graph analysis, there is a counterpart of multiplex visibility graphs \cite{Lacasa2015b}, which will be reviewed in Section~\ref{sec:multiplexVG}. For now, let us consider an $M$-dimensional multivariate time series $\{{\vec{x}}_i \}_{i=1}^{N}$, with ${\vec{x}}_i = (x^{[1]}_i, x^{[2]}_i, \dots, x^{[M]}_i) \in \mathbb{R}^M$ for any $i$. Then, the RN of the $\alpha$-th component of ${\vec{x}}(t)$ is created and forms the associated layer $\alpha$ of the multiplex network. For an $M$-dimensional multivariate time series, we can hence create $M$ different RNs which have the same number of nodes and each node is labeled by its associated time index $i$. These networks will form the different layers of a multilayer network. The layers are connected each other exclusively via those nodes with the same time labels. Furthermore, this procedure requires that the time points are the same for all component time series. We note that networks transformed from multivariate time series are generally compatible with the definition of multiplex networks, because each node is uniquely assigned to a certain time point of the multivariate time series, i.e., we find equally time-labeled nodes in all layers.

		More specifically, we denote the adjacency matrix of the $\alpha$-th layer as $\mathbf{A}^{[\alpha]} = \left(A_{ij}^{[\alpha]}\right)$ and $A_{ij}^{[\alpha]} = 1$ if nodes $i$ and $j$ are connected in layer $\alpha$, $A_{ij}^{[\alpha]}= 0$ otherwise. Then, the entire multiplex network $\mathcal{A}$ can be represented by the vector formed by the adjacency matrices of its layers, $\mathbf{\mathcal{A}} = (\mathbf{A}^{[1]}, \mathbf{A}^{[2]}, \dots, \mathbf{A}^{[M]})$, which can be alternatively expressed in matrix form as
		\begin{equation}
\mathbf{\mathcal{A}} = \left[ \begin{array}{cccc}
\mathbf{A}^{[1]} & \mathbf{1}_{N} & \ldots             & \mathbf{1}_{N}\\
\mathbf{1}_{N} & \mathbf{A}^{[2]} & \ddots             & \vdots   \\
\vdots                & \ddots                & \ddots            & \mathbf{1}_{N} \\
\mathbf{1}_{N} & \ldots                 & \mathbf{1}_{N} & \mathbf{A}^{[m]}\\
\end{array} \right]_{NM \times NM},
		\end{equation}
where $\mathbf{1}_{N}$ is again the identity matrix of size $N\times N$.

		Two different measures have been proposed to quantify the similarity between layer $\alpha$ and $\beta$ of a multiplex network \cite{Eroglu2018,Lacasa2015b}. The first is the \emph{interlayer mutual information} $I^{\alpha\beta}$,
		\begin{equation} \label{eq:RNmultiplex}
		I^{\alpha\beta} = \sum_{k^{[\alpha]}} \sum_{k^{[\beta]}} p(k^{[\alpha]}, k^{[\beta]}) \log \frac{p(k^{[\alpha]}, k^{[\beta]})}{p(k^{[\alpha]}) p(k^{[\beta]}) },
		\end{equation}
where $p(k^{[\alpha]}, k^{[\beta]}) $ is the joint probability of the existence of nodes with degree $k^{[\alpha]}$ in layer $\alpha$ and $k^{[\beta]}$ in layer $\beta$, and $p(k^{[\alpha]})$ and $p(k^{[\beta]}) $ are the degree distributions of the RNs in layer $\alpha$ and $\beta$, respectively. Since $I^{\alpha\beta}$ is computed based on the degree sequences, instead of the original time series, the mutual information $I^{\alpha\beta}$ considers the topological recurrence structures in phase space.

		A second measure to quantify the coherence of the original multivariate system is the average edge overlap \cite{Eroglu2018,Lacasa2015b},
		\begin{equation} \label{eq:RNmultiplexW}
			\omega = \frac{\sum_i\sum_{j>i} \sum_{\alpha}A_{ij}^{[\alpha]}}{M \sum_i\sum_{j>i}(1-\delta_{0, \sum_{\alpha}A_{ij}^{[\alpha]}})},
		\end{equation}
		where $\delta_{ij}$ is the Kronecker delta. This measure represents the average number of identical edges over all layers of the multiplex network \cite{Lacasa2015b}. Like the interlayer mutual information (Eq.~\ref{eq:RNmultiplex}), $\omega$ estimates the similarity and coherence via the averaged existence of overlapping links between nodes $i$ and $j$ in all layers $\alpha$ and $\beta$.

		We note that the interlayer similarity measures (Eqs.~\ref{eq:RNmultiplex} and \ref{eq:RNmultiplexW}) are computed for each pair of layers. The giant adjacency matrix $\mathbf{\mathcal{A}}$ of the multiplex network can be projected onto one weighted network representation encompassing the interlayer information only. In other words, we consider each single layer of the multiplex as a node and weighted edges between nodes $\alpha$ and $\beta$ are determined by the quantity $I^{\alpha\beta}$, which yields a weighted projection network of size $M \times M$. The conversion of multilayer systems to weighted network structures is a computationally very efficient approach, which allows further characterization by some traditional structural measures, i.e., clustering coefficient  $\mathcal{C}_w$ and average shortest path length $\mathcal{L}_w$, where the subscripts $w$ indicate that the measures are computed from weighted networks.

		It has been demonstrated that all measures of $I^{\alpha\beta}$, $\omega$, $\mathcal{C}_w$ and $\mathcal{L}_w$ capture similarities in the linking structures of the multiplex recurrence networks \cite{Eroglu2018}. In particular, high values for $I^{\alpha\beta}$, $\omega$ and $\mathcal{C}_w$ have been observed for periodic systems, while lower values correspond to more chaotic systems. However, the opposite holds for $\mathcal{L}_w$ because the diameter of a denser network is in general smaller. The discriminative power of these measures has been illustrated by both a numerical model of a coupled map lattices an real-world paleoclimate time series \cite{Eroglu2018}.

	\subsection{Inter-system recurrence networks} \label{sec:IntSRN}
	In the last decade, two different widely applicable bi- and multivariate extensions of RPs and RQA have been proposed \cite{marwan2007}: cross-recurrence plots \cite{marwan2002,marwan2002pla,Zbilut1998} and joint recurrence plots~\cite{romano2004}. In the following, we discuss some possibilities for utilizing these approaches in a complex network framework, following previous considerations in~\cite{Feldhoff2011,Feldhoff2012,Feldhoff2013}. For this purpose, let us consider $M$ (possibly multivariate) time series $\{\vec{x}_i^{[\alpha]}\}_{i=1}^{N_\alpha}$ with $\vec{x}_i^{[\alpha]}=\vec{x}^{[\alpha]}(t^{[\alpha]}_i)$ sampled at times $\{t^[{\alpha]}_i\}$ from dynamical systems $X^{[\alpha]}$ with $\alpha=1,\dots,M$.

		\subsubsection{From cross-recurrence plots to cross-recurrence networks}
\begin{figure}
	\centering
	\includegraphics[width=0.8\columnwidth]{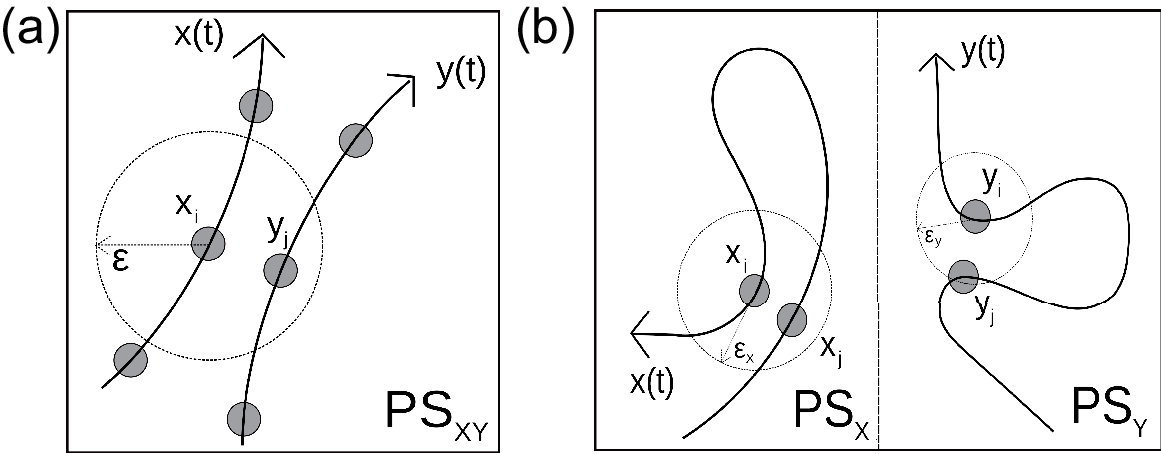}
	\caption{Schematic representation of cross-recurrence (a) and joint recurrence (b) between the trajectories $\vec{x}(t)$ and $\vec{y}(t)$ of two systems $X$ and $Y$. $PS_X$ and $PS_Y$ denote the individual phase spaces of systems $X$ and $Y$, respectively, whereas $PS_{XY}$ indicates the joint phase space of $X$ and $Y$. Modified from \cite{Feldhoff2011}. }
\label{fig:sketch_cr_jr}
\end{figure}

        		One way of extending recurrence analysis to the study of multiple dynamical systems is looking at \emph{cross-recurrences}, \textit{i.e.}, encounters of the trajectories of two systems $X_\alpha$ and $X_\beta$ sharing the same phase space, where $\vec{x}_i^{[\alpha]} \approx \vec{x}_j^{[\beta]}$~\cite{marwan2002,Zbilut1998} (see Fig.~\ref{fig:sketch_cr_jr}(a) for some illustration). It is important to realize that cross-recurrences are not to be understood in the classical sense of Poincar{\'e}'s considerations, since they do not indicate the return of an isolated dynamical system to some previously assumed state. In contrast, they imply an arbitrarily delayed close encounter of the trajectories of two \emph{distinct} systems. The elements of the cross-recurrence matrix $\mathbf{CR}^{[\alpha\beta]}$ are defined as
\begin{equation}
CR_{ij}^{[\alpha\beta]}(\varepsilon_{\alpha\beta})=\Theta(\varepsilon_{\alpha\beta} - \| \vec{x}_{i}^{[\alpha]} - \vec{x}_{j}^{[\beta]} \|),
\end{equation}
where $i=1,\dots,N_\alpha$, $j=1,\dots,N_\beta$, and $\varepsilon_{\alpha\beta}$ is a prescribed threshold distance in the joint phase space of both systems. As in the single-system case, $\varepsilon_{\alpha\beta}$ determines the number of mutual neighbors in phase space, quantified by the \emph{cross-recurrence rate}
\begin{equation}
RR^{\alpha\beta}(\varepsilon_{\alpha\beta})=\frac{1}{N_\alpha N_\beta}\sum_{i=1}^{N_\alpha} \sum_{j=1}^{N_\beta} CR_{ij}^{[\alpha\beta]}(\varepsilon_{\alpha\beta}),
\label{eq:crr}
\end{equation}
\noindent
which is a monotonically increasing function of $\varepsilon_{\alpha\beta}$ (i.e., the larger the distance threshold in phase space, the more neighbors are found). Notably, $RR^{\alpha\beta}$ corresponds to a cross-edge density ${\rho}^{\alpha\beta}$ (Eq.~\ref{eq:globrho_cross}) of a coupled network representation (see below). Furthermore, $\mathbf{R}^{[\alpha]}$ and $\mathbf{R}^{[\beta]}$ are symmetric for individual subsystems, but the cross-recurrence matrix $\mathbf{CR}^{[\alpha\beta]}$ is asymmetric, since we typically have $\|\vec{x}_i^{[\alpha]} - \vec{x}_j^{[\beta]}\|\neq\|\vec{x}_i^{[\beta]} - \vec{x}_j^{[\alpha]}\|$. Even more, it can be non-square if time series of different lengths ($N_\alpha\neq N_\beta$) are considered.

		Due to the aforementioned characteristics, $\mathbf{CR}^{[\alpha\beta]}$ cannot be directly interpreted as the adjacency matrix of a network with similar properties as single-system RNs. This is because the indices $i$ and $j$ label two distinct sets of state vectors belonging to systems $X^{[\alpha]}$ ($i$) and $X^{[\beta]}$ ($j$), respectively. In turn, we can interpret the state vectors $\{\vec{x}^{[\alpha]}_i\}$ and $\{\vec{x}^{[\beta]}_j\}$ as two distinct groups of vertices, and $\mathbf{CR}^{[\alpha\beta]}$ as being an adjacency matrix of a \emph{cross-recurrence network (CRN)} providing a binary encoding of the presence of edges between vertices belonging to different groups. This is the defining property of a bipartite graph~\cite{Newman2003}.

		Bipartite networks can be found in a wide range of fields~\cite{Guimera2007,Kitsak2011} and can be understood as a generic way for describing arbitrary complex networks~\cite{Guillaume2004,Guillaume2006}. The large variety of applications of bipartite graphs has triggered great interest in models describing their properties in an appropriate way. Particular attention has been spent on the problem of community detection \cite{Fortunato2010}, involving new definitions for the modularity function~\cite{Barber2007,Guimera2007,Murata2009,Suzuki2009} and the development of proper algorithms for community detection~\cite{Barber2007,Du2008,Lehmann2008,Sawardecker2009}, partially relating to the spectral properties of the networks. However, their specific structure renders some traditional definitions of network-theoretic measures non-applicable, calling for generalizations or even re-definitions of quantities such as the clustering coefficient~\cite{Lind2005,Zhang2008PhysA}. This is why we do not further consider here the explicit quantification of the properties of the bipartite CRN, but follow a different approach detailed below.

		\subsubsection{Coupled networks framework for $M$ subsystems}
        		As mentioned above, there is a lack of appropriate measures for characterizing explicit bipartite network structures as compared with the rich toolbox of general-purpose complex network characteristics~\cite{Boccaletti2006,Costa2007}. Therefore, instead of explicitly investigating the bipartite structure of the CRN, it is more useful to combine the information contained in the single-system recurrence matrices $\mathbf{R}^{[\alpha]}(\varepsilon_\alpha)$ and the cross-recurrence matrices $\mathbf{CR}^{[\alpha\beta]}(\varepsilon_{\alpha\beta})$ to construct an inter-system recurrence matrix~\cite{Feldhoff2012}
\begin{equation}
\mathbf{IR}(\mathbf{\varepsilon})=\left( \begin{array}{cccc} \mathbf{R}^{[1]}(\varepsilon_{11}) & \mathbf{CR}^{[12]}(\varepsilon_{12}) & \hdots & \mathbf{CR}^{[1M]}(\varepsilon_{1M}) \\
\mathbf{CR}^{[21]}(\varepsilon_{21}) & \mathbf{R}^{[2]}(\varepsilon_{22}) & \hdots & \mathbf{CR}^{[2M]}(\varepsilon_{2M}) \\
\vdots & \vdots & \ddots & \vdots \\ \mathbf{CR}^{[M1]}(\varepsilon_{M1}) & \mathbf{CR}^{[M2]}(\varepsilon_{M2}) & \hdots & \mathbf{R}^{[M]}(\varepsilon_{MM}) \end{array} \right).
\label{isrm}
\end{equation}
Here, $\mathbf{\varepsilon}=(\varepsilon_{\alpha\beta})_{\alpha\beta}$ is an $M \times M$ matrix containing the single-system recurrence thresholds $\varepsilon_{\alpha\alpha}=\varepsilon_\alpha$ and (cross-recurrence) distance thresholds $\varepsilon_{\alpha\beta}$. The corresponding \emph{inter-system recurrence network (IRN)}~\cite{Feldhoff2012} is fully described by its adjacency matrix
\begin{equation}
\mathbf{A}(\mathbf{\varepsilon})=\mathbf{IR}(\mathbf{\varepsilon}) - \mathbf{1}_N,
\end{equation}
where $N=\sum_{\alpha=1}^M N_\alpha$ is the number of vertices and $\mathbf{1}_N$ the $N$-dimensional identity matrix. As in the case of single-system RNs, the IRN is an undirected and unweighted simple graph, which additionally obeys a natural partition of its vertex and edge set (see Section~\ref{sec:irn_measures}). Specifically, for the ``natural'' partition of an IRN, the $G_\alpha$ correspond to the single-system RNs constructed from the systems $X^{[\alpha]}$, whereas the cross-recurrence structure is encoded in $E_{\alpha\beta}$ for $\alpha \neq \beta$. Vertices represent state vectors in the phase space common to all systems $X^{[\alpha]}$ and edges indicate pairs of state vectors from \emph{either} the same \emph{or} two different systems that are mutually close, whereby the definition of closeness can vary between different pairs of systems. To this end, we briefly mention two specific choices that may be convenient:

\begin{itemize}

\item Since we assume the considered systems to share the same phase space, it can be reasonable to measure distances in a way disregarding the specific membership of vertices to the different systems under study. This would imply choosing $\varepsilon_{\alpha\beta}=\varepsilon$ as equal values for all $\alpha,\beta=1,\dots,M$. In such a case, we can (modulo embedding effects) reinterpret the IRN as the RN constructed from the concatenated time series
$$\{\vec{y}_i\}_{i=1}^N=(\vec{x}_1^{[1]},\dots,\vec{x}_{N_1}^{[1]},\vec{x}_1^{[2]},\dots,\vec{x}_{N_2}^{[2]},\dots,\vec{x}_1^{[M]},\dots,\vec{x}_{N_M}^{[M]}).$$
In this situation, we reconsider the general framework of single-system RN analysis as discussed above for studying the geometric properties of the combined system as reflected in a RN. Note, however, that in this case it is hardly possible to explicitly exploit the given natural partitioning of the concatenated data. One corresponding strategy could be utilizing methods for community detection in networks~\cite{Fortunato2010}, such as consideration of modularity~\cite{Newman2004}. Notably, such idea has not yet been explored in this context, and it is unclear to what extent the inferred possible community structure of an IRN could exhibit relevant information for studying any geometric signatures associated with the mutual interdependences between different dynamical systems. To this end, we leave this problem for future research. In contrast, all state vectors are treated in exactly the same way.

\item An alternative choice of recurrence and distance thresholds is based on considering that the individual single-system RNs are quantitatively comparable. Since some of the network measures discussed in Section~\ref{sec:basictheoryCN} explicitly depend on the number of existing edges in the network, this requirement calls for networks with the same edge density $\rho_\alpha=\rho$ for all $\alpha=1,\dots,M$. In other words, the recurrence thresholds $\varepsilon_{\alpha\alpha}$ ($\alpha=1,\dots,M$) could be chosen such that the (single-system) recurrence rates are equal $(RR^1=\dots=RR^\alpha=RR)$. Given the natural partitioning of the IRN vertex set, such network can be viewed and statistically analyzed as a network of networks (see Section~\ref{sec:irn_measures}). In this case, in order to highlight the interconnectivity structure of the individual RNs, it is beneficial to choose the distance thresholds $\varepsilon_{\alpha\beta}$ for $\alpha\neq \beta$ such that the resulting cross-recurrence rates $RR^{\alpha\beta}$ yield $RR^{\alpha\beta}<RR^\alpha=RR^\beta=RR$ and possibly also take the same values $RR^{\alpha\beta}=CRR<RR$ for all $\alpha\neq \beta$. A further note is that, for an IRN, ${\rho}^{\alpha\beta}(\varepsilon_{\alpha\beta})$ equals the (cross-) recurrence rate $RR^{\alpha\beta}(\varepsilon_{\alpha\beta})$ (for $\alpha=\beta$, it gives the corresponding single-system recurrence rate $RR^\alpha(\varepsilon_\alpha)$).

\end{itemize}

As already stated above, the meaningful construction and analysis of IRNs requires time series $\{\vec{x}_i^{[\alpha]}\}_{i=1}^{N_\alpha}$ that share the same phase space and, hence, describe the same observables with identical physical units (Table \ref{tab:multivariate_rns}). However, time series under study can in principle be sampled at arbitrary times $\{t^{[\alpha]}_i\}_{i=1}^{N_\alpha}$ and have different lengths $N_\alpha$, because the method discards all information on time and focuses exclusively on neighborhood relationships in phase space. This type of geometric information is what can be exploited for studying coupling structures between different dynamical systems as reflected by the spatial arrangement of state vectors in the joint phase space (see Section~\ref{sec:coupling}).
\begin{table}[tb]
\centering
\setlength{\tabcolsep}{0.2cm}
\begin{tabular}{lll}
\hline
& IRN & JRN \\
\hline
Length & arbitrary & identical \\
Sampling & arbitrary & identical \\
Physical units & identical & arbitrary \\
Phase space dimension & identical & arbitrary \\
\hline
\end{tabular}
\caption[Multivariate generalizations of recurrence network analysis]{Comparison of inter-system and joint RNs regarding the principal requirements on the time series to be analyzed. \emph{Identical} means that a specific property must be the same for all involved time series, while \emph{arbitrary} implies that this does not need to be the case.}
\label{tab:multivariate_rns}
\end{table}%

		\subsubsection{Analytical description}
		In the same spirit as for the single-system RNs (Section~\ref{sec:analyticRNtheory}), we can consider the graph-theoretical measures for studying the interconnections between subnetworks within IRNs (Section~\ref{sec:irn_measures}) as discrete approximations of more general geometric properties \cite{Donges2012PhD}. Let $S_\alpha \subset Y$ be a subset of an $m$-dimensional compact smooth manifold $Y$ and $p^{[\alpha]}(\vec{x})$ represent its invariant density for all $\alpha=1,\dots,M$, where $\vec{x}\in S_\alpha$. In the following, the $S_\alpha$ and $p^{[\alpha]}$ are assumed to fulfill the same requirements that are stated for $S$ and $p$ in Section~\ref{sec:analyticRNtheory}. Notably, the $S_\alpha$ are assumed to have a considerable non-empty pairwise intersections. We will use the abbreviation $\int d\mu^{[\alpha]}(\vec{x})=\int_{S_\alpha} d^m\vec{x}\,p^{[\alpha]}(\vec{x})$, where $\mu_\alpha$ is a probability measure on $S_\alpha$. For simplicity, only a single recurrence threshold $\varepsilon=\varepsilon_{\alpha\beta}$ for all $\alpha, \beta$ will be used in the following. The generalization to different values of $\varepsilon_{\alpha\beta}$ is straightforward.

\paragraph{Local measures}

The \emph{continuous $\varepsilon$-cross-degree density}
\begin{equation}
\rho^{\alpha\beta}(\vec{x};\varepsilon) = \int_{B_\varepsilon(\vec{x}) \cap S_\beta} d\mu^{[\beta]}(\vec{y}) = \int d\mu^{[\beta]}(\vec{y}) \Theta(\varepsilon - \|\vec{x}-\vec{y}\|)
\end{equation}
\noindent
measures the probability that a randomly chosen point in $S_\beta$ is found in the neighborhood $B_\varepsilon(\vec{x})$ of $\vec{x}\in S_\alpha$. Its discrete version is the cross-degree density $\hat{\rho}_i^{\alpha\beta}(\varepsilon)$ (Eq.~\ref{eq:locrho_cross}).

The \emph{continuous local $\varepsilon$-cross-clustering coefficient}
\begin{equation}
\mathcal{C}^{\alpha\beta}(\vec{x};\varepsilon) = \frac{\int\!\!\!\int_{B_\varepsilon(\vec{x}) \cap S_\beta} \,d\mu^{[\beta]}(\vec{y})\,d\mu^{[\beta]}(\vec{z})\, \Theta(\varepsilon-\|\vec{y}-\vec{z}\|)}{\rho^{\alpha\beta}(\vec{x};\varepsilon)^2}
\end{equation}
gives the probability that two randomly chosen points $\vec{y},\vec{z}\in S_\beta$ are $\varepsilon$-close to each other ($\|\vec{y}-\vec{z}\|<\varepsilon$) if they both lie in the neighborhood of $\vec{x}\in S_\alpha$. The estimator of $\mathcal{C}^{\alpha\beta}(\vec{x};\varepsilon)$ is approximated by the discrete local cross-clustering coefficient $\hat{\mathcal{C}}_i^{\alpha\beta}(\varepsilon)$ (Eq.~\ref{eq:locclustering_cross}).

Considering the mutual global geometry of the sets $S_\alpha,S_\beta$, we furthermore introduce \textit{continuous $\varepsilon$-cross-closeness centrality}
\begin{equation}
c^{\alpha\beta}(\vec{x};\varepsilon) = \left( \int d\mu^{[\beta]}(\vec{y}) \, \frac{g(\vec{x},\vec{y})}{\varepsilon} \right)^{-1}
\end{equation}
quantifying the closeness of $\vec{x}\in S_\alpha$ to all points of the set $S_\beta$ along geodesics together with the related harmonic \textit{continuous local $\varepsilon$-cross-efficiency}
\begin{equation}
e^{\alpha\beta}(\vec{x};\varepsilon) = \int d\mu^{[\beta]}(\vec{y}) \, \left( \frac{g(\vec{x},\vec{y})}{\varepsilon} \right)^{-1}.
\end{equation}
Here, geodesics are defined with respect to the union of all involved systems' attractors $S=\bigcup_{\alpha=1}^M S_\alpha$ and $g(\vec{x},\vec{y})$ is a suitable distance metric on such geodesics (Section~\ref{sec:analyticRNtheory}). The discrete estimators of these two local path-based measures for interdependent networks are respectively given by $\hat{c}_i^{\alpha\beta}(\varepsilon)$ (Eq.~\ref{eq:closeness_cross}) and $\hat{e}_i^{\alpha\beta}(\varepsilon)$ (Eq.~\ref{eq:locefficiency_cross}).

Finally, we define the \textit{continuous $\varepsilon$-cross-betweenness centrality}
\begin{equation}
b^{\alpha\beta}(\vec{x};\varepsilon)=\int \int d\mu^{[\alpha]}(\vec{y})\; d\mu^{[\beta]}(\vec{z})\; \frac{\sigma(\vec{y},\vec{z}|\vec{x};\varepsilon)}{\sigma(\vec{y},\vec{z};\varepsilon)}.
\end{equation}
\noindent
As in the single network case, $\sigma(\vec{y},\vec{z}|\vec{x};\varepsilon)$ denotes the number of times $\vec{x}\in S$ (i.e., from any arbitrary subnetwork) lies on a geodesic between $\vec{y}\in S_\alpha$ and $\vec{z}\in S_\beta$, and $\sigma(\vec{y},\vec{z};\varepsilon)$ denotes the total number of such geodesics. Regarding the appropriate parametrization of $\sigma(\vec{y},\vec{z}|\vec{x};\varepsilon)$, we refer to our discussion for the single network case in Section~\ref{sec:analyticRNtheory}. The discrete estimator $\hat{b}_i^{\alpha\beta}(\varepsilon)$ is given in Eq.~(\ref{eq:betweenness_cross}).

\paragraph{Global measures}

The simplest continuous global property describing the geometric overlap between the sets $S_\alpha$ and $S_\beta$ is the \textit{continuous $\varepsilon$-cross-edge density}
\begin{equation}
\rho^{\alpha\beta}(\varepsilon) = \int\!\!\!\int d\mu^{[\alpha]}(\vec{x}) d\mu^{[\beta]}(\vec{y}) \Theta(\varepsilon - \|\vec{x} - \vec{y}\|)) = \rho^{\beta\alpha}(\varepsilon)
\end{equation}
that is empirically estimated by the discrete cross-edge density $\hat{\rho}^{\alpha\beta}(\varepsilon)$ (Eq.~\ref{eq:globrho_cross}).

The expectation value of the continuous local $\varepsilon$-cross-clustering coefficient $\mathcal{C}^{\alpha\beta}(\vec{x};\varepsilon)$ is referred to as the \textit{continuous global $\varepsilon$-cross-clustering coefficient}
\begin{equation}
\mathcal{C}^{\alpha\beta}(\varepsilon) = \int d\mu^{[\alpha]}(\vec{x})\, \mathcal{C}^{\alpha\beta}(\vec{x};\varepsilon),
\end{equation}
which is approximated by the discrete global cross-clustering coefficient $\hat{\mathcal{C}}^{\alpha\beta}(\varepsilon)$ (Eq.~\ref{eq:globclustering_cross}). Moreover, designed for quantifying transitivity in the cross-recurrence structure, the \textit{continuous $\varepsilon$-cross-transitivity}
\begin{equation}
\mathcal{T}^{\alpha\beta}(\varepsilon) = \frac{\int\!\!\!\int\!\!\!\int d\mu^{[\alpha]}(\vec{x}) d\mu^{[\beta]}(\vec{y}) d\mu^{[\beta]}(\vec{z}) \Theta(\varepsilon-\|\vec{x} - \vec{y}\|) \Theta(\varepsilon-\|\vec{y} - \vec{z}\|) \Theta(\varepsilon-\|\vec{z} - \vec{x}\|)}{\int\!\!\!\int\!\!\!\int d\mu^{[\alpha]}(\vec{x}) d\mu^{[\beta]}(\vec{y}) d\mu^{[\beta]}(\vec{z}) \Theta(\varepsilon-\|\vec{x} - \vec{y}\|) \Theta(\varepsilon-\|\vec{x} - \vec{z}\|)}
\end{equation}
gives the probability that two randomly chosen points $\vec{y},\vec{z}\in S_\beta$ which are $\varepsilon$-close to a randomly chosen point $\vec{x}\in S_\alpha$ are also $\varepsilon$-close with respect to each other. $\mathcal{T}^{\alpha\beta}(\varepsilon)$ is approximated by the discrete cross-transitivity $\hat{\mathcal{T}}^{\alpha\beta}(\varepsilon)$ (Eq.~\ref{eq:transitivity_cross}). As in the case of the discrete estimators, the two latter quantities are in general not symmetric, i.e., $\mathcal{C}^{\alpha\beta}(\varepsilon) \neq \mathcal{C}^{\beta\alpha}(\varepsilon)$ and $\mathcal{T}^{\alpha\beta}(\varepsilon) \neq \mathcal{T}^{\beta\alpha}(\varepsilon)$.

While the two former measures depend only on the local overlap structure between $S_\alpha$ and $S_\beta$ together with the invariant densities $p^{[\alpha]}(\vec{x})$ and $p^{[\beta]}(\vec{x})$, path-based measures contain information on the global geometry of both sets. The \textit{continuous $\varepsilon$-cross-average path length}
\begin{equation}
\mathcal{L}^{\alpha\beta}(\varepsilon) = \int\!\!\!\int d\mu^{[\alpha]}(\vec{x}) d\mu^{[\beta]}(\vec{y}) \frac{g(\vec{x},\vec{y})}{\varepsilon} = \mathcal{L}^{\beta\alpha}(\varepsilon)
\end{equation}
gives the average length of geodesic paths starting in $S_\alpha$ and ending in $S_\beta$ or vice versa. Similarly, we define the \textit{continuous global $\varepsilon$-cross-efficiency}
\begin{equation}
\mathcal{E}^{\alpha\beta}(\varepsilon) = \left( \int\!\!\!\int d\mu^{[\alpha]}(\vec{x}) d\mu^{[\beta]}(\vec{y}) \left( \frac{g(\vec{x},\vec{y})}{\varepsilon} \right)^{-1} \right)^{-1} = \mathcal{E}^{\beta\alpha}(\varepsilon)
\end{equation}
which is the harmonic mean geodesic distance between $S_\alpha$ and $S_\beta$. Discrete approximations of these global path-based quantifiers are provided by the cross-average path length $\hat{\mathcal{L}}^{\alpha\beta}(\varepsilon)$ (Eq.~\ref{eq:apl_cross}) and global cross-efficiency $\hat{\mathcal{E}}^{\alpha\beta}(\varepsilon)$ (Eq.~\ref{eq:globefficiency_cross}), respectively. As for their discrete estimators, the path-based characteristics $\mathcal{L}^{\alpha\beta}(\varepsilon)$ and $\mathcal{E}^{\alpha\beta}(\varepsilon)$ are invariant under an exchange of $S_\alpha$ and $S_\beta$.

		\subsubsection{Geometric signatures of coupling}\label{sec:coupling}
		The new class of statistical network measures designed for investigating the topology of networks of networks discussed in the previous subsections is readily applicable for analyzing the interdependency structure of multiple complex dynamical systems. For the special case of two coupled systems $X$ and $Y$, we have demonstrated numerically that in an IRN, the asymmetry intrinsic to the global measures cross-transitivity $\hat{\mathcal{T}}^{XY}$ and global cross-clustering coefficient $\hat{\mathcal{C}}^{XY}$ can eventually be exploited to reliably detect the direction of coupling between chaotic systems over a wide range of coupling strengths, requiring only a relatively small number of samples $N_{X,Y}\sim\mathcal{O}(10^2\dots 10^3)$~\cite{Feldhoff2012}. For this purpose, we make again use of the fact that transitivity-based characteristics quantify subtle geometric properties which can be easily evaluated both analytically and numerically. Note, however, that this finding has been purely heuristic so far and lacks a precise characterization under which conditions the corresponding considerations do apply.

		In order to see how cross-transitivities and global cross-clustering coefficients capture dynamical signatures of asymmetric vs. symmetric coupling configurations, let us assume a diffusive coupling with positive sign (i.e., an attractive interaction) as in Eq.~(\ref{eq:coupled_roessler}). In the uncoupled case, cross-triangles arise randomly according to the sampling from the systems' respective invariant densities. In this case, eventual asymmetries between $\hat{\mathcal{T}}^{XY}$ and $\hat{\mathcal{T}}^{YX}$ (or, equivalently, $\hat{\mathcal{C}}^{XY}$ and $\hat{\mathcal{C}}^{YX}$) originate from the geometry of the respective sets $S_X$ and $S_Y$ and the associated $p^X(\vec{x})$ and $p^Y(\vec{x})$, which should already be reflected in the single-system RN transitivities and global clustering coefficients. In turn, if both systems are represented by the same set of state variables (a prerequisite for the application of IRNs) and obey similar values of $\hat{\mathcal{T}}^{X}$ and $\hat{\mathcal{T}}^{Y}$ ($\hat{\mathcal{C}}^{X}$ and $\hat{\mathcal{C}}^{Y}$), it is likely that also $\hat{\mathcal{T}}^{XY}$ and $\hat{\mathcal{T}}^{YX}$ ($\hat{\mathcal{C}}^{XY}$ and $\hat{\mathcal{C}}^{YX}$) take similar values. Note that minor asymmetries in the interdependent network characteristics can already occur if both systems are only weakly non-identical, e.g., when considering uncoupled identical R\"ossler systems with just a small detuning of their natural frequencies~\cite{Feldhoff2012}.

\begin{figure}
	\centering
	\includegraphics[scale=0.6]{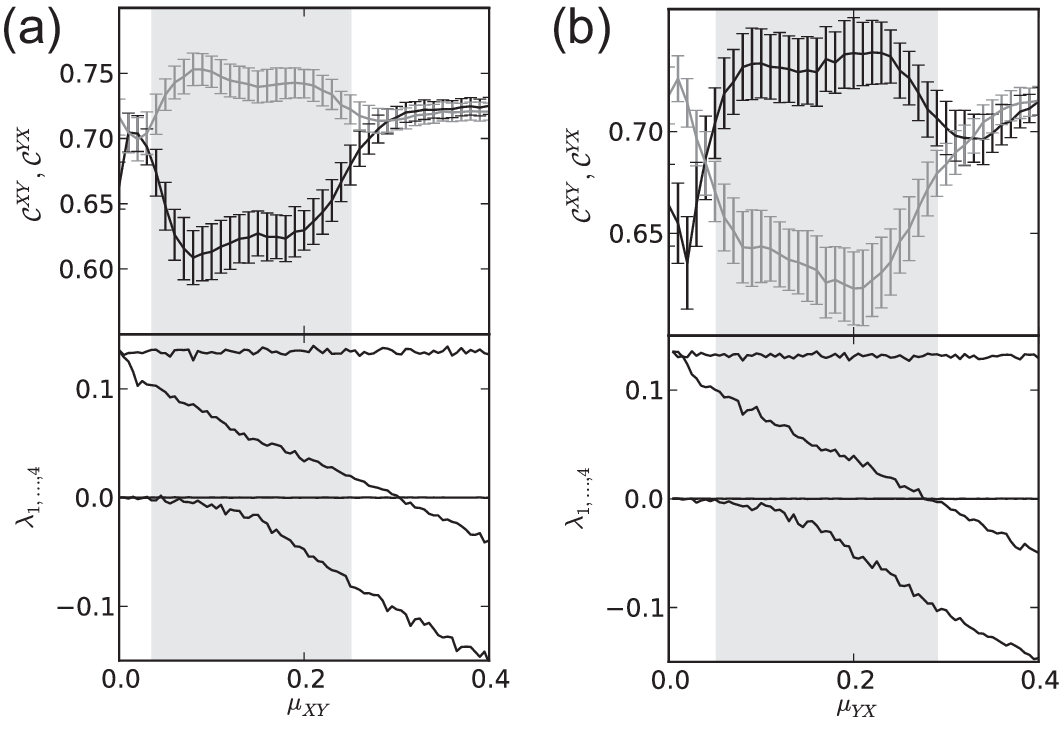}
\caption{Global cross-clustering coefficients (Eq.~\ref{eq:globclustering_cross}) $\hat{\mathcal{C}}^{XY}$ (black), $\hat{\mathcal{C}}^{YX}$ (gray) and the four largest Lyapunov exponents $\lambda_{1,\dots,4}$ estimated using the Wolf algorithm \cite{wolf1985} for two R\"ossler oscillators (Eqs.~\ref{eq:coupled_roessler}) subject to unidirectional coupling $X\to Y$ (a) and $Y\to X$ (b). The shaded regions mark the values of the coupling strength for which a correct identification of the coupling direction is achieved. Error bars represent mean values and standard deviations taken from an ensemble of 200 independent network realizations (with $N=1,500$ data points per system). Modified from \cite{Feldhoff2012}. }
\label{fig:roessler_coupling}
\end{figure}

		Let us suppose now that there is a unidirectional coupling $X\to Y$. In this case, the trajectory of the driven system $Y$ is attracted by that of the driver $X$ due to the considered form of coupling. As a result, it is likely to find more states in $Y$ that are close to mutually connected pairs of states in $X$ than in the uncoupled case. This implies that $\hat{\mathcal{T}}^{YX}$ ($\hat{\mathcal{C}}^{YX}$) increases since $X$ is ``pulling'' the trajectory of $Y$ and, hence, the number of triangles having their baseline in system $X$ increases relatively to those having their baseline in $Y$. Consequently, we expect to have $\hat{\mathcal{T}}^{YX}>\hat{\mathcal{T}}^{XY}$ and $\hat{\mathcal{C}}^{YX}>\hat{\mathcal{C}}^{XY}$, which is confirmed by numerical studies as shown in Fig.~\ref{fig:roessler_coupling} \cite{Feldhoff2012}.

		Moderate unidirectional coupling (below the threshold strength characterizing the onset of synchronization) increases the driven system's dimension \cite{Romano2007,zou2011} (we will numerically demonstrate this behavior in Section~\ref{sec:JRNSsync}), so that former neighbors of pairs of recurrent states in $X$ are not mutually close in $Y$ anymore. In this case, the number of ``cross-triangles'' with baseline in $Y$ decreases in comparison with those having their baseline in $X$. In fact, a corresponding decrease in $\hat{\mathcal{T}}^{XY}$ ($\hat{\mathcal{C}}^{XY}$) and an increase in $\hat{\mathcal{T}}^{YX}$ ($\hat{\mathcal{C}}^{YX}$) can often be observed in parallel.

		Figure~\ref{fig:roessler_coupling} shows an illustrative example of global cross-clustering coefficients $\hat{\mathcal{C}}^{XY}$ (Eq.~\ref{eq:globclustering_cross}) for two unidirectionally coupled R\"ossler systems (Eqs.~\eqref{eq:coupled_roessler}) in the rather complex funnel regime with the same parameters $a$, $b$ and $c$, but a weak detuning of $\nu=0.02$, following the setting of \cite{Feldhoff2012}. The obtained results are consistent with our above heuristic explanation for the emergence of asymmetries between the interdependent network characteristics in the presence of unidirectional coupling. Specifically, for a wide range of moderate coupling strengths, the difference between the two global cross-clustering coefficients $\hat{\mathcal{C}}^{XY}$ and $\hat{\mathcal{C}}^{YX}$ allows to correctly identify the direction of the imposed coupling. At large coupling strengths (i.e., close to and beyond the onset of generalized synchronization, which is indicated by the second largest Lyapunov exponent of the system approaching zero as shown in Fig.~\ref{fig:roessler_coupling}), both $\hat{\mathcal{C}}^{XY}$ and $\hat{\mathcal{C}}^{YX}$ become statistically indistinguishable, which is consistent with the fact that the behavior of the driven system is completely locked to the dynamics of the driver (cf.\, Section~\ref{sec:JRNSsync}). In turn, the indistinguishability of both coupling directions at very low coupling strengths is most likely due to the fact that the geometric deformations of the driven system's attractor are too small to be detected by the given finite values of $\varepsilon_X$, $\varepsilon_Y$ and $\varepsilon_{XY}$ and the chosen network size. We expect that for larger IRNs and smaller distance thresholds, the lower boundary of the interval of coupling strengths for which the two global cross-clustering coefficients differ statistically significantly from each other will shift towards zero.

		We emphasize that the same results can be obtained using the cross-transitivity replacing the global cross-clustering coefficient. Moreover, it is notable that the reported distinction can already be obtained at comparably small network sizes of some hundred vertices \cite{Feldhoff2012}.

		Furthermore, as one of successful applications of inter-system recurrence network approaches, Gao {\textit{et al.}} characterize different oil-water flow patterns by reconstructing networks from multi-channel measurements \cite{Gao2013,Gao2016b,Gao2016c}. In this series of works, Gao {\textit{et al.}} constructed multivariate RNs based on cross~recurrence plots. Further sufficient examples include the detection of coupling direction between the Indian and East Asian monsoon branches based on two representative paleoclimate records from Oman and China \cite{Feldhoff2012}.

	\subsection{Joint recurrence networks}
      		\subsubsection{Joint recurrence plots}
		Besides cross-recurrences, another possible multivariate generalization of RPs is studying joint recurrences of different systems in their individual (possibly different) phase spaces. Here, the basic idea is that the simultaneous occurrence of recurrences in two or more systems $X^{[\alpha]}$ (see Fig.~\ref{fig:sketch_cr_jr}(b)) contains information on possible interrelationships between their respective dynamics, for example, the emergence of generalized synchronization (GS)~\cite{romano2004,romano2005}. Consequently, based on time series $\{\vec{x}_i^{[\alpha]}\}$, the joint recurrence matrix $\mathbf{JR}$ with elements
\begin{equation}
JR_{ij}(\varepsilon_1,\dots,\varepsilon_M)=\prod_{\alpha=1}^M R_{ij}^{[\alpha]}(\varepsilon_\alpha)
\end{equation}
is defined as the element-wise product of the single-system recurrence matrices $\mathbf{R}^{[\alpha]}$ with the elements
\begin{equation}
R_{ij}^{[\alpha]}(\varepsilon_\alpha)=\Theta(\varepsilon_\alpha - \|\vec{x}_i^{[\alpha]} - \vec{x}_j^{[\alpha]} \|),
\end{equation}
where $(\varepsilon_1,\dots,\varepsilon_M)$ is the vector of recurrence thresholds that can be selected for each time series individually, typically such as to yield the same global recurrence rates $RR_\alpha=RR$ for all $\alpha=1,\dots,M$.

		\subsubsection{Network interpretation} \label{sec:jrn_classic}
		Analogously to single-system RN analysis, we take a graph-theoretical perspective by defining a \emph{joint recurrence network (JRN)} by its adjacency matrix
\begin{equation}
\mathbf{A}(\varepsilon_1,\dots,\varepsilon_M) = \mathbf{JR}(\varepsilon_1,\dots,\varepsilon_M) - \mathbf{1}_N,
\end{equation}
where $\mathbf{1}_N$ again denotes the $N$-dimensional identity matrix. Hence, the edges $(i,j)$ of a JRN indicate joint recurrences occurring simultaneously in \emph{all} $M$ time series under study. Alternatively, $\mathbf{A}(\varepsilon_1,\dots,\varepsilon_M)$ may be viewed as the element-wise product of the single-system recurrence networks' adjacency matrices $\mathbf{A}^{[\alpha]}(\varepsilon_\alpha)$.

		As single-system RN and IRN, the JRN describes an undirected and unweighted simple graph. However, due to the temporal simultaneity condition of the joint recurrence concept, vertices $i$ are explicitly associated with points in time $t^{[\alpha]}_i=t^{[\beta]}_i$ common to the $M$ considered time series (cf.~Tab.~\ref{tab:multivariate_rns}). This is conceptually different from RNs and IRNs where time information is not taken into account so that network characteristics are invariant under permutations of the state vectors (i.e., the -- possibily embedded -- observations). More specifically, it is not possible to relabel the observations in the underlying time series prior to the computation of the JRN, whereas the JRN vertices can be shuffled again without altering the resulting network properties.

		By construction, the time series $\{\vec{x}_i^{[\alpha]}\}$ used for constructing a JRN need to be sampled at identical times $\{t^{[\alpha]}_i\}$ and have to have the same length, \textit{i.e.}, $N_1=N_2=\dots=N_M=N$. However, since recurrences are compared instead of state vectors, the advantage of JRN is that the $\{\vec{x}_i^{[\alpha]}\}$ neither have to represent the same physical quantity measured in identical units, nor need they reside in the same phase space (Tab.~\ref{tab:multivariate_rns}).

		From a conceptual perspective, a JRN can be regarded as a RN for the combined system $(X^{[1]}\otimes\cdots\otimes X^{[M]})$ in its higher-dimensional phase space spanned by all state variables. However, recurrences are defined here in some non-standard way by taking distances in the subspaces associated with the individual systems $X^{[\alpha]}$ separately into account. This implies that the properties of JRNs can be studied in essentially the same way as those of single-system RNs (but with possibly more subtle geometric interpretations of the respective network characteristics). In turn, comparing the same properties for JRN and single-system RNs provides important information about the similarity of neighborhood relationships in the combined phase space and projections on the individual systems' subspaces. Specifically, we can gain insights about the effective degrees of freedom of the combined system, which may be reduced in comparison with the sum of the degrees of freedom of the uncoupled systems due to dynamical interdependences between its components. Note that there is a close analogy to multiplex recurrence networks, with the exception that JRNs do not exhibit any inter-layer linkages.

		\paragraph{$f$-joint recurrence networks}
		Equivalently to their interpretation outlined in Section~\ref{sec:jrn_classic}, we can also consider JRNs as the reduction of a generalized multiplex RN, where the vertices correspond to time points $t_i$, which can be connected by at most $M$ different types of (labelled) edges representing the mutual closeness of states in the $M$ different systems. In this viewpoint, the reduction towards the JRN follows from the requirement that for a given pair of vertices, in the multiplex RN \emph{all} $M$ possible labelled edges must be present. With other words, in terms of Boolean logics the entries of the binary recurrence matrices $\mathbf{R}^{[\alpha]}$ are connected by a logical AND for defining the elements of $\mathbf{JR}$.

		Notably, the presence of a joint recurrence becomes increasingly unlikely as the number of interacting systems $M$ increases. Even in the case of very strong interdependences, there may be stochastic fluctuations in the individual systems (e.g., observational noise) that mask recurrences in individual systems and, thus, subsequently reduce the \emph{joint recurrence rate}
\begin{equation}
JRR(\varepsilon_1,\dots,\varepsilon_M) = \frac{2}{N(N-1)} \sum_{i=1}^{N-1} \sum_{j=i+1}^N JR_{ij} (\varepsilon_1,\dots,\varepsilon_M)
\end{equation}
\noindent
aka JRN edge density $\rho_J$.

		One possibility to circumvent the problem sketched above is relaxing the requirement of having simultaneous recurrences in all sub-systems (i.e., the logical AND operation connecting the recurrence matrices of the individual systems in a component-wise way), but considering the case where at least a fraction $f \in(0,1]$ of all systems exhibit recurrences (the standard JRN follows for $f = 1$) \cite{Donner2015RPBook}. This point of view allows defining a hierarchy of networks, which we call \textit{$f$-joint recurrence networks ($f$-JRN)}. Starting from the union of single-system RNs (respectively, the multiplex RN) providing a network with $M$ different edge types corresponding to recurrences of the individual systems, we require that there exist at least $\lceil f M\rceil$ edges between two specified vertices (i.e., time points). In the specific case of $M=2$ systems and $f \in (0,0.5]$ (or, more generally, for $f \in(0,1/M]$), we can rewrite this requirement with a simple logical (Boolean) operation connecting the single-system recurrence matrices in a component-wise way as $JR^{f}_{ij}(\varepsilon_1,\varepsilon_2) = R^{[1]}_{ij}(\varepsilon_1)\ \mbox{OR}\ R^{[2]}_{ij}(\varepsilon_2)$.

		For the more general case, in order to mathematically formulate the requirement of $\lceil f M \rceil$ simultaneous recurrences, it is convenient to start from a practically equivalent re-definition of the joint recurrence matrix,
\begin{equation}
JR^*_{ij}(\varepsilon_1,\dots,\varepsilon_M) = \Theta\left( \sum_{\alpha=1}^M R_{ij}^{[\alpha]}(\varepsilon_\alpha) -M+\delta \right),
\end{equation}
\noindent
with the usual Heaviside function $\Theta(\cdot)$ and $\delta\to 0^+$ being infinitesimally small (to ensure $JR^*_{ij}=1$ if $\sum_{\alpha=1}^M R_{ij}^{[\alpha]} = M$), and set
\begin{equation}
JR^{f}_{ij}(\varepsilon_1,\dots,\varepsilon_M) = \Theta\left( \sum_{\alpha=1}^M R_{ij}^{[\alpha]}(\varepsilon_\alpha) - f M+\delta \right),
\end{equation}
\noindent
to be the \emph{$f$-joint recurrence matrix}. We can use the latter definition to define $f$-joint recurrence plots as well as $f$-JRNs in full analogy to the classical case $f =1$.

		Trivially, the number of edges in an $f$-JRN decreases monotonically for increasing $f$ if all single-system recurrence thresholds $\varepsilon_\alpha$ are kept fixed. We note that a similar relaxation of the strict requirement of a conjecture (AND relation) between the (Boolean) entries of different recurrence matrices has been recently discussed in the framework of symbolic recurrence plots~\cite{Donner2008}. Moreover, it might be interesting (but has not yet been explored) to use concepts from fuzzy logic as the basis for somewhat weaker requirements than in the rather restrictive definition of the original JRN.

		The conceptual idea of $f$-JRNs has not yet been further developed and studied elsewhere. One possible field of application could be finding proper values of $f$ (for example, in dependence on the magnitude of some observational noise) for which results commonly obtained using ``normal'' JRNs become stable in the case of real-world time series. To this end, we only emphasize the possibility of defining $f$-JRNs and studying the properties of these entities (e.g., the scaling of network characteristics as a function of $f$), but leave a corresponding investigation as a subject for future research.

		\subsubsection{Network properties and synchronization}\label{sec:JRNSsync}
		The concept of joint recurrence plots (JRPs) has been found very useful for studying the otherwise hard to detect emergence of generalized synchronization (GS) between two coupled chaotic systems $X$ and $Y$ \cite{romano2005}. GS describes the presence of a general functional relationship between the trajectories of both systems, $\vec{y}(t)=f(\vec{x}(t))$, which can arise at sufficiently large coupling strengths in both uni- and bidirectional coupling configurations. Most available methods for identifying GS from time series data have been developed for driver-response relationships, but a few approaches are also suitable for studying GS in the presence of symmetric couplings~\cite{Feldhoff2013}. Among the latter, JRPs have recently attracted specific interest.

		Romano~\textit{et~al.}~\cite{romano2005} argued that in case of GS, recurrences in the two coupled systems need to occur simultaneously (or with a given fixed time lag in the special case of lag synchronization, $y(t)=f(x(t-\tau))$). Hence, comparing the joint recurrence rate $JRR$ with the recurrence rates of the individual single-system RPs (taken to be the same for both systems) should show convergence of both values. The latter fact is quantified in terms of the \textit{joint probability of recurrence (JPR) index}
\begin{equation}
JPR = \max_{\tau} \frac{\Omega(\tau)-RR}{1-RR}
\label{eq:jpr}
\end{equation}
\noindent
with the lagged joint recurrence rate ratio
\begin{equation}
\Omega(\tau)=\frac{1}{N^2\, RR}\sum_{i,j=1}^N \Theta(\varepsilon_X-\|\vec{x}_i-\vec{x}_j\|)\, \Theta(\varepsilon_Y-\|\vec{y}_{i+\tau}-\vec{y}_{j+\tau}\|)
\end{equation}
\noindent
and $RR$ being the recurrence rate taken equal for both considered systems. Since for GS, we can expect that $\Omega(\tau)\to 1$ for some $\tau$, $JPR\to 1$. However, the latter measure has some disadvantages. On the one hand, testing for the significance of a specific value of $JPR$ usually requires complex surrogate data approaches for properly approximating the distribution of the underlying null hypothesis (no synchronization) adapted to the specific time series under study~\cite{thiel2006b}. On the other hand, comparing the single-system and joint recurrence rates may be insufficient since due to the complexity of fluctuations or the presence of stochastic components (observational noise), we can hardly ever capture all single-system recurrence in the JRP. Consequently, a solely $RR$-based characterization does not necessarily lead to the expected ``optimum'' value of the synchronization index ($JPR=1$) in case of fully developed GS.

		As an alternative, it has been suggested that looking at higher-order characteristics (specifically, three-point instead of two-point relationships) may improve the results~\cite{Feldhoff2013}, especially when relying on probabilistic arguments. One possible way is utilizing again the concept of transitivities from RN and JRN. The exploitation of alternative higher-order characteristics might be possible, but has not yet been explored. JRNs can be analyzed by standard statistical measures from complex network theory~\cite{Newman2003,Donner2010b}, which, however, need to be reinterpreted in terms of the underlying systems' joint recurrence structure~\cite{Feldhoff2011,Feldhoff2012,Donner2012}. Indeed, transitivity properties of joint recurrence networks have been shown to reveal complex synchronization scenarios, notably including the detection of the onset of GS, in coupled chaotic oscillators such as R\"ossler systems~\cite{Feldhoff2012}. Notably, the specific requirements on the time series data render JRNs a promising approach for detecting intricate interconnections between qualitatively distinct observables in observational or experimental real-world data.

		As a heuristic indicator for the presence of GS, one may use the \emph{transitivity ratio}~\cite{Feldhoff2013}
\begin{equation}
\hat{Q}_{\mathcal{T}}=\frac{\hat{\mathcal{T}}^J}{(\hat{\mathcal{T}}^X+\hat{\mathcal{T}}^Y)/2},
\label{eq:qt}
\end{equation}
\noindent
i.e., the ratio between the JRN transitivity and the arithmetic mean of the single-system RN transitivities. The rationale behind this definition is that for systems exhibiting GS, all degrees of freedom are completely locked, implying that both approach the same effective (fractal) dimension and should thus have the same RN transitivities, which approximately equal the JRN transitivity. Alternatively, we could also use other means of $\hat{\mathcal{T}}^{X}$ and $\hat{\mathcal{T}}^{Y}$, such as the geometric or harmonic means, for obtaining an appropriate ratio. However, numerical experiments show that using the arithmetic mean provides values of $\hat{Q}_{\mathcal{T}}$ that are mostly confined to the interval $[0,1]$ with only minor exceedances in the fully developed GS regime~\cite{Feldhoff2013}. One reason for this could be systematic biases of the different transitivity estimators in comparison with their analytical expectation values, which are to be further explored in future work. Since the arithmetic mean is always larger than the geometric one, normalizing with respect to the geometric mean $\sqrt{\hat{\mathcal{T}}^X \hat{\mathcal{T}}^Y}$ would lead to even larger values of $\hat{Q}_{\mathcal{T}}$ and, hence, an even stronger violation of the desired normalization of the transitivity ratio. However, even when considering the normalization by the arithmetic mean of single-system RN transitivities, the thus defined transitivity ratio has two major drawbacks:

		On the one hand, if the single-system RN transitivities are essentially different (a case that has not been studied in~\cite{Feldhoff2013}), the contribution of the lower-dimensional system (higher transitivity) dominates the arithmetic mean in the denominator of Eq.~(\ref{eq:qt}) and, hence, the transitivity ratio itself irrespective of a possible well-defined driver-response relationship.

		On the other hand, there is no rigorous theoretical justification for $\hat{Q}_{\mathcal{T}}$ being a good indicator of GS (as there is no such for $JPR$ either). Notably, the definition of the transitivity ratio is based on the idea that the transitivities are related with the effective dimensions of the individual systems \cite{Donner2015RPBook}. In the uncoupled case, the degrees of freedom of both systems are independent; hence, the effective dimension of the composed system $X\otimes Y$ just reads $D^{X\otimes Y}=D^X + D^Y$ (notably, due to the logarithmic transform between RN transitivity and transitivity dimension, this additivity does \emph{not} apply to the RN transitivities). In turn, in case of GS, the degrees of freedom of both systems become mutually locked, leading to $D^{X\otimes Y}=D^X=D^Y$ (i.e., one system can be viewed as a -- possibly nonlinear -- projection of the other), with $D^X$ and $D^Y$ eventually differing from their values in the uncoupled case depending on the specific coupling configuration (e.g., uni- versus bidirectional coupling). Taking the estimated transitivity dimensions $\hat{D}_{\mathcal{T}^{X,Y}}$ as proxies for $D^{X,Y}$ and the \emph{pseudo-dimension} $\hat{\Delta}_{\mathcal{T}^J}=\log(\hat{\mathcal{T}}^J)/\log(3/4)$ as an approximation of the true dimension $D^{X\otimes Y}$ of the composed system $X\otimes Y$, the latter case would translate into $\hat{Q}_{\mathcal{T}}=1$, which is approximately attained in numerical studies for coupled R\"ossler systems in different dynamical regimes~\cite{Feldhoff2013}. Note that the transitivity dimension of the RN obtained for $X\otimes Y$ reads as $\hat{D}_{\mathcal{T}^{X\otimes Y}}=\log(\hat{\mathcal{T}}^{X\otimes Y})/\log(3/4)$, which is in general not identical to the pseudo-dimension $\hat{\Delta}_{\mathcal{T}^J}$ due to the different metrics used for the definition of recurrences of $X\otimes Y$ and joint recurrences of $X$ and $Y$.

		In order to circumvent both problems, the authors of \cite{Donner2015RPBook} suggested utilizing an alternative indicator, which is directly based on the concept of effective dimensions (degrees of freedom) of the individual systems. In analogy with the mutual information (sometimes also called redundancy \cite{Palus1995,Prichard1995}) frequently used in nonlinear time series analysis, we define the \emph{transitivity dimension redundancies} \cite{Donner2015RPBook}
\begin{eqnarray}
\hat{\tilde{RD}}_{\mathcal{T}}&=&\hat{D}_{\mathcal{T}^X}+\hat{D}_{\mathcal{T}^Y}-\hat{\Delta}_{\mathcal{T}^J}, \\
\hat{RD}_{\mathcal{T}}&=&\hat{D}_{\mathcal{T}^X}+\hat{D}_{\mathcal{T}^Y}-\hat{D}_{\mathcal{T}^{X\otimes Y}},
\end{eqnarray}
which should assume zero values in the uncoupled case and exhibit $\hat{D}_{\mathcal{T}^X}=\hat{D}_{\mathcal{T}^Y}=\hat{D}_{\mathcal{T}^{X\otimes Y}}\approx\hat{\Delta}_{\mathcal{T}^J}$ in case of GS. In order to obtain a normalized measure for the presence of GS, we further define the \emph{dimensional locking index (DLI)}
\begin{eqnarray} \label{eq:DLItilde}
\widehat{\widetilde{DLI}} &=& \frac{\hat{\tilde{RD}}_{\mathcal{T}}}{\hat{\Delta}_{\mathcal{T}^J}}, \\
\widehat{DLI} &=& \frac{\hat{RD}_{\mathcal{T}}}{\hat{D}_{\mathcal{T}^{X\otimes Y}}}.
\end{eqnarray}
\noindent
Notably, this index is tailored to the dimensionality interpretation of RN transitivity. In a strict sense, this argument only applies if using the single-system RN transitivity dimension of the composed system $X\otimes Y$ instead of the JRN transitivity pseudo-dimension $\hat{\Delta}_{\mathcal{T}^J}$. However, at this point, the latter may be used as an approximation. A detailed comparison between the two definitions will be subject to future research.

\begin{figure}
	\centering
	\includegraphics[scale=0.6]{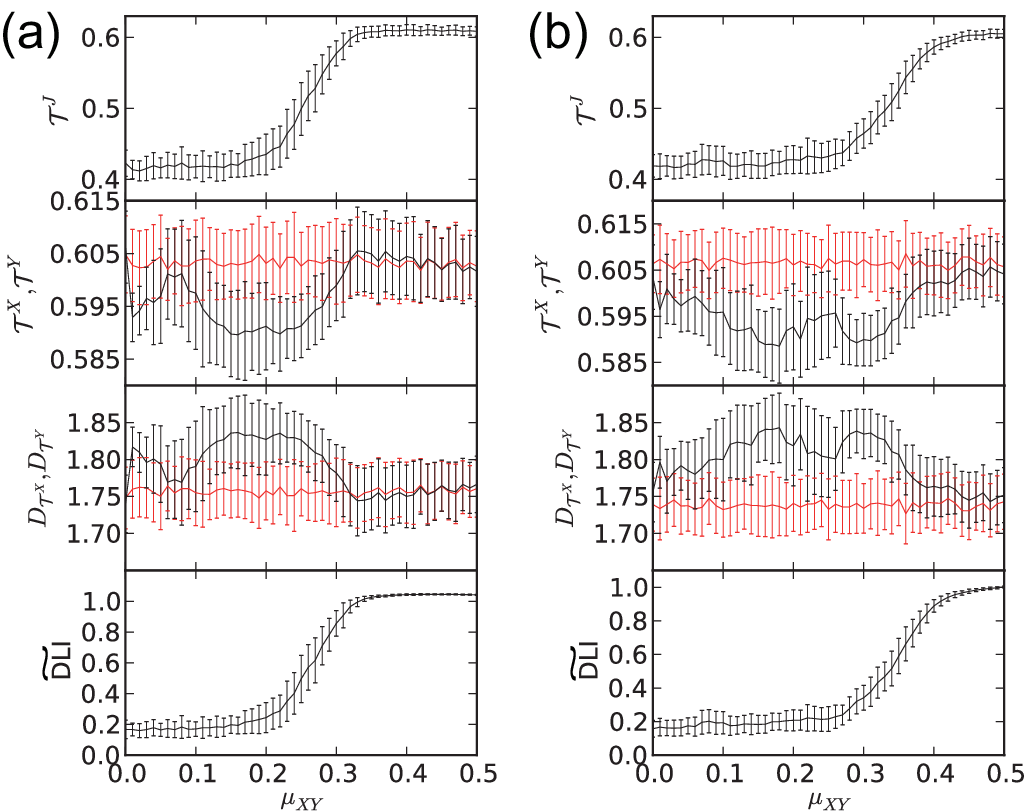}
\caption{Joint transitivity $\hat{\mathcal{T}}^J$, single-system RN transitivities $\hat{\mathcal{T}}^{X,Y}$ (Eq.~\ref{eq:transitivity}), corresponding transitivity dimensions $\hat{D}_{\mathcal{T}^X}$, $\hat{D}_{\mathcal{T}^Y}$ (Eq.~\ref{eq:dtu}) and derived dimensional locking index $\widehat{{\widetilde{DLI}}}$ (Eq.~\eqref{eq:DLItilde}) (from top to bottom) for two unidirectionally coupled R\"ossler systems ($X\to Y$, Eqs.~\ref{eq:coupled_roessler}) with $\nu=0.02$ (a) and $\nu=-0.02$ (b). The error bars indicate mean values and standard deviations estimated from 100 independent network realizations for each value of the coupling strength $\mu_{XY}$. For transitivities and transitivity dimensions, red (black) lines correspond to the values for system $X$ ($Y$). Modified from \cite{Feldhoff2013}. }
\label{fig:roessler_sync}
\end{figure}

		In order to further illustrate the behavior of the (J)RN-based characteristics for detecting the emergence of GS, we reconsider the example of two unidirectionally coupled identical but slightly detuned R\"ossler systems from Section~\ref{sec:coupling}. In contrast to \cite{Feldhoff2013}, \cite{Donner2015RPBook} studied different settings for uni- and bidirectional configurations with single realizations of the same system, we present here results obtained from ensembles of realizations. The results shown in Fig.~\ref{fig:roessler_sync} demonstrate that the estimated values of $\mathcal{T}^J$ and $\widetilde{DLI}$ exhibit a marked increase at the onset of GS. Specifically, the $DLI$ index approaches one (with little overshooting) in the synchronized regime as expected, but takes values of only about $0.2$ or lower in the non-synchronous case (in comparison with values of about $0.7$ exhibited by $Q_{\mathcal{T}}$, cf. Fig.~2B in \cite{Feldhoff2013}).

		As a second important observation, we find a systematic and significant decrease in the RN transitivity of the driven system at moderate coupling strengths before the onset of GS, which corresponds to an increase of the associated transitivity dimension. This behavior is precisely what was claimed in the context of coupling analysis in Section~\ref{sec:coupling} for providing an explanation of the numerically observed asymmetry between the transitivity-based coupled network characteristics. These results underline that some integrated utilization of single-system, inter-system and joint recurrence networks can eventually provide deep insights into the coupling regime and strength from bivariate observations.

	\subsection{Other types of proximity networks} \label{sec:otherproxnets}
		\subsubsection{Cycle networks}
In one of the first works on applying complex network methods to the analysis of time series, Zhang {\textit{el al.}} \cite{Zhang2006} suggested to study the topological features of pseudo-periodic time series. In such case, the underlying dynamical system possesses pronounced oscillations (examples are the well-known Lorenz and R\"ossler systems). Unlike for RNs and their algorithmic variants, we identify the individual cycles $p,q$ contained in a time series of this system with the vertices of an undirected network. Edges between pairs of vertices are established if the corresponding segments of the trajectory behave very similarly. For quantifying the proximity of cycles in phase space, different measures have been proposed. In \cite{Zhang2006b}, Zhang {\textit {et al.}} introduced a generalization of the correlation coefficient applicable to cycles of possibly different lengths. Specifically, this correlation index is defined as the maximum of the cross correlation between the two signals when the shorter of both is slid relative to the longer one. That is, if the two cycles being compared are $C_p=\{\vec{x}_1,\vec{x}_2,\ldots,\vec{x}_m\}$ and $C_q=\{\vec{y}_1,\vec{y}_2,\ldots,\vec{y}_n\}$ with (without loss of generality) $m \leq n$, then we compute
\begin{equation}
r(C_p,C_q)=\max_{i=0,\ldots (n-m)} \left<(\vec{x}_1,\vec{x}_2,\ldots,\vec{x}_m),(\vec{y}_{1+i},\vec{y}_{2+i},\ldots, \vec{y}_{n+i})\right> = r_{pq},
\label{cyclecorr}
\end{equation}
where $\left<\cdot,\cdot\right>$ denotes the standard correlation coefficient of two $\alpha$-dimensional vectors, and set
\begin{equation}
A_{pq}=\Theta(\rho(C_p,C_q) - r_{max})-\delta_{pq}.
\end{equation}
where $\rho_{max}$ is a properly chosen threshold value and $\delta_{pq}$ is again the Kronecker delta necessary in order to obtain a network without self-loops. As an alternative, the phase space distance~\cite{Zhang2006b}
\begin{equation}
D(C_p,C_q)=\min_{i=0,\ldots (n-m)} \frac{1}{m} \sum_{j=1}^{\alpha} \|\vec{x}_j-\vec{y}_{j+i}\| = D_{pq}
\label{psd}
\end{equation}
has been suggested, leading to the following definition:
\begin{equation}
A_{pq}=\Theta(D_{max}-D(C_p,C_q))-\delta_{pq}.
\end{equation}
\noindent
Of course, there are other possible similarity measures that one could use here as well.

		The advantage of cycle networks is that explicit time-delay embedding is avoided. In addition, the method is more robust against additive noise, given a small enough noise magnitude to allow a clear identification of the individual cycles from the time series. Moreover, cycle networks are invariant under reordering of the cycles (this is precisely the same property that was also exploited for cycle-shuffled surrogate methods \cite{Theiler1996} but not the pseudo-periodic surrogate method \cite{Small2001}). However, for chaotic and nonlinear systems in a near-periodic regime, we typically observe significant orderly variation in the appearance of individual cycles. For systems that are linear or noise driven, that orderly variation will be less pronounced. As a consequence, the networks constructed with these methods will have characteristic and distinct properties: linear and periodic systems have cycle networks that appear randomly, while chaotic and nonlinear systems generate highly structured networks \cite{Zhang2006,Zhang2008e}. Therefore, the vertex and edge properties of the resultant networks can be used to distinguish between distinct classes of dynamical systems. Moreover, in \cite{Zhang2006b}, authors used meso-scale properties of the networks --- and in particular the clustering of vertices --- to locate unstable periodic orbits (UPOs) within the system. This approach is feasible, since a chaotic system will exhibit a dense hierarchy of unstable periodic orbits, and these orbits act as accumulation points in the Poincar\'e section. Hence, the corresponding vertices form clusters in the cycle network.

		For an implementation of the cycle network approach, the time series must be divided into distinct cycles. In \cite{Zhang2006,Zhang2008} the preferred method for defining cycles is splitting the trajectory at peaks (or equally troughs). In order to quantify the mutual proximity of different cycles, different measures can be applied depending on the specific application. On the one hand, the cycle correlation index $r_{pq}$ (Eq.~\ref{cyclecorr}) can be properly estimated without additional phase space reconstruction (embedding), which has advantages when analyzing noisy and non-stationary time series, e.g., experimental data~\cite{Zhang2006}. Moreover, this choice effectively smoothes the effect of an additive independent and identically distributed noise source~\cite{Zhang2006b}. On the other hand, the phase space distance $D_{pq}$ (Eq.~\ref{psd}) is physically more meaningful~\cite{Zhang2008}. For example systems as well as some real-world clinical electrocardiogram recordings studied in \cite{Zhang2006,Zhang2008}, both methods have been found to perform reasonably well. However, whether the previously considered approaches also lead to feasible results for other cases has to be further investigated in future research.

		In general, the construction and quantitative analysis of cycle networks requires a sufficiently high sampling rate, i.e., we require that both cycle lengths $m$ and $n$ in Eqs.~(\ref{cyclecorr}) and (\ref{psd}) are reasonably large. The main reason for this requirement is that even two cycles that are fully identical but sampled in a different way may have rather different cycle correlation indices (and phase space distances) depending on the exact values of the observed quantity. Hence, for a very coarse sampling, it is possible that two cycles that are actually close in phase space may not be connected in the cycle network. However, for large sampling rates, the variance of this measure decreases, resulting in a more reliable network reconstruction.

		Instead of computing correlation coefficients to quantify the linear correlation between two cycles, the mutual information could be used to capture nonlinear effects in a time series, as it may provide more accurate estimates of the similarity of nonlinear time series. In this context, in \cite{Emmert2011} the authors defined a node in the constructed network as an \emph{episode}, i.e., a time interval of a given series that may consist of $n_e \ge 1$ consecutive cycles. This means that such an episode is $n_e$ times longer than a cycle. The extended length of an episode, compared to a cycle, has the advantage of increasing the accuracy of statistical estimates of  mutual information. Note that a cycle does not need to have a certain minimal length to qualify as a cycle. However, it is clear that very short cycles convey less information about the time series than long cycles. Due to the fact that the notion of a ``cycle" is parameter free, one cannot adjust for this shortcoming. For this reason we extend the general idea behind the usage of a cycle in the construction of a network \cite{Zhang2006} in terms of an episode. Moreover, the corresponding network construction algorithm is a parametric method because an episode is a function of $n_e$, the number of consecutive cycles. This gives us a parameter $n_e$ that can be optimized to result in the ``best" network for a given time series.

		The choice of the threshold $r_{max}$ (alternatively, of $D_{max}$) influences the link density of the resulting network, which could be discussed in a similar framework as for constructing recurrence networks. One solution is to study the dependence of network characteristics on $r_{max}$ explicitly \cite{Zhang2006}.

        In terms of possible applications of this approach, Zhang {\textit {et al.}} constructed cycle networks for sinus rhythm electrocardiogram recordings of coronary care unit patients and healthy volunteers. It has been demonstrated that the degree distributions of the resulting networks for affected patients show more prominent variations in comparison to those of healthy volunteers which vary rather smoothly. Other network measures including clustering coefficients and average path length also show significant differences between healthy and coronary care patients. Furthermore, the cycle network has been applied for characterizing electrical signals of acupuncture \cite{Men2011}, showing different network topologies when the control parameter is in different regimes, for instance, either twisting or lifting and thrusting conditions.

		Most of the vertex and edge properties of cycle networks have been explained by UPOs of the underlying chaotic systems \cite{Zhang2006}. Since UPOs are crucial to the understanding of chaotic systems, Kobayashi {\textit{et al.}} performed a network analysis of UPOs \cite{Kobayashi2017}. By means of the Poinar{\'e} map, they first numerically extracted a large number of UPOs, which were considered as vertices of the network. Note that most of the existing algorithms can only detect UPOs of lower orders, which have been sufficient for characterizing the properties of the underlying chaotic system \cite{Cvitanovic1988,Grebogi1988}. The edges between two UPOs are established by a transition process of a typical chaotic orbit. More specifically, if a typical chaotic orbit $\{ \vec{x}_i \}$ travels close to UPO$_p$ at time $i$ and later shifts to the neighborhood of UPO$_q$ at time $i+1$, we consider a connection between these two UPOs. Due to the chaotic nature of the orbit, the transitions between different UPOs are irregular. The resulting network presents SW and SF features, which confirm the results reported earlier in \cite{Zhang2006}.

		\subsubsection{Correlation networks}\label{sec:correlationnetworks}
		Generalizing the idea of cycle networks to arbitrary time series, individual state vectors $\vec{x}_i$ in the $m$-dimensional phase space of the embedded variables can be considered as vertices of an undirected complex network. In contrast to the standard RN approach, let us consider the case of $m$ being very large (i.e., the time series has been over-embedded). Specifically, if the Pearson correlation coefficient $r_{pq} = \left<\vec{x}_p,\vec{x}_q\right>$ between two sequences $p$ and $q$ is larger than a given threshold $r^*$, the corresponding vertices $p$ and $q$ are considered to be connected \cite{Yang2008,Gao2009}:
\begin{equation}
A_{pq}=\Theta(r^*-r_{pq})-\delta_{pq}.
\end{equation}
Interpreting $1-r_{pq}$ as a proximity measure, the condition $r_{pq}\geq r^*$ corresponds to the definition (Eq.~\ref{eq:rn_definition}) of a recurrence with $\varepsilon=1-r^*$. The consideration of correlation coefficients between two phase space vectors usually requires a sufficiently large embedding dimension $m$ for a proper estimation of $r_{pq}$. This high value of $m$ often includes several oscillation periods as compared to a cycle network. Hence, information about the short-term dynamics might get lost. Moreover, since embedding is known to induce spurious correlations \cite{thiel2006}, the results of the correlation method for network construction may suffer from related effects.

The correlation network method has been applied to stock price series, unveiling Gaussian distributions for the degree sequences that are constructed from return and amplitude series \cite{Yang2008}. Furthermore, different two-phase (gas-liquid) flow patterns have been well characterized by correlation network approaches \cite{Gao2009}.

		Statistical concerns regarding the Pearson correlation coefficient arise when smaller values of $m$ are used, say $m = 10$, which requires more statistical robust measures. In \cite{Hou2014}, Hou {\textit{et al.}} proposed to use the inner composition alignment measure (IOTA), which is a permutation based measure that was originally introduced to identify couplings from rather short gene-expression data \cite{Hempel2011}, to quantify the connectivity strength between two embedding vectors. As compared to the standard symmetric (undirected) correlation network, a directed correlation network is obtained, which has been further applied to characterize pathological changes in the cardiovascular system from short-term heartbeat time series \cite{Hou2014}.

%% file: Chapter04_VisibilityGt/Chapter04_VisibilityGt.tex
\section{Visibility graphs}\label{sec:VisibilityGt}
	
	Another approach for transforming time series into complex network representations, which has recently attracted great interest, is the visibility graph (VG) algorithm. Originally, this concept has been introduced for the analysis of mutual visibility relationships between points and obstacles in two-dimensional landscapes in the framework of computational geometry, with applications ranging from robot motion planning to architectural design and topographic descriptions of geographical space \cite{Lozano1979,Nagy1994,Floriani1994,Turner2001}. Lacasa \emph{et al.}\cite{Lacasa2008} adopted the VG approach to the analysis of structures in scalar, univariate time series. Since this seminal work, different algorithmic variants of the original VG algorithm have been proposed, which we will summarize in course of this section. In general, visbility algorithms constitute a family of geometric and ordering criteria for scalar real-valued time series, providing a combinatorial representation of the underlying dynamical system \cite{Lacasa2018}.
	
	Some mini-review of VGs has already been presented in \cite{Nunez2012,Luque2016}. In particular, the application of this approach to geophysical time series has been focused on in \cite{Donner2012}, which has attempted to link the complete variety of different network properties describing the structure of VGs with specific structural features of geophysical processes in some more detail. Here, we summarize the recent developments of the method and discuss some practical issues which pose considerable challenges to VG analysis for experimental time series, such as missing data, homo- and heteroscedastic uncertainty of observations, and time-scale uncertainty. These practical problems for VG analysis have not yet fully answered since the early work of \cite{Donner2012}. In addition, we also discuss some successful applications of VGs and related methods to testing time-irreversibility of nonlinear time series. 
	
	\subsection{Algorithmic variants of visibility algorithms}
	In VG analysis, individual observations are considered as vertices. For instance, given a univariate time series $\{x_i \}_{i=1}^{N}$ with $x_i = x(t_i)$, the binary adjacency matrix $\mathbf{A}$ has size ${N \times N}$. Depending on the particular visibility conditions in defining the edges of the resulting graph, we can distinguish different versions of VGs. 
	
		\subsubsection{Natural visibility graphs}
\begin{figure}
		  \centering
		  \includegraphics[width=0.8\columnwidth]{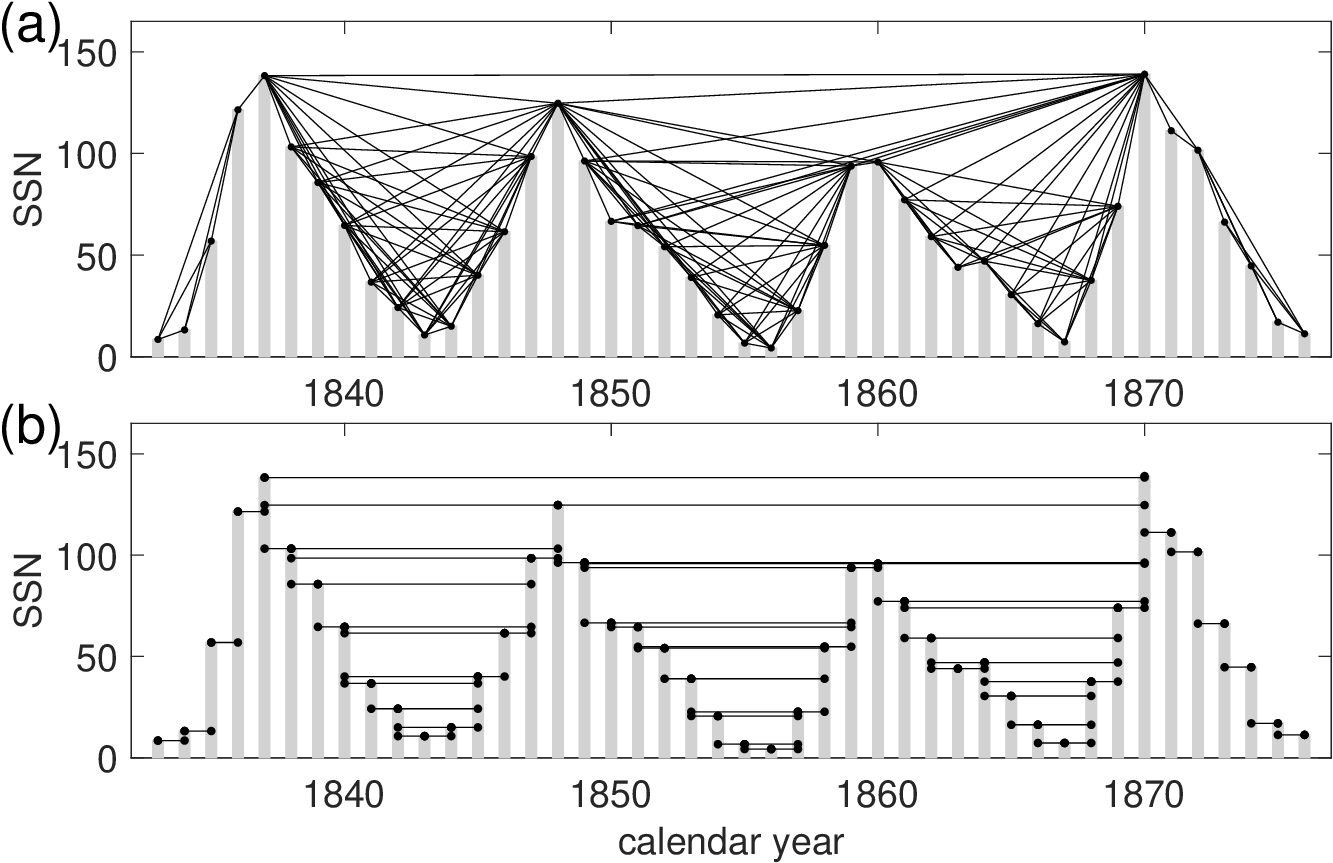}
		  \caption{Schematic illustration of the algorithm for constructing (a) natural visibility graphs and (b) horizontal VG for an excerpt of the time series of sunspot numbers. Reproduced from \cite{Zou2014}. \label{fig_chap04:timeseriesSS}}
		\end{figure}

		First, in the framework of the standard visibility graph (VG), the non-zero entries of $\mathbf{A}$ correspond to two time points $t_i$ and $t_j$ which are mutually connected vertices if the criterion 
\begin{equation} \label{vis_cond}
	\frac{x_i-x_k}{t_k - t_i} > \frac{x_i - x_j}{t_j - t_i}
\end{equation}
is fulfilled for all time points $t_k$ with $t_i < t_k < t_j$~\cite{Lacasa2008}. Therefore, the edges of the network take into account the temporal information explicitly. In Fig.~\ref{fig_chap04:timeseriesSS}(a), we illustrate the algorithm of constructing natural VGs for an example of a the sunspot time series. More detailed information on recent results obtained by VGs analysis of the sunspot series will be provides in Section~\ref{sec:appVGs}. 

By default, two consecutive observations are always connected in a VG, so that the graph forms a completely connected component without disjoint subgraphs. However, unlike for other types of time series networks, there are potentially relevant boundary effects, for instance, the first time point can only be visible to points that are in the future of this observation (Fig.~\ref{fig_chap04:timeseriesSS}(a)), see below for a detailed discussion. In turn, as an advantage, the VG is not affected by the choice of any algorithmic parameters -- in contrast to most other methods of constructing complex networks from time series data which depend on the choice of some parameters (e.g., the threshold $\varepsilon$ of recurrence networks, see more details in Section \ref{sec:RecurrenceNt}). 
		
		\subsubsection{Horizontal visibility graphs}
		As a notable modification of the standard VG algorithm, Luque \textit{et al.} \cite{Luque2009,Lacasa2010} proposed utilizing a simplified criterion of horizontal visibility for transforming a time series into a complex network. Specifically, they considered two observations made at times $t_i$ and $t_j$ to be connected in a horizontal visibility graph (HVG) if and only if
\begin{align}
x_k < \min \left\{ x_i, x_j \right\} \label{eq:hvc}
\end{align}
for all $t_k$ with $t_i<t_k<t_j$. 

		The algorithmic difference between HVG and VG is illustrated in Fig.~\ref{fig_chap04:timeseriesSS}(b). Note that the geometric criterion defining the HVG algorithm is more ``visibility restrictive" than its analogous for the standard VG. That is to say, the nodes within a HVG will have ``less visibility" than their counterparts within a VG. It is easily seen that the edge set of the HVG associated with a given time series is a subset of the edge set of the associated VG, which means that if the horizontal visibility criterion in Eq.~(\ref{eq:hvc}) is fulfilled, then also Eq.~(\ref{vis_cond}) holds, but not necessarily vice versa. In addition, VGs are invariant under affine transformations of the entire time series, whereas HVGs are not. One notable advantage of HVGs is that they provide an even higher degree of algorithmic simplicity than standard VGs, resulting in the observation that for certain simple stochastic processes and the quasiperiodic transition route to chaos, some basic graph properties can be calculated analytically \citep{Luque2009,Luque2013a,Luque2013d}. On the other hand, the fact that HVGs typically contain a lower number of edges increases the demands regarding the time series length relative to those of the standard VG when using this approach in applications, such as tests for time-reversal asymmetry \citep{Donges2013} (see Section~\ref{sec:timeIRvg}). 
		
        To further account for the differences between natural and horizontal visibility graphs, Zhu \textit{et al.} \cite{Zhu2014b} introduded the concept of \emph{difference visibility graphs (DVGs)}, which are based on the node and edge sets of the natural VG with the edges of the HVG being removed, i.e., $E_{DVG}=E_{VG}\backslash E_{HVG}$ and $V_{DVG}=V_{VG}=V_{HVG}$. This construction immediately implies that the degree sequence of the DVG is defined as the (strictly non-negative) difference series between the degree sequences of the VG and HVG, respectively. Wang \textit{et~al.} \cite{Wang2017} recently demonstrated that the degree properties of DVGs obtained from EEG data can be employed to differentiate between different types of epileptic seizures.
        
		\subsubsection{Other variants of (H)VG}
		Throughout this section, we will mostly focus on the discussion of the standard VG and horizontal VG algorithms and their applications. However, there are some other generalizations of these two algorithms that will be briefly summarized in the following. Recent applications of the (H)VG algorithms and their variants to experimental time series of various origins will be reviewed in Section \ref{sec:appVGs}. 
		
		Given the definitions of VG and HVG, the resulting graphs are undirected and unweighted. One straightforward generalization of (H)VGs to \emph{directional (H)VGs} is to introduce directed edges between vertices, i.e., from the cause at $t_i$ to the effect at $t_j > t_i$. As it will be shown in Section~\ref{sec:timeIRvg}, such directed graphs provide information on time-reversal asymmetry of the considered time series. 
        
        Since an edge represents the visibility between two time points $t_i$ and $t_j$ that can be either consecutive in time or be separated by various other observations, another generalization of the original ideas of (H)VGs is to construct a \emph{weighted (H)VG} which takes into account the time distance between $t_i$ and $t_j$ \cite{Bianchi2017}. More specifically, the weight $w_{ij}$ is defined as $w_{ij} = 1 / \sqrt{(t_j - t_i)^2 + (x_j - x_i)^2} \in [0, 1]$. Thereby, these weights capture time distance information $(t_j - t_i)$ as well as amplitude differences $(x_j-x_i)$ of two data points connected by the respective visibility rule. The corresponding approach has been applied to characterize heterogeneity of recurrent neural network dynamics \cite{Bianchi2017}. 
		
		There are several further algorithmic variants of (H)VG that address specific properties of time series. For instance, given a binary series, in \cite{Ahadpour2012}, a simplified VG has been developed yielding a \emph{binary VG}, in which the visibility condition (Eq.~\ref{vis_cond}) is reduced to $x_i + x_j > x_k$ for all $t_k$ such that $t_i < t_k < t_j$. The resulting VG from binary series is always connected and undirected and more easily tractable than the standard VGs. 
		
		\emph{Parametric VGs} have been proposed in \cite{Bezsudnov2014,Snarskii2013a}, introducing ``viewing angle'' $\alpha$. When this angle $\alpha = \pi$, the parametric VG and the standard VG are the same. However, the angle $\alpha = \pi / 2$ does not turn into HVG because $\alpha$ introduces a direction of links in the resulting graph. By means of this algorithm, it is possible to study the dependence of network structural measures on the parameter $\alpha$. 
		
		The concept of \emph{limited penetrable VG} (LPVG) \cite{Zhou2012,Gao2013e} can be regarded as a continuous construction of a HVG based on a properly coarse-grained time series. In the original HVG, two time points $t_i$ and $t_j$ are connected if no other intermediate points $x_k$ are larger than $\min{x_i, x_j}$. Now we use a less restrictive criterion, allowing one of $x_k$ to be larger than $\min{x_i, x(_j}$, as represented by the new parameter of $L=1$. Similarly, we allow two points $x_k$ larger than $\min{x_i, x_j}$ if $L = 2$, etc. When increasing $L$, there are more edges in the resulting LPHVGs as compared to the standard HVG. The standard VG is recovered from the LPVG if $L \to \infty$. A straightforward extension in terms of the multi-scale \emph{limited penetrable HVG algorithm} (LPHVG) has been proposed in \cite{Gao2016,Pei2014a} and successfully applied to the analysis of EEG time series \cite{Wang2016} and electromechanical signals \cite{Wang2016b}. Recently, the limited penetrable VG algorithm has been combined with a parametric VG in \cite{Li2018}. Moreover, the key topological characteristics of LP(H)VGs have been studied in great detail in a series of papers \cite{Wang2018,Wang2018b,Wang2018c}.
				
		An extension of (H)VGs from a univariate time series to scalar fields has been recently reported in \cite{Xiao2014a,Lacasa2017}, which is conceptually closer to the original idea of visibility graphs. In addition, one may reconstruct (H)VGs for a set of ordered data (either in descending or increasing order) \cite{Wang2015a}. 
						
		It is also possible to combine the concepts of VGs with transition networks (Section~\ref{sec:TransitionNt}), e.g., by using the \emph{visibility graphlet} approach proposed in \cite{Mutua2015,Mutua2016a}. In this case, VGs are constructed for sliding windows, and each window is regarded as a network node. The connection between two nodes represents their temporal succession. The resulting transition network is regular for periodic dynamics while showing more complex structural properties for chaotic systems \cite{Mutua2016a}. The computation of network measures (including the average path length and network diameter) have successfully tracked the different bifurcations from period doubling and intermittency to chaos in the logistic map. 
		
		In all above cases, (H)VGs have been constructed from a given time series. In \cite{Tsiotas2018}, Tsiotas {\textit {et al.}} expanded the VG algorithm to analyze node attributes of a given graph. More specifically, let us consider a graph $G(V, E)$ and a node attribute $Y$ that might be any of the node-wise network measures, e.g., local clustering coefficient or betweenness centrality. The secondary VG analysis for node-wise attributes $Y$ shows a specific capability in pattern recognition \cite{Tsiotas2018}. 
				
		Other properties of combinatorics have been used to characterize HVGs successfully in \cite{Gutin2011}. Recently, some analytic results have been obtained for independent and identically distributed random noise, which has an exponential degree distribution \cite{Wang2018}. 

		The time complexity of the basic natural VG algorithm is $O(N^2)$, which means that it takes a lot of time when dealing with long time series. A faster transform algorithm has been proposed in \cite{Lan2015a} to reduce the computation time, showing much more efficient time complexity $\sim O(N \log N)$. Note that for the HVG algorithm, it is not possible to further improve the computational efficiency because it has already reached the lower bound $\sim O(N)$. 
		
	\subsection{Visibility graph properties}
		\subsubsection{Degree distributions}
		A vast body of early works on (H)VGs has mainly concentrated on the properties of the degree distribution $p(k)$ resulting from different kinds of processes. Specifically, VGs obtained from periodic signals appear as a concatenation of a finite number of network motifs (given that the basic period is an integer multiple of the sampling rate), i.e., have a regular structure with only a few distinct values of the vertex degree. The opposite extreme case, white noise, yields VGs appearing as exponential random graphs, i.e., random networks characterized by an exponential degree distribution. For example, exponential degree distribution have been reported for wind speed records measured in central Argentina \cite{Pierini2012}. 
		
		In fractal processes, numerical results suggest that $p(k)$ exhibits a power law \cite{Lacasa2008}, $p(k) \sim k^{-\gamma}$. Taking this empirical observation, VG analysis has been suggested to characterize fractional Brownian motions and $f^{\beta}$-noise, finding some heuristic relationship between $\gamma$ and the process Hurst exponent $H$ as $\gamma = 3 - 2H$, and $\gamma = 5 - 2H$ for fractional Gaussian noise \cite{Lacasa2009,Ni2009}. Depending on the fractal properties of the underlying process, recently a resampling algorithm for constructing VGs from segmented time series has been proposed \cite{Ahmadlou2012}, which estimates power-law exponents reflecting SF properties quite well. This improved algorithm has been applied to diagnose Autism spectrum disorders \cite{Ahmadlou2012}. Since there are many concerns regarding the statistical justification of the power laws of VGs, one can directly analyze the degree sequence (instead of the distribution) by detrended fluctuation analysis \cite{Czechowski2016}, which quantifies the multifractal properties better than standard VGs analysis.  
	
		Some exact results of $p(k)$ for HVGs associated with generic uncorrelated random series have been obtained in \cite{Luque2009}. More specifically, for a bi-infinite time series created from a random variable $X$ with the probability distribution $p(x)$ and $x\in [0, 1]$, it has been proven that the degree distribution of the graph has an exponential form 
		\begin{equation}\label{pkhvg}
		p(k) = \frac{1}{3} \left(\frac{2}{3}\right)^{k-2}. 
		\end{equation}
Interestingly, for every probability distribution $p(x)$ of uncorrelated random series, we find the same exponential form for the HVG's degree distribution. Numerical results for $p(x)$ with a uniform, Gaussian and power-law form (e.g., $p(x) \sim x^{-2}$) show perfect agreements with this theoretical prediction \cite{Luque2009}. A general diagrammatic theory has been proposed in \cite{Lacasa2014b} to compute $p(k)$ for any given dynamical process with well-defined invariant measure. Taking into account the time information explicitly as in so-called directed HVG (as will be explained below), the outgoing degree distribution $p_{out}(k) = (1/2)^{k}$. Further solvable examples include Markovian processes with an integrable invariant measure $p(x)$, for instance, the stationary Ornstein-Uhlenneck process, and one-dimensional chaotic and quasi-periodic maps of with smooth invariant measure. In addition, the mean degree $\left< k \right>$ of the HVG associated with an uncorrelated random process is then given by 
\begin{equation}
\left< k \right> = \sum k p(k) = \sum_{k=2}^{\infty} \frac{k}{3}\left(\frac{2}{3}\right)^{k-2} = 4.
\end{equation}
For an infinite periodic series of period $T$, the mean degree is $\left< k \right> = 4 ( 1 - \frac{1}{2T})$. 
	
		Similar to VGs, HVGs have been successfully applied to studying time series from various fields of sciences. We particularly notice a recent paper \cite{Yu2012} which studied the multifractal properties of some solar flare index in terms of HVG characteristics. Among others, the properties of HVGs have been studied in river flows \cite{Braga2016}, showing exponential degree distributions. 
		
		In order to understand the hypothetical SF property of $p(k)$ for VGs from fractal records, one has to note that typically, maxima of the time series have visibility contact with more other vertices than other points, i.e., hubs of the network often form at maximum values of the recorded observable. Put it differently, the degree of a vertex in the VG characterizes the maximality property of the corresponding observation in comparison with its neighborhood in the time series. Although locally large time series values have better visibility than other small values, hub nodes of large degrees of VGs do not necessarily correspond to higher values, especially when there is some sort of periodic trend in the given data sequence, for instance, in wind speed records \cite{Pierini2012,Zou2014a}. Therefore, the relationship between maxima time series points and hubs of VGs is not completely general, since there can be specific conditions (e.g., a concave behavior over a certain period of time) which can lead to highly connected vertices that do not coincide with local maxima, for example, in case of a Conway series \cite{Lacasa2008}. 
				
		In addition, studying the minima of a given time series provides complementary insights for the understanding of the particular process, for instance, in case of sunspot series \cite{Zou2014a}. In standard VGs, the contributions of local minimum values are largely overlooked by the degree distribution $p(k)$ because minimum values are basically represented by non-hubs. One simple solution is to study the negative counterpart of the original time series, $\{-x_i\}$, the VG of which highlights the properties of the local minima \cite{Zou2014a}. For convenience, we use $k^{-x}_i$ and $P(k^{-x})$ to denote the degree sequence and distribution of VG resulting from $\{-x_i\}$. Here, we remark that this simple inversion of the time series allows us to create an entirely different complex network. This technique has demonstrated to be useful to understand the long-term behavior of strong minima of solar activity \cite{Zou2014a}. We will review these results in Section~\ref{subsec:sunnum}. 
		
		The graph entropy is often computed as $\mathcal{S} = - \sum_{k} p(k) \log p(k)$ and is used as the approximation to the Shannon entropy $S$ of the corresponding time series $\{x_i\}$ \cite{Luque2016,Goncalves2016}. Furthermore, based on the degree sequences, the VG aggregation operator has been proposed in \cite{Chen2014,Jiang2016}. This operator includes temporal information in the weights of the aggregation, showing computational simplicity comparing to other traditional aggregation operators.  
					
		\subsubsection{Distinguishing stochastic and deterministic dynamics}
		For the HVG, exponential functional forms, $p(k) \sim e ^{-\lambda k}$, have been found for many random processes. A scaling factor of $\lambda_c = \ln (3/2)$ has been found in the case of uncorrelated noise (white noise), which has been further proposed to separate stochastic from chaotic dynamics in the following senses \cite{Lacasa2010,Lacasa2014b,Ravetti2014}: (i) correlated stochastic series are characterized by $\lambda > \lambda_c$, slowly tending to an asymptotic value of $\ln (3/2)$ for very weak correlations, whereas (ii) chaotic series are characterized by $\lambda_{chaos} < \lambda_c$ for decreasing correlations or increasing chaos dimensionality, respectively \cite{Lacasa2010}. In \cite{Zhang2017}, Zhang {\textit{et al.}} have provided some further examples supporting argument (i). Meanwhile, some peculiar results have been found indicating that $\lambda_c$ should not be interpreted as a general critical value separating chaos from noise \cite{Ravetti2014}. 
		
		Let us focus on applying (H)VG analysis to auto-regressive (AR) stochastic processes, which often describe a general model for colored noise as an idealization of time-varying processes in nature, economics, etc. The AR model specifies that the output variable depends linearly on its own previous values and on a stochastic term (Eq.~\ref{def:AR}). More specifically, we perform both VG and HVG analysis for stationary AR(1) processes, i.e., with $|\varphi_1| < 1$ where $\varphi_1$ is the characteristic parameter of the AR(1) model. It is known that $\varphi_1 > 0$ corresponds to positive correlations and the correlation length increases when $\varphi_1$ increases from 0 to 1. In contrast, anti-correlation is observed for negative coefficient $\varphi_1$. Similar H(VG) analysis for the AR(2) model has been reported in \cite{Zhang2017}. 

\begin{figure}
	\centering
	\includegraphics[width=\columnwidth]{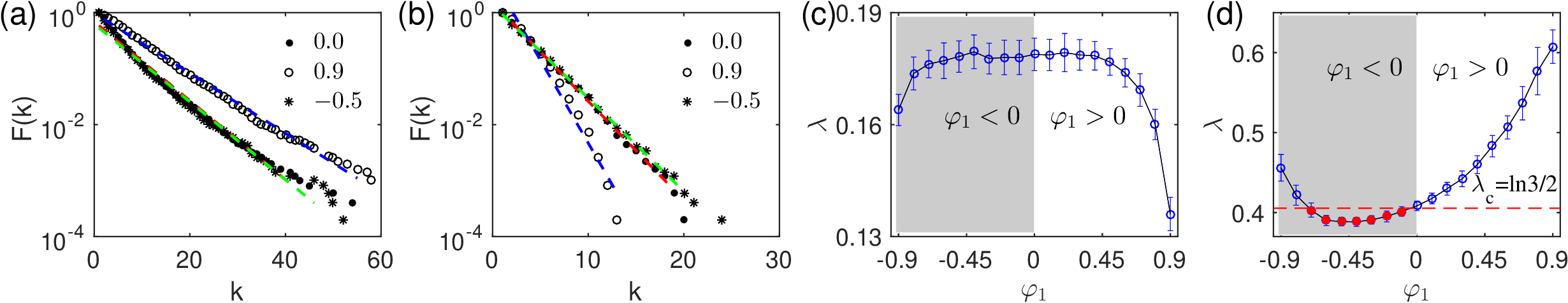}
\caption{(color online) (a, b) Estimates of $\lambda$ for approximately exponential degree distributions of the AR$(1)$ process. (c, d) $\lambda$ versus $\varphi_1$. (a, c) VG, and (b, d) HVG. Each dot in panels (c) and (d) represents an average over 50 independent random realizations of 5000 data points. In (d), $\lambda$ values smaller than $\ln 3/2$ are highlighted by red color. Reprinted from \cite{Zhang2017} with permission of the original publisher. \label{fig:lambda_AR1}}
\end{figure}

		In the case of $\varphi_1 > 0$, we find that $p(k)$ approximately follows an exponential distribution. To illustrate this finding, the complementary cumulative degree distributions $F(k)$ for $\varphi_1 = 0.3 $, $0.9$ and $-0.5$ are shown in Fig. \ref{fig:lambda_AR1}(a) and (b), where clear scaling regimes are present in the semi-logarithmic plots. Furthermore, when increasing $\varphi_1$, in the VG, the exponent $\lambda$ shows a monotonically decreasing trend (Fig.~\ref{fig:lambda_AR1}(c)). In contrast, the value $\lambda$ for the HVG is increased (Fig.~\ref{fig:lambda_AR1}(d)). The result of Fig.~\ref{fig:lambda_AR1}(d) confirms the hypothesis stated in \cite{Lacasa2010} that all $\lambda$ should be larger than $\lambda_c = \ln (3/2)$ as the correlation length is increased in the case of positively correlated increments.

		In turn, when $\varphi_1 < 0$, we observe some peculiar results that seem to contradict the original hypothesis of \cite{Lacasa2010} that $\lambda$ should be larger than $\lambda_c$ ($\lambda > \lambda_c$) in stochastic processes. Notably, the results of Fig.~\ref{fig:lambda_AR1}(d) do not support this claim when $\varphi_1$ is negative in the AR(1) model. Instead, we find a region where the slope of the exponential degree distribution is smaller than $\ln (3/2)$ (as highlighted in Fig.~\ref{fig:lambda_AR1}(d)). This suggests that the critical value of $\ln (3/2)$ should not be understood as a general law for separating correlated stochastic from chaotic processes, which requires further investigation. 

		Working with correlated stochastic time series, further results in \cite{Manshour2015} do not adequately support the arguments of exponential degree distributions as reported in \cite{Lacasa2010}. More specifically, they have constructed (H)VGs for fractional time series with three different methods, a generic $1 = 1 / f^{\beta}$ noise constructed by Fourier filtering, a deterministic fBm process of the Weierstrass-Mandelbrot function, and a stochastic fBm process generated with a successive random addition method. Numerical analysis shows that the VG algorithm may not provide a good method to extract correlation information of a time series and its statistics is not essentially the same as that of the HVG. The degree distributions of HVGs are shown to have parabolic exponential forms with the fitting parameter depending on the Hurst exponent \cite{Manshour2015}. 
		
        \subsubsection{Degree sequences of horizontal visibility graphs}
        
        Due to their specific construction procedure, the degree sequences of HVGs carry essential information on the properties of the system under study. Notably, it can be shown that there exists a bijection between the degree sequence and the associated adjecency matrix of HVGs \cite{Luque2017b}. Even more, the degree sequence of a HVG can be interpreted as a symbolic discretization of the underlying time series (i.e., a transformation from real into integer-valued observations). By studying the scaling of the associated block entropies of degree subsequences, Lacasa and Just \cite{Lacasa2018} demonstrated the convergence of these block entropies towards the Kolmogorov-Sinai entropy of the system, indicating that the degree sequence asymptotically fulfills the properties of a generating partition of the system under study and, hence, provides an encoding without information loss.

		\subsubsection{Local network properties}
		\paragraph{Local clustering coefficient $\mathcal{C}_i$} 
		In the case of VGs, the local clustering coefficient $\mathcal{C}_i$ and its relationship with the degree $k_i$ have been numerically studied recently in human heartbeat data \cite{Shao2010}. Particularly, it has been observed that $\mathcal{C}(k) \sim k^{-\gamma}$ and $\gamma = 1$, pointing to a hierarchical organization of the network \cite{Albert2002}, since vertices $i$ with high $\mathcal{C}_i$ and low $k_i$ (which are most abundant) form densely connected subgraphs, indicating a strong modular structure of the VG. 

		In the case of HVGs associated an uncorrelated random series, $\mathcal{C}_i$ can be easily deduced by means of geometrical arguments. For a given vertex $i$, $\mathcal{C}_i$ denotes the fraction of nodes connected to $i$ that are connected between each other. In other words, we have to calculate from a given vertex $i$ how many vertices from those visible to $i$ have mutual visibility (triangles), normalized with the cardinality of the set of possible triangles, $\binom{k}{2}$, where $k$ is the degree of vertex $i$. Based on a general rule between the degree $k$ and the local clustering coefficient
		\begin{equation}
		\mathcal{C}_v(k) = \frac{k-1}{\binom{k}{2}} = \frac{2}{k}, 
		\end{equation}
one obtains the distribution $p(\mathcal{C})$ as 
		\begin{equation}
		p(\mathcal{C}) = \frac{1}{3} \left(\frac{2}{3}\right)^{2/\mathcal{C} -2}. 
		\end{equation}
The above theoretical result has been numerically confirmed for uncorrelated random series \cite{Luque2009}. In addition, for HVG of a binary sequence, $p(\mathcal{C})$ has a simplified expression as reported in \cite{Ahadpour2012}. 
			
		\paragraph{Betweenness $b_i$}
		In many cases local maxima of the underlying time series are expected to have large values of betweenness because high values often correspond to hubs in VGs which separate different parts of the series without mutual visibility contact and, thus, act as bottlenecks in the network structure, bundling a large number of shortest paths between vertices at $t < t_i$ and $t > t_i$, respectively. However, in contrast to $k_i$, $b_i$ is additionally affected by the vertex' position in the underlying time series due to a simple combinatorial effect: Considering that the majority of shortest paths that cross a vertex $i$ connecting observations before and after $i$ with each other, there are more possible combinations of such points for $i$ being close to the middle of the time series than for vertices close to the edges of the record. In this respect, in a VG betweenness centrality of a vertex mixes information on the local maximality of the corresponding observation and its position within the time series.
		
		\paragraph{Closeness centrality $c_i$}
		The position of a vertex in the time series is even more important for the closeness centrality $c_i$. Specifically, $c_i$ is strongly determined by the number of vertices to its left and right, respectively. In this spirit, it can be argued that in the middle of the time series, high values $c_i$ are more likely than at its ends. As argued above, a similar (but weaker) effect contributes to betweenness and - close to the edges of the record - also to the degree (consequently, the highest degree and betweenness values can be taken by other vertices than that corresponding to the global maximum). In contrast, $\mathcal{C}_i$ is almost unaffected except for vertices very close to the beginning and end of the time series, since direct connectivity is mainly established between vertices that correspond to observations that are not very distant in time. 
		
		From the above discussion, we note that boundary effects play an important role in the computation of the local centrality measures as that has been numerically reported in \cite{Donner2012}. Except for $\mathcal{C}_i$, the impact of boundaries on the estimated vertex properties is particularly strong for short records, which is typical for many observational time series. In particular, degree and other centrality properties of observations close to both ends of a time series are systematically underestimated, which may artificially alter the interpretation of the corresponding results in their specific context. Hence, a careful treatment and interpretation of the results of VG analysis is necessary in such cases \cite{Donner2012}.
				
		\subsubsection{Global network properties}
		The edge density $\rho$ of a (H)VG is a true network characteristic rather than a parameter of the method, which is in contrast to other approaches to complex network based time series analysis (for instance recurrence networks in Section \ref{sec:RecurrenceNt}). Specifically, a maximum edge density of 1 would be present if the underlying time series is globally convex (e.g., of regular parabolic shape), whereas low values indicate a strong fragmentation of the VG and, hence, irregularity of fluctuations of the underlying observable. 
		
		For a holistic characterization of a (H)VG, $\mathcal{C}$ and $\mathcal{L}$ have attracted particular interest, since their common behavior gives rise to a mathematical evaluation of the SW phenomenon, i.e., the emergence of real-world networks with a high degree of clustering $\mathcal{C}$ and a short average path length $\mathcal{L}$ \cite{Watts1998}. The corresponding characterizations of HVGs reconstructed from fractional Brownian motions (fBm) with different Hurst indices $H \in (0, 1)$ have been reported in \cite{Xie2011}. It was found that the clustering coefficient $\mathcal{C}$ decreases when $H$ increases. 
		
		It can be expected that the value of $\mathcal{L}$ is large when there are only few edges in the VG (low edge density $\rho$) and low for a high edge density $\rho$. Hence, $\mathcal{L}$ and $\rho$ capture essentially similar properties of the underlying time series. For uncorrelated random series, the value of $\mathcal{L}$ of a HVG has a logarithmic scaling relationship with the length of time series $N$, in particular, $\mathcal{L}(N) = 2 \ln N + 2 (\gamma - 1) + O(1/N)$, where $\gamma$ is the Euler-Mascheroni constant \cite{Luque2009}. In the case of fBm with different Hurst indices $H \in (0, 1)$, and for a fixed length of time series of $N$ points, $\mathcal{L}$ increases exponentially with $H$. In addition, $\mathcal{L}$ increases linearly with respect to $N$ when $H$ is close to 1 and in a logarithmic form when $H$ is close to 0 \cite{Xie2011}. 
	
		Besides studies on the SW effect, the assortativity coefficient $\mathcal{R}$ of VGs has recently attracted considerable interest. Specifically, the presence of assortative behavior ($\mathcal{R} > 0$) implies so-called hub attraction, whereas disassortative behavior ($\mathcal{R} < 0$) relates to hub repulsion. It has been shown that the latter is a necessary condition for the emergence of fractal structures in networks \cite{Song2006}. For example, for the particular case of fBm, hub repulsion is not present, and the resulting VGs are non-fractal, but show a scaling of $\mathcal{L}$ with network size $N$ as $\mathcal{L}(N)\sim \log N$, which is typical for SW networks. In contrast, for the Conway series (a deterministic fractal) one finds hub repulsion and $\mathcal{L}(N)\sim N^{-\beta}$, which implies the presence of fractal properties as reported in \cite{Lacasa2008}. In this respect, $\mathcal{R}$ or, more specifically, the scaling of the degree correlation determines the fractality of a VG \cite{Song2006}, which is an interesting and potentially relevant property when studying fractal time series. The interrelationship between the properties of fractal time series and that of the resulting (H)VGs needs more careful numerical validations. 	
	
    A final global H(VG) characteristic that has been studied in several recent works \cite{Ahmadlou2010,Tang2013,Nasrolahzadeh2018} is the so-called \emph{graph index complexity}, practically a rescaled version of the largest eigenvector $\lambda_{max}$ of the network's adjacency matrix \cite{Kim2008},
    \begin{equation}
    GIC=4c(1-c)\ \mbox{with}\ c=\frac{\lambda_{max}-2\cos(\pi/(n+1))}{n-1-2\cos(\pi/(n+1))}.
    \label{gic}
    \end{equation}
    \noindent
    Among others, this measure has been successfully applied in conjunction with VGs to discriminating between different conditions of heart rate variability,
    
		Of course, beyond the aforementioned characteristics there are multiple other measures one could also consider for describing the properties of VGs. This includes also measures characterizing the properties of individual edges as well as the distributions of small subgraphs (motifs). For example, four-node subgraphs show different dominant motifs rankings in the VGs of human ventricular time series, which distinguishes ventricular fibrillations from normal sinus rhythms of a subject \cite{Li2011,Li2012}. Furthermore, the profiles of sequential $n$-node motifs of (H)VGs appear with characteristic frequencies which have been computed analytically for certain deterministic and stochastic dynamics \cite{Iacovacci2016}. 
		
		\subsection{Practical considerations} \label{secsec:VGpractical}
		Many recent publications on (H)VG analysis of time series have particularly made use of data from model systems, which are characterized by rather ideal conditions for statistical analysis. We note that the numerical implementation of the (H)VG algorithms is rather straightforward without extensive precautions. However, when operating with data obtained from experiments, specific features challenging basically any kind of time series analysis are often present, including missing data, heteroscedastic ``noise", or even uncertainties in the time domain. The explicit treatment of the resulting effects on (H)VG properties has not yet been properly addressed in the literature. The associated practical problems that have been discussed in \cite{Donner2012} remain to be further explored. In the following, we summarize these issues in the following. 
		
		Here, some of the practical issues will be discussed using a Gaussian white noise process which serves as a simple, but still illustrative example. It has to be emphasized that for ``real" data characterized by a non-Gaussian probability distribution function, serial dependences, or even (multi-)fractal behavior, the resulting effects could well be much stronger than in this example. A detailed study of the interdependences between such features of the data and the resulting effects of missing data and uncertainties on (H)VG properties is, however, beyond the scope of this review. Anyway, the considerations below provide a first attempt when performing (H)VG analysis and do not cover all relevant aspects of the methodological issues for real time series analysis. Furthermore, all discussions should be performed separately for the natural VG, horizontal VG and their variants, while we mainly focus here on the traditional VG algorithm for the sake of brevity. 

		\subsubsection{Missing data}
		One important problem of many observational time series is the presence of missing data. Since existing methods of time series analysis typically require a uniform spacing in time, this problem is most often addressed by means of interpolation or sophisticated imputation of the missing observations. In general, there is a great variety of possible approaches for such gap filling, which shall not be further discussed here. Anyway, we emphasize that it is not always a priori clear which method performs the best under the specific conditions of the data studied. From the view point of (H)VG analysis, it would be an interesting task to evaluate which effects that interpolation or imputation methods exert on the resulting network structures. 
		
		Unlike many other approaches of time series analysis, VGs do not explicitly require uniform sampling. Hence, missing data could be ignored when performing a corresponding analysis. However, if it is known that there must have been an observation at a given time, it could be conceptually problematic to neglect this information in the analysis. From a broader perspective, it can be argued, however, that this argument applies to all kinds of time series, since values of the considered observable (with a continuous-time variability) taken in between two subsequent observations remain always unknown, but could have a certain impact on the results of the analysis. 
		
		From the viewpoint of complex network studies, it is also possible to interpret the problem of missing values as an attack at or a failure of the complex network represented by the visibility graph. In complex network theory, the impact of such attacks on various types of networks has been intensively studied in terms of safety and robustness of infrastructures \cite{Albert2002}. In this context, one has to distinguish random failures (corresponding to randomly missing values) from intentional attacks, which typically affect the network hubs. Since hubs of a VG correspond to the maxima of the underlying time series, this effect of intentional attacks is particularly relevant for certain types of censored data, e.g., in case of measurement failures due to the limited detection range of a measurement device. Since attacks on hubs typically have a more severe effect on the network architecture than other vertices, censoring can strongly alter the properties of the resulting VGs. However, Donner {\textit{et al.}} have shown in \cite{Donner2012} that even a random removal can have notable consequences for the VG properties on both global and local scale. 
		
\begin{figure}
		  \centering
		  \includegraphics[width=\columnwidth]{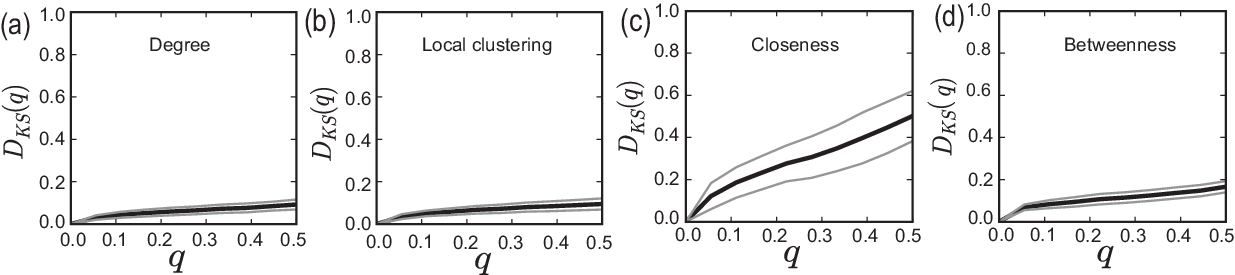}
		  \caption{The Kolmogorov-Smirnov (KS) test statistics $D_{KS}$ versus the fraction $q$ of randomly removed single time series values for the case of a Gaussian white noise time series with $N=100$. (a) Degree $k_i$, (b) local clustering coefficient $\mathcal{C}_i$, (c) closeness $c_i$, and (d) betweenness $b_i$. Missing values are neglected. Black and gray lines correspond to mean values and $\pm 1$ standard deviation levels obtained from $1000$ realizations of the removal process. Modified from \cite{Donner2012}. \label{fig_chap04:missingD}}
		\end{figure}

		The effects of missing data on the properties of VGs have been discussed by two different types of treatment, which have been presented in \cite{Donner2012} and can be considered as opposite extreme cases. On the one hand, missing data will be simply neglected in the generation of the VG. On the other hand, since there is no information about the magnitude of the missing values, it can be a more honest solution to consider the VG as being fragmented into pieces corresponding to times before and after the missing observation, i.e., decomposing the VG into mutually disconnected subgraphs. It should be noted, however, that the latter approach results in the emergence of additional boundary effects. In this case, some sophisticated gap filling by means of interpolation or imputation will be most likely a better strategy in many practical applications. In order to quantify the corresponding effect, the Kolmogorov-Smirnov (KS) test has been proposed in \cite{Donner2012} to quantify the distance (similarity) between the two distributions of a given local vertex property (including $k_i$, $\mathcal{C}_i$, $c_i$, and $b_i$) for both the original and perturbed time series. The corresponding results are shown in Fig.~\ref{fig_chap04:missingD} and demonstrate that ignoring missing values has a considerable effect mainly on closeness $c_i$ (Fig. \ref{fig_chap04:missingD}(c)), whereas the changes to the distributions of $k_i$, $\mathcal{C}_i$, and $b_i$ are in a certain tolerable range. 
				
		In real-world applications, missing data often occur not randomly independent of each other, but as blocks. Some further numerical results in this regard can be found in \cite{Donner2012}.

		\subsubsection{Homo- and heteroscedastic uncertainties} 
\begin{figure}
		  \centering
		  \includegraphics[width=\columnwidth]{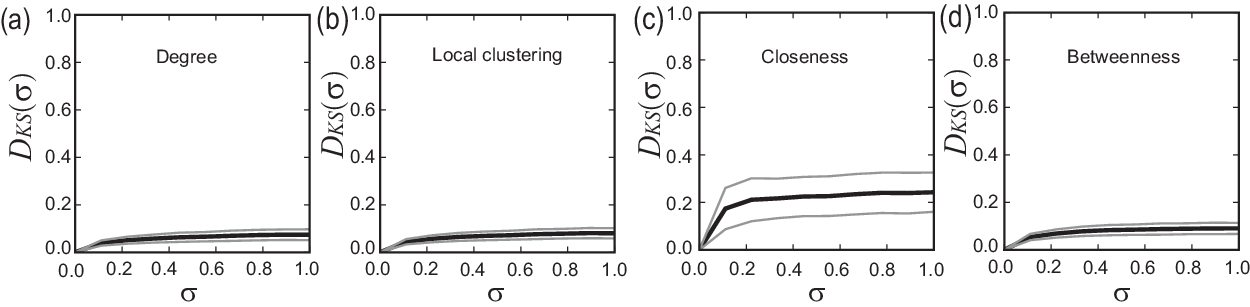}
		  \caption{As in Fig.~\ref{fig_chap04:missingD} for quantifying the effect of additive Gaussian white noise with variance $\sigma^2$ on the VGs for one uncorrelated Gaussian random noise of $N = 100$. Black and gray lines correspond to mean values and $\pm 1$ standard deviation levels obtained from $M=1000$ realizations of the additive noise. Modified from \cite{Donner2012}. \label{fig_chap04:homouncertain}} 
		\end{figure}

		The influence of measurement uncertainties on the resulting VG properties can be studied in a similar way as for the case of missing values. In particular, Donner {\textit{et al.}} \cite{Donner2012} considered homoscedastic uncertainties as an additional additive Gaussian white noise component, whereas the heteroscedastic case is studied by multiplicative noise with a reasonable, simple analytical distribution. As shown in Fig.~\ref{fig_chap04:homouncertain}, it is found that in both cases the signal-to-noise ratio has a considerable effect by systematically shifting the distributions of vertex properties obtained for the original data towards those expected for the noise process. Note that since in the considered numerical example both signal and noise originated from mutually independent Gaussian white noise processes, there is a saturation of the KS statistics for moderate noise at values corresponding to the variance of VG properties for independent realizations of the same signal process. Further examples of heteroscedastic uncertainties can be found in \cite{Donner2012}. 

		\subsubsection{Uneven and irregular timings}
\begin{figure}
		  \centering
		  \includegraphics[width=\columnwidth]{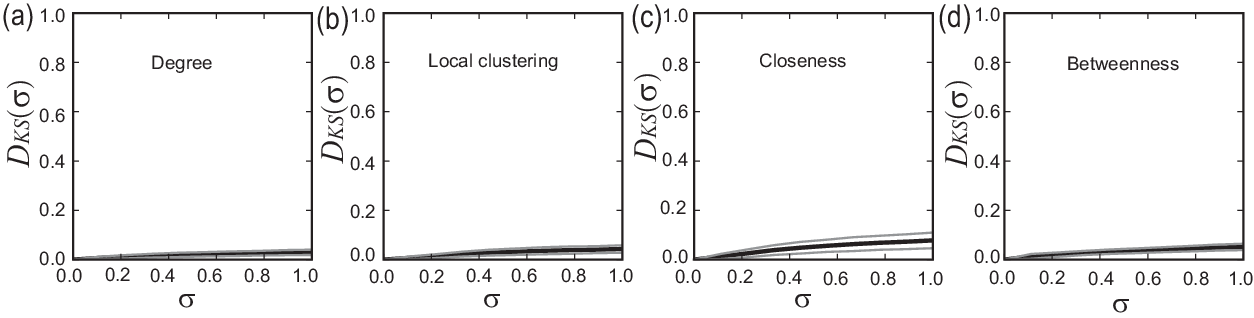}
		  \caption{As in Fig. \ref{fig_chap04:homouncertain} for fixed data, but uncertain timing of observations. Uncertainty in the time domain has been modeled by considering new times $\tilde{t_i} = t_i + \Delta t (| 1 - 2\eta_i | - 0.5)$, where $\eta_i$ are independent realizations of a random variable with uniform distribution in $[0,1]$, and $\Delta t$ is the spacing between subsequent observations in the original data set.  Reproduced from \cite{Donner2012}. \label{fig_chap04:uncertimings}}
		\end{figure}

		In a full analogy to the case of uncertainties in the observable $\{x_i \}_{i=1}^N$ with $x_i = x(t_i)$, one can study the impact of uncertain timings $t_i$ on the properties of the resulting VGs. The issue of uncertain timings is a wide-spread problem particularly in the analysis of paleoclimate time series \cite{Telford2004}. Since in the construction of VGs both observable and sampling time enter in terms of an inequality defined by a linear relationship (Eqs.~\ref{vis_cond} and \ref{eq:hvc}), it is not surprising that uncertain timing can indeed have a similar effect on the VG properties as uncertainties in the measurement itself. Figure~\ref{fig_chap04:uncertimings} displays the corresponding KS test statistic results for a realization of Gaussian white noise originally observed with regular spacing, with the timings being corrupted. Note that this specific form of the time-scale corruption, which allows preserving the temporal order of observations, has been inspired by the tent map as a paradigmatic nonlinear mapping often used as an illustrative example in complex systems sciences \cite{Donner2012}. These results demonstrate that the distribution of local vertex properties of a VG are indeed affected by modifications of the time-scale. However, comparing Fig.~\ref{fig_chap04:uncertimings} with Fig.~\ref{fig_chap04:homouncertain}, the changes are considerably smaller than for noisy corruptions of the measurements themselves. The reason for this is that the modification used here has been restricted by the normal sampling interval, whereas the changes induced by additive and multiplicative noise allowed for comparably larger modifications in the data.
				
		Uneven timings are paradigmatic for point processes, which are ubiquitous in geoscientific processes, for instance, seismic magnitude series \cite{Telesca2012}. A point process is often characterized by a random time occurrence of the events and these events are typically clustered because they are neither Poissonian nor regularly distributed over time. To check the effects of irregular timing on the degree distributions $p(k)$, Telesca {\textit{et al.}} \cite{Telesca2012} constructed two VGs from (1) the seismic series of the original random occurrence times, and the series (2) that has been substituted by regular conventional time unit. Interestingly, almost identical results have been obtained for $p(k)$, which suggests that the effects of irregular timing are not crucial. However, as we demonstrated in Fig.~\ref{fig_chap04:uncertimings}, network measures are affected by the level of uncertainties in the sampling timings. 

		The trivial connection of neighboring points in time in the (H)VG enhances the signature of structures due to autocorrelations in the record under study. Although this might be desirable for (H)VGs, since some of their respective network properties are explicitly related with the presence of serial dependences (e.g., the typical scale of the degree distribution of HVGs, cf.\, \cite{Luque2009}), there could be situations in which one is interested in removing the corresponding effects. In such cases, it is possible to introduce a minimum time difference for two observations to be connected in the network for removing the effect of slowly decaying auto-dependences, which would correspond to the Theiler window in other concepts of nonlinear time series analysis \cite{Theiler1986}.
		
		We note that other than for VGs or related methods, the uncertain timings or irregular time spacings can have substantial effects on the construction of ordinal pattern transition networks \cite{Kulp2016a,McCullough2016,Sakellariou2016}, which will be reviewed in Section \ref{sec:TransitionNt}. Furthermore, the current discussion is limited to the case of a Gaussian white noise process and more generalizations to non-Gaussian assumptions are necessary to be explored, which are much closer to the situations of real time series.

	\subsection{Multivariate visibility graph methods}
	Despite their success, the range of applicability of (H)VGs methods has been mainly limited to univariate time series, although the most challenging problems in the area of nonlinear science concern systems that are described by multivariate time series. Synchronization analysis is one of traditional topics when generalizing (H)VG analysis from a univariate to bivariate time series \cite{Ahmadlou2012a,Mitra2012}. We notice that there are several different ways of extending the ideas from a single time series to multivariate time series, for instance, multiplex VGs, cross-VGs and joint VGs. For instance, the cross-visibility algorithm has been proposed to understand coupling and information transfer between two time series \cite{Mehraban2013}. Here, we summarize some of the different approaches to characterize bivariate time series using (H)VGs. 
	
		\subsubsection{Multiplex visibility graphs} \label{sec:multiplexVG}
		Based on the definition of HVGs, Lacasa \textit{el al.} \cite{Lacasa2015b} proposed to transform a multidimensional time series into an appropriately defined multiplex visibility graph. New information can be extracted from the original multivariate time series, with the aims of describing signals in graph-theoretical terms or to construct novel feature vectors to feed automatic classifiers in a simple, accurate and computationally efficient way. Multiplex VG analysis has been applied to characterize the heterogeneity of recurrent neural network dynamics \cite{Bianchi2017} and different conditions of resting state human fMRO data \cite{Sannino2017}. 
		
		Consider an $M$-dimensional real valued time series $\{ \vec{x}_i \} = (\{ x_i^{[1]}\}, \{x_i^{[2]} \}, \dots, \{x_i^{[M]} \})$ with $\vec{x}_i = \vec{x}(t_i)  \in \mathbb{R}^M$ for any value of $t_i \in [1, N]$, measured empirically or extracted from an $M$-dimensional, either stochastic or deterministic dynamical system. In full analogy to the notation used for describing multiplex recurrence networks in Section~\ref{sec:mrn}, the superscript index represents here the $\alpha$-th variable. An $M$-layer multiplex visibility graph $\mathcal{M}$ is then constructed, where layer $\alpha$ corresponds to the HVG associated to the time series $\{ x_i^{[\alpha]}\}_{t_i=1}^{N}$ of state variable $X^{[\alpha]}$. Note that $\mathcal{M}$ is represented by the vector of adjacency matrices of its layers $\mathbf{\mathcal{A}} = \{\mathbf{A}^{[1]}, \mathbf{A}^{[2]}, \dots, \mathbf{A}^{[M]}\}$, where $\mathbf{A}^{[\alpha]} = \{A_{ij}^{[\alpha]} \}_{ij}$ is the adjacency matrix of layer $\alpha$. Such a mapping builds a bridge between multivariate series analysis and recent developments in the theory of multilayer networks \cite{Boccaletti2014}, making it possible to employ the structural descriptors introduced to study multiplex networks as a toolbox for the characterization of multivariate signals. 
		
		Two measures have been proposed to capture, respectively, the abundance of single edges across layers and the presence of inter-layer correlations of node degrees \cite{Lacasa2015b}, which help to characterize information shared across variables (layers) of the underlying high dimensional system. Simply speaking, we compute the two measures defined by Eqs.~(\ref{eq:RNmultiplex},\ref{eq:RNmultiplexW}), which are based on the adjacency matrices of HVGs. More specifically, the first measure is the \textit{average edge overlap} $\omega$ (Eq.~\ref{eq:RNmultiplexW}),  	
which computes the expected number of layers of the multiplex on which an edge is present. Note that $\omega$ takes values in $[1/M, 1]$ and in particular $\omega = 1/M$ if each edge $(i,j)$ exists in exactly one layer, i.e. if there exist a layer $\alpha$ such that $A_{ij}^{[\alpha]} = 1$ and $A_{ij}^{[\beta]} = 0$\, $\forall\, \beta \neq \alpha$, while $\omega = 1$ only if all the $M$ layers are identical. As a consequence, the average edge overlap of a multiplex VG can be used as a proxy of the overall coherence of the original multivariate time series, with higher values of $\omega$ indicating high correlation in the microscopic structure of the signal. 

		The second measure proposed in \cite{Lacasa2015b} allows to quantify the presence of \textit{interlayer correlation} between the degrees of the same node at two different layers. More specifically, given a pair of layers $\alpha$ and $\beta$ of $\mathcal{M}$, respectively characterized by the degree distributions $p(k^{[\alpha]})$ and $p(k^{[\beta]})$, the interlayer correlation is defined by the mutual information $I^{\alpha\beta}$ (Eq.~\ref{eq:RNmultiplex}) between $p(k^{[\alpha]})$ and $p(k^{[\beta]})$. 
		The higher $I^{\alpha\beta}$ the more correlated are the two layers and therefore the structures of the associated time series. Then, the average of $I^{\alpha \beta}$ over every pair of layers of $\mathcal{M}$ gives a scalar variable $I = \left < I^{\alpha\beta}\right>_{\alpha, \beta}$, which captures the amount of information flow in the multivariate time series. 
		
		Note that the interlayer correlation gives a weighted correlation matrix of size $M \times M$ and each entry is represented by $I^{\alpha\beta}$. That means that the original $M$-dimensional time series is transformed to a weighted graph of $M$ nodes, where each node represents one layer and the weights of the edges denote the magnitude of mutual information computed from the associated (H)VG degree distributions.

		\subsubsection{Joint and excess degrees} \label{subsec:jointdegreeVG}
		Some network-theoretic quantities to quantify asymmetries in bivariate time series have been introduced in \cite{Zou2014}. Following the notations as described for multiplex (H)VG, we restrict here our considerations to two-dimensional time series which results in two layers $\alpha$ and $\beta$. For $\{x^{[\alpha]}_i\}_{i=1}^{N}$ and $\{x^{[\beta]}_i\}_{i=1}^{N}$, we again consider two (H)VGs with adjacency matrices $\mathbf{A}^{[\alpha]}$ and $\mathbf{A}^{[\beta]}$, respectively. Note that the sets of vertices are the same for both subgraphs, with differences exclusively in the set of edges.

Based on the thus obtained (H)VGs, we proceed as follows:
\begin{enumerate}
\item From the two (H)VGs, we have two sets of neighbors, $\mathcal{N}_i^{[\alpha],[\beta]} = \left\{ A_{ij}^{{[\alpha]}, {[\beta]}} \equiv 1, j \in \left\{1,\ldots, N \right\}/\left\{i\right\} \right\}$ for each time $i \in \left\{1,\ldots, N \right\}$. The \textit{degree sequences} are then defined as
\begin{align}
&k_i^{{[\alpha]}, {[\beta]}} = \# \mathcal{N}_i^{{[\alpha]}, {[\beta]}} = \sum_{j} A_{ij}^{{[\alpha]}, {[\beta]}}.
\end{align}

\item The \textit{joint degree}
\begin{align} \label{eq:jointK}
&k_i^\textrm{joint} = \# \left( \mathcal{N}_i^{{[\alpha]}} \cap \mathcal{N}_i^{{[\beta]}} \right) = \sum_{j} A_{ij}^{{[\alpha]}}\cdot A_{ij}^{{[\beta]}}
\end{align}
gives the number of common neighbors of a vertex corresponding to time $i$ in both sequences. Notably, we can define $\{k_i^{joint}\}$ as the degree sequence of a \textit{joint} (H)VG combining the visibility criteria for two distinct time series. Here, the adjacency matrix is defined by the point-wise multiplication of the individual (H)VGs' adjacency matrices. This idea is conceptually related to the concept of joint recurrence plots encoding the simultaneous recurrence of two dynamical systems in their respective phase spaces \cite{romano2004}. One simple idea is to compare the cross correlation or mutual information between the two degree sequences of $\{k_i^{[\alpha]}\}$ and $\{k_i^{[\beta]}\}$ \cite{Lacasa2015b,Ahmadi2018}. 

\item In a similar spirit as the joint degree sequence, we can quantify the number of edges associated with time $i$, which connect to vertices contained in $\mathcal{N}_i^{{[\alpha]}}$ but \textit{not} in $\mathcal{N}_i^{{[\beta]}}$, or vice versa. More specifically,
\begin{align}
k_i^{{o,[\alpha]}} & = \# \left( \mathcal{N}_i^{{[\alpha]}} \cap \overline{\mathcal{N}_i^{{[\beta]}}} \right)  =  \sum_{j} A_{ij}^{{[\alpha]}}\cdot \left( 1-A_{ij}^{{[\beta]}} \right) \\
k_i^{{o,[\beta]}} & = \# \left( \mathcal{N}_i^{{[\beta]}} \cap \overline{\mathcal{N}_i^{{[\alpha]}}} \right) = \sum_{j} A_{ij}^{{[\beta]}}(t)\cdot \left(1-A_{ij}^{{[\alpha]}} \right)
\end{align}
where $\overline{\mathcal{N}_i^{{[\alpha]}, {[\beta]}}(t)}=\{1,\ldots,T\}\cap \left\{ \mathcal{N}_i^{{[\alpha]}, {[\beta]}}\cup i\right\}$ is the complementary set of $\mathcal{N}_i^{{[\alpha]}, {[\beta]}}$, which measures the number of neighbors that belong \textit{only} to $\mathcal{N}_i^{{[\alpha]}}$ or $\mathcal{N}_i^{{[\beta]}}$, respectively. Therefore, we refer $k_i^{o,{[\alpha]}, {[\beta]}}$ as the \textit{conditional degree sequences}. By definition,
\begin{align}\label{eq:nad}
k_i^{o,{[\alpha]}, {[\beta]}}=k_i^{{[\alpha]}, {[\beta]}}-k_i^\textrm{joint}.
\end{align}
\end{enumerate}

Based on the latter definitions (Eq.~\ref{eq:nad}), we compute the following properties:
\begin{align} \label{eq:deltaK}
\Delta k_i &= k_i^{o,{[\alpha]}} - k_i^{o,{[\beta]}} = k_i^{{[\alpha]}} - k_i^{{[\beta]}} \\ \label{eq:deltaReK}
\Delta_{rel} k_i &= \Delta k_i/\left(k_i^{{[\alpha]}} + k_i^{{[\beta]}} \right).
\end{align}
The \textit{excess degree} $\Delta k_i$ quantifies how much ``more convex'' the fluctuations of $\mathbf{A}^{[\alpha]}$ are in comparison with $\mathbf{A}^{[\beta]}$ around a given time $i$ (i.e., how many more or less visibility connections the observation of $\mathbf{A}^{[\alpha]}$ at time $i$ obeys in comparison with $\mathbf{A}^{[\beta]}$). By additionally considering the \textit{relative excess degree} $\Delta_{rel} k_i$ normalized by the sum of the individual degrees, we obtain a measure that does not exhibit marked sensitivity with respect to the actual degrees $k_i^{{[\alpha]}, {[\beta]}}$, which may considerably vary over time according to the statistical and dynamical characteristics of the data.

In Section~\ref{subsec:sunspotsAsym}, we will summarize the applications of $\Delta k_i$ and $\Delta_{rel} k_i$ to characterize the north--south asymmetry of solar activity \cite{Zou2014}, which provide many nonlinear properties that have not been studied by other methods. Notably, this approach is conceptually related with recently developed (H)VG-based tests for time series irreversibility, which compare (among others) degree distributions obtained when considering edges to past and future observations separately \citep{Donges2013}. We will address these methods in the following.

        \subsubsection{Visibility graph similarity}
						
As an early alternative to studying inter-layer similarities in multiplex (H)VGs, Ahmadlou and Adeli \cite{Ahmadlou2012a} introduced the concept of visibility graph similarity. Here, a combination of ideas from time delay embedding, recurrence analysis and visibility graphs is used to quantify the statistical relationship between two observed sequences. For this purpose, each sequence is first subjected to classical time delay embedding. Next, a certain state vector from the previously constructed set is chosen as a reference, and a time series of distances with respect to this reference point is constructed. For the resulting time series of distances, the corresponding VG is constructed. Finally, by pair-wise comparison between the thus obtained VG degree sequences obtained for the two time series of interest in terms of their correlation coefficient, a VG-based similarity measure is constructed, which has been used for capturing signatures of generalited synchronization in paradigmatic model systems \cite{Ahmadlou2012a}, but also for constructing time-dependent functional network representations of human brain actvity from multi-channel EEG measurements \cite{Sengupta2013}.

	\subsection{Decomposition of visibility graphs} \label{sec:timeIRvg}
		\subsubsection{Time-directed visibility graphs and characterizations}
		So far, visibility graphs have mostly been considered as undirected, since visibility does not have a predefined temporal arrow. However, (H)VGs be made directed by again assigning to the links a time arrow, which result in the so called directed VGs and HVGs. Accordingly, a link between $i$ and $j$ (where time ordering yields $i< j$) generates an outgoing link for $i$ and an incoming link for $j$. Therefore, the degree $k_i$ of the node $i$ is now split into an in-degree $k_i^{in}$, and an out-degree $k_i^{out}$, such that $k_i= k_i^{in}+k_i^{out}$. The in-degree $k_i^{in}$ is defined as the number of links of node $i$ with other past nodes associated with data in the series (that is, nodes with $j < i$). Conversely, the out-degree $k_i^{out}$, is defined as the number of links with future nodes. Then we define the in- and out-degree distributions of a directed VG (HVG) as $p(k^{out})$ and $p(k^{in})$, respectively. An important property at this point is that the ingoing and outgoing degree sequences are interchangeable under time reversal. 
		
		More specifically, given the adjacency matrix $\mathbf{A}$ of a (H)VG, the degree {$k_i =\sum_{j} A_{ij}$} measures the number of edges incident to a given vertex {$i$}. Then, Donges \textit{et~al.} \cite{Donner2012Nolta,Donges2013} adapted a somewhat different yet equivalent notation for in- and out-degree sequences. More specifically, they decomposed the quantity $k_i$ for a vertex corresponding to a measurement at time $i$ into contributions due to other vertices in the past and future of $i$,
\begin{eqnarray} \label{eq:kvin}
k_i^r &= \sum_{j<i} A_{ij},\\ \label{eq:kvout}
k_i^a &= \sum_{j>i} A_{ij}
\end{eqnarray}
with $k_i=k_i^r+k_i^a$, being referred to as the \emph{retarded} and \emph{advanced degrees}, respectively. Here, $k_i^r$ and $k_i^a$ correspond to the respective in- and out-degrees of time-directed (H)VGs \cite{Lacasa2012}. While the degrees of an individual vertex can be significantly biased due to the finite time series length~\cite{Donner2012}, the resulting frequency distributions of retarded and advanced degrees are equally affected. 

		The local clustering coefficient $\mathcal{C}_i = {k_i \choose 2}^{-1} \sum_{j,k} A_{ij} A_{jk} A_{ki}$ is another vertex property of higher order characterising the neighbourhood structure of vertex $i$~\cite{Newman2003}. Here, for studying the connectivity due to past and future observations separately, \cite{Donges2013} defined the \emph{retarded} and \emph{{advanced local clustering coefficients}}
\begin{eqnarray} \label{eq:cvin}
\mathcal{C}_i^r &= {k_i^r \choose 2}^{-1} \sum_{{j<i,k<i}} A_{ij} A_{jk} A_{ki},\\ \label{eq:cvout}
\mathcal{C}_i^a &= {k_i^a \choose 2}^{-1} \sum_{{j>i,k>i}} A_{ij} A_{jk} A_{ki}.
\end{eqnarray}
Hence, both quantities measure the probability that two neighbours in the past (future) of observation $i$ are mutually visible themselves. Note that the decomposition of $\mathcal{C}_i$ into retarded and advanced contributions is not as simple as for the degree and involves degree-related weight factors and an additional term combining contributions from the past and future of a given vertex.

		Finally, {we note that} other measures characterising complex networks on the local (vertex/edge) as well as global scale could be used for similar purposes as {those} studied in this work. However, {since} path-based network characteristics {(e.g.,} closeness, betweenness, or average path length{)} cannot be easily decomposed into retarded and advanced contributions, the approach followed here is mainly restricted to neighbourhood-based network measures like degree, local and global clustering coefficient, or network transitivity. As a possible solution, instead of decomposing the network properties, the whole edge set of a (H)VG could be divided into two disjoint subsets that correspond to visibility connections forward and backward in time, as originally proposed by Lacasa~\textit{et~al.}~\cite{Lacasa2012}. For these directed (forward and backward) (H)VGs, also the path-based measures can be computed separately and might provide valuable information. 
		
		\subsubsection{Tests for time series irreversibility}
		Testing for nonlinearity of time series has been of great interest. Various approaches have been developed for identifying signatures of different types of nonlinearity as a necessary precondition for the possible emergence of chaos. Since linearity of Gaussian processes directly implies time-reversibility \cite{Weiss1975,Lawrance1991,Diks1995}, nonlinearity results (among other features) in an asymmetry of certain statistical properties under time-reversal~\cite{Theiler1992}. Therefore, studying reversibility properties of time series is an important alternative to the direct quantitative assessment of nonlinearity~\cite{Voss1998}. In contrast to classical higher-order statistics requiring surrogate data techniques~\cite{Theiler1992}, most recently developed approaches for testing irreversibility have been based on symbolic dynamics~\cite{Daw2000,Kennel2004,Cammarota2007} or statistical mechanics concepts~\cite{Costa2005,Porporato2007,Roldan2010}. 
		
		The time series reversibility has the following definition: a time series $\Sigma = \{ x_1, x_2, \dots, x_n \}$ is called statistically time reversible if the time series $\Sigma^{\ast} = \{x_{-1}, x_{-2}, \dots, x_{-n} \}$ has the same joint distribution as $\Sigma$. Therefore, time reversibility implies stationarity \cite{Lawrance1991}. By this definition, time series reversibility reduces to the equivalence between forward and backward statistics and hence, nonstationary series are infinitely irreversible and therefore $\Sigma$ and $\Sigma^{\ast}$ have different statistics that increase over time \cite{Weiss1975}. More specifically \cite{Lawrance1991}, time-{ir}reversibility of a stationary stochastic process or time series $\{x_i\}$ requires that for arbitrary $n$ and $m$, the tuples $(x_n,x_{n+1},\dots,x_{n+m})$ and $(x_{n+m},x_{n+m-1},\dots,x_n)$ have the same joint probability distribution. It is practically unfeasible in most situations due to the necessity of estimating high-dimensional probability distribution functions from a limited amount of data. Instead of testing this condition explicitly, for detecting time series irreversibility it can be sufficient to compare the distributions of certain statistical characteristics obtained from both vectors (e.g.,~\cite{Tong1990}). In the following, we will review the corresponding VG-based approaches.

		\paragraph{Kullback-Leibler divergence}
		In many applications, one can actually quantify different kinds of time asymmetries in the underlying dynamics on nonstationary processes. Following previous work \cite{Lacasa2012}, the topological properties of (H)VGs associated to several types of nonstationary processes have been proposed to quantify different degrees of irreversibility of several nonstationary processes \cite{Lacasa2015}. Furthermore, they take advantage of the fact that the topological properties of these graphs are effectively invariant under time shift for large classes of nonstationary processes, which allows to introduce the concept of visibility graph stationarity. This in turn allows to extract meaningful information on the time asymmetry of nonstationary processes. 

		Lacasa \textit{et al.} \cite{Lacasa2015} defined time series reversibility in terms of (H)VGs in the following way: a time series $\Sigma = \{ x_i \}_{i=1}^N$ is said to be (order $p$) (H)VG reversible if and only if, for large $N$, the order $p$ block ingoing and outgoing degree distribution estimates of the directed (H)VG associated to $\Sigma$ are asymptotically identical, i.e., 
\begin{equation}
p(k_1^{in}, k_2^{in}, \dots, k_p^{in}) = p(k_1^{out}, k_2^{out}, \dots k_p^{out}).
\end{equation}
This property yields that the ingoing and outgoing degree sequences of the original and time reversed series have the same distribution, namely, 
\begin{equation}
p(k^{in}) [\Sigma] = p(k^{out}) [ \Sigma^{\ast}]; p(k^{out}) [\Sigma] = p(k^{in})[\Sigma^{\ast}],  
\end{equation}
where $\Sigma^{\ast} = \{ x_{n+1-i} \}_{i=1}^{n}$ represents the time reversed series. For time series, we assess how close the system is to reversibility by quantifying the distance between $p(k^{in})$ and $p(k^{out})$.  

		The distance between the \textit{in} and \textit{out} degree distributions has been calculated by the Kullback-Leibler divergence (KLD) \cite{Lacasa2012,Lacasa2015}, which is used in information theory as a measure to distinguish between two probability distributions. More specifically, the KLD between these two distributions is
\begin{equation}\label{eq:KLD}
D_{k l d}(in \| out) = \sum_{k} p(k^{in}) \log \frac{p(k^{in})}{p(k^{out})}. 
\end{equation}
This measure $D_{k l d}(in\|out)$ is a semi-distance which vanishes if and only if the outgoing and ingoing degree probability distributions of a time series are identical, but is positive otherwise. Then the (H)VG reversibility is redefined if the following expression holds:
\begin{equation}
\lim_{N \to \infty} {D_{k l d}(in\|out)} = 0. 
\end{equation}		
Truly irreversible process have positive values of Eq.~\eqref{eq:KLD} in the limit of large $N$ \cite{Lacasa2012}. The consistency of this test for HVGs has been demonstrated numerically for chaotic dissipative systems (HVG-irreversible), long-range dependent stochastic processes (HVG-reversible) \cite{Lacasa2012} and non-Markovian random walks (HVG-irreversible) \cite{Lacasa2015}.

		Lacasa \textit{et al.}\cite{Lacasa2012, Lacasa2015} conjectured that the information stored in the in- and out-degree distributions takes into account the amount of time irreversibility of the associated series. More precisely, they suggested that this can be measured, in a first approximation, as the distance (in a distributional sense) between the in- and out-degree distributions ($p(k^{in})$ and $p(k^{out})$). This claim was recently supported by quantitative numerical analysis of $D_{k l d}$ for a modified Langevin equation and some simple population growth model with tunable long-range dependence \cite{Telesca2018c}. Specifically, it was found that both persistence and anti-persistence increase the value of the Kullback-Leibler divergence, with the effect of positive correlations being somewhat stronger than that of negative ones.
        
        Instead of degree based characteristics, higher order measures can be used as well if necessary, such as the corresponding distance between the in and out degree-degree distributions ($p(k^{in}, k'^{in})$ and $p(k^{out}, k'^{out})$). These are defined as the in and out joint degree distributions of a node and its first neighbors, describing the probability of an arbitrary node whose neighbor has degree $k'$ to have degree $k$. That is, we compare the out-degree distribution in the actual (forward) series $p(k^{out})$ with the corresponding probability in the time-reversed (or backward) time series, which is equal to the probability distribution of the ingoing degree in the actual process. 
		
\begin{figure}[htbp]
		\centering
			\includegraphics[width=\columnwidth]{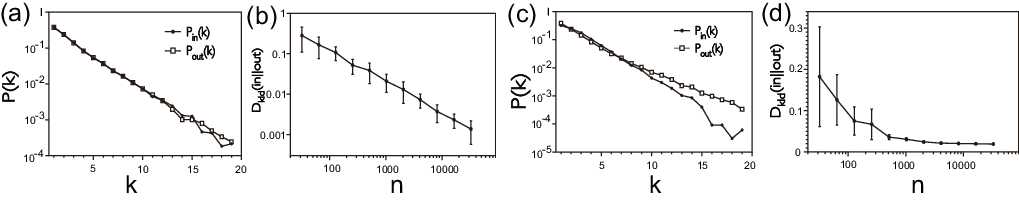}
			\caption{Time irreversibility testing for white noise (a,b) and the chaotic logistic map (c,d). (a,c) Semi-logarithmic plots of the in- and out-degree distributions of the VG. (b,d) Time irreversibility measure $D_{k l d}(in\|out)$ versus series size $N$. Modified from \cite{Lacasa2015} with permission by American Physical Society. } \label{fig:chapter4timeReverse}
		\end{figure}

		Therefore, by calculating Eq.~\eqref{eq:KLD}, Lacasa \textit{et al.}\cite{Lacasa2012} have shown that one can correctly distinguish between reversible and irreversible stationary time series, including analytical and numerical studies of its performance for: (i) reversible stochastic processes (uncorrelated and Gaussian linearly correlated), (ii) irreversible stochastic processes, (iii) reversible (conservative) and irreversible (dissipative) chaotic maps, and (iv) dissipative chaotic maps in the presence of noise. Numerical examples of time irreversibility tests have been provided in Fig.~\ref{fig:chapter4timeReverse}. In particular, they first considered time series of length $2^{15}$ from white noise of $i.i.d.$ uniformly random uncorrelated variables $\sim U[0,1]$ (Fig. \ref{fig:chapter4timeReverse}(a,b)). In this case, Fig.~\ref{fig:chapter4timeReverse}(a) shows that both distributions are identical up to finite-size effects fluctuations. In addition, the irreversibility measure $D_{k l d}(in\|out)$ vanishes asymptotically as $1/N$, showing that finite irreversibility values for finite size are due to statistical fluctuations that vanish asymptotically. Therefore, Fig.~\ref{fig:chapter4timeReverse}(a, b) suggest that the underlying process is time reversible. In contrast, both distributions are clearly different for the chaotic logistic map (Fig.~\ref{fig:chapter4timeReverse}(c)), which is further confirmed by the convergence of $D_{k l d}(in\|out)$ to a finite value with $N$ (Fig.~\ref{fig:chapter4timeReverse}(d)). Therefore, the logistic map is time irreversible. 
		
		Notice that the majority of previous methods to estimate time series irreversibility generally proceed by first making a (somewhat ad hoc) local symbolization of the series, coarse-graining each of the series data into a symbol (typically, an integer) from an ordered set ~\cite{Daw2000,Kennel2004,Cammarota2007,Costa2005,Porporato2007,Roldan2010}. The method based on directed (H)VGs here lacks an ad hoc symbolization process, which may in principle take into account multiple scales. The unnecessary requirement of symbolization is desirable if we want to tackle complex signals and, hence, it can be applied directly to any kind of real-valued time series.  

		We note that (H)VG reversibility varies depending on the detailed properties of particular processes \cite{Lacasa2015,Telesca2018c}, which calls for careful interpretations. For instance, both analytical calculations and numerical simulations show that unbiased additive random walks, while nonstationary, are both (H)VG  stationary and (H)VG time reversible. On the other hand, biased memoryless additive random walks are HVG irreversible with finite irreversibility measures that quantify the degree of time asymmetry, while these are still VG reversible, as VGs are invariant under superposition of linear trends in the original data. Numerical examples suggest that HVGs can capture, for both finite and infinite series size, the irreversible nature of non-Markovian additive random walks, whereas VGs are only able to do so for finite series. For multiplicative random walks, the processes are HVG reversible if the process is akin to an unbiased additive process in logarithmic space, and time irreversible if the process reduces to a biased additive process in logarithmic space. Finally, the VGs capture the time irreversible character of multiplicative random walks, yielding finite values in the unbiased case and asymptotically diverging quantities in the biased case. Furthermore, these conclusions are based on the limit of infinitely long time series $N \to \infty$ and finite size time series always yield finite, non-zero values of HVG and VG irreversibility \cite{Xiong2018}, which needs a proper test justifying the statistical significance. 
		
		\paragraph{Kolmogorov-Smirnov (KS) test for irreversibility}
		While the results of the KLD measure for reversible and irreversible dynamics quantitatively differ in several orders of magnitude, a statistical test is required. Lacasa \textit{et al.} \cite{Lacasa2012} proposed to address the statistical significance by surrogate techniques as follows: one first proceeds to shuffle the series under study in order to generate a randomized resampled data set with the same underlying probability density. This resampled series, whose irreversibility measure is asymptotically zero, is considered as the null hypothesis of the test. Recently, a combination of Kullback-Leibler distance between the ingoing and outgoing degree sequences and the so-called inversion number of the permutation of the original time series has been proposed to characterize asynchronous patterns of time irreversibility \cite{Yang2018}. Taking a slightly different algorithm, Donges \textit{et~al.} \cite{Donner2012Nolta,Donges2013} have extended this idea and proposed to utilize some standard statistics for testing the homogeneity of the distribution of random variables between two independent samples, which can be formulated for both standard and horizontal VGs and for different network properties. 
		
		More specifically, following the decomposition of vertex properties into time-directed contributions proposed above (including network degrees and local clustering coefficients, Eqs.~(\ref{eq:kvin})-(\ref{eq:cvout})), Donges {\textit{et al.}} \cite{Donges2013} utilized the frequency distributions {$p(k^{r})$ and $p(k^{a})$ ($p(\mathcal{C}^{r})$ and $p(\mathcal{C}^{a})$)} of retarded and advanced {vertex properties} as representatives for the statistical properties of the time series when viewed forward and backward in time. In the case of time-reversibility, they conjecture that both sequences {$\{k_i^{r}\}$ and $\{k_i^{a}\}$} (or {$\{\mathcal{C}_i^{r}\}$ and $\{\mathcal{C}_i^{a}\}$}) should be drawn from the same probability distribution, because the visibility structure towards the past and future of each observation has to be statistically equivalent. In turn, for an irreversible (i.e., nonlinear) process, we expect to find statistically significant deviations between the probability distributions of retarded and advanced characteristics. In other words, rejecting the null hypothesis that {$\{k_i^{r}\}$ and $\{k_i^{a}\}$} ({$\{\mathcal{C}_i^{r}\}$ and $\{\mathcal{C}_i^{a}\}$}) are drawn from the same probability distribution, respectively, implies rejecting the null hypothesis that the time series under investigation is reversible. 
		
		{Since for sufficiently long time series (representing the typical dynamics of the system under study), the available samples of individual vertex properties approximate the underlying distributions sufficiently well, we can (despite existing correlations between subsequent values)} consider the Kolmogorov-Smirnov (KS) test for testing this null hypothesis. {Specifically}, a small $p$-value of the KS test statistic (e.g., $p<0.05$) implies that the time series has likely been generated by an irreversible stochastic process or dynamical system. Even more, these $p$-values are distribution-free in the limit of $N\to\infty$. {Neglecting possible effects of the intrinsic correlations between the properties of subsequent vertices on the estimated $p$-values (which shall be addressed in future research), this} implies that we do \textit{not} need to construct surrogate time series for obtaining critical values of our test statistics as in other {ir}reversibility tests. Note that other (not network-related) statistical properties sensitive to the time-ordering of observations could also be exploited for constructing similar statistical tests for time series irreversibility \cite{Donges2013}. 
		
\begin{figure}[htbp]
		\centering
			\includegraphics[width=\columnwidth]{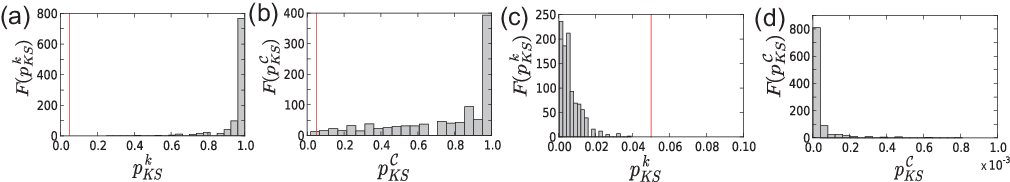}
			\caption{(color online) Frequency distributions of $p$-values of the KS statistic for comparing the distributions of retarded/advanced (a,c) degree {$k_i^{r}$, $k_i^{a}$} and (b,d) local clustering coefficient {$\mathcal{C}^{r}_i$, $\mathcal{C}^{a}_i$} of standard VGs from an ensemble of $M=1,000$ realizations of model system time series of length $N=500$: (a,b) AR(1) process, (c,d) H\'enon map. Vertical red lines indicate the typical significance level of 0.05 {where appropriate (note the different scale in panel (d))}.  Modified from \cite{Donges2013}. } \label{fig:chapter4KSReverse}
		\end{figure}

		The KS test for irreversibility has been demonstrated by model time series of AR(1) process ($p = 1, \varphi_1 = 0.5$ in Eq.~\eqref{def:AR}) and the chaotic H\'enon map (Eq.~\eqref{eq:henon}) as shown in Fig. \ref{fig:chapter4KSReverse}. For the linear (reversible) AR(1) process, the empirical distributions of retarded/advanced {vertex properties} collapse onto each other \cite{Donges2013}. Consequently, the null hypothesis of reversibility is never rejected by the test based on the degree (Fig.~\ref{fig:chapter4KSReverse}(a)), and only rarely rejected by the clustering-based test well below the expected false rejection rate of 5\% (Fig.~\ref{fig:chapter4KSReverse}(b)). In contrast, for the irreversible H\'enon map, the null hypothesis of reversibility is nearly always (degree, Fig.~\ref{fig:chapter4KSReverse}(c)) or always (local clustering coefficient, Fig.~\ref{fig:chapter4KSReverse}(d)) rejected. To further evaluate the performance of the tests for varying series size $N$, the receiver operating characteristics (ROC curves) have been applied to quantify the false positive rate versus true positive rate when testing irreversibility \cite{Donges2013}. 		

		Utilizing standard as well as horizontal VGs for discriminating between the properties of observed data forward and backward in time has at least two important benefits: (i) {Unlike for some classical tests (e.g.,~\cite{Theiler1992}),} the reversibility properties are examined without the necessity of constructing surrogate data. Hence, the proposed approach saves considerable computational costs in comparison with {such methods} and, more importantly, avoids the problem of selecting a particular type of surrogates. Specifically, utilizing the KS test statistic or a comparable two-sample test for the homogeneity (equality) of the underlying probability distribution functions directly supplies a $p$-value for the associated null hypothesis that the considered properties of the data forward and backward in time are statistically indistinguishable. (ii) The proposed approach {is applicable} to data with non-uniform sampling {(common in areas like} paleoclimate~\cite{Donner2012,Donner2012Nolta} or astrophysics) {and} marked point processes (e.g., earthquake catalogues~\cite{Telesca2012}). For {such} data, constructing surrogates for {non}linearity tests in the most common way {using} Fourier-based techniques is a challenging task, {which is avoided by} {(H)}VG-based methods. 

		We emphasize that this method exploits the time-information explicitly used in constructing (H)VGs. Other existing time series network methods (e.g., recurrence networks \cite{Marwan2009,Donner2010a,Donner2011}) not exhibiting this feature cannot be used for the same purpose. Furthermore, there are methodological questions such as the impacts of sampling, observational noise, and intrinsic correlations in vertex characteristics as well as a detailed comparison to existing methods for testing time series {ir}reversibility that need to be systematically addressed in future research. Furthermore, {(H)}VG-based methods are generally faced with problems such as boundary effects and the ambiguous treatment of missing data \cite{Donner2012}, which call for further investigations, too. 
		
		It may be noted that unlike neighborhood-based network measures like degree and local clustering coefficients, path-based measures of (H)VGs are known to be strongly influenced by boundary effects \cite{Donner2012}, so that they could possibly lose their discriminative power if employed in a similar fashion in the context of {ir}reversibility tests. 
        
        Irreversibility tests have been conducted for various real valued time series. Examples include neuro-physiological EEG recordings \cite{Donges2013}, mean temperature anomaly series \cite{Xie2014}, financial time series \cite{Flanagan2016}, oil-water two-phase flows \cite{Meng2016a}, meteorological stream flow fluctuations \cite{Serinaldi2016}, correlated fractal processes \cite{Xiong2018}, and seismic sequences of a Mexican subduction zone \cite{Telesca2018}.

%% file: Chapter05_TransitionNt/Chapter05_TransitionNt.tex
\section{Transition networks} \label{sec:TransitionNt}

A third main group of transformations of time series into complex networks is commonly referred to as transition networks. Unlike recurrence (and other proximity-based) networks or the numerous algorithmic variants of visibility graphs, these networks are directed by definition and, hence, trace the succession of dynamical states as time proceeds. Specifically, the nodes of a transition network correspond to certain discrete states or patterns, and directed links are established if one of these nodes is followed by the other with non-zero (empirical) probability along the observed trajectory of the system under study. Mathematically, this corresponds to a Markov chain with given transition probabilities between discrete states, which can be conveniently used for constructing a weighted and directed graph \cite{Schnakenberg1976}. In this regard, the history of transition networks is closely tied to the development of the mathematical theory of Markov chains, while explicitly exploiting the topological properties of these complex networks constructed from the transition probabilities between states or patterns in dynamical systems has probably only been started in the last about 15 years, initiated by the seminal work by Nicolis \emph{et~al.}~\cite{Nicolis2005}. 

While it is most common to formulate transition networks as directed and weighted graphs, it should be noted that an adjoint unweighted network representation can be easily obtained by omitting the explicit information on transition probabilities and considering those pairs of nodes (i.e., states or patterns) as being linked via an unweighted edge which exhibit non-zero mutual transition probabilities. Alternatively to considering all non-zero probabilities, one may also choose some threshold for the transition probabilities to exclude rare transitions (e.g., due to noise in deterministic dynamical systems).

In general, there are various ways to define the nodes of a transition network upon a given data set. The simplest situation is when the data themselves take only discrete values. In this case, each possible value or, more generally, each possible $m$-tuple of such values can be directly used to define a node. This results in two parameters that should be selected taking the number $K$ of discrete states, the total length $N$ of the underlying time series, and the available computational resources into account. Notably, these algorithmic parameters -- the number $m$ of states contained in each tuple and their mutual time distance $\tau$ -- are conceptually equivalent to the parameters of time delay embedding (i.e., embedding dimension and delay) in classical nonlinear time series analysis \cite{Packard1980,kantz1997}. However, unlike in most applications of time delay embedding, in the context of transition network approaches one is not necessarily interested in having statistically independent states in the considered tuple, so that $\tau=1$ is often a convenient choice even in case of serially dependent data sequences. It should be noted that the idea of considering tuples of discrete data allows a straightforward generalization of the transition network approach to multivariate series by combining, e.g., simultaneous values of the individual univariate component time series for defining a certain node.

In the more common case of variables with continuous distributions, obtaining discretized states requires some initial transformation from continuous to discrete data, which is commonly referred to as symbolic encoding. Although this encoding results in a loss of information on the detailed state of the system under study, proper encodings minimize this loss (up to zero if using so-called generating partitions of the system under study). After having performed this discretization, one may proceed as in the case of originally discrete-valued time series.

In summary, transforming a time series into a transition network representation is a (possibly two-step) process of mapping the temporal information into a Markov chain to obtain a compressed or simplified representation of the original dynamics. In the following, the individual steps for constructing and analyzing different types of transition networks will be described in full detail.

\subsection{Symbolic encoding}\label{sec:symbolic}

Symbolic encoding transforms a time series into a set of $K$ discrete states or patterns (``symbols") $\{\pi_1, \dots , \pi_K \}$. Therefore, when analyzing time series of a variable with continuous distribution, the first step towards constructing a transition network representation is applying a suitable discretization of the considered series. 

In general, there is a well-developed theory of using symbolic sequences in the context of nonlinear time series analysis \cite{Daw2003,Finn2003,Amigo2010}, which draws upon methodological concepts like entropies and complexity measures derived upon them, commonly originating from the field of statistical mechanics and information theory. Notably, there is generally a large freedom of defining different kinds of symbolizations for an underlying time series. For deterministic discrete-time dynamical systems, there exist so-called generating partitions, in which case there is a direct correspondence between the observed trajectory and the resulting symbolic sequence that is unique up to a set of measure zero \cite{Grassberger1985,Christiansen1996,Christiansen1997}. While using such generating partitions would be desirable in many applications, their estimation from real-world data commonly poses a challenging problem \cite{Kennel2003,Hirata2004,Buhl2005}. For this purpose, in practice simplified symbolization strategies are commonly employed:

\begin{itemize}

\item Coarse-graining of phase space, sometimes also referred to as \emph{static encoding} \cite{Donner2008}, classifies the data into $K$ different groups based on intervals defined by a set of pre-defined threshold values $\{\xi_1,\dots,\xi_{K-1}\}$. In this case, symbols $\pi_q$ with $q=1,\dots,K$ are assigned according to the group membership of the respective data value (i.e., $\pi_p=q$), and
\begin{equation}
\xi_i=\left\{
\begin{array}{ll}
1 & \quad \mbox{if} \quad x_i < \xi_1, \\
q & \quad \mbox{if} \quad \xi_{q-1} \leq x_i < \xi_q \quad \mbox{for} \quad 1<q<K-1), \\
k & \quad \mbox{if} \quad x_i \geq \xi_{K-1}.
\end{array} \right.
\end{equation}

\item While the coarse-graining approach relies solely on the amplitudes of each individual data value, different types of \emph{dynamic encodings} can be defined as alternatives. One simple way consists of thresholding the difference-filtered time series of order $p$, whereby the difference filtering operator is recursively applied as
\begin{eqnarray}
\Delta^{1}x_i &=& x_{i+1}-x_i \\
\Delta^{p}x_i &=& \Delta^{p-1}x_{i+1}-\Delta^{p-1}x_i \quad (p>1).
\end{eqnarray}
\noindent
Most commonly, one uses this specific approach for obtaining some binary encoding, i.e., $\xi_t=\Theta(\Delta^{p}x_t)\in\{0,1\}$ with $\Theta(\cdot)$ being again the Heaviside function. One important variant of this approach is considering order relationships among subsequent data values in subsets of $m$ observations defined based upon the first-order difference filtered series as described above. This approach results in a symbolic encoding into $2^{m-1}$ different states representing coarse-grained local dynamical patterns. In the following, we will call this type of symbolization \emph{order-pattern based encoding}.

\item Conceptually related to the aforementioned encoding based on the \emph{pair-wise} inter-comparison between subsequent values is \emph{group-wise} ordering of values within sets of $m$ subsequent observations. Neglecting possible ties, we assign each value $x_i$ within such a sequence its local rank order $r_i\in\{1,\dots,m\}$. Notably, there will be $m!$ possible permutations of such rank order sequences, which are enumerated to obtain a symbolic encoding. This strategy allows utilizing statistical techniques like permutation entropies \cite{Bandt2002} commonly referred to as \emph{ordinal time series analysis} methods \cite{bandt2005}. Accordingly, we will refer to such permutation-based symbolizations as \emph{ordinal pattern based encoding} or simply \emph{ordinal encoding}.

\end{itemize}

While the three general strategies for symbolic encoding of time series values described above are quite common in nonlinear time series analysis, they are not exhaustive. Other types of encodings, as well as mixed strategies \cite{Donner2008}, are possible as well, depending on the specific aspects of dynamics one wishes to highlight when performing the symbolization. In this regard, there is commonly no ``optimal'' symbolization strategy for a given time series \cite{Bollt2001}, but rather great flexibility according to the dynamical features or time-scales of interest as well as the available computational resources. In general, symbolic encoding performs a natural coarse-graining of the dynamics of the system under study, which preserves essential dynamical information. Such a strategy naturally addresses common issues in many real-world time series, such as the presence of noise or intrinsic multi-stability. 

Based upon the accordingly discretized time series, a variety of relevant dynamical properties can be conveniently estimated, including symbolic correlation functions \cite{Lee1997}, mutual information \cite{Li1992}, permutation entropy \cite{Bandt2002}, or transfer entropy \cite{Staniek2008}. In most cases, the choice of the symbolization strategy and its possible algorithmic parameters (e.g., number and location of thresholds, length of considered subsets, etc.) affects the resulting estimates. For example, mutual information estimators obtained from groups with equal probabilities are commonly more reliable than such from groups of equal interval size. Since defining symbols with equal probability of occurrence can be challenging in case of other than threshold-based encodings, this observation poses natural restrictions to the interpretation of quantitative values of many statistical characteristics of symbolic sequences. 

Instead of considering statistics based upon the (joint) probability distributions of individual symbols, it is often useful to combine sequences of symbols into ``words'' of a given length $m$, which allow not only for looking at the coarse-grained state of the system under study, but rather a succession of several of such states, thereby conserving some information on its dynamics. Among the quantities that can be estimated from statistics upon such sequences, the source entropy of the underlying system is often approximated by the limit of the \textit{conditional block entropies} $s_m=S_{m+1}-S_m$ for $m\to\infty$ where
\begin{equation}
S_m=-\sum_{p=1}^{K^m} p_p^{(m)}\log p_p^{(m)}
\end{equation}
\noindent
are the block entropies, i.e., the Shannon entropies of symbolic sequences of length $m$, with $p_q^{(m)}$ being the probabilities of occurrence of all possible subsequences (words) of this length (enumerated by the index $p$) within the time series \cite{Daw2003,Ebeling1992}. Moreover, the statistical properties of the sequence distributions can be used for defining various measures of complexity \cite{Grassberger1986,Rosso2007}. Last but not least, investigating the properties of forbidden symbols and words (i.e., values or patterns that are not observed during the system's dynamical evolution) have recently attracted particular attention \cite{Amigo2007,Carpi2010}.

\subsection{Markov chains}

After having defined suitable symbols or words, which are taken from an alphabet $\mathcal{A}$ of $\omega$ discrete values, the next step towards constructing a transition network representation upon a given time series is to explicitly use the temporal order of the accordingly coarse-grained observations to represent the dynamics of the observed system. 
For this purpose, we consider the transition probabilities $w_{pq} = p (\xi_{i+1} = \pi_q | \xi_i = \pi_p)$ between subsequent symbols (words) to define a weighted and directed transition network with the weight matrix $\mathbf{W} = \{w_{pq} \}, p,q \in [1, \dots, \omega]$. Note that because of $\sum_{q=1}^{\omega} w_{pq}=1$ for all $p\in\mathcal{A}$ by definition (conservation of probability), $\mathbf{W}$ is a column-stochastic matrix.

In the terminology of stochastic processes \cite{Nicolis2005}, the resulting transition networks describe a Markov chain with the nodes representing some set of states that encode the time series' amplitudes or local variations, and directed and weighted edges indicating the temporal succession of such states or patterns. Specifically, Markov chains are memoryless stochastic processes, implying that the state of the process at some time $i$ depends solely on its previous state at $i-1$. If approximating the coarse-grained dynamics of a time series as such a Markov process, this implies that the $n$-step transition probabilities are simply given by the entries of $\mathbf{W}^n$. It should be emphasized, however, that trajectories of dynamical systems commonly exhibit serial correlations, so that this simplifying approximation is commonly not suitable for fully describing the longer-term dynamical evolution of the system. Nonetheless, the Markov chain analogy allows reconsidering terms like absorbing or recurrent states or stationary densities in terms of transition network properties. Specifically, absorbing states of a Markov chain can be identified as transition network nodes with zero out-degree (respectively, zero out-strength), since $w_{pq}=\delta_{pq}$. Recurrent states are characterized by their membership in loops of length greater than one, while the stationary density is given as the eigenvector of $\mathbf{W}$ with eigenvalue 1, which is unique in case of non-degenerated Markov chains. This implies a close connection with the network measure of eigenvector centrality.

It should be emphasized that the duality between Markov chains and transition networks holds for any Markov chain irrespective of the existence of an underlying time series.

	\subsection{Coarse-graining based transition networks}\label{sec:cgtn}

	The construction of coarse-graining based transition networks draws upon a proper phase space partition. For instance, we first mesh the $d$-dimensional phase space with boxes of equal size following the traditional idea of fractal dimension computations \cite{Grassberger1983PRL,Grassberger1983PLA} or complexity measures \cite{Crutchfield1989}. When working with univariate time series, $d=m$ (i.e., the length of the considered symbolic sequences). An alternative, mathematically preferable yet harder to construct alternative would be a separation into boxes with equal probability \cite{Liu2016}. Then, each box $p$ is labeled with the symbol $\pi_p$ and regarded as a vertex in the network. The connectivity between two boxes (nodes) $\pi_p$ ($p$) and $\pi_q$ ($q$) is then represented by the empirical transition frequency following the temporal order of observations. 

	The transition probability approach is well suited for identifying such ``states" (i.e., regions in phase space) that have a special importance for the dynamical evolution of the studied system, for example, in terms of their betweenness centrality $b_p$ or similar measures. Moreover, the resulting networks do not only depend on a single parameter, but on the specific definition of the full set of classes. Note, however, that coarse graining might be a valid approach in case of noisy real-world time series, where extraction of dynamically relevant information hidden by noise can be supported by grouping the data. In contrast to the other approaches for constructing complex networks from time series, the topology of transition networks depends on the specific choice of discretization. 

	For a trajectory that does not leave a finite volume in phase space, there is only a finite number of discrete ``states" $\pi_p$ with a given minimum size in phase space. This implies the existence of absorbing and/or recurrent states in the associated Markov chain. Specifically, in case of dissipative dynamics, the phase space segments corresponding to absorbing and recurrent nodes provide a coarse-grained description of the system's attractor(s).
		
		It should be noted that unlike in visibility graphs and related methods, detailed temporal information is lost at the network level after the coarse graining since the transition frequency matrix $\mathbf{W}$ is estimated over the entire time series. Therefore, the resulting transition network is a static representation of the system's dynamics, which requires the system to be stationary and ergodic. Violation of the stationarity assumption could imply the transition matrix $\mathbf{W}$ being explicitly time-dependent, which however would make a proper estimation of its coefficients challenging if only a single time series is available as a realization of the non-stationary system dynamics. In this context, Weng {\emph {et al.}} \cite{Weng2017} proposed constructing a temporal network from time series by unfolding temporal information into an additional topological dimension. More specifically, a transition from node $p$ to $q$ is established whenever the trajectory flow performs a transition from $p$ to $q$ at time $t_i$ which is denoted as ($p \to q; i$). By adding the additional time axis to the transition route, the consecutive memory network is constructed by introducing a memory factor. Weng {\textit{et al.}} further proposed memory entropy analysis to characterize the memory effect in the observed time series. The identified memory effect can accurately differentiate between various types of time series including white noise, $1/f$ noise, AR model, periodic and chaotic time series. 

In general, there is a close analogy between coarse-graining based transition networks and Lagrangian flow networks \cite{Lindner2017,Donner2019} used for describing structural characteristics of flows. The latter type of network representations has been originally introduced for studying geophysical flows in the atmosphere and ocean \cite{Rossi2015,Ser-Giacomi2015}, but can also be applied for investigating the behavior of dynamical systems in their underlying phase space. The latter type of network representations of nonlinear maps or ordinary differential equations has also been termed \emph{Ulam networks} \cite{Shepelyansky2010,Ermann2010,Chakhmakhchyan2013,Ermann2015,Frahm2018} in parts of the literature, referring to the estimation of transition probabilities of trajectories between finite boxes being known as Ulam's method. While the main difference with respect to the transition networks used for the purpose of time series analysis is that Lagrangian flow networks and Ulam networks encode transition probabilities between volume elements that are commonly based on the observation of the trajectories of ensembles of tracer particles that are passively advected within a given flow field, in case of stationary and ergodic systems, results obtained for flow networks can be directly translated to transition networks in the asymptotic limit. Specifically, it has been demonstrated that node properties like degree, eigenvector centrality, or cutoff closeness have a close correspondence with spatial patterns of finite-time Lyapunov exponents or entropies (highlighting the positions of invariant manifolds of hyperbolic trajectories of the system under study) \cite{Ser-Giacomi2015,Lindner2017}. Other dynamically relevant structures like elliptic fixed points and periodic orbits can be identified by different network properties like the local clustering coefficient \cite{Rodriguez-Mendez2017}.

		\subsubsection{Ordinal pattern transition networks for univariate time series} \label{sec:OPtransition}
		As an alternative to phase space partitioning, one may define the node set of a transition network based on some different symbolic representation of the studied time series, for instance, ordinal patterns \cite{Li2008,McCullough2015}. This strategy has been followed recently in a growing number of studies \cite{Small2013,McCullough2015,Kulp2016b,McCullough2017b,Small2018}. Among others, a series of systematic investigations of ordinal methods has been conducted on irregularly sampled time series \cite{Kulp2016a,McCullough2016,Sakellariou2016}, which shows great potentials for studies of experimental observation data from climate sciences. 

\begin{figure}[ht]
	\centering
	\includegraphics[width=\columnwidth]{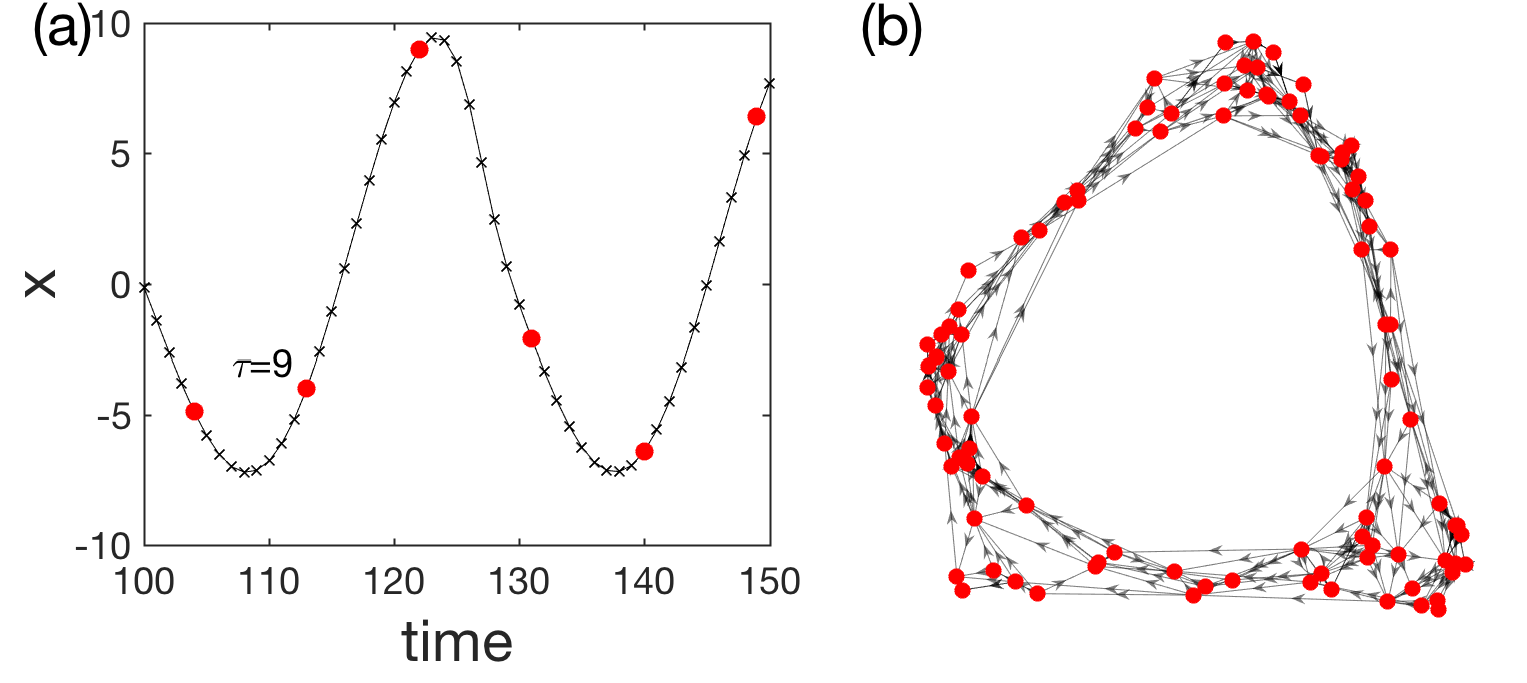}
\caption{(a) Illustration of permutation symbols from a time series of the R\"ossler attractor ($a = 0.165$ in Eq.~\eqref{eq:roessler}). Assume $\tau = 9$ and $m = 6$. One embedded state vector $\vec{x}_{104} = \{x_{104}, x_{113}, x_{122}, x_{131}, x_{140}, x_{149} \}$ is highlighted by red color, and its corresponding pattern is defined by the rank ordering $\pi_{104} = \{5, 1, 2, 4, 6, 3\}$. (b) Resulting OPTN (isolated vertices and self-loops are excluded in the visualization). Directed edges are indicated by arrows. Reproduced from \cite{McCullough2015}. \label{fig:rosTNm}}
\end{figure}

		To construct an ordinal pattern transition network (OPTN), the first step is to embed the given one-dimensional time series $\{x_i\}$ by using traditional time delay embedding with a proper choice of embedding dimension $m$ and time delay $\tau$. Then, embedded points in phase space are mapped to nodes in the network space according to the sequence of rank orders, and links are allocated between nodes based on temporal succession on the trajectory. In Fig.~\ref{fig:rosTNm}, we show an example of an OPTN using the algorithm of \cite{McCullough2015}. 
		
		In \cite{McCullough2015}, McCullough {\textit{et al.}} illustrated the construction algorithm in detail for the R\"ossler system and found that periodic dynamics translates to ring structures whereas chaotic time series translate to band or tube-like structures -- thereby indicating that this algorithm generates networks whose structure is sensitive to qualitatively different system dynamics. Furthermore, it has been demonstrated that simple network measures (including the mean out-degree and variance of out-degrees) can track changes in the dynamical behavior in a manner comparable to the largest Lyapunov exponent \cite{McCullough2015}. Therefore, topological characteristics of OPTNs have the potential to provide useful indicators for dynamical discrimination of different states and the detection of change points. 
		
		Measures of transitional complexity have been further proposed in \cite{McCullough2017b} to quantify the resulting OPTNs. Based on the transition matrix $\mathbf{W}$ of an OPTN that excludes the possibility of self-loops, McCullough {\textit{et al}} \cite{McCullough2017b} considered the local out-link transition frequency between the ordinal patterns $\pi_p$ and $\pi_q$
		\begin{equation} \label{eq:localTp}
			p_{pq} = \left \{ \begin{aligned}
							& 0,  \;\;\;\;\;\;\;\;\;\;\; \text{if} \; p = q\\
							& \frac{w_{pq}}{\sum_{q, q \neq p} w_{pq}} \;\;\;\;\;\; \text{if} \; p \neq q.\\
							\end{aligned}
					    \right.
		\end{equation}
		Note that the transition frequency of Eq.~\eqref{eq:localTp} is pattern (row-) wise normalized. Then, the Shannon entropy of each row of Eq.~\eqref{eq:localTp} gives the \emph{local node out-link entropy} $s_{p}^{L} = -\sum_{q} p_{pq} \log p_{pq}$. By averaging over the network, we obtain the expected value of the transitional complexity, which is called \emph{global node out-link entropy} 
		\begin{equation}\label{eq:globalTp}
			S^{GNE} = \sum_{p} p_p s_{p}^{L}. 
		\end{equation}
		The lower bound $S^{GNE} = 0$ implies that there is no uncertainty in the system, which corresponds to time series that are strictly monotonic or have a strictly periodic series of ordinal patterns. Further, an infinitely long time series of uniform independent and identically distributed noise yields the upper bounds for $S^{GNE}$ \cite{McCullough2017b}. 
		
		Embedding dimension $m$ and time delay $\tau$ are two important parameters in the construction of OPTNs, having crucial impacts on the appearance of forbidden order patterns \cite{Kulp2016b,McCullough2016,Sakellariou2016}. The selection of time delay $\tau$ must be performed in relation to the sampling rate for continuous systems. Sun \emph{et al.}~\cite{Sun2014} proposed using a time delay $\tau > 1$. In \cite{McCullough2015}, the authors recommended to select these two parameters by traditional methods used for time series embedding, for instance, the first root of the autocorrelation function of the underlying time series, because it provides a sufficiently good phase space reconstruction of a deterministic dynamical system. While there does not yet exist a robust metric for determining the correct choice of $m$, for time series of the R\"ossler system they found that networks with the most visually intuitive structure often coincide with maxima of the degree variance in dependence on $m$. In addition, it was demonstrated that the range $6 \leq m \leq 10$ was the most useful when using simple network measures to track changes in the dynamics \cite{McCullough2015}. Note that the choice of $m$ also determines the level of simplification of the original phase space by ordinal partitions. 
		
Based on similar considerations, Masoller \emph{et~al.}~\cite{Masoller2015} introduced an alternative set of entropic quantifiers. Specifically, they assigned each node of an OPTN from a univariate time series a weight $w_q$ according to the probability of occurrence of the corresponding pattern, and studied the Shannon entropy of the distribution of these node weights,
\begin{equation}
\mathcal{S}^P=-\sum_{p=1} p_p \log p_p
\end{equation}
\noindent
which corresponds to the classical permutation entropy, together with the local per-node entropy and network-averaged node entropy,
\begin{equation}
\mathcal{S}_p=-\sum_{q} w_{pq}\log w_{pq} \ \mbox{and} \ \mathcal{S}^N = \frac{1}{\omega} \sum_{p=1}^\omega \mathcal{S}_p.
\end{equation}
\noindent
In addition, they studied the \emph{asymmetry coefficient}
\begin{equation}
a=\frac{\sum_p \sum_{q\neq p} \left| w_{pq}-w_{qp} \right|}{\sum_p \sum_{q\neq p} \left( w_{pq}+w_{qp} \right)},
\end{equation}
\noindent
which takes values between $a=0$ (for a completely symmetric network) and $a=1$ (in a fully directed network where each link between any pair of nodes is completely unidirectional). Masoller \emph{et~al.}~\cite{Masoller2015} demonstrated that the aforementioned quantities trace well the succession of qualitatively different dynamical states in bifurcation scenarios of paradigmatic model systems like logistic map or tent map, as well as in real-world data from semiconductor laser experiments.

		\subsubsection{Order pattern transition networks for multivariate time series}
		Most recent works have focused only on univariate time series $\{x_i\}$, while the generalization to multivariate time series has remained largely untouched. However, many phenomena in the empirical sciences are of a multivariate nature. For instance, different assets in stock markets are often observed simultaneously and their joint development is analyzed to better understand tendencies. In climate science, multiple observations (temperature, pressure, precipitation, human activities, etc., from different locations) are the basis of reliable predictions of upcoming meteorological conditions. 
		
		Zhang \emph{et~al.}~\cite{Zhang2017b} proposed constructing OPTNs from multivariate (high dimensional) data. Recall that given a scalar time series $\{x_i\}$ which is produced by a deterministic dynamical system, the order structure depends on the embedding dimension $m$ and delay $\tau$. Let us start with embedding dimension $m = 2$. Neglecting equality, we have two relations between $x_i$ and $x_{i+\tau}$, namely, two symbol sequences representing order patterns $\pi_{x}$:
\begin{equation} \label{piX2D_eq}
\pi_{x,i} =  \begin{cases}
 1 & \text {if} \; x_i < x_{i+\tau}, \\
 0 & \text {if} \; x_i > x_{i+\tau}. 
\end{cases}
\end{equation}
In \cite{Small2013}, Small {\textit{et al.}} used a fixed lag $\tau = 1$ for embedding, and we adopt this idea in the following. In this case, the order pattern $\pi_x^{1}=1$ captures an increasing trend, respectively, $\pi_x^{0}=0$ corresponds to a decreasing trend of the time series. This definition is equivalent to considering the signs of the increments $\Delta x_i = x_{i+1} - x_i$ by a first-order difference of the original series. 

		Generalizing the aforementioned idea to the case of three-dimensional time series $(x_i, y_i, z_i)$, we first obtain the increment series $(\Delta x_i, \Delta y_i, \Delta z_i)$. Then the order patterns are defined by the combinations of signs of $\Delta x_i$, $\Delta y_i$ and $\Delta z_i$. In particular, the ordinal patterns $\Pi_i \in (\pi_1, \cdots, \pi_K)$ (with $K = 8$) of a three-dimensional time series are summarized in Tab.~\ref{tab:3D}. 
\begin{table}
\centering
\begin{tabular}{|c|c|c|c|c|c|c|c|c|}
\hline
$\Pi$      & $\pi_1$ & $\pi_2$ & $\pi_3$ & $\pi_4$ & $\pi_5$ & $\pi_6$ & $\pi_7$
& $\pi_8$\\
\hline
$\Delta x$ & $\pi_x^{1}, + $ & $\pi_x^{1}, + $ & $\pi_x^{1}, +$ &
$\pi_x^{1}, +$ & $\pi_x^{0}, - $ & $\pi_x^{0}, - $ &
$\pi_x^{0}, -$ & $\pi_x^{0}, -$\\
\hline
$\Delta y$ & $\pi_y^{1}, + $ & $\pi_y^{1}, + $ & $\pi_y^{0}, -$ &
$\pi_y^{0}, -$ & $\pi_y^{1}, + $ & $\pi_y^{1}, + $ &
$\pi_y^{0}, -$ & $\pi_y^{0}, -$\\
\hline
$\Delta z$ & $\pi_z^{1}, + $ & $\pi_z^{0}, - $ & $\pi_z^{1}, +$ &
$\pi_z^{0}, -$ & $\pi_z^{1}, + $ & $\pi_z^{0}, - $ &
$\pi_z^{1}, +$ & $\pi_z^{0}, -$\\
\hline
\end{tabular}
\caption{Order patterns in three-dimensional time series $(x_t, y_t, z_t)$. 
\label{tab:3D}}
\end{table}
More generally, the size of the alphabet describing the order patterns $\Pi_i$ for an $n$-dimensional time series $(\{x_{1,i}\}, \ldots, \{x_{n,i}\})$ is $m = 2^{n}$ since each component has either increasing or decreasing trend at time $i$. 

\begin{figure}[ht]
	\centering
	\includegraphics[width=0.7\columnwidth]{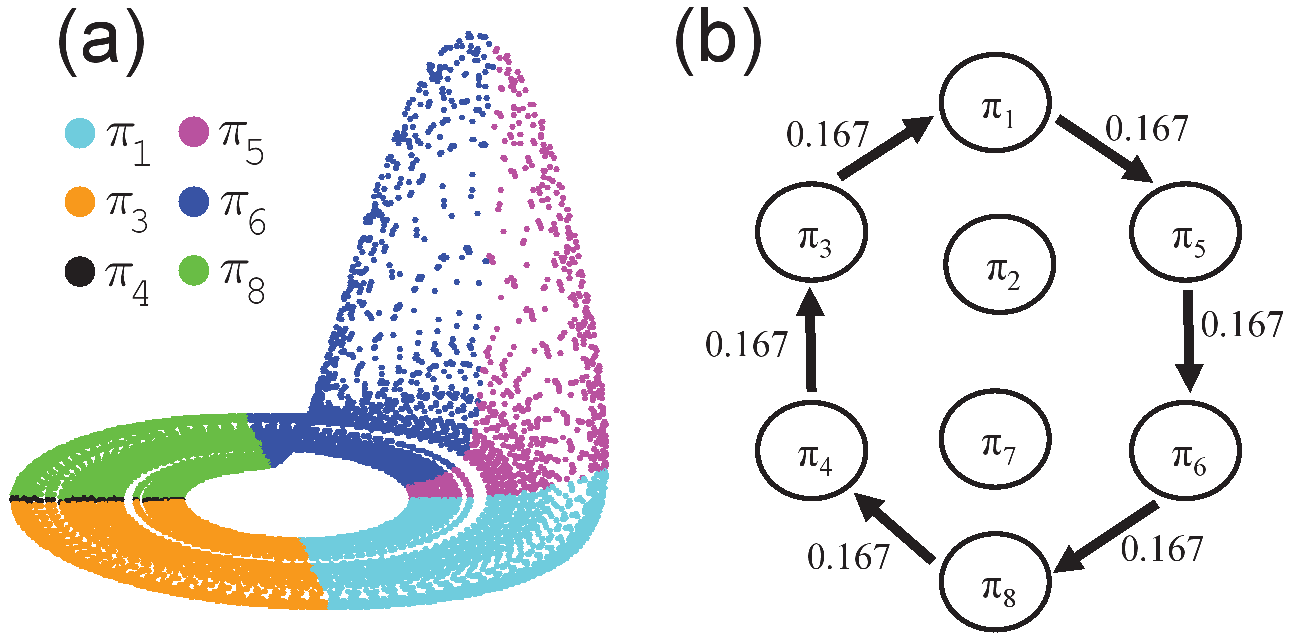}
\caption{(a) R\"ossler attractor in phase space color coded by the order patterns defined in Tab.~\ref{tab:3D} (for $a = 0.165$ in Eq.~\eqref{eq:roessler}), (b) OPTN and self-loops are excluded. Reproduced from~\cite{Zhang2017b}. \label{fig:rosslerTwoA}}
\end{figure}

		We emphasize that multivariate OPTNs consider the increments between two consecutive time points of each one-dimensional measurement series in the space of multivariate measurements, which captures the dynamical properties of the multivariate time series in its associated ``velocity space'' (difference space). Therefore, the time delay $\tau$ in the order pattern definition (Eq.~\eqref{piX2D_eq}) has a rather different interpretation than the time delay that is often used in classical time delay embedding. The above discussion can be directly generalized to the case of $\tau>1$ and embedding dimension $m>2$ for each variable. The resulting OPTN utilizes discretized approximations of the local nullclines (i.e., $\Delta x=0$, etc.) to obtain a symbolic partitioning of the systems. In Fig. \ref{fig:rosslerTwoA}(a), the chaotic R\"ossler system is color-coded by the ordinal pattern partitions and the corresponding transition network in shown in Fig. \ref{fig:rosslerTwoA}b. 

In order to highlight the importance of non-self transitions between ordinal patterns, one straightforward modification of the OPTN definition would consist of removing any self-loops, which is a typical step in many applications of complex networks \cite{Costa2007}. Specifically, we can remove self-loops before computing the weight matrix $\mathbf{W}$ to keep the normalization $\sum_{p,q}^{2^n} w_{pq} = 1$. Note that in stochastic processes, self-loops can be expected not to contribute with large probabilities. 
		
		Given the empirical observations of different occurrence frequencies of ordinal patterns $p(\pi_p)$ and their mutual transition frequencies $w_{pq}$, two Shannon entropies are defined as 
		\begin{align} \label{eq:Ho}
		\mathcal{S}_O &= - \sum_{p=1}^{2^n} p(\pi_p) \log_2 p(\pi_p) , \\ \label{eq:Ht}
		\mathcal{S}_T &= - \sum_{p,q=1}^{2^{n}} w_{pq} \log_2 w_{pq}. 
		\end{align}
		In the terminologies of \cite{McCullough2017b}, $\mathcal{S}_{O}$ characterizes the vertex (node) complexity (i.e., is defined as the Shannon entropy of node weights, which is equivalent to the classical permutation entropy) and $\mathcal{S}_{T}$ the edge (link) transitional complexity, both of which have been found useful for characterizing synchronization transitions \cite{Zhang2017b}. Note that Eq.~\eqref{eq:Ht} measures the transitional complexity of the ordinal patterns, but with a slightly different normalization than that was used in \cite{McCullough2017b,Small2018}. 
		
\subsubsection{Ordinal pattern transition networks for synchronization transitions}

\begin{figure}
	\centering
	\includegraphics[width=0.6\columnwidth]{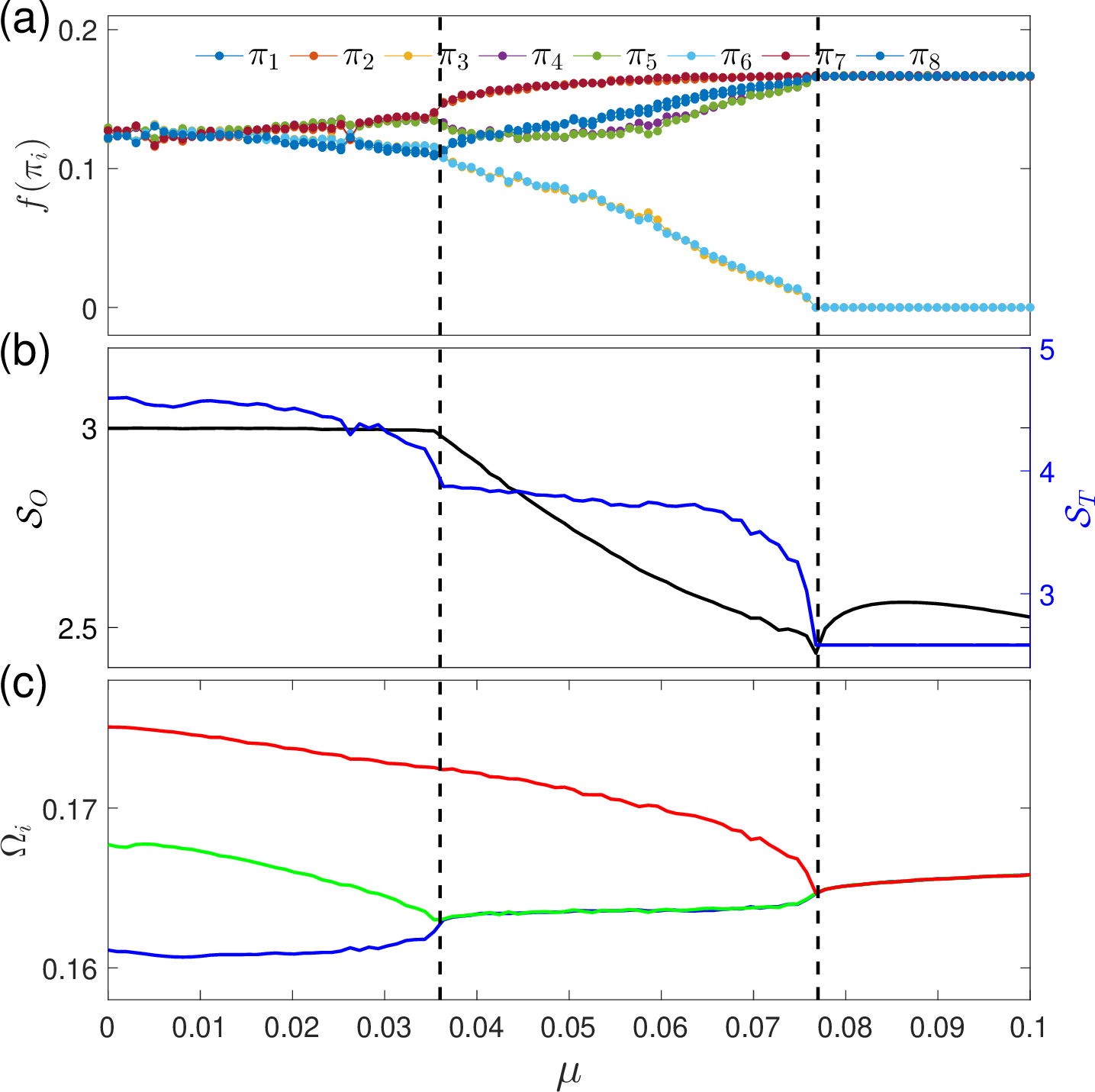}
\caption{Phase synchronization transitions of three coupled R\"ossler systems. (a) Frequency of each ordinal pattern $\pi_i$, (b) entropy values $\mathcal{S}_O$ (Eq.~\eqref{eq:Ho}) and $\mathcal{S}_T$ (Eq.~\eqref{eq:Ht}), (c) mean rotation frequency $\Omega_i$ of each oscillator. Subsystem $k_1$ and $k_2$ become synchronized at $\mu_{c,1}=0.036$, and $k_3$ joins the synchronization only at a stronger coupling strength $\mu_{c.2}=0.077$. Both critical coupling values are highlighted by vertical dashed lines. Modified from~\cite{Zhang2017b}. \label{fig:rosslerSync}}
\end{figure}

As a numerical application, Zhang {\textit{et al.}} \cite{Zhang2017b} constructed OPTNs to identify dynamical regimes shifts and characterize routes to phase synchronization. In particular, they considered three R\"ossler systems diffusively coupled via their $x$-components \cite{Nawrath2010} (see Eqs.~\eqref{threeRosPRL}), where $k = 1, 2, 3$ are indices for the different subsystems and $\mu$ is the coupling strength. For illustrative purposes, let us consider non-identical oscillators by choosing $\omega_1 = 0.98, \omega_2 = 1.02, \omega_3 = 1.06$ in Eqs.~\eqref{threeRosPRL}. The oscillator $k = 2$ is bidirectionally coupled to both $k=1$ and $k=3$, whereas there is no direct coupling between $k=1$ and $k = 3$. The Eqs.~\eqref{threeRosPRL} are numerically integrated by a fourth-order Runge-Kutta method with integration step $h = 0.01$. We construct OPTNs from the $x$ components, i.e., $(x_1, x_2, x_3)$, and use the definitions of patterns as summarized in Tab.~\ref{tab:3D}. The results are shown in Fig.~\ref{fig:rosslerSync}, which have been averaged over 50 random initial conditions when integrating Eqs.~\eqref{threeRosPRL}. 

\begin{figure}
	\centering
	\includegraphics[width=0.8\columnwidth]{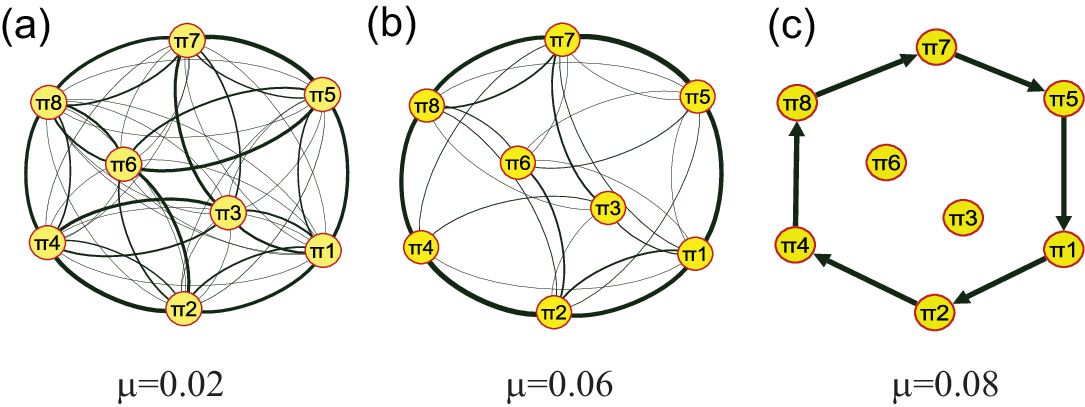}
\caption{Ordinal transition networks on the path to phase synchronization. (a) Non-synchronized regime, $\mu = 0.02 < \mu_{c,1}$, (b) regime in which the oscillators $k=1$ and $k=2$ are phase-synchronized, but not with $k=3$, $\mu = 0.06 \in [\mu_{c,1}, \mu_{c,2}]$, (c) regime in which all three oscillators are phase locked, $\mu = 0.08 > \mu_{c,2}$. The thickness of links indicates the corresponding transition frequencies. In (a) and (b), link arrows are suppressed. Reproduced from~\cite{Zhang2017b}. \label{fig:rosslerSyncNet}}
\end{figure}

In the absence of synchrony ($\mu < \mu_{c,1}=0.036$), the three oscillators evolve almost independently such that all ordinal patterns have the same frequencies of $0.125$. There are rather small gradual changes only when $\mu$ approaches $\mu_{c,1}$ (Fig.~\ref{fig:rosslerSync}(a)). The entropy value $\mathcal{S}_T$ is sensitive to these gradual changes by showing a pronounced downward trend, while $\mathcal{S}_O$ stays about constant (Fig.~\ref{fig:rosslerSync}(b)). The average rotation frequencies $\Omega_k$ of all three oscillators are shown in (Fig.~\ref{fig:rosslerSync}(c)), which confirms the absence of synchrony in this weak coupling regime.

Increasing the coupling strength, phase synchronization first appears between oscillators $k=1$ and $k=2$, but not with $k=3$ ($\mu \in [\mu_{c1}, \mu_{c2}] =[0.036, 0.077]$). In this regime, we observe monotonic increases in the frequencies of the order patterns $\pi_1$, $\pi_2$, $\pi_7$, and $\pi_8$ (Fig.~\ref{fig:rosslerSync}(a)). In addition, we find slower increases for the frequencies of patterns of $\pi_4$ and $\pi_5$, whereas those of $\pi_3$ and $\pi_6$ systematically decrease. The changes in the frequencies of order patterns are captured by both entropy values $\mathcal{S}_O$ and $\mathcal{S}_T$, showing gradual downward trends (Fig.~\ref{fig:rosslerSync}(b)) indicating a reduction in dynamical complexity of the coupled system. The average rotation frequencies $\Omega_k$ shown in Fig.~\ref{fig:rosslerSync}(c) indicate that $k=1$ and $k=2$ are phase locked to the same rotation frequency, but $k=3$ still evolves independently.

Finally, in the regime with all oscillators being phase-synchronized ($\mu > \mu_{c2} = 0.077$), we find that the frequencies of patterns $\pi_1$, $\pi_2$, $\pi_4$, $\pi_5$, $\pi_7$ and $\pi_8$ converge to the same value of $p(\pi_q) = 1/6$, while $\pi_3$ and $\pi_6$ are absent (Fig.~\ref{fig:rosslerSync}(a)). In other words, the patterns $\pi_3$ and $\pi_6$ are forbidden if all oscillators are synchronized. The entropy $\mathcal{S}_O$ shows some parabolic trend (first increasing and then decreasing slowly), while $\mathcal{S}_T$ stays constant at a value of about $2.585$ (Fig.~\ref{fig:rosslerSync}(b)). All mean rotation frequencies converge to the same value since three oscillators are phase locked (Fig.~\ref{fig:rosslerSync}(c)).

Across the transition from asynchrony to phase synchronization, the transition networks experience rather random transitions between all possible pairs of patterns to finally approach a state of transitions between a limited number of ordinal patterns as shown in Fig.~\ref{fig:rosslerSyncNet}. Specifically, as already discussed above, $\pi_3$ and $\pi_6$ are forbidden patterns if all three oscillators are synchronized.

		\subsection{Cross and joint ordinal transition networks} 
		The method of \cite{Zhang2017b} has been further generalized to construct cross and joint ordinal partition transition networks for two coupled systems \cite{Guo2018}. For this purpose, we start with an example of a single chaotic R\"ossler system as represented by three variables $(x_{1,i}, y_{1,i}, z_{1,i})$. The OPTN is constructed based on the signs of the increments of each variable, $(\Delta x_{1,i}, \Delta y_{1,i}, \Delta z_{1,i})$, where $\Delta x_{1,i} = x_{1,i+1} - x_{1,i}$, $\Delta y_{1,i} = y_{1,i+1} - y_{1,i}$, and $\Delta z_{1,i} = z_{1,i+1} - z_{1,i}$. The definition of corresponding patterns $\Pi_i \in (\pi_1, \cdots, \pi_K)$ with $K=8$ follows again the setting in Tab.~\ref{tab:3D}. 
        
\begin{figure}
	\centering
	\includegraphics[width=\columnwidth]{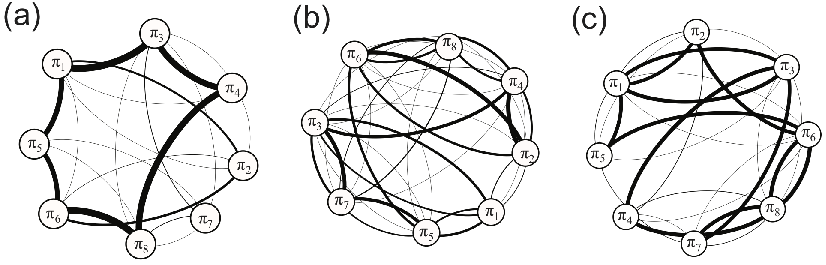}
	\caption{\small{Cross and joint OPTNs, which are reconstructed from two coupled R\"ossler systems (Eqs. \eqref{eq:coupled_roessler}) in the non-synchronized regime \cite{Guo2018}. (a) Normal cross ordinal pattern transition network (COPTN), (b) alternative version of the COPTN, and (c) joint ordinal pattern transition network (JOPTN). Link directions have been omitted in the  visualization. Reproduced from \cite{Guo2018}. } \label{fig:rosCOPT}}
\end{figure}

		For two coupled systems, we have additional time series from the other system as represented by $\{(x_{2,i}, y_{2,i}, z_{2,i})\}$. A cross-ordinal pattern transition network (COPTN) compares the relative rates of changes between the two systems by the signs of $(\Delta x_{1,i} - \Delta x_{2,i}), (\Delta y_{1,i} - \Delta y_{2,i})$ and $(\Delta z_{1,i} - \Delta z_{2,i})$. The resulting pattern definitions of a COPTN are summarized in Tab.~\ref{tab:3DCOPT}. An example of a COPTN constructed from two coupled R\"ossler systems in a non-synchronized regime \cite{Guo2018} is shown in Fig. \ref{fig:rosCOPT}(a). 
\begin{table}[htb]
\centering
\begin{tabular}{|c|c|c|c|c|c|c|c|c|}
\hline
$\Pi$      & $\pi_1$ & $\pi_2$ & $\pi_3$ & $\pi_4$ & $\pi_5$ & $\pi_6$ & $\pi_7$
& $\pi_8$\\
\hline
$\Delta x_1 - \Delta x_2$ & $+ $ & $+ $ & $+$ & $+$ & $ - $ & $ - $ & $-$ & $ - $\\
\hline
$\Delta y_1 - \Delta y_2$ & $ + $ & $ + $ & $ -$ & $ -$ & $ + $ & $ + $ & $ -$ & $ -$\\
\hline
$\Delta z_1 - \Delta z_2$ & $ + $ & $ - $ & $ +$ & $ -$ & $ + $ & $ - $ & $+$ & $ -$\\
\hline
\end{tabular}
\caption{Pattern definitions of a COPTN. $``+"$ means a positive value while $``-"$ stands for a negative value.  \label{tab:3DCOPT}}
\end{table}

Considering the effects of the different magnitudes of the three variables, in \cite{Guo2018} the authors defined an {\emph{alternative}} COPT by replacing $\Delta x_{1,i} - \Delta x_{2,i}$ by $\Delta x_{1,i} / x_{1,i} - \Delta x_{2,i} / x_{2,i}$, respectively, $\Delta y_{1,i} - \Delta y_{2,i}$ by $\Delta y_{1,i} / y_{1,i} - \Delta y_{2,i} / y_{2,i}$, and $\Delta z_{1,i} - \Delta z_{2,i}$ by $\Delta z_{1,i} / z_{1,i} - \Delta z_{2,i} / z_{2,i}$. An example of this alternative COPTN is shown in Fig.~\ref{fig:rosCOPT}(b). Comparing Figs.~\ref{fig:rosCOPT}(a) and \ref{fig:rosCOPT}(b), this alternative COPT reflects better the non-coherent transitions between ordinal patterns since the coupling strength is in the non-synchronized regime ($\nu = 0.02$, $\mu_{21} = 0$ and $\mu_{12} = 0.01$ in Eqs. \eqref{eq:coupled_roessler}). We note, however, that normalizing by the local value of each component may result in numerical problems close to the roots of each component time series. To better account for amplitude effects in the different variables, other normalizations (like with respect to the individual components' variances or ranges) are possible but have not yet been explored systematically.

Similar to the COPTN, a joint ordinal pattern transition network (JOPT) compares the relative rates of changes between two systems by the signs of $\Delta x_{1,i} \cdot \Delta x_{2,i}$, $\Delta y_{1,i} \cdot \Delta y_{2,i}$ and $\Delta z_{1,i} \cdot \Delta z_{2,i}$. The definitions of the resulting patterns are summarized in Tab.~\ref{tab:3DJOPT}. An example of a JOPTN is shown in Fig.~\ref{fig:rosCOPT}(c).
\begin{table}[htb]
\centering
\begin{tabular}{|c|c|c|c|c|c|c|c|c|}
\hline
$\Pi$      & $\pi_1$ & $\pi_2$ & $\pi_3$ & $\pi_4$ & $\pi_5$ & $\pi_6$ & $\pi_7$
& $\pi_8$\\
\hline
$\Delta x_1 \cdot \Delta x_2$ & $+ $ & $+ $ & $+$ & $+$ & $ - $ & $ - $ & $-$ & $ - $\\
\hline
$\Delta y_1 \cdot \Delta y_2$ & $ + $ & $ + $ & $ -$ & $ -$ & $ + $ & $ + $ & $ -$ & $ -$\\
\hline
$\Delta z_1 \cdot \Delta z_2$ & $ + $ & $ - $ & $ +$ & $ -$ & $ + $ & $ - $ & $+$ & $ -$\\
\hline
\end{tabular}
\caption{Pattern definitions of a JOPT. $``+"$ means a positive value while $``-"$ stands for a negative value.  \label{tab:3DJOPT}}
\end{table}
In contrast to cross ordinal patterns, we notice that the joint ordinal patterns represent whether the respective variables of two systems show the same trend of changes or not, regardless of the magnitudes of the respective variables.

The ideas of both COPTN and JOPTN have been applied to analyze synchronization transitions \cite{Guo2018}. Note that COPTN and JOPTN provide two different ways to construct networks from multivariate time series, providing complementary information. The ordinal patterns of a COPTN are defined by considering the signs of the difference $\Delta \vec{x}_1 - \Delta \vec{x}_2$ between two subsystems. In contrast, the ordinal patterns of a JOPTN are defined by the signs of the product of $\Delta \vec{x}_1 \cdot \Delta \vec{x}_2$. Notably, the amplitudes of oscillations of different variables influence directly the definition of a COPTN. However, amplitudes become practically irrelevant for a JOPTN because only the signs of the product are considered. In addition, it is straightforward to generalize the ideas of JOPTNs from two to three (or even $n$) coupled subsystems with an extended number of pattern definitions. In turn, it remains a challenge to construct COPTNs for three or more coupled subsystems.

\subsection{Other related approaches}

		For one-dimensional symbolic sequences, one may construct a directed symbolic transition network \cite{Emmert2012}. Working with experimental data of inter-beat ($RR$) intervals of the human heart from cardiac regulations, Makowiec~\emph{et~al.} proposed to construct transitions networks from the subsequent increment series $\Delta RR_i = RR_i - RR_{i-1}$ \cite{Makowiec2013,Makowiec2013b,Makowiec2014b,Makowiec2015,Makowiec2015b,Makowiec2016}. In this series of work, they have demonstrated that transition network approaches are a powerful tool for quantifying the unique properties of the $RR$-interval time series of patients after heart transplant surgery. 
		
		When constructing OPTNs by a sliding window scheme \cite{Small2013}, the ordinal pattern of the windowed sequence corresponds to one node of the network. Since amplitude information is neglected, this approach may be combined with a transition network, where the nodes of the network are the binned amplitudes of the time series \cite{Sun2014}. More specifically, each time step $i$ is associated with a pair of symbols containing both, amplitude information $a_i$ and ordinal pattern $o_i$. The former is calculated by binning the time series in the interval $[\min_i y_i, \max_i y_i]$ into $K$ subintervals of equal size. $a_i$ is then simply the bin number associated with $y_i$. On the other hand, $o_i$ is the ordinal pattern of the embedded vector $(y_i, y_{i+\tau}, \dots, y_{i+(m-1)\tau})$. The symbol pair at step $i$, $\xi_i=(a_i,o_i)$, is then one node of the network, and it is connected by a directed link to the symbol pair $(a_{i+1},o_{i+1})$ of the successive time step. Furthermore, this algorithm has been combined with recurrence networks and surrogate networks, which has proven powerful in detecting weak nonlinearities in time series~\cite{Laut2016}. 
		
		By coarse graining of financial time series, one may focus on the particular up-down behavior associated with volatility in stock index series \cite{Li2006b,Li2007a}. More specifically, the authors of the former papers symbolize time series by using the parameter $\theta = \arctan \Delta x / \Delta t$, which characterizes the local rate of increase or decrease of the observations. Then, these local rates are coarse-grained into four states ($R, r, d, D$), which correspond to violent-up meta, common-up meta, common down meta, and violent down meta patterns, respectively. It was found that the topologically relevant nodes of the resulting transition networks play important roles in both information control and transport at the stock market \cite{Li2007a}. 
		
		In a series of works by Gao \textit{et al.}~\cite{Gao2014a,Gao2014,Gao2015}, the authors proposed a linear regression pattern transmission algorithm which captures the evolution of linear regression of bivariate time series. This algorithm has been demonstrated to be a useful tool to show the correlation mode transmission in crude oil spot price and future price \cite{Huang2015}. Based on reduced autoregressive models generated from time series, a directed transition network reconstruction algorithm has been used in \cite{Nakamura2012a}, such that the delay information has been successively captured by the transition behavior of the resulting network. In combination with proper surrogate methods, extending these ideas from a univariate to multivariate time series analysis is possible~\cite{Nakamura2016}. 
		
        Another alternative way to approach a symbolization based upon multivariate time series has been proposed by Gao \emph{et~al.}~\cite{Gao2015b}. In that paper, the authors studied the behavior of two-phase (gas--fluid) flows based on experimental observations by four conductance sensors within the two media. Starting from the corresponding acquired time series, they used a sliding window approach to estimate time-dependent correlation between all pairs of time series. A symbolic encoding has been achieved by rank-ordering the resulting six pairwise correlation coefficients. Based on the corresponding symbolization by ordinal patterns of correlation values, the authors constructed an OPTN as described above and demonstrated that corresponding network properties like weighted local clustering coefficients and closeness centralities were able to trace qualitative changes in the resulting dynamics in dependence on the gas' superficial velocity.
		
        Finally, a recent modification of transition network approaches relieves the restriction to sequences or ordinal patterns of the same degree, but rather utilizes an encoding based on optimal symbols representing sequences or patterns of variable length that form a unique alphabet for a loss-free compression of the underlying symbolic sequence \cite{Walker2018}. In the original paper, the authors employed this idea by making use of Lempel-Ziv-Welch like compression algorithm \cite{Welch1984} together with coarse-graining based transition networks constructed upon single threshold (i.e., a binary encoding). As a statistics of interest, they used the fraction of unused codewords in these \emph{coarse-graining based compression networks} and demonstrated the discriminative skills of this measure for different stochastic and chaotic model systems as well as real-world EEG data involving epileptic seizures. A more recent paper by the same authors combined the same compression algorithm with the idea of ordinal pattern based encoding and showed that considering the minimal cycle basis of the resulting \emph{ordinal compression networks} allows for testing for time irreversibility of the underlying time series.

%% file: Chapter06_Applications/Chapter06_Applications.tex
\section{Real-world applications}\label{sec:Applications}
In this section, we illustrate the application of the above discussed methods on selected  examples from theory and real-world research questions. We focus on recurrence network approaches, (horizontal) visibility graphs, and transition network approaches, since those methods have found much wider and deeper applications in diverse fields.

	\subsection{Recurrence networks}
		Although recent work on RNs and multivariate generalizations thereof has been focused on the development of the theoretical framework and its numerical exploration using simple low-dimensional model systems, there have already been several successful applications to characterizing system's properties from experimental or observational time series. For example, the successful application of RN to predict protein structural classes has been reported in \cite{Olyaee2016}. Here we summarize some main applications to various disciplines and choose one successful application to understanding climate regularities.

		\paragraph{Applications in Earth sciences}
		One important field of recent applications is paleoclimatology, which has already been taken as an illustrative example in the seminal paper by Marwan \textit{et~al.} \cite{Marwan2009}. The corresponding study was later extended to a systematic investigation of the temporal variability profile of RN-based complexity measures for three marine sediment records of terrigenous dust flux off Africa during the last 5 million years. Donges \textit{et~al.} \cite{Donges2011,Donges2011a} argued that RNs can be used for characterizing dynamics from non-uniformly sampled or age-uncertain data, since this methodological approach does not make explicit use of time information. In turn, due to the necessity of using time-delay embedding, there is implicit time information entering the analysis, which has been recognized but widely neglected in previous works. Notably, disregarding age uncertainty and sampling heterogeneity appears a reasonable approximation only in cases where the distribution of instantaneous sampling rates remains acceptably narrow. In fact, more recent findings point to a strong dependence of the validity and robustness of the results obtained for paleoclimate time series on the specific archive and proxy variable \cite{Lekscha2018}. As an alternative, the letter study suggested utilizing derivative embedding instead of time delay embedding for phase space reconstruction and discussed several approaches for numerically estimating the numerical derivatives of the time series values required for this procedure. Another approach for tackling the problem of potentially non-robust individual significance of RN properties for individual time series integrates information from multiple paleoclimate time series and explicitly propagates age uncertainty to RN-measure uncertainties. This multi-archive approach has been used to investigate non-linear regime shifts in Asian summer monsoon variability during the Holocene and its potential impacts on human societies, conflicts and migration \cite{donges2015nonlinear}.

		As another methodological step towards better understanding climatic mechanisms, Feldhoff {\textit{et al.}} have used two speleothem records for studying interdependencies between the two main branches of the Asian summer monsoon (the Indian and East Asian summer monsoon) by means of inter-system recurrence network (IRN) approaches \cite{Feldhoff2012,Marwan2012Nolta}. For this purpose, they selected two data sets of oxygen isotope anomalies from speleothems obtained from two caves in China and the Oman, respectively, which can be considered as proxies for the annual precipitation and, hence, the overall strength of the two monsoon branches over the last about 10,000 years. The asymmetries of the IRN cross-transitivities and global cross-clustering coefficients provided clear evidence for a marked influence of the Indian summer monsoon on the East Asian branch rather than vice versa, which is in good agreement with existing climatological theories. As a subsequent extension of this work, Feldhoff {\textit{et al.}} emphasized the possibility of repeating the same kind of analysis in a sliding windows framework, thereby gaining information on possible temporal changes of the associated climatic patterns during certain time periods as recently revealed using correlation-based complex network analysis applied to a larger set of speleothem records from the Asian monsoon domain \cite{Rehfeld2012}.

		In order to characterize dynamical complexity associated with more recent environmental variability, Lange and B\"ose \cite{Boese2012,Lange2013Book} used RQA as well as RN analysis for studying global photosynthetic activity from remote sensing data in conjunction with global precipitation patterns. Specifically, they studied 14-years long time series (1998-2011) of the fraction of absorbed photosynthetically active radiation with a spatial resolution of 0.5$\degree$ around the Earth and a temporal sampling of about ten days, providing time series of $N=504$ data points. Their results revealed very interesting spatial complexity patterns, which have been largely, but not exclusively determined by the amplitude of the annual cycle of vegetation growth in different ecosystems.

Finally, RN analysis -- in combination with more traditional RQA characteristics -- has been employed recently for studying temporal variations in the dynamical complexity of geomagnetic field fluctuations at time-scales between hours and weeks associated with sequences of quite-time magnetic field episodes and geomagnetic storms \cite{Donner2018}. It has been shown that especially the RN transitivity allows a unique discrimination between the corresponding ``physiological'' and ``pathological'' states of the Earth's magnetosphere. A follow-up work investigated the corresponding changes in RN and RQA characteristics in greater detail for the same geomagnetic activity index (Dst) together with corresponding changes in solar wind variables that could particularly affect the stability of the Earth's magnetic field \cite{Donner2018b}.

		\paragraph{Applications in fluid dynamics}
		In a series of papers, Gao \textit{et~al.} investigated the emerging complexity of dynamical patterns in two-phase gas-liquid or oil-water flows in different configurations using RN techniques. Bifurcation scenarios from slugs to bubbles of a two phase flow of water-air occurring in a circular horizontal mini-channel have been recently analyzed by RPs and RN approaches in \cite{Gorski2015}. Similarly, the RN transitivity of pressure drop fluctuation time series has been used to distinguish between different dynamical patterns in two-phase flows \cite{Mosdorf2015}. In general, multiple sensors measuring fluctuations of electrical conductance have been used for obtaining signals that are characteristic for the different flow patterns. For gas-liquid two-phase upward flows in vertical pipes, different types of complex networks generated from observational data have been proposed, among which the degree correlations (assortativity) of RNs was proven to be particularly useful for distinguishing between qualitatively different flow types \cite{Gao2009,Gao2009a,Gao2010a}. One may also construct a directed weighted RN \cite{Gao2012,Gao2012a,Gao2013b,Gao2013d,Zhang2013b}. For oil-water two-phase upward flows in a similar configuration, the global clustering coefficient of RNs reveals a marked increase in dynamical complexity (detectable in terms of a decreasing $\hat{\mathcal{C}}$) as the flow pattern changes from slug flow over coarse to very finely dispersed bubble flow \cite{Gao2013,Gao2013c}. In case of oil-water two-phase flows in inclined pipes \cite{Gao2010}, the motif distributions of RNs (specifically, the frequency distributions of small subgraphs containing exactly four vertices) revealed an increasing degree of heterogeneity, where the motif ranking was conserved in all experimental conditions, whereas the absolute motif frequency dramatically changed. The corresponding results were independently confirmed using some classical measures of complexity, which indicated increasing complexity in conjunction with increasing heterogeneity of the RN motif distributions. Finally, for characterizing horizontal oil-water flows \cite{Gao2013,Gao2013d}, RN and inter-system RN analysis were combined for studying conductance signals from multiple sensors. Specifically, cross-transitivity was found a useful measure for tracing the transitions between stable stratified and unstable states associated with the formation of droplets. Furthermore, Gao {\textit{et al.}} \cite{Gao2015a,Gao2016,Gao2016b,Gao2016c} further extended these ideas to construct multivariate weighted recurrences networks from multi-channel measurements from different oil-water flow patterns.
        
        In a similar context, Charakopoulos \emph{et~al.} \cite{Charakopoulos2014} combined the $k$-nearest neighbor version of RNs and visibility graphs for studying temperature time series obtained from different parts of a turbulent heated jet, which allowed distinguishing dynamically different regions within the jet and attributing them to distinct physical mechanisms.

		\paragraph{Applications in electrochemistry}
		Zou \textit{et~al.} \cite{Zou2012b} studied the complexity of experimental electrochemical oscillations as one control parameter of the experiments (temperature) was systematically varied. By utilizing a multitude of complementary RN characteristics, they could demonstrate a systematic rise in dynamical complexity as temperature increased, but an absence of a previously speculated phase transition \cite{Wickramasinghe2010} separating phase-coherent from noncoherent chaotic oscillations. The latter results were independently confirmed using other classical indicators for phase coherence, as well as studies of a corresponding mathematical model of the specific electrochemical processes.

		\paragraph{Applications in medicine}
		Finally, there have been a couple of successful applications in a medical context. Marwan \textit{et~al.} \cite{Marwan2010c} demonstrated that the global clustering coefficients of RNs obtained from heartbeat intervals, diastolic and systolic blood pressure allowed a reliable identification of pregnant women with pre-eclampsia, a cardiovascular disease during pregnancy with a high risk of fetal and maternal morbidity. Their results were further improved by Ram\'{i}rez \textit{et~al.} \cite{Ramirez2012,Ramirez2013} who considered combinations of various RN-based network characteristics. In a similar spirit as for cardiovascular diseases, recent results point to the capability of RN characteristics for discriminating between the EEG signals of healthy and epileptic patients or to identify pre-seizure states in epilepsy patients \cite{Subramaniyam2013,Subramaniyam2015,ngamga2016,gao2018}.

		\paragraph{Understanding climate regularity transitions by RN analysis}
		The results of Donges \textit{et~al.}~\cite{Donges2011a} pointed to the existence of spatially coherent changes in the long-term variability of environmental conditions over Africa, which have probably influenced the evolution of human ancestor species. Specifically, RN transitivity and average path length have been interpreted as indicators for ``climate regularity'' (i.e., the complexity of fluctuations as captured by the transitivity dimensions) and ``abrupt dynamical changes'', respectively. By identifying three time intervals with consistent changes of the RN properties obtained from spatially widely separated records, it has been possible to attribute the corresponding long-term changes in the dynamics to periods characterized by known or speculated mechanisms for large-scale climate shifts such as changes in the Indian ocean circulation patterns, the intensification of the atmospheric Walker circulation, or changes in the dominant periodicity of Northern hemispheric glacial cycles. Moreover, Donges \textit{et~al.}\cite{Donges2011} demonstrated a good robustness of the results of RN analysis obtained in a sliding windows framework when varying the corresponding parameters (e.g., window size or embedding delay) over a reasonable range.

\begin{figure}[htbp]
		\centering
			\includegraphics[width=\textwidth]{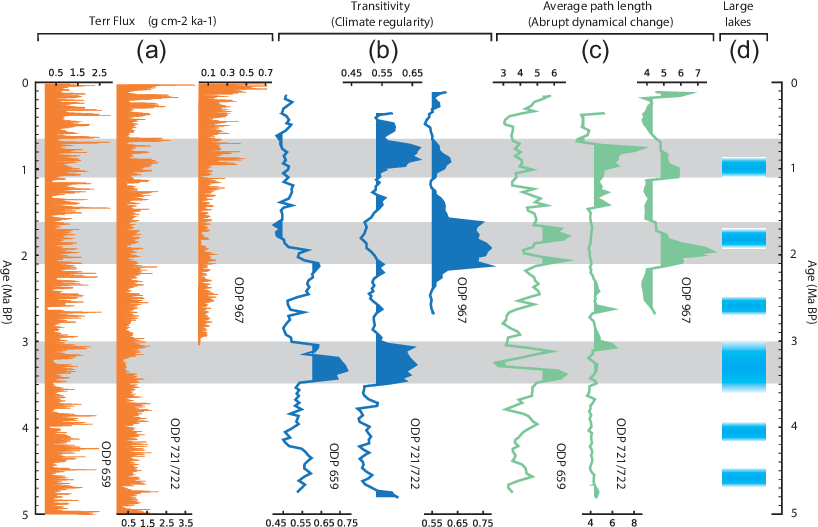}
		\caption{(a) Terrestrial dust flux records from the three considered ODP sites distributed around Africa covering the Plio-Pleistocene. (b, c) Results of RN analysis of the three dust flux records including 90\% confidence bands (vertical shadings) of a stationarity test. Comparing both measures transitivity (interpreted as a measure of climate regularity) and average path length (interpreted as a measure of abrupt dynamical change in climate) for all records reveals significant and synchronous large-scale regime shifts in dust flux dynamics (horizontal shadings). (d) Time intervals with geological evidence for large lakes in East Africa, comprising collected information from different areas in the EARS and additional results from the Afar basin. Modified from \cite{Donges2011a}. } \label{fig:appl_recurrence_network}
		\end{figure}

		More specifically, the paleoclimate variability transitions in East Africa have been demonstrated by analyzing marine sediment paleoclimate records from ocean drilling program (ODP) sites 659 in the East Atlantic, 721/722 in the Arabian Sea, and 967 in the Eastern Mediterranean Sea. The time series of three chosen sites are shown in Fig. \ref{fig:appl_recurrence_network}(a). To date, these marine sediments provide the only archive that allows the study of the Plio-Pleistocene African climate on all relevant time scales. However, earlier analyses of terrigenous dust flux records using traditional time series analysis techniques to detect important transitions in the African climate yielded partly contradictory results with respect to the signature and timing of these events \cite{deMenocal2004,Trauth2009}. Difficulties like these are to be expected when applying linear methods to the highly nonlinear climate system underlying paleoclimate proxy records. To circumvent this problem and explore the vast remainder of nonlinear phenomena, RNs can be used \cite{Donges2011a}.

		To this end, Donges \textit{et~al.} rely on two established measures of RNs: transitivity $\mathcal{T}$ and average path length $\mathcal{L}$. Furthermore, transitivity $\mathcal{T}$ has been interpreted as a climate regularity index since noisy or chaotic dynamics gives rise to low values, whereas (almost) periodic or laminar behavior induces high values. The average path length $\mathcal{L}$ shows much sensitivity on abrupt dynamical changes between different dynamical regimes when extreme values have been observed for $\mathcal{L}$. Both $\mathcal{T}$ and $\mathcal{L}$ together provide double-evidence points at a particularly relevant feature in the data since they are responsive to different nonlinear aspects of the time series data and do not necessarily show transitions at the same epochs.

		The results of $\mathcal{T}$ and $\mathcal{L}$ for the considered terrigenous dust record time series are shown in Fig. \ref{fig:appl_recurrence_network}(b, c). More specifically, $\mathcal{T}$ reveals surprisingly similar long-term change in short-term fluctuations before about 1.5 Ma B.P. (in contrast to the Mediterranean ODP site 967), although both sites are strongly geographically separated and, hence, characterized by distinct wind systems and dust sources (Fig.~\ref{fig:appl_recurrence_network}(b)). This overall picture indicates that changes in $\mathcal{T}$ during the Pliocene and early Pleistocene are robust manifestations of long-term variations in the dynamics of large-scale African dust mobilization and transport. More importantly, three transition periods can been identified which can be clearly related to distinct and known climatic mechanisms (Fig.~\ref{fig:appl_recurrence_network}) \cite{Donges2011a}.

		The first transition period is identified by a pronounced maximum of $\mathcal{T}$ between 3.5 and 3.0 Ma B.P. at both ODP sites 659 and 721/722 signaling a period of exceptionally regular dust flux dynamics (Fig.\ref{fig:appl_recurrence_network}(b)). The transition epochs highlighted by average path length $\mathcal{L}$ support these findings (Fig.~\ref{fig:appl_recurrence_network}(c)). The time interval 3.5 -- 3.0 Ma B.P. is characterized by three distinct and highly significant extrema (two maxima and one minimum in $\mathcal{L}$) in the ODP site 659 record, indicating shifts between regimes of higher and lower regularity in the variations of environmental conditions \cite{Donges2011a}. This identified long period during which global mean temperatures have been consistently higher than present day, which is thus considered as an analog for the future climate of the late 21st century if anthropogenic emissions of greenhouse gases continue to rise. At the same time, RN analysis reveals an enhanced $\mathcal{T}$ of African dust flux variations in both the Arabian Sea and Atlantic Ocean. At ODP site 721/722, this observation is predominantly caused by a well-pronounced epoch of relatively weak and approximately constant dust flux between about 3.36 and 3.17 Ma B.P. A similar --- but shorter --- feature is found at ODP site 659 between about 3.25 and 3.19 Ma B.P. (Fig. \ref{fig:appl_recurrence_network}(d)) as well as at ODP sites 661 and 662 in the Eastern Equatorial Atlantic. The presence of such very similar features with a clearly different timing suggests the presence of either one common climatological mechanism influencing the Arabian Peninsula much earlier than Northwest Africa or two distinct (but eventually interrelated) factors, where the first affected only the Northeast African and/or Arabian dust flux dynamics.

		The second identified transition is represented by an extended and highly significant maximum of $\mathcal{T}$ between about 2.25 and 1.6 Ma B.P. and of $\mathcal{L}$ between 2.2 and 1.7 Ma B.P. The latter one roughly coincides with the observations made at ODP site 659. The timing of this transition period in the Early Pleistocene (2.25 -- 1.6 Ma B.P.) well coincides with known large-scale changes in atmospheric circulation associated with an intensification and spatial shift of the Walker circulation.

		Moreover, the third transition period is between about 1.1 and 0.7 Ma B.P., when both $\mathcal{T}$ and $\mathcal{L}$ show significant maxima for ODP sites 967 and 721/722 but not for ODP site 659. This interval corresponds to the Middle Pleistocene transition (MPT) characterized by a change from glacial cycles predominantly related to obliquity variations of Earth's orbit (approximately 41 ka period) to such with an approximately 100 ka periodicity. The timing of this transition and its underlying mechanisms have been extensively studied elsewhere. That the MPT is not detected in the record from ODP site 659 by RN analysis does not imply that it did not have any climatic impact in the corresponding dust-source areas in Northwest Africa. Instead it shows that our technique is not sensitive to the local signature of the transition, if present, e.g., if it manifests itself in some change of trend \cite{Donges2011a}. Alternatively, the locally available data may be insufficient in quality and/or resolution to reveal the subtle type of events RN analysis is focusing on.

		In summary, this analysis identifies three main epochs of interest: 3.5 -- 3.0, 2.25 -- 1.6, and 1.1 -- 0.7 Ma B.P., as shown in Fig.\ref{fig:appl_recurrence_network}. All three are characterized by statistically significant extrema of $\mathcal{T}$ and/or $\mathcal{L}$ in at least two of the analyzed records. In addition, there are further shorter time periods of considerably increased or decreased $\mathcal{T}$ or $\mathcal{L}$ observed in the different records which coincide with environmental changes too (lake level high stands, Fig.~\ref{fig:appl_recurrence_network}(d)). Based on results from a meta-anaylsis using event coincidence analysis, tt was furthermore proposed that these detected large-scale changes in climate regularity may have acted as drivers of human evolution in Africa during the Plio-Pleistocene \cite{Donges2011a}.

	\subsection{Visibility graphs} \label{sec:appVGs}

	Very similar as RN approaches, (H)VGs have been applied to experimental time series from various fields, for instance, energy dissipation rates in fully developed turbulence \cite{Liu2010,Manshour2015,Manshour2015a}, financial data \cite{Ni2009,Yang2009,Qian2010,Wang2012,Flanagan2016}, physiological time series \cite{Lacasa2009,Shao2010,Dong2010,Ahmadlou2010,Jiang2013,Hou2014}, seizure detections by EEG signals \cite{Bhaduri2015,Liu2017a,Zhang2018}, cardiorespiratory interaction signals \cite{Long2014}, and alcoholism identification by EEG signals \cite{Zhu2014}. In the geoscientific context, \cite{Elsner2009} studied the time series of annual US landfalling hurricane counts. Subsequently, further studies investigated daily streamflow series from the US and China \cite{Tang2010}, air temperature data from China \cite{Wang2009}, wind speed records from central Argentina \cite{Pierini2012}. In the field of paleoclimatology, VG-based tests for time-reversal asymmetry were used to detect indications for a North Atlantic ocean circulation regime shift at the onset of the Little Ice Age \cite{schleussner2015indications}. VGs were also used for studying seismic activity in Italy \cite{Telesca2012}  and the  Corinth rift in western central Greece \cite{Hloupis2017}. Nonlinear features of seismic time series have been recently reviewed in \cite{Telesca2018b}. Both $k$ nearest neighbors network and VG analysis show almost the same qualitative behavior and allow to reveal the underlying system dynamics in turbulent heated jets \cite{Charakopoulos2014}. Furthermore, the VG approach has been used to disclose the fractal properties of the event-by-event fluctuations of multiparticle production in Nucleus-Nucleus collisions \cite{Mondal2018}.

	 Motif structures and subgraphs played important roles in forming VGs, which have been used to human ventricular fibrillation (ECG) time series \cite{Li2011,Li2012}, ECG diagnosis of epilepsy \cite{Tang2013}, and air traffic flow data \cite{Liu2018}. Simple topological measures such as the diameter, average path length, modularity, clustering coefficient, density and hierarchical organizations of networks have been used to characterize different dynamic properties between atmospheric and oceanic variables \cite{CHARAKOPOULOS2018}. In addition, the (H)VGs have been proposed to predict catastrophes of a non-autonomous network which derived from a marine system \cite{Zhang2018b}, which demonstrates that the topological characteristics like average degrees of the networks do show pronounced signatures at the onset of catastrophes. Fractal characteristics of (H)VGs have been reported in fractional Brownian motions, which has been recently extended to multiparticle emission data in high energy heavy-ion collisions \cite{Mali2018}, which shows consistent power law degree distributions as compared to the results as obtained by the traditional sandbox algorithm. In \cite{Gao2016}, a slight modification of HVG algorithm has been proposed to extract the multiscale properties of time series from oil-water two phase flow signals. In the case of intermittent time series, some phenomenology theories have been obtained in order to link the laminar episodes and chaotic bursts with the connectivity of the resulting HVGs \cite{Nunez2013,Nunez2014}.

		Next, we illustrate two specific examples showing the applications of (H)VG analysis to identify nonlinear ocean circulation regime shifts by paleoclimate time series of ocean sediment cores and time series of sunspots.

		\paragraph{VG analysis of nonlinear ocean circulation regime shifts at the onset of the Little Ice Age} \textit{Schleussner et al.} applied HVG-based tests for time-reversal asymmetry to paleoclimate records to detect indications for a North Atlantic ocean circulation regime shift at the onset of the Little Ice Age \cite{schleussner2015indications}.

		The transition from the Medieval Climate Anomaly (MCA) to the Little Ice Age (LIA) primarily in the  Northern Hemisphere is one of the most important climatic shifts during the pre-industrial last millennium. Although recent paleoclimatic reconstructions reveal no coherent global-scale cooling at the onset of the LIA, they agree on a generally colder period from the 16th to the 19th century (e.g. \cite{Pages2K2013}). Alternatively, \textit{Masson-Delmotte et al.} \cite{IPCC_AR5_WG5_Ch5} give a period between 1450 and 1850. In Europe, the regional expression of the LIA is associated with a spatially and temporally heterogeneous cooling, being most pronounced in central and northern Europe \cite{buentgen_tegel11,Pages2K2013}.

		Besides uncertainties in timing and extent, also the origin of this climate shift is still a subject of debate. Since the LIA coincides with several minima in the total solar irridiance (TSI), solar activity has been proposed as a possible driver already by \textit{Eddy et al.}\cite{eddy76}. The impact of TSI changes on the coupled ocean-atmosphere system in the North Atlantic has been investigated in a variety of different model studies \cite{crowley00,zorita_storch04,swingedouw_terray12}. As an alternative hypothesis, volcanic eruptions have been suggested as the origin of the regional cooling \cite{robock79,crowley00}. Despite the short life-time of volcanic aerosol loadings, they have been found to influence North Atlantic climate variability on multi-decadal time scales \cite{ottera_bentsen10,fischer_luterbacher07,zanchettin_timmreck11,goosse_crespin12}. Decadally-paced volcanic eruptions have been reported to trigger coupled sea-ice oceanic feedbacks leading to a sustained slow-down of the Atlantic Meridional Overturning Circulation (AMOC) and persistent hemispheric cooling in modelling studies of the last millennium \cite{miller_geirsdottir12,schleussner_feulner13}.

		\textit{Schleussner et al.} \cite{schleussner2015indications} present additional evidence for such a non-linear regime shift in the North Atlantic circulation dynamics during the MCA-LIA transition based on an analysis of two fossil diatom based high-resolution August sea surface temperature series from two ocean sediment cores from the central subpolar basin (Rapid 21-COM) and the Norwegian Sea (CR 948/2011). They find robust signatures of time-irreversibility in both records during the MCA-LIA transition using a test based on horizontal visibility graphs. Comparison with simulations with the climate model of intermediate complexity CLIMBER-3$\alpha$ reveals good agreement between the sediment cores and model outcome. Paleo reconstructions as well as model results support the hypothesis of a non-linear oceanic regime shift at the onset of the LIA.

\begin{figure}
		\noindent\includegraphics[width=\columnwidth]{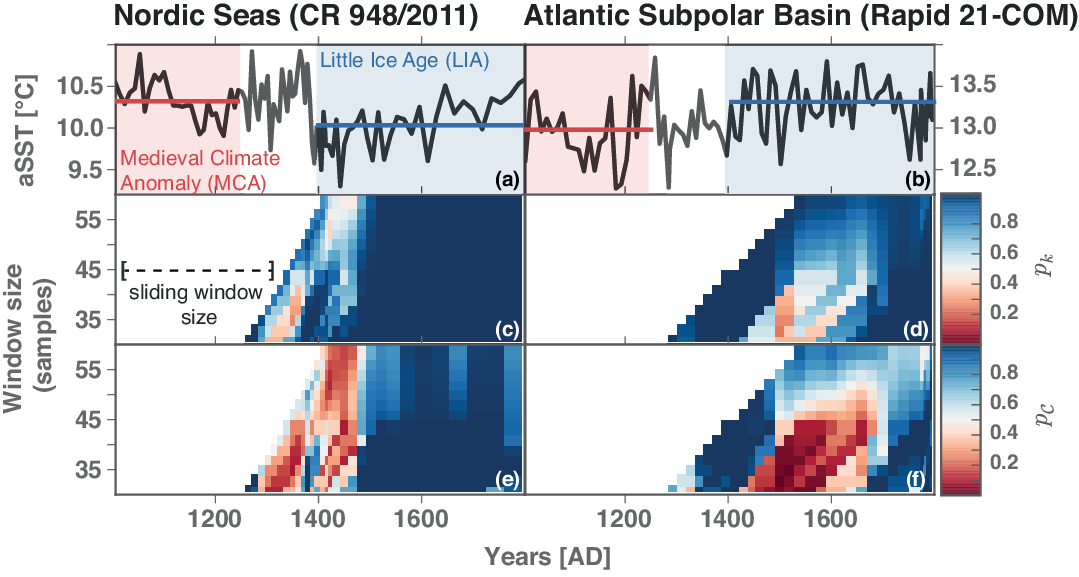}
		\caption{Results of sliding window HVG-based tests for time-series irreversibility to detect nonlinear regime shifts in Atlantic paleooceanic dynamics during the past millennium. Reconstructed aSST time series from (a) the Nordic Seas (CR 948/2011) and (b) the Atlantic subpolar basin (Rapid 21-COM). The MCA (until 1250) and the LIA period (1400--1850) are shaded in red and blue, respectively, and the means over the MCA and LIA periods are depicted by solid lines coloured accordingly. (c, d) Results of the degree - based HVG time series irreversibility tests ($p_k$) for different window sizes. $p$-values close to unity (blue) indicate full reversibility, whereas close to zero (red) point towards time-irreversibility. (e, f)) Results of the clustering - based HVG time-irreversibility tests ($p_\mathcal{C}$). Figure modified from \cite{schleussner2015indications}.}
		\label{fig:lia_vg_analysis}
		\end{figure}

		Despite a basin-wide cooling in the whole North Atlantic, the Rapid 21-COM time series exhibits a warming during the LIA. On the contrary, CR 948/2011 shows an abrupt cooling after 1400, preceding the Rapid 21-COM warming by about 50 years (compare Fig.~\ref{fig:lia_vg_analysis} a (b) for CR 948/2011 (Rapid 21-COM)). A warming signal in the subpolar North Atlantic in contrast to a cooling in the Nordic Seas is also found in sub-decadal ocean sediment records from North Iceland and North East Newfoundland \cite{sicre_wekstroem14}. \textit{Andrews et al.}\cite{andrews_jennings14} report signatures of a major environmental shift at the MCA-LIA transition in two calcite and quartz based sediment records from the Denmark Strait.

		\textit{Schleussner et al.} perform a sliding window test for time series irreversibility as described above for the Rapid 21-COM as well as CR 948/2011 record over the pre-industrial last millennium from 1000 to 1800 AD (Fig. \ref{fig:lia_vg_analysis} c--f). Results for the degree and local clustering based tests are depicted in the middle and bottom panel for different window sizes. The window size is varied between 30 and 60 data points, which comprises between 240 and 480 years given an average sampling rate of 8 years for both cores. The choice of small window sizes comes at the cost of an increased rate of false positives \cite{Donges2013}, but allows to detect regime shifts that occur within time scales of a few decades.

		The authors detect a clear signature of time-irreversibility using the clustering based test with $p$-values $p_C < 0.05$ for Rapid 21-COM and window sizes below 45 data points between 1450 and 1550 and $p_C \leq 0.1$ for CR 948/2011 and all window sizes around 1400 (see Fig. \ref{fig:lia_vg_analysis}, e,f). The $p$-values for the degree based test are somewhat higher (about 0.2, see Fig. \ref{fig:lia_vg_analysis}, c, d), which shows that the degree based test alone does not imply a rejection of the NH at a high significance level. Still, the timing of the signatures of NH rejection for the degree based test matches very well with the clustering based test, thus giving additional confidence in the results.

		\paragraph{VG analysis for sunspot numbers} \label{subsec:sunnum}
		Solar activity is characterized by complex dynamics, showing the famous 11 years cycle. Zou {\textit{et al.}} \cite{Zou2014a} performed the VGs analysis on both the daily and monthly sunspot series. The natural VGs focus on the effects of the local maxima on the resulting graphs. In the particular case of sunspot series, local minima play important roles in forming the increasing and decreasing phases of the solar cycles. In order to disclose the contributions of local minima to the VGs, they proposed two ways to construct the network: one is from the original observable measurements and the other is from a negative-inverse-transformed series.

		More specifically, let us discuss the results of VGs for the International Sunspot Number (ISN)~\cite{sidcDataBelgium} (see \cite{Zou2014a} for more consistent results that are based on the sunspot area series). The VG analysis have been performed for both monthly and daily sunspot series, which yields, respectively, month-to-month and day-to-day correlation patterns of the sunspot activities. The degree sequence $k_i = \sum_j A_{i,j}$ and its distribution $p(k)$ reflects the maximal visibility of the corresponding observation in comparison with its neighbors in the time series. In the case of sunspot time series,
the contributions of local minimum values to the network is of interest -- something that has been largely overlooked by the traditional VGs. One simple solution is to study the negatively inverted counterpart of the original time series, namely, $-x(t_i)$, which quantifies the properties of the local minima. Therefore, we use $k_{-x}$ and $p(k_{-x})$ to denote the case of $-x(t_i)$. This simple inversion of the time series allows us to create an entirely different complex network.

\begin{figure}
  		\centering
			\includegraphics[width=0.8\columnwidth]{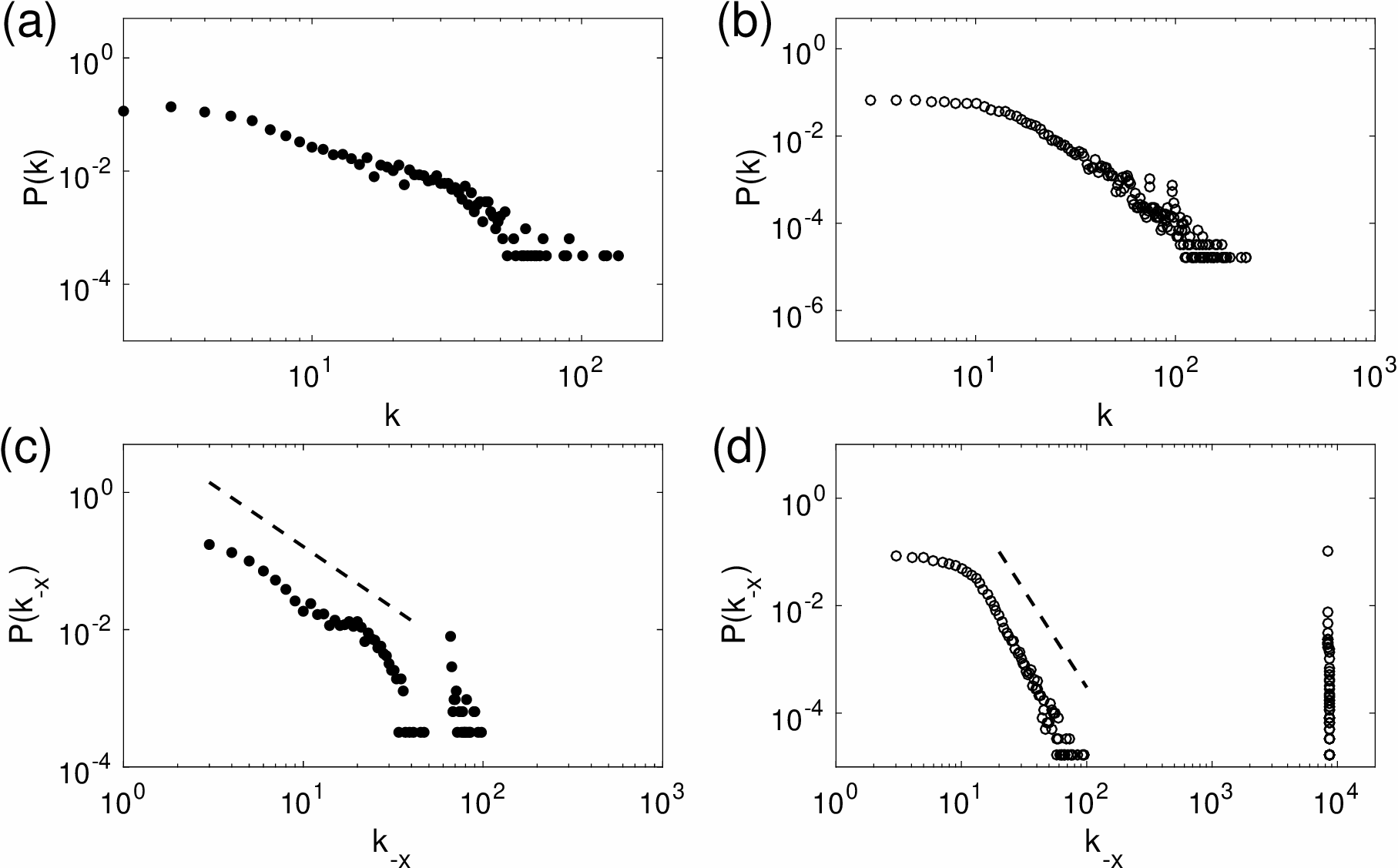}
		\caption{$p(k)$ of VGs from monthly (a,c) and daily data (b,d). (a,b) is for $\{x_i\}$, and (c,d) $\{-x_i\}$. One would suspect a fit to the first part of $p(k^{-x})$ yields that the slope of dashed line in (c) is 1.79, and that of (d) is 3.61, {\emph{but}} all $p$-values are $0$, rejecting the hypothetical power laws. Modified from \cite{Zou2014a}. } \label{sn_sa_data}
		\end{figure}

		Figure~\ref{sn_sa_data}(a,b) show $p(k)$ of the VGs derived from the ISN series $\{x_i\}$ with heavy-tails corresponding to hubs of the graph, which clearly deviates from Gaussian properties. In contrast, $p(k^{-x})$ of the negatively inverted sunspot series $\{-x_i\}$ shows a completely different distribution, consisting of a bimodal property (Fig.~\ref{sn_sa_data}c,d), extra large degrees are at least two orders of magnitude larger than most of the vertices (Fig.~\ref{sn_sa_data}(d)).
		Since well-defined scaling regimes are absent in either $p(k)$ or $p(k^{-x})$ (nor do they appear in the cumulative distributions, see more details of the statistical tests in \cite{Zou2014a}), the hypothesis of power laws is rejected.

\begin{figure}[htbp]
  		\centering
   			\includegraphics[width=\columnwidth]{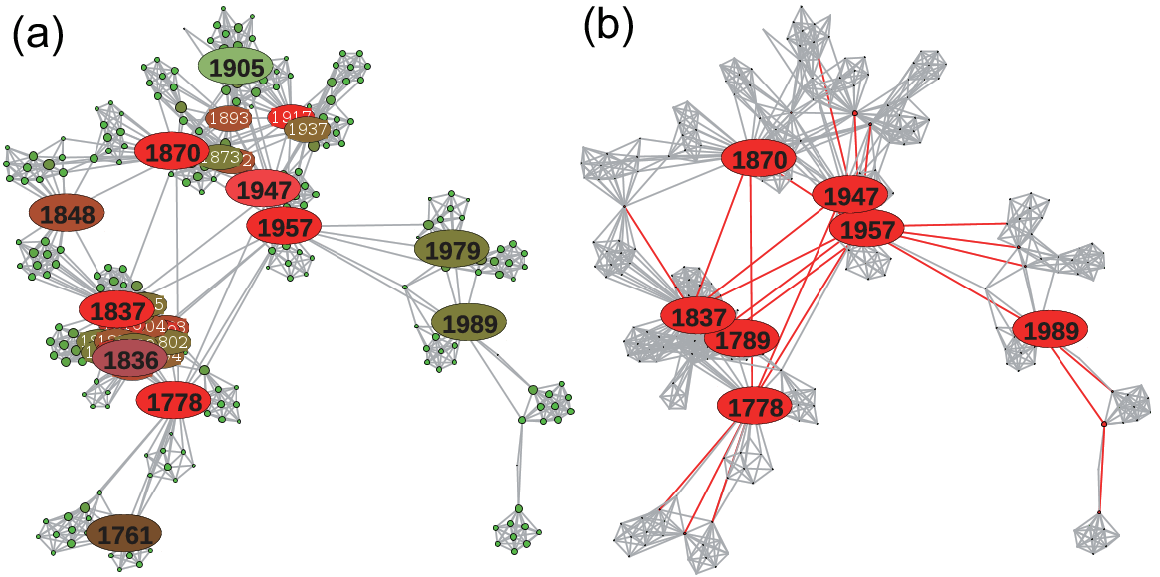}
			\caption{Network representations of the VG constructed from the annual sunspot numbers of the entire series. Highlighted visible nodes are: (A) large degrees ($k_i>15$), and (B) high betweenness centrality ($b_i>0.2$). Modified from \cite{Zou2014a}. } \label{year_sspnDegBet}
		\end{figure}

		Based on the degree sequences $k_x$ and $k_{-x}$, we further investigate the long term variations of local maxima/minima of the sunspot series. We find that the positions of strong maxima are largely homogeneously distributed over the time domain, while that of the strong minima are much more clustered in the time axis. These results of the difference between maxima and minima could be used for evaluating models for solar activity because they reflect important properties that are not included in other measures reported in the literature. Furthermore, VGs for sunspot series show rich community structures, each of which mainly consists of the temporal information of two consecutive solar cycles. The solar cycle of approximately 11-years yields that most of the temporal points of the decreasing phase of one solar cycle are connected to those points of the increasing phase of the next cycle in the resulting VGs \cite{Zou2014a}. When the sunspot number reaches a stronger but more infrequent extreme maximum, we have inter-community connections, since they have a better visibility contact with more neighbors than other time points -- hence, forming hubs in the graph. The inter-community connections extend over several consecutive solar cycles encompassing the temporal cycle-to-cycle information. In Fig.~\ref{year_sspnDegBet}(a), some hubs of large degrees ($k_i > 15$) are highlighted, which have been suggested to identify solar cycles~\cite{Zou2014a}. In addition, there are strong positive correlations between large degrees $k_i$ and high betweenness $b_i$, which further characterizes the node's ability to transport information from one place to another along the shortest path (Fig.~\ref{year_sspnDegBet}(b)).

		\paragraph{Asymmetry of sunspots} \label{subsec:sunspotsAsym}
		Another important feature of sunspots is the presence of a marked, time-varying hemispheric asymmetry, which have not yet been completely resolved \cite{Newton1955,zolotova2009,donner2007}. The hemispheric asymmetry of solar activity manifests itself in the statistical properties of a variety of activity indicators such as sunspot numbers, areas and spatial distribution, the numbers of flares and coronal mass ejections, solar radio and X-ray flux, etc., and has been recognized to vary on multi-decadal time scales (see, e.g.,\cite{Newton1955,Carbonell1993,zolotova2006,donner2007,donner2008a,Li2008MNRAS,li2008a,zolotova2009}, and references therein). Notably, it is commonly believed that the observed distinct hemispheric asymmetry is an intrinsic property associated with the underlying solar magnetic field dynamics, which in turn serves as the driver of solar activity responsible for particle and electromagnetic emissions directly affecting the Earth. However, even despite these methodological advances, properly quantifying the North--South asymmetry is a challenging problem by itself. Specifically, the complex dynamics of the entire solar activity cycles calls for replacing traditional linear statistical approaches by methods originated in the field of nonlinear dynamics \cite{zolotova2006,donner2007}. In \cite{Zou2014}, Zou {\textit{et~al.}} proposed (H)VGs analysis to study the asymmetric distributions of the sunspots over the solar surface. They have argued that (H)VGs provide complementary information on hemispheric asymmetries in dynamical properties.

		More specifically as we discussed in Sec.\ref{subsec:jointdegreeVG}, the excess degree $\Delta k(t)$ (Eq. \eqref{eq:deltaK}) and the relative excess degree $\Delta_{rel} k(t)$ (Eq. \eqref{eq:deltaReK}) have been proposed to characterize the possible asymmetric properties for (H)VGs that are reconstructed from bivariate time series, which are resulted from two interacting layers $\alpha$ and $\beta$. These two measures are based on the computations of joint degree $k^{joint}(t)$ (Eq. \ref{eq:jointK}) and the conditional degree sequences $k_{{[\alpha]}, {[\beta]}}^{O}(t)$ (Eq. \ref{eq:nad}). We emphasize that the absolute excess degree can be easily interpreted in terms of inter-hemispheric differences, whereas the relative excess degree partially corrects for the skewness effect and allows quantitatively assessing the relevance of differences between the degree sequences of both hemispheres. When analyzing sunspot time series, it has been demonstrated in \cite{Zou2014} that absolute and relative excess degrees exhibit qualitatively the same long-term variability. Therefore, we review some results based on $\Delta k(t)$ only and further results of $\Delta_{rel} k(t)$ can be found in \cite{Zou2014}.

		First we construct the (H)VGs for monthly hemispheric sunspot area series $A_{N}(t)$ and $A_{S}(t)$, yielding the degree sequences $k_\mathrm{N}(t)$ and $k_\mathrm{S}(t)$, respectively. The long-term asymmetric distribution behavior of the sunspots has been captured by utilizing a sliding window technique that averages the degree sequence over some time period. In all following considerations the window size has been chosen as $w = 270$ months, with a mutual overlap of 12 months between subsequent time windows. This specific choice of the window size covers about one full period of the solar magnetic field polarity cycle (approximately 22 years). There are no marked changes in the long-term variability of the (H)VG-based characteristics for $w$ being between about 180 and 400 months.
\begin{figure}[htbp]
	\centering
	\includegraphics[width=0.8\columnwidth]{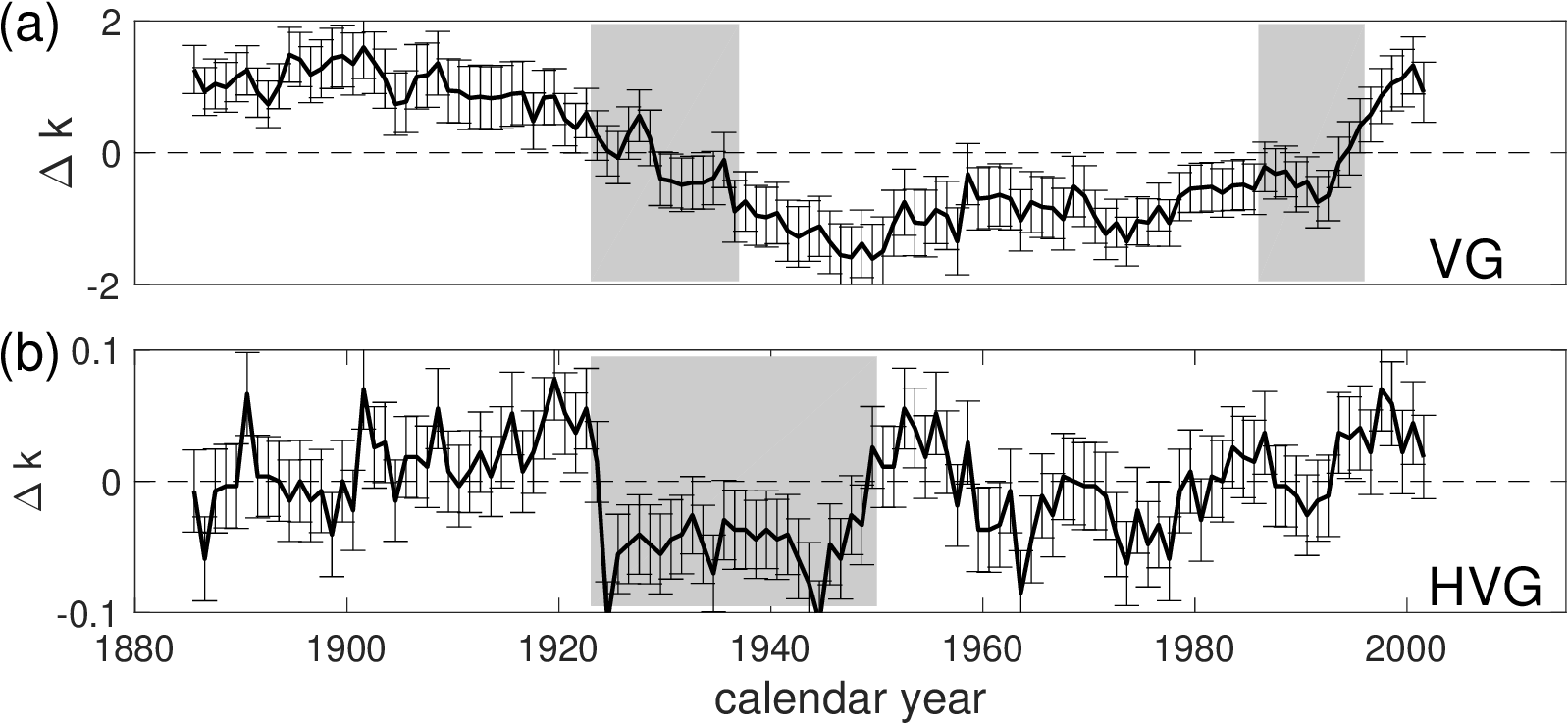}
\caption{\small {(a) Absolute excess degrees $\Delta k_i$ obtained from the VGs of $A^{N,S}$ computed over the sliding windows with a width of $w = 270$ months and a mutual overlap of $12$ months. Error bars display mean values and standard deviations within a given time window centered at the respective point in time. Gray areas mark those time intervals where the sign of the excess degree changes. (b) $\Delta k_i$ from HVG. Reproduced from \cite{Zou2014}. }
\label{degreeComX_ns_area270}}
\end{figure}

		Figure~\ref{degreeComX_ns_area270} shows the mean features associated with the degree sequences for our sliding windows, together with the associated window-wise standard deviations. In the VG analysis (Fig.~\ref{degreeComX_ns_area270}(a)), our results reveal two transitions between periods of positive and negative mean absolute excess degrees, which take place at about 1925--1935 (from higher degrees in the Northern Hemisphere to those in the Southern one) and 1985--1995 (vice versa). Furthermore, positive (negative) excess degrees imply higher mean degrees in the Northern (Southern) Hemisphere. These observations have been partially explained by the strong asymmetry of the probability distributions of sunspots over the north and south hemispheres, for instance, very high positive skewness \cite{Zou2014}. The corresponding analysis by mean of HVGs (Fig.~\ref{degreeComX_ns_area270}(b)) reveals some interesting facts: first of all, all degree-related quantities obey considerably lower values and weaker overall variability than for the VG. This is to be expected since the HVG is a subgraph of the VG. However, while the absolute degree values in the HVG typically reduce by a factor of about 2--4 in comparison with the VG, the absolute excess degrees are by more than one order of magnitude smaller (Fig.~\ref{degreeComX_ns_area270}(b)).

		Moreover, for the HVG-based excess degree $\Delta k_i$ we do not find comparably clear indications for transitions between time periods with clear hemispheric predominance as for the VG (Fig.~\ref{degreeComX_ns_area270}(a)). The only notable exception is the time period between about 1925 (corresponding to the formerly identified first transition in the VG) and 1950, where the excess degree of the HVG is significantly negative (as also observed before for the VG). Specifically, the transition in the hemispheric predominance reflected by the VGs' conditional degree sequences coincides with a sharp drop in the corresponding series for the HVG at about 1925, whereas the end of the period of significantly negative excess degrees in the HVG at about 1950 accompanies the termination of the gradual downward trend of the excess degree obtained from the VGs (Fig.~\ref{degreeComX_ns_area270}(b)). Taken together, we interpret these findings such that the effect of the asymmetry of the hemispheric sunspot area values mostly dominates possible variations in dynamical characteristics. However, to this end we tentatively conclude that parts of the observed long-term changes of the VG-based excess degree cannot be explained by combining the corresponding changes in skewness and HVG-based excess degree (i.e., distribution and dynamics, respectively). One possible reason for this could be complex changes in the PDF of the sunspot areas, which go beyond fluctuations in skewness, but yet have a significant effect on the resulting VGs' properties.

		In summary, we conclude that temporal changes in the hemispheric predominance of the graph properties lag those directly associated with the total hemispheric sunspot areas. These findings open a new dynamical perspective on studying the North--South sunspot asymmetry, which needs to be further explored in future work.

	\subsection{Transition networks}
	Depending on the particular symbolic representations of time series, there are various applications of transition network approaches to real time series. For instance, co-movement time series of economic growth and high-end talent development efficiency \cite{Zhang2018b}. Here we focus on the application of ordinal pattern transition network approach as proposed by McCullough {\textit{et al.}} in \cite{McCullough2015}, where they applied this analysis to experimental time series generated by a diode resonator circuits. They argue that the network size, mean shortest path length, and network diameter are highly sensitive to the interior crisis captured in this particular data set. Meanwhile, the ordinal pattern partition networks have been reconstruct from Electrocardiogram (ECG) data from patients with a variety of heart conditions \cite{Kulp2016b}. Network measures of mean degrees, entropies of the set of ordinal patterns and the number of non-occurring ordinal patterns have been computed for the resulting transition networks, showing statistically significant difference between healthy patients and several groups of unhealthy patients with varying heart conditions.

	\paragraph{Ordinal pattern networks for externally driven diode resonator circuit}
	In \cite{McCullough2015}, McCullough {\textit{et al.}} constructed ordinal pattern transition networks for experimental time series from an externally driven diode resonator circuit. In this experiment, each time series of the circuit output voltage $U_R$ were recorded for evenly spaced values, which consists of 65536 observations. The amplitude of the driving sinusoidal voltage $U_0$ serves as a control parameter. When this control parameter $U_0$  is changed systematically in the range of $3V \leq U_0 \leq 5V$, the system presents rich bifurcation scenarios from periodic to chaotic dynamics. Our motivation of this example is to show that structural measures of the reconstructed ordinal pattern transition networks can track different routes to chaos. More visualizations of attractors in the corresponding phase space and their associated ordinal networks have been well demonstrated in \cite{McCullough2015}. In addition, we focus on three main network measures, the mean out-degree $\left < k_{out} \right >$, the mean shortest path length $\mathcal{L}$ and network diameter $\mathcal{D}$ (maximum shortest path length).
	\begin{figure}
	\centering
		\includegraphics[width=0.8\columnwidth]{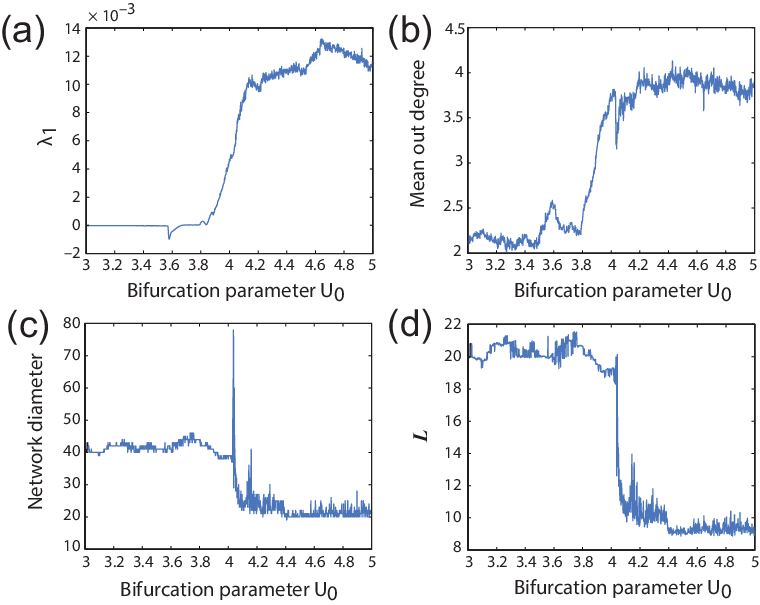}
	\caption{Bifurcation diagram of the diode resonator data which are characterized by (a) the largest Lyapunov exponent $\lambda_1$ for the range $3V \leq U_0 \leq 5V$, (b) the mean out degree, (c) network diameter, and (d) mean shortest path length. Networks were generated for each time series with $\tau = 8$ and $D=8$. Modified from \cite{McCullough2015} with permission by AIP Publishing. } \label{fig:appl_transition_network}
	\end{figure}

	The full bifurcation spectrum of the data set is characterized by the largest Lyapunov exponent $\lambda_1$ and network measures (Fig.~\ref{fig:appl_transition_network}). The system begins in period-3 oscillations and undergoes a period doubling bifurcation into period-6 when the control parameter approaches $U_0 \approx 3.6$. The period doubling cascade to chaos is observed for the control parameter approximately $0.38 \leq U_0 \leq 0.405$, and further undergoes a step change at the interior crisis, reflecting the filling of the attractor. Each ordinal network is generated for each time series with $\tau = 8$ and $D = 8$. The period-3 and period-6 time series are mapped to ring structures. The network measures capture the bifurcation scenarios successfully (Figs.\ref{fig:appl_transition_network}(b-d)).

	More specifically, the periodic regime is captured by zero Lyapunov exponent $\lambda_1$. Note that $\lambda_1$ is computed by the {\tt{lyapk}} function of the TISEAN package \cite{kantz1997}. The small range of the control parameter $U_0$ for which $\lambda_1$ becomes negative around the first period doubling bifurcation is a numeric error due to poor parameter selection for those particular corresponding time series \cite{McCullough2015}. The size of the network exhibits sensitivity to both the period doubling bifurcation at $U_0 =3.6$, the period doubling cascade to chaos for approximately $0.38 \leq U_0 \leq 0.405$, and undergoes a step change at the interior crisis, reflecting the filling of the attractor. Furthermore, the mean out-degree $\left < k_{out} \right>$ provides robust tracking of dynamical change similar to $\lambda_1$, and also appear sensitive to the period doubling bifurcation and the interior crisis. The mean shortest path length $\mathcal{L}$ and network diameter $\mathcal{D}$ both undergo a clearly discernible step change at the interior crisis, with the latter also exhibiting a peak value at the change point. Both of these results are easily understood in terms of the relationship between the networks and phase space as follows: additional nodes and edges are created immediately after the crisis, corresponding to the intermittent chaotic trajectories that begin to fill the space between the bands of the pre-crisis attractor in phase space. These new nodes and edges become shortcuts in the network. The spike in diameter corresponds to the small number of time series which have only a limited number of trajectories in between the bands of the pre-crisis attractor because they are in the immediate vicinity of the crisis and we are dealing with finite non-stationary data. These trajectories will form new strands in the network which are only connected to the main structure where they leave and rejoin the bands of the pre-crisis attractor, and hence these trajectories will have a significant impact on the network diameter. Moreover, these strands or subgraphs will have a far lower degree and degree variance than the remainder of the network, hence why the value for mean out degree and degree variance also dips at the interior crisis.

	In summary, this set of results demonstrates that while $\left < k_{out} \right>$, mean shortest path length $\mathcal{L}$ and diameter $\mathcal{D}$ all share the deficiency that they do not provide an absolute criteria for discriminating between periodic and chaotic dynamics, they have the potential to be useful as an indicator for dynamical discrimination in a relative sense, and for detecting change points \cite{McCullough2015}.

	\paragraph{Ordinal pattern networks for electrocardiograms}
	In \cite{McCullough2017b}, McCullough {\textit{et al}} have introduced to compute both local and global out-link entropies of ordinal transition networks to quantify the complexity of temporal structure in the networks from time series. The numerical comparative investigation in the R\"ossler system has demonstrated that these complexity measures track dynamical changes through period doubling and periodic windows over a range of the bifurcation parameter. Furthermore, the analysis has been applied to time series of electrocardiograms (ECGs). More specifically, complexity measures are able to capture the unique properties, discriminating between short-time ECG recordings characterized by normal sinus rhythm (NSR), ventricular tachycardia (VT) and ventricular fibrillation (VF). The global node out-link entropy of each time series is computed for both a short and a long time embedding lag and the resulting two-dimensional vector constitutes a measure of multiscale complexity description.

	More specifically, the dataset comprises $81$ ECGs that were measured to observe different cardiac dynamics, including 31 records of NSR, 30 records of VT, and 20 records of VF \cite{McCullough2017b}. Each time series consists of $10000$ points in length and have been sampled at 500 Hz with 10 bits resolution. Then, the global node out-link entropy (Eq. \eqref{eq:globalTp}) has been used to quantify transitional complexity of the resulting ordinal pattern network which has been reconstructed for each ECG record.

	As we have discussed in Section \ref{sec:OPtransition}, the choice of embedding delay $\tau$ has certain effects on the resulting ordinal pattern networks. For the particular dataset of ECGs, McCullough {\textit{et al}} used a simple assumption that the mean resting heart rate is $80$ beats per minute, which is about 375 samples per cycle. Therefore, the short embedding lag is chosen as $\tau = 20$ which is approximately a quarter period of the complete cycle. Furthermore, a long embedding lag is chosen to be one order of magnitude larger $\tau = 200$ which will capture dynamics over segments of two to six cycles and encoding inter-cycle variability of the ECG in the resulting ordinal networks. The embedding dimension $m$ has been suggested as $5 \leq m \leq 10$. The transitional complexity $S^{GNE}$ has been computed for both a short and a long time embedding lag and the resulting two-dimensional figure constitutes a measure of multiscale complexity.
\begin{figure}[ht]
	\centering
	\includegraphics[width=\columnwidth]{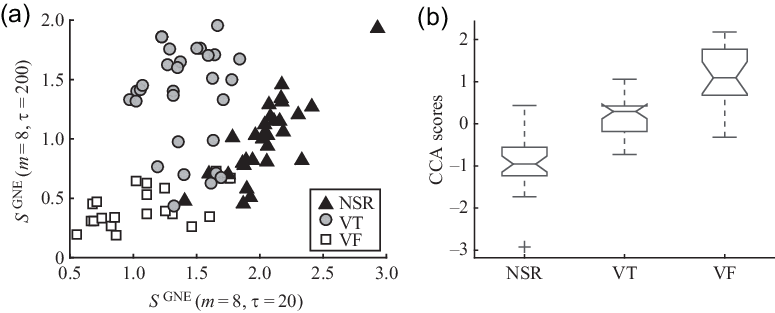}
\caption{(a) Scatter plot of global node out-link entropy $S^{GNE}$ for the $81$ ECG dataset. The $x$-and $y$-axes
correspond to $S^{GNE}$ computed with embedding $\tau =20$ and $\tau = 200$, respectively. (b) Box plot for the scores from a canonical correlation analysis of the data. Modified with permission from Ref. \cite{McCullough2017b}. Courtesy of M. Small. \label{fig:rieECG}}
\end{figure}

	Figure \ref{fig:rieECG} illustrates the two dimensional multiscale plot of $S^{GNE}$ and the corresponding box plot, which shows clear discrimination between the pathological groups. Therefore, ordinal networks present different level of transitional complexity corresponding to different networks.

		In addition, the ordinal network analysis is performed to characterize age-related effects in interbeat interval dynamics from ECGs. In \cite{McCullough2017b}, the authors further showed that the standard permutation entropy is unable to discriminate between age groups. In contrast, the global node out-link transitional complexity $S^{GNE}$ generally is higher for elderly subjects on short time scales and lower on long time scales, and $S^{GNE}$ has significantly greater variability than that can be observed for young subjects.

%% file: Chapter07_Software/Chapter07_Software.tex
\section{Software implementation -- \texttt{pyunicorn} }\label{sec:Software}


	In this chapter, we briefly introduce the Python software package \texttt{pyunicorn}, which implements methods from both complex network theory and nonlinear time series analysis, and unites these approaches in a performant, modular and flexible way \cite{Donges2015}. Here, we mainly present a brief introduction of \texttt{pyunicorn} and a discussion of software structure and related computational issues. More details of the illustrative examples have been presented in \cite{Donges2015}. Although in the tutorial of \cite{Donges2015}, the work flow of using \texttt{pyunicorn} is mainly illustrated drawing upon examples from climatology, we have to emphasize that the package is applicable to all fields of study where the analysis of (big) time series data is of interest, e.g. in neuroscience \cite{Bullmore2009,Subramaniyam2014,Subramaniyam2015}, hydrology \cite{sun2015global} or economics and finance \cite{Wang2012}.

	\texttt{pyunicorn} is intended to serve as an integrated container for a host of conceptionally related methods which have been developed and applied by the involved research groups for many years. Its aim is to establish a shared infrastructure for scientific data analysis by means of complex networks and nonlinear time series analysis and it has already greatly taken advantage from the backflow contributed by users all over the world. The code base has been fully open sourced under the BSD 3-Clause license.

	First of all, we emphasize that \texttt{pyunicorn} covers rather general topics of complex network studies. The \texttt{pyunicorn} library consists of five subpackages: (1) {\it core}, which contains the basic building blocks for general network analysis and modeling. For instance, it is capable for analyzing and modeling general complex networks, spatial networks, networks of interacting networks or multiplex networks and node-weighted networks. (2) {\it funcnet}, which contains advanced tools for construction and analysis of general functional networks \cite{Bullmore2009,Donges2012PhD}. For instance, this module calculates cross-correlation, mutual information, mutual sorting information and their respective surrogates for large arrays of scalar time series. (3) {\it climate}, which focus on the construction and analysis of climate network \cite{Tsonis2004,Yamasaki2008,Donges2009,Donges2009b} and coupled climate network analysis \cite{Donges2011b}. (4) {\it timeseries}, which provides various tools for the analysis of non-linear dynamical systems and uni- and multivariate time series. This subpackage covers all aspects of RNs (Section \ref{sec:RecurrenceNt}) and (H)VGs (Section \ref{sec:VisibilityGt}) that have been reviewed in this report, except ordinal pattern transition networks. Furthermore, \texttt{pyunicorn} also presents methods for generating surrogate time series \cite{Schreiber2000}, which are useful for both functional networks and network-based time series analysis. (5) {\it utils}, which includes \texttt{MPI} parallelization support and an experimental interactive network navigator.

	\texttt{pyunicorn} is conveniently applicable to research domains in science and society as different as neuroscience, infrastructure and climatology. Most computationally demanding algorithms are implemented in fast compiled languages on sparse data structures, allowing the performant analysis of large networks and time series data sets. The software's modular and object-oriented architecture enables the flexible and parsimonious combination of data structures, methods and algorithms from different fields. For example, combining complex network theory and RPs yields RN analysis (Section \ref{sec:RecurrenceNt}) \cite{Donges2015}.

	Along these lines, \texttt{pyunicorn} has the potential to facilitate future methodological developments in the fields of network theory, nonlinear time series analysis and complex systems science by synthesizing existing elements and by adding new methods and classes that interact with or build upon preexisting ones. Nonetheless, we urge users of the software to ensure that such developments are theoretically well-founded as well as motivated by well-posed and relevant research questions to produce high-quality research.

	Besides \texttt{pyunicorn}, some other software packages are available, for instance, the MATLAB toolbox \texttt{CRP Toolbox} that allows to get the adjacency matrix of a RN (i.e., the recurrence matrix). Furthermore, this toolbox provides the computation of several network measures as well, for instance, the node degree $k$ (corresponds to the recurrence rate), the clustering coefficients $\mathcal{C}$ and transitivity $\mathcal{T}$. Some basic elements of recurrence network analysis have been published as Mathematica demonstrations \cite{Zech2010a,Zech2010b}. While we are not aware of further existing comprehensive software packages in the style of \texttt{pyunicorn} or \texttt{CRP Toolbox}, other groups have sometimes published code implementing specific methods of network-based nonlinear time series analysis, such as for example visibility graph analysis developed by the group of Lacasa et al. (\url{http://www.maths.qmul.ac.uk/~lacasa/Software.html}). In addition, we recommend some network visualization toolboxes, including Gephi (\url{http://gephi.org}) or Networkx \url{http://networkx.github.io} which can be used to analyze the networks once the adjacency matrix is available.

%% file: Chapter08_Discussion/Chapter08_Discussion.tex
\section{Conclusions and future perspectives} \label{sec:Discussion}

	\subsection{Conclusions}
Time series analysis by means of complex networks is an innovative and powerful
approach with many ramifications and applications. In this report, we have reviewed several major algorithms for transforming a time series into a network representation, depending on the definitions of vertices and edges. The network approach makes use of different established methods, such as Markov chains or recurrences, but also of more abstract concepts, as visibility graphs, which form three main classes of methods, namely, recurrence network (RN), visibility graphs (VG) and transition network (TN) that have been discussed in detail throughout this report.

These methods complement available approaches with alternative measures, e.g., describing geometrical properties of the system under study in its phase space, but also broadening the applicability of time series analysis to short, complex, and multivariate data. As such, network based time series analysis can be used to characterize systems dynamics from a single time series, to distinguish different dynamics, to identify regime shifts and dynamical transitions, to test for time series reversibility, or to predict the future system states.

%
%
%
%
%

	\subsection{Future perspectives}
	As shown in this review, applying complex network approaches in the context of time series analysis has already gained a number of valuable insights from both, a theoretical dynamical systems and/or stochastic processes perspective and in the context of various types of applications. However, as for any emerging field, there is a large body of relevant questions and upcoming developments that may further increase the relevance of the discussed frameworks. For example, in the particular case of RNs, there are some evident questions, the most relevant being about the invariance of findings under variation of the threshold value $\varepsilon$ since this value determines the link density of the network and all network characteristics become trivial in the limit of full connectivity \cite{Bradley2015c}. Therefore, future work needs to further demonstrate the wide applicability of existing as well as newly developed methods for transforming time series into complex networks by considering additional applications from various disciplines, particularly regarding the methods' capabilities to provide deeper insights beyond the already existing knowledge of the respective topics. Depending on the particular working subject, it will be crucial to use network analysis to extract some features that are not easily captured by most of standard methods of linear and nonlinear time series analysis, thereby demonstrating the added value of the network methods.

	To this end, we would like to highlight some particularly prospective directions for future research, being aware of the fact that this selection will be necessarily incomplete and subjective.

\subsubsection{Evolving and temporal network analysis for time series}

In their basic formulations as discussed in this review, most existing approaches for transforming and analyzing time series from a complex network perspective have been designed primarily to cope with stationary systems. However, generalizations are desirable that account for the fact that real-world time series often exhibit changing dynamical patterns as a hallmark of nonstationarity.

While many complex network approaches traditionally assume static network structures, there are natural extensions of evolving network analysis, i.e., the consideration of complex networks that change as a function of time. For proximity networks as well as visibility graphs and related concepts, a generalization to evolving networks appears straightforward if we consider both node and edge sets being time-dependent. In turn, for transition networks, one may keep the underlying node set but consider the transition frequencies between patterns encoded in the weights of the directed edges as changing with time.

In this context, the most common way to generate evolving time series networks would be employing a sliding windows analysis. Here, evolving networks can be understood as successive snapshots of static networks obtained for individual, mutually overlapping time windows. As a result, we may trace changes in the resulting network properties over time and use them as proxies for dynamical changes in the underlying time series, thereby revealing non-stationarity of the system or even distinct episodic events such as regime shifts. However, the sliding windows approach brings about some natural limitation, that is, the necessity of making empirical choices for the temporal window lengths and the mutual window overlap, for which there are no optimal strategies but rather heuristics depending on the individual case study. In general, using extremely small window sizes between two consecutive snapshot networks allows for a high resolution of tracked changes in the network properties, but could obscure slower trends which only become visible over longer time-scales. Conversely, using larger window sizes integrates information over considerably large time intervals and thus loses both, resolution and information on the effects of individual events within each window. Therefore, objective strategies for choosing an appropriate time-scale for dividing the evolution of a network into static snapshots are required, which are likely to have positive effects on proper interpretations of the obtained results \cite{Donges2011,Zou2014,schleussner2015indications,Franke2017}.

Even with such optimal and objective choices of time windows, a sliding window technique as described above by definition cannot cover all potentially relevant aspects associated with the temporal structures in the underlying time series that should be captured by their network representations \cite{Holme2012}. For this purpose, we need to include an additional time dimension to take the detailed information on the temporal succession of network structures (emergence and/or disappearance of nodes and links) into account in the context of quantitative analyses. In this spirit, there have been attempts to analyse visibility graphs as temporal networks \cite{Mutua2015}. Furthermore, there might be cases in transition networks where transitions between patterns do exist at certain times but not at others, i.e., times with active versus inactive links (so-called blinking links \cite{Gozolchiani2008}). In such a situation, the time ordering of observations in the underlying time series can have important effects that cannot be captured by static network representations. In the context of time series analysis, Weng {\textit{et al.}} \cite{Weng2017} proposed to transform time series into temporal networks \cite{Holme2012} by encoding temporal information into an additional topological dimension of the graph, which captures the ``lifetime" of edges. We note that a proper modification of the ordinal pattern transition network approach (for instance, considering short-term transition networks) may provide the necessary temporal information for this problem since the transition matrix describes the probability of future evolution directions of the observed trajectory.

\subsubsection{Multilayer and multiplex network analysis for multiple time scale time series}

In this work, we have provided a review on existing methods for reconstructing multilayer and multiplex networks from time series, for instance, multiplex recurrence networks, multiplex visibility graphs, inter-system recurrence networks, and joint recurrence networks. However, most of these methods are only appropriate for stationary time series as we have discussed in Section~\ref{subsec:practicalRN}. When monitoring complex physical systems over time, one often finds multiple phenomena in the data that work on different time scales. For example, observations are collected on a minimal (short) time scale, but also reflect the time series' behavior over larger time scales, which is rather typical for real-world climate data. Another prominent example from neuroscience is the recording of spiking activity of individual neurons (discrete event series) and local field potentials (time continuous measurement). Higher-frequency variability of such data can obscure the time series behavior of the data at larger scales, making it more difficult to identify the associated trends. If one is interested in analyzing and modeling these individual phenomena, it is crucial to recognize the multiple time scales in the construction of multilayer and multiplex networks from time series. One corresponding way could be applying successive scale-sensitive filters prior to network generation, with the choice of a particular method depending on the specific data set and research question, such as empirical mode decomposition \cite{gao2018} or some type of wavelet transform \cite{chen2012}.

\subsubsection{The inverse problem of time series regeneration from networks}

		Most of the existing works focus on investigating proper transformation methods for mapping time series into network representations. To study the inverse question of how much information is encoded in a given network model of a time series \cite{Wiedermann2017}, some studies have been undertaken to recover the original time series from the network, to use the network to reconstruct the phase space topology of the original system, or to generate new time series from the networks and compare these with the original \cite{Campanharo2011,hirata2008,Hirata2016,McCullough2017}. This inverse problem of getting back from the network adjacency matrix to time series of the underlying dynamical system remains a big challenge, which certainly has many applications \cite{Lancaster2018}. In general, transformations of complex networks to time series are not straightforward. Specifically, without having additional node labels informing about the temporal succession of vertices, the order of vertices in a complex network can be arbitrarily exchanged without affecting the network topology. In turn, for reconstructing the trajectory from a network representation, the temporal order of the nodes needs to be known.

A few algorithms have been proposed so far to reconstruct time series from networks. For instance, under a certain condition for reconstructability, Thiel {\textit{et al.}} proposed an algorithm to reconstruct time series from their recurrence plots \cite{thiel2004b,Robinson2009}. In this case, the reconstructed attractor shows topological equivalence with the original attractor \cite{Zhao2014}. Furthermore, based on recurrence plots with fixed
number of recurrences per state (equivalent to $k$-nearest neighbor networks), the topological properties of the underlying time series have been reconstructed by multidimensional scaling \cite{hirata2008}. Recently, it has been shown that $k$-nearest neighbor and $\varepsilon$-recurrence networks can be viewed as identical structures under a change of (equivalent) metrics \cite{Khor2016}. Based on this fact, an improved inversion algorithm has been proposed in \cite{Khor2016}, which further supports the use of complex networks as a means of studying dynamical systems, while also revealing an equivalence between $\varepsilon$-recurrence and $k$-nearest neighbor classes of complex networks. In addition, algorithms based on random walks have been proposed in the literature. For instance, a random walk algorithm has been used in \cite{Hou2015}, which further compares the performance of RNs and adaptive $k$-nearest neighbor networks. The performances of these algorithms have been compared in \cite{Liu2013c}. Recently, a constrained random walk algorithm has been proposed to regenerate time series from ordinal transition networks \cite{McCullough2017}.

For all these different algorithms, there are several important algorithmic parameters that have to be chosen empirically in order to guarantee consistent topology between the reconstructed time series and the original system. The general performance and applicability of each algorithm has to be evaluated in future work. One application of such regeneration algorithms is to perform surrogate analysis, for example, to test for the statistical significance of the results obtained from analyzing the original time series. Therefore, we also have to take into account the proper choice of null hypothesis while proposing algorithms for regenerating time series from networks.

\subsubsection{Combining data mining tools with time series network approaches}
In the framework of time series mining \cite{FU2011}, some fundamental tasks include dimension reduction by introducing proper indexing mechanisms, similarity comparison between time series subsequences and segmentation. The final goal of mining tools is to discover hidden information or knowledge from either the original or the transformed time series, for instance, using a proper clustering method to identify patterns. Note that the interesting pattern to be discovered here relate to rather general categories, including patterns that appear frequently versus such that occur rather surprisingly in the datasets \cite{FU2011,Aghabozorgi2015}. From the perspective of time axis, time series clustering can be classified into three categories, whole time series clustering, subsequence clustering and time point clustering \cite{Aghabozorgi2015}. Several algorithms have been proposed to perform time series clustering based on shapes of raw time series, feature vectors of dimension reduced time series, and distances between parametric model outputs and raw time series. These conventional mining algorithms have found various applications to time series of different origins, which, however, are challenged by practical issues like high dimensionality, very high feature correlations, and large amount of noise.

Despite the rapid increase in size and complexity of datasets in the era of big data, a proper combination of data mining tools with the complex network approaches for time series analysis has largely remained untouched so far. Such a joint research effort should combine methodologies and techniques from different fields, such as statistics, data mining, machine learning and visualization. It has been recently demonstrated that complex network approaches and data mining tools can indeed be integrated to provide novel insights for the understanding of complex systems \cite{Zanin2016}. From the viewpoint of nonlinear time series analysis, both sides of data mining and nonlinear time series analysis can benefit from each other, which will be one important topic for future research.

\subsubsection{Building network models for time series prediction}

Time series modeling and forecasting has attracted a great number of researchers' attention and provides the core of nonlinear time series analysis \cite{kantz1997,Bradley2015c}. To this end, we can build a proper model to forecast the system's future behavior, given a sequence of observations of one or a few time variable characteristics. Most existing methods originating from nonlinear dynamics are state-space models, which build local models in ``patches'' of a reconstructed state space and then use these models to predict the next point on the system's trajectory, which remains an active area of research \cite{Bradley2015c}. We note that most existing network approaches to nonlinear time series analysis have been focusing on characterizing network features of phase space (and, hence, diagnosing rather than forecasting the observed dynamics). Examples for using network approaches for time series modeling and prediction have not been reported in the literature yet to out best knowledge, but could provide another exciting future research avenue.

%% file: Chapter09_Appendix/Chapter09_Appendix.tex
\appendix
\section{Mathematical models}
Here we list the mathematical models used in the examples of this paper: 

\begin{enumerate}
\item The Lorenz system \cite{Lorenz1963}
\begin{equation}
\left(\begin{array}{c}\dot{x}\\ \dot{y}\\ \dot{z}\end{array}\right)=\left(\begin{array}{c}\sigma(y-x) \\ x(r-y)\\ xy-\beta z \end{array}\right), 
\label{eqlorenz}
\end{equation}
with the parameters $r=28$, $\sigma=10$ and $\beta=8/3$. 

\item The R\"ossler system \cite{Roessler1976}
\begin{equation}
\left( \begin{array}{c}\dot{x}\\ \dot{y}\\ \dot{z}\end{array} \right) = \left( \begin{array}{c} -y-z\\ x+ay\\ b+z(x-c) \end{array} \right),
\label{eq:roessler}
\end{equation}
where $a, b$, and $c$ are parameters. 

\item Auto-regressive process $p$ order
\begin{equation} \label{def:AR}
x_t = \sum_{j=1}^{p}\varphi_j x_{t-j} + \varepsilon_t, 
\end{equation}
where $\varphi_j, j \in [1, p],$ are real-valued coefficients of the model, and $\varepsilon_t$ is white noise. We further assume that the error terms $\varepsilon_t$ follow a Gaussian distribution with zero mean and unit variance. 

\item The H\'enon map
\begin{equation} \label{eq:henon}
 \left \{ \begin{aligned}
x_t &= A - x_{t-1}^2 + B y_{t-1}, \\
y_t &= x_{t-1},
\end{aligned}
\right.
\end{equation}
with $A=1.4$ and $B=0.3$. 

\item Two R\"ossler systems that are diffusively coupled via the second $y$-component:
\begin{equation}
\begin{split}
\dot{x}^{(1)} &= -(1+\nu)y^{(1)} - z^{(1)} \\
\dot{y}^{(1)} &= (1+\nu)x^{(1)} + ay^{(1)} + \mu_{21}(y^{(2)}-y^{(1)}) \\
\dot{z}^{(1)} &= b+z^{(1)}(x^{(1)}-c) \\
\dot{x}^{(2)} &= -(1-\nu)y^{(2)} - z^{(2)} \\
\dot{y}^{(2)} &= (1-\nu)x^{(2)} + ay^{(2)} + \mu_{12}(y^{(1)}-y^{(2)}) \\
\dot{z}^{(2)} &= b+z^{(2)}(x^{(2)}-c), 
\end{split}
\label{eq:coupled_roessler}
\end{equation}
where $\nu$ is the frequency mismatch and $\mu_{12}$ and $\mu_{21}$ are coupling strength. Symmetric coupling is achieved if $\mu_{12} = \mu_{21}$. 

\item Three diffusively coupled R\"ossler systems via the $x$ component \cite{Nawrath2010}:
\begin{equation} \label{threeRosPRL}
\left ( \begin{aligned}
& \dot{x}_k \\ 
& \dot{y}_k \\
& \dot{z}_k
\end{aligned} \right )
= \left ( \begin{aligned}
& -\omega_k y_k - z_k + \sum_{l\neq k} \mu_{k,l}(x_l - x_k) \\ 
& \omega_k x_k + 0.165 y_k \\
& 0.4 + z_k(x_k - 8.5)
\end{aligned} \right ),
\end{equation}
where $k = 1, 2, 3$ and $\mu$ is the coupling strength.

\item The standard map
\begin{equation} \label{std_map_book}
\mathbf{v}(t): \left \{ \begin{aligned}
& y_{n+1} = y_{n} + \frac{\kappa}{2 \pi}\sin(2 \pi x_{n}),  \\
& x_{n+1} = x_{n} + y_{n+1},
\end{aligned} \;\;\text{mod}\;\; 1
\right.
\end{equation}
with $\kappa $ denoting the system's single control parameter. 

\end{enumerate}